\documentclass[12pt,twoside,openright]{report}
\usepackage[dvips]{epsfig}
\usepackage[german,english]{babel}
\usepackage{amsmath,amssymb,amsfonts,exscale,mathrsfs,latexsym}
\begin{document}
\begin{titlepage}
\begin{center}
{\bf \huge Topological Defects and Gravity in}\\
\vspace{5mm}
{\bf \huge  Theories with Extra Dimensions} \\
\vspace{5.5cm}
\large{PhD~Thesis} \\
\vspace{2.5cm}
\large{Ewald Erich R\"o\ss l} \\
\vspace{1.5cm}
\large{Supervisor: Prof. Mikhail Shaposhnikov} \\
\vspace{1.5cm}
August 2004 \\
\vspace{1cm}
{\it Institut de Th\'eorie des Ph\'enom\`enes Physiques, \\
Laboratoire de Physique des Particules et Cosmologie (LPPC), \\
Sciences de Base, Ecole polytechnique f\'ed\'erale de Lausanne, BSP-Ecublens, 1015~Lausanne, Switzerland}
\end{center}
\end{titlepage}

\thispagestyle{empty}
\cleardoublepage
\thispagestyle{empty}
\vspace*{18cm}
\hfill {\large\em To my beloved wife and children}
\clearpage
\thispagestyle{empty}
\clearpage
\tableofcontents
\clearpage
\thispagestyle{empty}
\chapter*{Abstract}
Recent proposals of large and infinite extra dimensions triggered a strong research activity 
in theories in which our universe is considered as a sub-manifold of some higher-dimensional 
space-time, a so-called $3$-brane. In this context, it is generally assumed that some mechanism 
is at work which binds Standard Model particles to the $3$-brane, an effect often referred to as 
the localization of matter on the brane. Gravity, however, is allowed to 
propagate in general also in the extra dimensions. As demonstrated by Randall and Sundrum in 1999,
it is also possible to localize gravity itself on a $3$-brane. In the setup they proposed, 
the $3$-brane is realized as a singular domain wall separating two patches of 5-dimensional anti-de-Sitter 
($AdS_5$) space-time. 
The potential between two test masses on the brane at distances larger than the $AdS_5$-radius was shown 
to be the usual $4$-dimensional Newtonian $1/r$ potential with strongly suppressed corrections. 

The model of Randall and Sundrum, usually referred to as the Randall-Sundrum II setup, constitutes 
the center of interest for this thesis. The main goal of this work is to find 
possible generalizations to higher dimensions of the simple setup considered by Randall and Sundrum.
One of the motivations for such a generalization is that a realistic theory should possibly be able to
explain the chiral nature of $4$-dimensional fermions on the brane. One way to explain chiral fermions from 
higher dimensions is to consider $3$-branes identified with the cores of topological defects located in 
a higher-dimensional transverse space. Naturally a richer topological structure of the 
field configuration in transverse space provides the possibility of a more realistic spectrum of chiral 
fermions localized on the $3$-brane.

After two introductory chapters on extra dimensions and non-factorizable geometries which are relevant 
for the Randall-Sundrum II model, we briefly discuss basics of topological defects in the following 
third chapter. In the rest of the third chapter we consider various solutions to higher-dimensional
Einstein equations coupled to a series of physically different sources and discuss their properties 
of localization of gravity. Due to their asymptotic nature, these solutions are only valid far from the 
cores of the defects in transverse space. Therefore, it seems reasonable to complement the considerations 
by  presenting a particular numerical 
example of a solution to the Einstein equations coupled to a set of scalar and gauge fields: this 
solution describes a $3$-brane realized as a 't~Hooft-Polyakov monopole residing in the 
$3$-dimensional transverse space of a $7$-dimensional space-time. 

The last chapter of this work is dedicated to the study of a  modification of the original 
Randall-Sundrum II model of another kind. The motivation is given by the geodesic incompleteness 
of the latter scenario with respect to time-like and light-like geodesics. We will describe 
a model which resembles the Randall-Sundrum II model with respect to its properties 
of gravity localization but with the advantage that the underlying space-time manifold 
is geodesically complete. Parts of the calculations related to the properties of gravity at low energies 
in this model are rather technical in nature and we therefore preferred to assemble them in several 
appendices. We finally add some concluding remarks and discuss possible further developments.

\chapter{Introduction} \label{cha1}
The discovery of the hidden $SU(2) \times U(1)$ gauge symmetry by Sheldon Glashow, Mohammed 
Abdus Salam and Steven Weinberg which provides a unified 
description of the physics of electromagnetic and weak interactions between elementary particles 
is commonly considered not only as one of the major 
contributions to our understanding of the physics of elementary particles but 
also as an outstanding achievement in theoretical physics in the $20^{\mbox{\small th}}$ century as a whole. 
Together with the theory of the 
strong interactions of elementary particles, quantum chromodynamics, based on the gauge group
$SU(3)$, the Glashow-Salam-Weinberg theory constitutes the so-called Standard 
Model of the interactions of elementary particles, the pride of particle physicists today due 
to its impressive agreement between observations and predictions.  

Despite of the enormous successes of the Standard Model, it is Weinberg himself who 
starts his by now famous review article \cite{Weinberg:1988cp} of the so-called 
cosmological constant problem with the lines:

\vspace{5mm}
{\em  \hspace{1cm} Physics thrives on crisis.} 
\vspace{5mm}

The crisis Weinberg describes is not only a result of the puzzling and yet unexplained mismatch between the 
observed and predicted values of the cosmological constant of around $120\, \!$ orders of magnitude, but 
also stands for numerous unsuccessful attempts to elucidate the quantum structure 
of the fourth fundamental interaction in nature, gravity.

Since its introduction by Einstein in 1916, the classical theory of general relativity remains  
the main theoretical framework for our understanding of the expanding universe and its history.
On the other hand, general relativity so far withstood all attempts to be combined with the principles
of quantum mechanics, hence we presently still lack a successful theory of quantum gravity.   

From today's point of view, string theory appears to be the most promising candidate for 
achieving the goal of a unified description of all fundamental interactions since the graviton is 
part of the field contents of all known theories of strings. A consistent formulation 
of string theory requires that the dimensionality of space-time
is equal to $10$ (or $11$ in the case of M-theory). The reduction of a $10$-dimensional
string theory to the $4$-dimensional Standard Model in the limit of low energies is complicated by the wealth 
of possibilities of compactifying the additional dimensions.

Even though the requirement of a $10$-dimensional space-time in string theory was unexpected, the idea of 
extra dimensions was not new. It was first proposed by Nordstr\"om~\cite{Nordstr} and further developed 
by Kaluza and Klein \cite{Kaluza:tu,Klein:tv}. 

Later, inspired by the success of Kaluza in the unification (on a classical level) of the 
gravitational and electromagnetic interactions (the only two known interactions of his times) many 
physicists applied the idea to non-abelian gauge theories, supergravity and string theories in the search for
a realistic Kaluza-Klein theory 
\cite{Scherk:fm,Cremmer:1975sj,Cremmer:1976zc,Witten:me,Salam:1981xd,Randjbar-Daemi:1983bx,Randjbar-Daemi:1983jz,Randjbar-Daemi:1984an}. 

The idea of extra dimensions also fell on fruitful grounds outside the usual Kaluza-Klein approach. 
During the last few decades, theoretical physics discovered the potential of extra dimensions in 
addressing many outstanding problems of physics: 
the cosmological constant problem \cite{Rubakov:1983bz,Wetterich:1984rv,Randjbar-Daemi:1985wg},
the origin of chirality
\cite{Rubakov:bb,Wetterich:1984uc,Witten:1983ux,Randjbar-Daemi:1982hi,Randjbar-Daemi:1983qa}, 
the strong CP-problem 
\cite{Khlebnikov:1987zg,Dienes:2001zz}, 
the hierarchy problem 
\cite{Antoniadis:1990ew,Arkani-Hamed:1998rs,Antoniadis:1998ig,Randall:1999ee,Cohen:1999ia},
and the horizon-problem 
\cite{Chung:1999xg} 
to mention only some of the most prominent ones.

In spite of the long history of extra dimensions, a boost in research in this area was triggered only
recently. In the search of a solution to the hierarchy problem 
it was realized that, from a phenomenological point of view, the existence of large extra dimensions from
about $1\, \mbox{mm}$ to $1\, \mbox{fm}$ is not excluded by observations 
\cite{Arkani-Hamed:1998rs,Antoniadis:1998ig}, as long as one 
assumes a mechanism of localization of the Standard Model fields on a $4$-dimensional sub-manifold 
in the extra dimensions. The reason is that the validity of Newton's law of gravity 
is established only at distances larger than about $0.1\, \mbox{mm}$ \cite{Hoyle:2000cv}. Although models 
with extra dimensions in the above distance range suffer generically from the effects of light Kaluza-Klein 
gravitons, the corresponding constraints from particle physics 
\cite{Giudice:1998ck}-\cite{Gupta:1999iy} 
and astrophysics and cosmology \cite{Arkani-Hamed:1998nn}-\cite{Hall:1999mk} 
are not stringent enough to rule them out.

Moreover, inspired by the discovery of solitonic solutions in string theory \cite{Polchinski:1995mt}, 
so-called D(irichlet)-branes, it was pointed out that even an infinite extra dimension can be in complete 
agreement with $4$-dimensional observations \cite{Randall:1999vf}.
In particular it was shown that under the assumption that Standard Model fields were bound to a 
so-called $3$-brane in an $5$-dimensional bulk space-time characterized by a non-factorizable geometry, it is 
possible to reproduce standard $4$-dimensional Newtonian gravity on the brane. 
Eventually, a similar setup involving two branes using the same idea of a warped geometry 
provided a solution to the hierarchy problem different from the one mentioned above \cite{Randall:1999ee}. 
The main advantage of this latter approach is that it does not predict light Kaluza-Klein 
excitations of the graviton.

The fact that these new ideas of large and infinite extra dimensions make use of the concept of 
a $4$-dimensional sub-manifold, a $3$-brane, embedded in a higher-dimensional space-time resulted in an 
immense effort in the study of properties of models of this kind. The investigations in this area of 
brane-world scenarios range from questions of stability 
\cite{Sasaki:1999mi,Ghoroku:2001pi,Kobayashi:2001jd}, 
general space-time structure 
\cite{Mukohyama:1999wi,Bowcock:2000cq}, 
properties of gravity
\cite{Sasaki:1999mi,Garriga:1999yh,Tanaka:2000er,Deruelle:2001ms},
black holes on the brane
\cite{Chamblin:1999by}-\cite{Frolov:2004wy}, 
cosmological aspects (related to a brane motion in the bulk) 
\cite{Binetruy:1999ut}-\cite{Langlois:2004kc}
including brane inflation 
\cite{Dvali:1998pa}-\cite{Nojiri:2000gb}, 
to gauge invariant formalism to study brane perturbations
\cite{Kodama:2000fa,Mukohyama:2000ui,Riazuelo:2002mi}, etc.

This work constitutes a contribution to the above research activity related to generalizations 
of the above mentioned Randall-Sundrum II setup \cite{Randall:1999vf} to either higher 
dimensions or to models with different spatial topology.


The structure of this work is the following: in chapter \ref{cha2} we review the 
subject of non-factorizable geometries and in particular the  
example of the Randall-Sundrum II setup \cite{Randall:1999vf} which lies in many respects at the
center of our considerations. The developments of the following chapters will be in one or another 
way related to the above mentioned setup.


The subject of chapter \ref{cha3} illustrates another example of a fruitful
symbiosis of the a priori unrelated subjects of topological defects and extra dimensional physics.
The physics of topological defects (or topological solitons) is extremely rich and it is clearly 
out of the scope of this report to provide a complete treatment. We refer here to excellent  
textbooks, relevant articles and reviews 
\cite{Kibble:1976sj}-\cite{Hindmarsh:1994re} and references therein
since our main goal will be to provide basics ideas and concepts that 
will be of relevance for this work. Based on the idea of identifying a $3$-brane with a topological 
defect residing in the transverse space of some higher-dimensional space-time \cite{Rubakov:bb}, we 
present a discussion of solutions to higher-dimensional Einstein equations with different 
kinds of source terms.
Attention will be paid in particular to the localization of gravity in models of this kind, where by 
localization we mean that (at least) the effective $4$-dimensional Newtonian theory of gravity 
is reproduced on the brane. The solutions for the space-time metric presented in this chapter will be of 
an asymptotic nature with the meaning that they will constitute solutions to the Einstein equations only 
for large distances from the topological defects. Under certain conditions the underlying field theory 
allows to consider a singular defect limit in which the size of the defect tends to zero in such 
a way that the above asymptotic solutions become complete solutions when supplemented by appropriate
fine-tuning relations. In order to support the analytical evidence for the existence of a particular set of 
these solutions (involving generic $p$-form fields in the bulk) we present a specific numerical example 
in chapter \ref{cha4}: a 't~Hooft-Polyakov monopole residing
in the $3$-dimensional transverse space of a $7$-dimensional space-time. After a short discussion 
of 't~Hooft-Polyakov monopoles in Minkowski space-time, 
we present numerical solutions of the $7$-dimensional case of interest. 
Moreover, we identify the regions in parameter
 space leading to the localization of gravity together with the corresponding fine-tuning relations 
linking various brane tension components.
Finally, we present a specific $5$-dimensional brane-world model in chapter \ref{cha5} with similar
phenomenological properties (in particular similar large distance corrections to Newton's law)
as the Randall-Sundrum II setup. The main difference to the latter 
is that the global topology of the spatial sections of its metric is compact. 
In contrast to the Randall-Sundrum II model in which the 3-brane is characterized by a flat, induced metric, 
the induced metric of the model we present is given by the metric of Einstein's static universe. The main 
advantage of this modification is that the given setup does not suffer from being geodesically 
incomplete, a problem of the Randall-Sundrum II model. 
We close by drawing conclusions in chapter~\ref{cha6}.

\vspace{2cm}
\noindent
This thesis is based on the following publications:

\vspace{4mm}
\noindent
T.~Gherghetta, E.~Roessl and M.~E.~Shaposhnikov,
``Living inside a hedgehog: Higher-dimensional solutions that localize
gravity'', 
Phys.\ Lett.\ B {\bf 491}, 353 (2000)
[arXiv:hep-th/0006251].

\vspace{4mm}
\noindent
E.~Roessl and M.~Shaposhnikov,
``Localizing gravity on a 't Hooft-Polyakov monopole in seven dimensions'', 
Phys.\ Rev.\ D {\bf 66}, 084008 (2002)
[arXiv:hep-th/0205320].

\vspace{4mm}
\noindent
A.~Gruppuso, E.~Roessl and M.~Shaposhnikov,
``Einstein static universe as a brane in extra dimensions'', 
[arXiv:hep-th/0407234]; accepted for publication in JHEP.
\chapter{Non-factorizable geometries} \label{cha2}

\section{Introduction}
In this introductory chapter we will discuss briefly the relevance of non-factorizable geometries to 
brane-world scenarios.
After collecting basic formulae for the metric, the connection coefficients, the Ricci tensor 
and the Ricci scalar of non-factorizable geometries and presenting some motivations 
for their study, we will briefly discuss the Einstein equations with a cosmological constant term in
vacuum. Then we will study the two examples of the Randall-Sundrum brane-world models.  
Our motives for the last choice are two-fold: first, as we already emphasized, it is to a good part 
the Randall-Sundrum models which renewed and reinforced the interest in non-factorizable geometries. 
Second, because it will serve as a reference for later setups studied in chapters \ref{cha3}--\ref{cha5}.
In appendix \ref{appDoubleWarp} we consider a simple generalization of the metric 
(\ref{WarpedMetric}) given in section \ref{2WarpedMetricAndConnections} below.

\section{Metric, connection coefficients and curvature of non-factorizable geometries} 
\label{2WarpedMetricAndConnections}
The concept of non-factorizable product manifolds was first introduced by Bishop and O'Neill in 1969
\cite{Bishop} and later applied to general relativity and semi-Riemannian geometry \cite{Beem,ONeill}. 
Many of the well known metrics of general relativity have this form. For example the simple 
Schwarzschild solution of general relativity in four dimensions can be considered as a non-factorizable 
product of a two-sphere $S^2$ and a $2$-dimensional space-time represented by the Carter-Penrose 
diagram for a black hole \cite{Carter}. For recent work in this area we refer to 
\cite{Carot,Schmidt:1997mq,Katanaev:1998ry}. We will point out later the 
relevance of non-factorizable geometries 
in theories with extra dimensions.

By a non-factorizable metric we mean a metric which locally can be written in the following 
form:\footnote{Throughout this report we will consistently use the sign 
conventions of ref. \cite{wheeler} for the 
metric and the Riemann and Ricci tensors. The metric signature is mostly
 plus $-+\ldots+$. The expression for 
the Riemann curvature tensor is 
$R^\alpha_{\beta \gamma \delta}=
\Gamma^{\alpha}_{\beta \delta, \gamma}-
\Gamma^{\alpha}_{\beta \gamma, \delta}+
\Gamma^{\alpha}_{\sigma \gamma} \Gamma^{\sigma}_{\beta \delta}-
\Gamma^{\alpha}_{\sigma \delta} \Gamma^{\sigma}_{\beta \gamma}$ and 
the Ricci tensor is defined by 
$R_{\alpha \beta}=R^{\sigma}_{\alpha \sigma \beta}$.}
\begin{equation} \label{WarpedMetric}
  ds^2=\hat{g}_{M N} \, dx^M \, dx^N \, = \, \sigma\left(x^a \right) \, g_{\mu \nu}\left(x^\rho\right) \, 
   dx^\mu \, dx^\nu +
  \tilde{g}_{a b}\left(x^c\right) dx^a \, dx^b \, .
\end{equation}
The function $\sigma\left(x^a \right)$ is supposed to depend only on the ``internal'' coordinates $x^a$
and is sometimes called a ``warp'' factor. Moreover, geometries described by metrics of the kind 
(\ref{WarpedMetric}) are often referred to as warped geometries. We will assume that the dimensionality of 
space-time is given by $d_1+d_2$, where $d_1$ and $d_2$ denote the dimensionalities of the spaces described
by $g_{\mu \nu}\left(x^\rho\right)$ and $\tilde{g}_{a b}\left(x^c\right)$, respectively. 
From the explicit expression (\ref{WarpedMetric}) it is straightforward to derive the following relations
for the connections $\hat{\Gamma}^L_{M N}$, $\Gamma^\rho_{\mu \nu}$, $\tilde{\Gamma}^c_{a b}$
\begin{align} \label{WarpedConnections}
  \hat{\Gamma}^c_{a b}=\tilde{\Gamma}^c_{a b} \, , \quad
  \hat{\Gamma}^c_{\mu \nu}=-\frac{1}{2} g_{\mu \nu} \tilde{\nabla}^c \, \sigma \, , \quad
  \hat{\Gamma}^\nu_{\mu b}=\frac{1}{2 \, \sigma}\delta^\nu_{\; \mu} \tilde{\nabla}_b \, \sigma \, , \quad
  \hat{\Gamma}^\rho_{\mu \nu}=\Gamma^\rho_{\mu \nu} \, ,
\end{align}
Ricci tensors $\hat{R}_{M N}$, $R_{\mu \nu}$, $\tilde{R}_{a b}$ 
\begin{align} \label{WarpedRicciTensor}
  \hat{R}_{\mu \nu} &=R_{\mu \nu} -\frac{1}{2} g_{\mu \nu} 
  \left[ \tilde{\nabla}_a \, \tilde{\nabla}^a \, \sigma + \frac{d_1-2}{2 \, \sigma} 
         \tilde{\nabla}_a \, \sigma \tilde{\nabla}^a \, \sigma \right] \, , \\
  \hat{R}_{a b} &=\tilde{R}_{a b}
  - \frac{d_1}{2 \,\sigma} \tilde{\nabla}_a \, \tilde{\nabla}_b \, \sigma  \, 
  + \frac{d_1}{4 \,\sigma^2} \tilde{\nabla}_a \, \sigma \tilde{\nabla}_b \, \sigma \, , \\
  \hat{R}_{\mu a} &=0 \, ,
\end{align}
and the Ricci scalars $\hat{R}$, $R$ and $\tilde{R}$ corresponding to the metrics 
$\hat{g}_{M N}$, $g_{\mu \nu}$ and $\tilde{g}_{a b}$:
\begin{align} \label{WarpedRicciScalar}
  \hat{R}=\tilde{R}+\frac{1}{\sigma} R -\frac{d_1}{\sigma} 
     \tilde{\nabla}_a \, \tilde{\nabla}^a \, \sigma  - 
    \frac{d_1\left(d_1-3\right)}{4\sigma^2} 
   \tilde{\nabla}_a \, \sigma \, \tilde{\nabla}^a \, \sigma \, .
\end{align}
In the upper formulae $\tilde{\nabla}_a$ denotes the covariant derivative associated with the metric
$\tilde{g}_{a b}\left(x^c\right)$. Note that in the case $\sigma=1$ we recover in 
eq.~(\ref{WarpedRicciScalar}) 
the well-known property of additive Ricci scalars for direct product spaces.

\section{Einstein equations with a cosmological constant term}
One of the early motivations of considering a metric of the form (\ref{WarpedMetric}) was to provide
a solution to the cosmological constant problem. The original hope was that the higher dimensional 
Einstein equations with a cosmological term might give a dynamical 
explanation for the smallness of the observed $4$-dimensional cosmological constant. 
The $d_1+d_2$ dimensional Einstein equations with a cosmological constant and an otherwise empty
space-time are given by:
\begin{equation} \label{FullEinstein}
  \hat{R}_{M N}-\frac{1}{2} \, \hat{g}_{M N} \, \hat{R}+\Lambda \, \hat{g}_{M N} =0 \, .
\end{equation}
Using the above relations (\ref{WarpedRicciTensor})-(\ref{WarpedRicciScalar}) in (\ref{FullEinstein}) gives
after some algebra:
\begin{align} \label{FullEinsteinExplicit1}
  &R_{\mu \nu}-\frac{1}{2} \, g_{\mu \nu} \, R+\Lambda_{\mbox{\small phys}} \, g_{\mu \nu} =0 \, , \\
  &\tilde{R}_{a b}=\frac{2}{d_1+d_2-2} \, \Lambda \, \tilde{g}_{a b}\label{FullEinsteinExplicit2}
   + d_1  \left(\frac{1}{2\, \sigma} \tilde{\nabla}_a \, \tilde{\nabla}_b \, \sigma
               -\frac{1}{4 \, \sigma^2} \tilde{\nabla}_a \, \sigma \, \tilde{\nabla}_b \sigma \right) \, , \\
  &\Lambda_{\mbox{\small phys}} = \frac{d_1-2}{d_1+d_2-2} \Lambda \, \sigma 
   + \frac{d_1-2}{4} \tilde{\nabla}_a \, \tilde{\nabla}^a \, \sigma \label{FullEinsteinExplicit3}
   + \frac{\left(d_1-2\right)^2}{8 \, \sigma} \, \tilde{\nabla}_a \, \sigma \, \tilde{\nabla}^a \, 
  \sigma \, . 
\end{align} 
These equations were first studied in 
\cite{Rubakov:1983bz,Wetterich:1984rv,Randjbar-Daemi:1985wg,Wetterich:1984uc}, 
in order to provide an alternative way to the usual Kaluza-Klein 
compactification of higher dimensions and, as we mentioned above, a way to address the cosmological 
constant problem.
In \cite{Rubakov:1983bz} solutions of the above equations 
(\ref{FullEinsteinExplicit1})--(\ref{FullEinsteinExplicit3}) are given in
the case of two extra dimensions $(d_2=2)$. Since $\Lambda_{\mbox{\small phys}}$ can 
be considered as an arbitrary integration constant, the solution of the cosmological constant problem
relies on an additional explanation why a vanishing $\Lambda_{\mbox{\small phys}}$ should be preferred.
In \cite{Randjbar-Daemi:1985wg} it is demonstrated that a similar continuous spectrum of solutions
exists in the case of more than two internal non-compact dimensions.

In the next two sections we discuss the two brane-world models of Randall and Sundrum 
\cite{Randall:1999ee},\cite{Randall:1999vf} as well known examples of 
warped geometries in physics of extra dimensions.

\section{The Randall-Sundrum I model}
In 1999 Randall and Sundrum proposed a solution to the hierarchy problem based on a
single spatial extra dimension \cite{Randall:1999ee}. In their model, the observable 
$4$-dimensional universe is considered to be a subspace of a $5$-dimensional space-time, referred to 
as a $3$-brane. The metric that characterizes the model takes the following form:
\begin{equation} \label{RSImetric}
  ds^2=g_{M N} \,dx^M \, dx^N = e^{-2 \, k \, r_c \, \phi } \, \eta_{\mu \nu} \, dx^\mu \, dx^\nu + r_c^2 \, d\phi^2 \, .
\end{equation}
Upper-case latin indices $M\, ,N$ range from $0$ to $4$. 
Greek indices denote the coordinates of the four familiar dimensions and range from $0$ to $3$ 
while $\phi \in [0,\,\pi]$ denotes the extra (compact) dimension (of ``size'' $\pi r_c$) 
given by the quotient $S^1/\mathbf{Z_2}$. $\eta_{\mu \nu}$ is the usual Minkowski-metric in 
$4$-dimensions with signature $-+++$.
The line element (\ref{RSImetric}) describes a certain region of $AdS_5$-space-time and 
$k$ can be identified with the inverse $AdS$-radius. 
There are actually two $3$-branes in this setup located at the boundaries of the interval $[0,\,\pi]$: 
a negative tension brane located at $\phi=\pi$ with the usual standard model fields localized on it 
and a positive tension brane at $\phi=0$ needed for overall consistency. The action functional can 
be written in the following way:
\begin{align} \label{RSIaction}
  S&=\int d^4x \int\limits_{-\pi}^\pi d\phi \, \sqrt{-g}  \left(-\Lambda+2 \, M^3 \, R\right) \\
   &+\int d^4x \sqrt{-g_{\mbox{\tiny vis}}} 
      \left(\mathcal{L}_{\mbox{\tiny vis}}-\sigma_{\mbox{\tiny vis}}\right)
   + \int d^4x \sqrt{-g_{\mbox{\tiny hid}}} 
     \left(\mathcal{L}_{\mbox{\tiny hid}}-\sigma_{\mbox{\tiny hid}}\right) \nonumber 
\end{align}
In the last expression $\Lambda$ and $M$ denote the $5$-dimensional cosmological constant in the bulk 
space-time and the fundamental scale of gravity.  $g_{\mbox{\tiny vis}}$ ($g_{\mbox{\tiny hid}}$),
 $\mathcal{L}_{\mbox{\tiny vis}}$ ($\mathcal{L}_{\mbox{\tiny hid}}$) and 
$\sigma_{\mbox{\tiny vis}}$ ($\sigma_{\mbox{\tiny hid}}$) represent the induced metric, 
Lagrangian and tension of the visible (hidden) brane.

As shown in detail in \cite{Randall:1999ee}, the metric (\ref{RSImetric}) solves the Einstein equations 
following from (\ref{RSIaction}) (in the sense of distributions) provided the following relations 
between the brane-tensions, the cosmological constant, the fundamental scale and the inverse 
$AdS$-radius hold:
\begin{align} \label{RSIFineTuning}
  \sigma_{\mbox{\tiny hid}} = -\sigma_{\mbox{\tiny vis}}=24 \, M^3 \, k\, ,\quad 
   \Lambda = -24\, M^3 \, k^2 \, .
\end{align}

From the action (\ref{RSIaction}) it is straightforward 
to relate the fundamental scale $M$ to 
the $4$-dimensional Planck-scale $M_{\mbox{\tiny Pl}}^2$ with the result:
\begin{equation} \label{RSIMMrelation}
  M_{\mbox{\tiny Pl}}^2=\frac{M^3}{k} \, \left(1-e^{-2 \, k \, r_c \, \pi}\right) \, ,
\end{equation}
which shows that $M_{\mbox{\tiny Pl}}$ stays finite in the limit $r_c \to \infty$ contrary to the
 case of a product-space compactification.

The way the model solves the hierarchy problem can be summarized as follows: physical mass parameters 
as measured by an observer on the visible brane are generically related to the fundamental 
mass parameters via a renormalization of the induced metric $g_{\mbox{\tiny vis}}$. For $k r_c \approx 50$ 
the exponential suppression of the metric at the position of the visible brane implies that any 
fundamental mass parameter of the order of the Planck mass corresponds to a physical mass in the 
TeV regime from the point of view of an observer on the visible brane. 
The predictions of the model are very different from the predictions of models with large 
extra dimensions introduced in \cite{Arkani-Hamed:1998rs,Antoniadis:1998ig}.
The constraints coming from particle physics 
\cite{Giudice:1998ck}-\cite{Gupta:1999iy}, astrophysics and cosmology 
\cite{Arkani-Hamed:1998nn}-\cite{Hall:1999mk} 
restricting large extra dimensions do not apply 
in the case of the Randall-Sundrum I model, since the splittings of the Kaluza-Klein spectrum of gravitons
are roughly given by the weak scale. Moreover, their coupling to matter is also expected to be set by the 
weak scale and not by the Planck scale. A somewhat unpleasant feature of this model is that the 
fields of the Standard model are localized on a $3$-brane with negative brane tension.

\section{The Randall-Sundrum II model}
In a related article \cite{Randall:1999vf}, Randall and Sundrum demonstrated that even 
an infinite extra dimension is not in contradiction with present day observations 
of gravity at low energy scales. The main motivation was to prove that the Newtonian force
law is effectively $4$-dimensional for two test masses placed on the $3$-brane. In order to achieve this 
``localization of gravity'', it is necessary to assume that the Standard Model fields are localized on the 
brane at $\phi=0$ (in the coordinates of the last section). The brane at $\phi=\pi$ serves merely as a
``regulator'' brane for imposing boundary conditions and will eventually be removed from the setup by taking 
the limit $r_c \to \infty$. 

\subsection{Localization of gravity} 
In order to work out the force between two test masses on the brane we need to determine the spectrum 
of Kaluza-Klein excitations of the $4$-dimensional graviton $h_{\mu \nu}$ defined by:
\begin{equation} \label{RSImetricfluct}
  ds^2=\left(e^{-2 \, k \, \vert y \vert} \, \eta_{\mu \nu} +h_{\mu \nu} \right) \, dx^\mu \, dx^\nu + dy^2
 \, ,
\end{equation}
where $y=r_c \phi$ in the notation of the last section.\footnote{Note that even thought in reference 
\cite{Randall:1999vf} Randall and Sundrum stick to the topology $S^1/\mathbf{Z_2}$ at the beginning, 
they allow $y$ to range from $(-\infty,\infty)$ later in the paper, in order to avoid a semi-infinite 
extra dimension.}
Following \cite{Randall:1999vf} we use the gauge $\partial^\mu h_{\mu \nu}=h^\mu_{\, \mu}=0$ in order 
to obtain a single perturbation equation for the metric fluctuations $h_{\mu \nu}$ from the 
linearized Einstein equations in the background (\ref{RSImetric}). After neglecting the tensor structure 
of $h_{\mu \nu}$ an expansion in $4$-dimensional plane waves gives:
\begin{equation} 
  \left[ -\frac{m^2}{2} e^{2 k \vert y \vert}-\frac{1}{2}\partial_y^2-2 k \delta(y) + 2k^2 \right]
  \psi(y)=0 \, .
\end{equation}
We now transform the above equation into a Schr\"odinger equation by using the change of 
independent and dependent variables $z=\mbox{sign}(y) \left(e^{k \vert y \vert}-1\right)/k$, 
$\hat\psi(z)=\psi(y) \, e^{k \vert y \vert/2}$ with the result:
\begin{equation} \label{RSSchroedinger}
  \left[-\frac{1}{2} \partial_z^2+\frac{15 \, k^2}{8 \left(k\,\vert z \vert+1\right)^2}-
    \frac{3 \, k}{2} \, \delta(z)\right] \hat\psi(z)=m^2 \, \hat\psi(z) \, .
\end{equation}
From the potential term in the above equation one can infer the main properties of the Kaluza-Klein spectrum 
in the limit $r_c \to \infty$.
The $\delta$-function allows for a single normalizable bound state 
$\hat\psi_0(z)=k^{-1} \left(k \vert z\vert+1\right)^{- 3/2}$ corresponding to $m^2=0$. Next, since the 
potential vanishes in the limit $z \to \infty$ there will be a continuous spectrum for all $m^2>0$. 
Finally, one expects a suppression of the wave-functions of the continuum in the vicinity of the 
brane ($z=0$) due to the potential barrier. The correct continuum wave-functions are given in terms of Bessel 
functions:
\begin{equation}
  \hat\psi_m\left(z\right)=
  \left(\vert z\vert+\frac{1}{k}\right)^\frac{1}{2} \left\{ 
   A Y_2\left[m\left(\vert z\vert+\frac{1}{k}\right)\right]+
   B J_2\left[m\left(\vert z\vert+\frac{1}{k}\right)\right] \right\} \, .
\end{equation}
The constants $A$ and $B$ can be determined by going one step back and using the 
regulator brane to impose the following boundary condition on 
$\hat\psi(z)$ at $z=0$ and $z=z_c=\left(e^{k \pi r_c}-1\right)/k$ :
\begin{equation}
  \partial_z \hat\psi(z)+\frac{3 \, k}{2\left(k z + 1\right)} \hat\psi(z)=0\, .
\end{equation} 
In this way the spectrum becomes discrete again and at least in the most interesting 
case of small masses $m \ll k$ the correct normalization of the modes can be deduced. As usual, 
masses in the range $m \ll k$ will determine the behavior of the potential in the distance regime 
$r\gg 1/k$. Going back to the infinite extra dimensions, we can now assemble all contributions coming 
from the zero-mode and the continuum modes in the potential between two test particles on the 
brane at $z=0$:
\begin{align} \label{RSCorrections}
 V(r) &\sim G_N \frac{m_1 \, m_2}{r} + \int\limits_0^\infty dm \, \frac{G_N}{k} \, 
   \frac{m_1 \, m_2 \, e^{- m r}}{r} \, \frac{m}{k} \nonumber \\ 
  &= G_N \frac{m_1 \, m_2}{r} \left(1+\frac{1}{r^2 \,  k^2}\right) \, ,
\end{align}
with $G_N=k/M^3$ following directly from (\ref{RSIMMrelation}).
We see that it is the zero mode that reproduces the conventional $4$-dimensional Newtonian behavior 
while the collective contributions of the higher modes is manifest through a 
correction term which is strongly suppressed at distances $r$ larger than $1/k$.

We conclude this section with a few remarks about the above calculation. First, the result 
(\ref{RSCorrections}) applies only in the regime  $r \gg 1/k$, since the normalization constant
has been worked out correctly only in the regime of small masses. Second, due to the incomplete 
treatment of the tensor structure of the gravitational perturbation we expect 
(\ref{RSCorrections}) to represent the correct potential only up to numerical factors in front of the 
$1/(r^2 k^2)$-correction term inside the parenthesis on the right hand side. Indeed several careful studies 
have shown that the factor of $1$ in (\ref{RSCorrections}) has to 
be replaced with $1/2$ in the case of the massless scalar in the Randall-Sundrum background, whereas a full
treatment accounting for the tensorial nature of the graviton and the effects of 
``brane-bending'' gives a factor of $2/3$, 
see for example \cite{Garriga:1999yh,Deruelle:2001ms,Giddings:2000mu,Kiritsis:2002ca} and 
\cite{Callin:2004py} and references therein for a recent discussion. We also refer to appendix
 \ref{appParallel} where we redo the above calculation bypassing the complications of 
the determination of the 
Kaluza-Klein spectrum and wave-function normalization by directly solving for the (scalar) Green's
function.

The singular treatment of branes in the models of Randall and Sundrum, manifest through the appearance 
of delta-functions in the 
stress-energy tensor components, constitutes a welcome simplification 
when it comes to solving Einstein equations. However, from a field theoretical point of view, branes 
can be considered as finite energy solutions to some field equations with an energy-density 
localized in some region of space of a given finite thickness $\delta$. On the one hand, in the 
case of a singular brane, a complete solution to Einsteins equations can be obtained 
in a straightforward way by imposing 
the Israel-junction condition \cite{Israel:rt} at the boundary between the two regions separated by the 
brane. On the other hand, the treatment of the non-singular, thick-brane case 
is in principle based on the complete solution of the coupled Einstein and matter field equations and is 
therefore in general a much more involved task with the result that analytic solutions are rare. As we will
see in the following two chapters, the idea of replacing a singular brane by a thick domain wall
can be generalized to higher-dimensional cases. Depending on the dimension of space-time, 
we will have to replace domain walls by various other topological defects like strings or monopoles. 
\thispagestyle{empty}
\chapter{Topological defect solutions of higher-dimensional Einstein equations} \label{cha3}
\section{Introduction} Topological defects like domain walls, strings and monopoles arise in a 
variety of different areas of physics 
starting from condensed matter physics, where particular examples of defects can be 
observed experimentally, over particle physics, astrophysics
and cosmology to even cosmology involving extra dimensions. Their physics is enormously rich 
and consequently the attention devoted to the study of topological defects in the past and present 
is considerable. Before embarking on a discussion
of the relevance of topological defects in theories with extra dimensions we would like to review briefly 
basic properties and examples in the next section. It is clear that our presentation is again 
not more than an introduction to this vast field and in no sense complete. We start by  giving a definition 
for what we will mean by topological defect (or topological solitons), enumerate the most prominent 
examples and discuss general common physical properties. 
After a short sketch of homotopy theory, together with the statement of the two fundamental theorems, we are ready to discuss the by now standard classification scheme of topological defects going back to 
Kibble \cite{Kibble:1976sj}. We will conclude this introduction by shortly discussing the role of particular 
topological defects in condensed matter physics, astrophysics and cosmology.

\subsection{Definitions and physical properties of topological defects}
Topological defects or equivalently topological solitons are particular cases of a larger group of 
objects called solitons or better solitary waves. The last two expressions are not used in a standardized 
way in the literature. We decide to use the definitions given in \cite{Raj}. On the one hand, a {\em solitary wave} is a 
classical solution
to any in general non-linear field equation (or sets of non-linear field equations) with the properties 
that 
\begin{enumerate}
  \item the total energy of the given field configuration is finite,
  \item the energy density $\epsilon(t,\vec x)$ is nonsingular and localized in space. 
  Moreover, for time-dependent solutions the shape of $\epsilon(t,\vec x)$ propagates undistorted 
  in time.
\end{enumerate}

A {\em soliton}, on the other hand, is a solitary wave with the additional property that the 
underlying nonlinear dynamics involved in a scattering process of two or more solitary wave packets 
asymptotically (for large times) restores the forms of the incoming wave packets.
Obviously, this definition is very restrictive and apart from a few examples the vast majority of 
known solitary waves does not satisfy this criterion.\footnote{It is mainly for this reason that it 
became common practice to use the term soliton instead of solitary wave even in cases where 
the above criterion is not satisfied.} 
With the above definitions, we can now come back to topological defects and identify them with solitary 
waves or solitons whose stability is guaranteed by means of a topological argument. More precisely this 
means that the underlying field theory allows for a topologically conserved (and generically quantized) 
charge in such a way that field configurations with a given charge $Q$ are forbidden to evolve into 
configurations characterized by charges different from $Q$.

As we will see in the next section, a mathematical classification of topological defects is based 
on specific topological properties of the vacuum manifold of the underlying field theory. A more 
phenomenological (and related) approach classifies defects according to their spatial dimensionality. 
Therefore, one distinguishes 
\begin{itemize}
  \item $0$-dimensional, point-like defects or monopoles,
  \item $1$-dimensional, line-like defects or strings,
  \item $2$-dimensional, surface-like defects or domain walls.
\end{itemize}
Somewhat out of this scheme lie textures which are not localized in space and hybrids of the 
above mentioned defects. For example one can consider theories in which monopoles appear to be 
attached to the endpoints of strings or in which domain walls are bounded by strings.

Some observed physical examples in condensed matter physics comprise magnetic flux lines in a type II
superconductor, quantized vortex lines in superfluid $^4 \mbox{He}$, textures in $^3 \mbox{He}$, 
line defects and dislocations in crystalline substances, various defects in different phases of 
liquid crystals and domain structures in ferromagnets. The formation of all kinds of defects is believed 
to be triggered via rapid phase transitions in connection with spontaneous symmetry breakdown by the 
so-called Kibble-mechanism \cite{Kibble:1976sj}. The main idea is that below a certain critical temperature
the effective Higgs potential will develop degenerate minima. Random quantum fluctuations of the
 Higgs field will decide which vacuum will be chosen in a given volume. If two neighboring regions in space
 turn out to fall into different minima, then a defect will form at the boundary between them.
The typical length scales of these structure will be set by the correlation length beyond which the 
Higgs phase is completely uncorrelated. An example for the above situation is the magnetic domain structure
in a ferromagnet. 

\subsection{Homotopy theory and classification of topological defects} \label{Homotopy}
In this section we intend to assemble basic definitions and results of homotopy theory 
as far as they are of relevance to the classification of topological defects. Our presentation
is based on \cite{Vilenkin}. For more complete discussions of the subject we refer to 
\cite{Hilton,Iyanaga,Mermin,Maunder}. We start with the definition of homotopy groups. Let us consider 
the set $\mathcal{L}_x \left(\mathcal{M}\right)$ of all closed loops with given base point $x \in \mathcal{M}$
in a given topological space $\mathcal{M}$, that is the set of all continuous maps $f$ from the circle 
$S^1$ into $\mathcal{M}$ with $x$ in the image of $f$. We can define an equivalence relation on $\mathcal{M}$
in the following way: $f$ and $g \in \mathcal{L}_x \left(\mathcal{M}\right)$ are said to be homotopic if they
can be continuously deformed into each other (within $\mathcal{M}$) keeping the base point $x$ fixed. Now 
we have $f \sim g$ if $f$ and $g$ are homotopic and we denote $[f]$ the equivalence class or homotopy class
corresponding to the loop $f$. This set of equivalence classes $\left\{ [f], [g], \ldots \right\}$ can be 
given a group structure:
the inverse of $[f]$ is defined by $[f]^{-1}\equiv [f^{-1}]$ with $f^{-1}$ being the loop obtained from
$f$ by reversing the order in which it is traversed. The group multiplication of $[f]$ and $[g]$ is realized 
by the operation of connecting two loops according to $[f]\cdot[g]=[f \circ g]$ and the identity class 
$[I]$ is the class of loops continuously contractible to the point $x$. The resulting  group is called the 
first fundamental group or the first homotopy group of the topological space $\mathcal{M}$ with base point
$x$ and is denoted by $\pi_1\left(\mathcal{M}\, , x\right)$. It can easily be seen that for a connected space $\mathcal{M}$ two base points $x, y \in \mathcal{M}$ 
give rise to isomorphic first homotopy groups  $\pi_1\left(\mathcal{M}\, , x\right) \cong 
\pi_1\left(\mathcal{M}\, , y\right)$ so that it is justified to drop the reference to the base point 
altogether and simply write $\pi_1\left(\mathcal{M}\right)$. 

In a similar manner it is possible to define homotopy groups of $\mathcal{M}$ of higher order as equivalent 
classes of maps from $n-$spheres into $\mathcal{M}$. For example the $n^{\mbox{\small th}}$ homotopy group of 
the topological space $\mathcal{M}$, denoted by $\pi_n \left(\mathcal{M}\right)$ is the 
set of homotopically equivalent classes of maps form the $n$-sphere into the 
manifold $\mathcal{M}$.\footnote{Note that
we immediately dropped the dependence on the base point which again is possible only for a connected space.}
 An important difference between the groups $\pi_1 \left(\mathcal{M}\right)$ and 
$\pi_n \left(\mathcal{M}\right)$ (for $n\geq2$) is that the former is not necessary abelian while the 
latter always are. Finally, it is useful to introduce $\pi_0 \left(\mathcal{M}\right)$ as 
the homotopy classes of maps from $S^0$ (a point) into $\mathcal{M}$. This set labels the 
disconnected components of the space $\mathcal{M}$. However, in general it cannot be given a group structure. 
A trivial $\pi_0\left(\mathcal{M}\right)$ is equivalent to the statement that $\mathcal{M}$ is a 
connected space. 

We are now prepared to state (without proof) the first two fundamental theorems of homotopy theory.

\vspace{5mm}
\noindent
{\em First Fundamental Theorem}

\noindent
Let $G$ be a connected and simply-connected Lie group ($\pi_0\left( G \right)=\pi_1\left( G \right)=I$),
and let $H$ be a subgroup of $G$. We then have
\begin{equation} \label{1stHomTheorem}
  \pi_1\left(G/H\right) \cong \pi_0\left(H\right) \, .
\end{equation}

\vspace{5mm}
\noindent
{\em Second Fundamental Theorem}

\noindent
Let $G$ be a connected and simply-connected Lie group ($\pi_0\left( G \right)=\pi_1\left( G \right)=I$),
and let $H$ be a subgroup of $G$. Let $H_0$ be the component of $H$ connected to the identity. We then have
\begin{equation} \label{2ndHomTheorem}
  \pi_2\left(G/H\right) \cong \pi_1\left(H_0\right) \, .
\end{equation}

The usefulness of the above two theorems follows directly from the classification schema for topological
defects which we set out to describe in the following. Topological defects are thought to be produced
during phase transitions in the early universe. Let us suppose that in the high temperature phase the 
theory possesses a symmetry described by a connected, simply-connected continuous Lie group $G$. At the 
onset of the phase transition, when the temperature drops below a critical value, the Higgs potential 
develops non-trivial minima and the symmetry is spontaneously broken down to the subgroup $H$ of $G$.
Let $\mathcal{M}$ be the vacuum manifold of the theory and $\Phi_0$ a given vacuum
expectation value in $\mathcal{M}$. Then $D(h) \Phi_0=\Phi_0$ for all $h \in H$ by definition of the 
unbroken subgroup $H$. On the other hand, any $g \in G$ with $g \notin H$ will give rise to another
possible choice of vacuum distinct from $\Phi_0$. More than that, all different distinct elements 
in the vacuum manifold $\mathcal{M}$ can be obtained by the action of distinct cosets $g \, H$ of $G$ 
with respect to the unbroken subgroup $H$ on $\Phi_0$. Therefore, the vacuum manifold can be identified with 
the coset space $\mathcal{M}=G/H$.

We are now in a position to state the classification of topological defects established by Kibble in 1976
 \cite{Kibble:1976sj}. Accordingly, domain walls occur during phases transitions accompanied by spontaneous symmetry 
breaking if the vacuum manifold has disconnected components, that is $\pi_0\left(G/H\right)\neq I$. 
Strings occur if the vacuum manifold contains unshrinkable loops that is 
$\pi_1\left(G/H\right)\neq I$. Analogously we find monopoles for $\pi_2\left(G/H\right)\neq I$ that is 
unshrinkable $2$-spheres. A somewhat special kind of defects that can also be incorporated in the above 
classification scheme is so-called textures. While the Higgs field of domain walls, strings and 
monopoles leaves the vacuum manifold inside and around the core of the defect, the Higgs field for textures
always takes values inside the vacuum manifold. Therefore, only gradient energy contributes 
to the total energy of textures. Textures arise in theories in which the  vacuum manifold allows for 
non-shrinkable $3$-surfaces that is for non-trivial $\pi_3\left(G/H\right)$. We summarize the classification of topological defects according to the topology of the vacuum manifold of the underlying field theory in the 
following table~\ref{TDClassification}.
\begin{table}[htbp]
  \begin{center}
  \begin{tabular}{|l|c|}
    \hline
    {\em Topological Defect} & {\em Classification} \\
    \hline\hline
    Domain walls & $\pi_0\left(\mathcal{M}\right)$ \\
    Strings & $\pi_1\left(\mathcal{M}\right)$ \\
    Monopoles & $\pi_2\left(\mathcal{M}\right)$ \\
    Textures & $\pi_3\left(\mathcal{M}\right)$ \\
    \hline
  \end{tabular}
\caption{\label{TDClassification} Homotopy classification of topological defects. $\mathcal{M}=G/H$.}
  \end{center}
\end{table}

By making use of the first fundamental theorem (\ref{1stHomTheorem}), we see that 
for connected and simply-connected groups $G$ the question of whether a given theory allows for 
topological string solutions can be reduced to the question of whether the unbroken subgroup 
$H$ has disconnected components or not. In a similar manner (and under the same assumptions 
about $G$) the condition for the existence of monopole solutions $\pi_2\left(G/H\right)\neq I$
is equivalent to $\pi_1\left(H_0\right)\neq I$, where we recall that $H_0$ was the component
of $H$ connected to the identity.

Since the discoveries of topological defect solutions around 30 years ago in simple 
relativistic field theoretical models 
\cite{Abrikosov:1956sx,Nielsen:1973cs,Dashen:1974cj,Goldstone:1974gf,'tHooft:1974qc,Polyakov:1974ek}, 
the field underwent an immense 
boost in research activities in different directions. Apart from their direct relevance in 
condensed matter physics where corresponding non-relativistic analogs of the above-mentioned models 
describe magnetic flux lines and vortex lines in superconductors and superfluids, the rich potential for
new physics was immediately recognized also in the areas of particle physics, astrophysics and cosmology. 
Research interests ranged from field theoretical, classical and quantum aspects of topological defects over 
revealing detailed defect structures, extension of the classes of known defects to more and more complicated
field theoretical models including supersymmetry versions, dynamics of interactions of defects accompanied by 
particle productions, the study of hybrid defects etc. Also the gravitational properties of defects 
were extensively studied for example the gravitational lensing effects of strings, emission of 
gravitational waves from oscillating string loops, formation of black holes, or the gravitational 
fields of collapsing textures to mention only a few.

Interest in cosmology ranged from the formation and evolution of domain structures or string networks 
in the universe in connection with structure formation, predicted power spectra, imprints in the 
temperature fluctuation of the cosmic microwave background radiation, etc.
Our list is by no means complete and we refer to excellent textbooks and review articles for 
more exhaustive descriptions 
\cite{Raj,Vilenkin,Durrer:1999na,Vachaspati:2000cq,Goddard:1977da,Preskill:1986kp,Coleman:1982cx,Hindmarsh:1994re} and also 
for lists of relevant articles.

We now turn to our main interest for studying topological defects, namely to point out 
the role they play in cosmological models in connections with extra dimensions. As we will see 
in the next section, there are numerous motivations for this symbiosis.

\section{Topological defects in higher dimensions}

The picture of identifying our $4$-dimensional universe with a topological defect in a 
higher dimensional space-time has emerged in an attempt to find alternatives to Kaluza-Klein
compactifications 
\cite{Rubakov:1983bz,Wetterich:1984rv,Randjbar-Daemi:1985wg,Rubakov:bb,Antoniadis:1990ew,Arkani-Hamed:1998rs,Antoniadis:1998ig,Randall:1999ee,Randall:1999vf,akama,Visser:1985qm,Dvali:1996xe}. 
In this connection, the $4$-dimensional sub-manifold of the higher-dimensional 
bulk associated with our world is usually referred to as a $3$-brane. While in theories of this 
type, Standard Model particles are thought to be confined to the brane, gravity is free to propagate in
the extra dimensions, being the dynamics of space-time itself. One way of achieving such a 
confinement is to use string theoretical arguments \cite{Polchinski:1995mt}. 
Other ways use various field theoretical effects.  
In \cite{Rubakov:bb} it was shown that in the 
absence of gravity a domain wall residing in a $5$-dimensional Minkowski space-time can explain the 
necessary trapping of Standard Model particles in the vicinity of the brane, making the extra dimension
inaccessible to particles below a certain threshold energy. Moreover, it was shown that 
this mechanism can give rise to $4$-dimensional chiral fermions on the brane.     
Similar mechanisms of confining 
particles to $4$-dimensions had also been envisaged in theories of pure gravity
\cite{Visser:1985qm}. In \cite{Dvali:1996xe} a possibility of obtaining massless gauge bosons confined to 
domain walls has been demonstrated. Lattice simulations on related topics can be found in 
\cite{Laine:2002rh,Laine:2004ji}.

As already discussed in chapter \ref{cha2}, also gravity itself can be localized on the brane 
in a simple (singular) domain wall setup in $5$ dimensions involving a warped 
geometry \cite{Randall:1999vf}, even 
in the case of an extra dimension of infinite size. 
This last work motivated also the generalization of the simple model of Randall and Sundrum to 
more involved possibilities and to higher-dimensional models. In \cite{Chamblin:1999cj} a 
non-singular supergravity version of the Randall-Sundrum II model was considered. Motivated by solving the 
hierarchy problem along similar lines as in the Randall-Sundrum I model and based on their 
earlier works in $4$-dimensions \cite{Cohen:1988sg}, the authors of \cite{Cohen:1999ia} considered 
a brane made from a global cosmic string defect residing in a six-dimensional space-time. The 
solution corresponds to a vanishing cosmological constant in the bulk and is characterized by a 
singularity at a finite distance from the string. A similar but singularity-free static solution with 
a negative bulk cosmological constant based on the compactification on a 
global string was found in \cite{Gregory:1999gv}. More general constructions in six dimensions have been 
considered in \cite{Chodos:1999zt}. Then, in \cite{Gherghetta:2000qi} it was shown that similar to the 
Randall-Sundrum II case in $5$-dimensions, gravity can be localized on a global string like defect in 
six dimensions. The corrections to Newton's law were also calculated and shown to be highly suppressed
at large distances. See also \cite{Tinyakov:2001jt} for a related discussion of the physical 
properties of the thin-string 
limit and \cite{Ponton:2000gi} for a more detailed study of 
the localization of gravity in this model. A discussion 
of the localization of matter fields of different spin in the above background of the local string 
can be found in \cite{Oda:2000zc}. 
Further generalizations of these scenarios to so-called ``hedgehog''
scalar and p-from field configurations were given in \cite{Olasagasti:2000gx},\cite{Olasagasti:2001hm} 
and \cite{Gherghetta:2000jf}. In the latter paper, 
the corrections to Newton's law were given for theories with $n$-transverse extra dimensions.
After finishing this short overview of existing literature, we will come back to \cite{Gherghetta:2000jf} 
in the rest of this section in more detail.
While \cite{Olasagasti:2000gx} and \cite{Gherghetta:2000jf} concentrate on the case of a spherical symmetric 
transverse spaces, in \cite{Randjbar-Daemi:2000ft} more general cases  
of Ricci-flat transverse manifolds or homogeneous spaces with topologically non-trivial gauge field
configurations were included. The motivation for considering more complex structures of 
transverse spaces lies within the possibilities of obtaining a more realistic spectrum 
of $4$-dimensional particles localized on the brane, see e.g. \cite{Frere:2000dc,Libanov:2000uf} for 
a construction involving three families of chiral fermions localized on the brane 
which in this setup is identified with an Abrikosov-Nielsen-Olesen vortex in 6 dimensions. 
Another attempt to construct a realistic theory of extra dimensions in the Kaluza-Klein approach 
can be found in \cite{Dvali:2001qr}, with the interesting observation that the identification of 
extra components of a gauge field with the Higgs field provide a possible realization 
of the spontaneous electroweak symmetry breaking. In the latter model, the existence of a topological
background in the extra dimensions is crucial for explaining both, the symmetry breaking of the 
Standard Model and chiral fermions in $4$ dimensions.
More recently (see \cite{Midodashvili:2004ce} and references therein) similar work realized a purely 
gravitational localization of three generations of fermions on a brane.

As already mentioned in the introductory chapter, it was realized early 
\cite{Rubakov:bb,Wetterich:1984uc,Witten:1983ux,Randjbar-Daemi:1982hi,Randjbar-Daemi:1983qa}
that the topological nature of the field configuration in transverse space is crucial for 
obtaining chiral fermions in $4$-dimensions. For a recent discussion of the 
localization of chiral fermions in brane-world models see 
\cite{Randjbar-Daemi:2000cr,Randjbar-Daemi:2002pq}.  

We should also mention another motivation for studying topological defects in connection
with brane-world scenarios. In field theories in flat space-time which allow for 
topological defect solutions, the conservation of a topological charge guarantees the stability of the 
configuration. When gravity is included, it will of course not be possible to conclude 
stability by topological arguments only, but at least in cases where gravity is weak
one can hope that ``gravitating''
topological defects can provide good potential candidates for stable configurations.   

Also some numerical work exists in this general context. 
See e.g.~\cite{Giovannini:2001hh} for a discussion of a gravitating abelian vortex in 6 dimensions and 
\cite{Roessl:2002rv}
for a gravitating 't~Hooft-Polyakov monopole in $7$ dimensions. The latter case will be the subject
of chapter \ref{cha4}. Other related works include for example 
the case of a global monopole in $7$ dimensions \cite{Benson:2001ac} 
or a texture in 5 dimensions \cite{Cho:2002ui}. In contrast to the corresponding 
$6$-dimensional case of the global string \cite{Cohen:1999ia}, the geometry of  
\cite{Benson:2001ac} is singularity-free but the graviton zero-mode is not normalizable.


As announced earlier in this section, we are now going to generalize the results 
of ref.~\cite{Gherghetta:2000qi} to the case of transverse spaces of dimensionality $n$
greater than $2$. We are restricting ourselves to transverse spaces with spherical
symmetry and as we shall see in the following, their properties are qualitatively different 
for $n\leq2$ and $n \geq 3$ extra dimensions, a fact 
which complicates the generalization.  In contrast to the case with $n \geq 3$, 
for  $n=1$ the extra space is flat while for $n=2$ the extra space can be curved, but is still
conformally flat. 

We will consider three different possibilities: we will call the first possibility
a strictly local defect and mean the situation where the stress-energy tensor 
of the defect is zero outside the core or, for the more realistic situation of a non-singular brane 
of finite thickness, exponentially vanishing outside of the defect core. 
In this case, Einstein's equations do not allow a simple generalization of the 
geometries for $n=1$ and $n=2$ leading to the localization of gravity.

The second possibility is related to the so-called global defects.
In this case one assumes a multiplet of scalar field, with a global symmetry 
e.g. $O(N)$ which is spontaneously broken. Outside the core of the defect this 
field may have a hedgehog type configuration, which
gives a specific contribution to the stress-energy tensor. 
As mentioned above, this case was studied in \cite{Cohen:1999ia,Gregory:1999gv} for $n=2$ and in
\cite{Olasagasti:2000gx} for higher dimensions, where the solutions with an
exponential warp factor were found. 
We compute the corrections to Newton's law in this case and 
define a generalization of these metric solutions.

The third possibility is related to configurations of the monopole
type. In this case, outside of the defect, the stress-energy tensor is 
dominated by the energy of a magnetic field or its
generalization to higher dimensions, $p$-form abelian gauge fields. We
consider different spherically symmetric configurations and define those 
which lead to gravity localization on the $3$-brane and a
regular geometry in the bulk. We also discuss the corrections to
Newton's law for these solutions. 

\section{Einstein equations with a 3-brane source}

In D-dimensions the Einstein equations with a bulk cosmological
constant $\Lambda_D$  and a stress-energy tensor $T_{AB}$ are
\begin{equation}
   R_{AB} - \frac{1}{2} g_{AB} R + \frac{\Lambda_D}{M_D^{n+2}} g_{AB} 
    = \frac{1}{M_D^{n+2}}  T_{AB}~,
\end{equation}
where $M_D$ is the reduced D-dimensional Planck scale.  We will
assume  that there exists a solution that respects 4d Poincar\'e
invariance. A D-dimensional metric with $D=4+n$ satisfying this ansatz for $n$
transverse, spherical coordinates with $0\leq \rho < \infty$,  
$0\leq \{\theta_{n-1},\dots,\theta_2\} < \pi$ and $0\leq \theta_1
< 2\pi$ is
\begin{equation}
\label{3metric}
    ds^2 = \sigma(\rho) g_{\mu\nu} dx^\mu dx^\nu 
    +d\rho^2+\gamma(\rho)d\Omega_{n-1}^2\, ,
\end{equation}
where as usual we follow the conventions of \cite{wheeler} with the metric signature 
of $g_{\mu\nu}$ given by $(-,+,+,+)$. 
$d\Omega_{n-1}^2$ is defined recursively as
\begin{equation}
     d\Omega_{n-1}^2 = d\theta_{n-1}^2 + 
     \sin^2\theta_{n-1} d\Omega_{n-2}^2~,
\end{equation}
with $d\Omega_0^2 = 0$. 
We assume in the following that the $3$-brane is located at the origin $\rho=0$
of the above coordinate system (\ref{3metric}) and that its 
source is described by a stress-energy tensor $T^A_{\;\; B}$ with nonzero
components
\begin{equation}
\label{3source}
     T^\mu_{\;\; \nu} = \delta^\mu_{\;\; \nu} f_0(\rho), \quad T^\rho_{\;\; \rho} = f_\rho(\rho),
     \quad {\rm and} \quad T^\theta_{\;\; \theta} = f_\theta(\rho)~.
\end{equation}
Here we introduced three source functions $f_0$, $f_\rho$, and
$f_\theta$ which depend only on the radial coordinate $\rho$ and by
spherical symmetry all the angular source functions are identical.
By definition we have $\theta\equiv \theta_{n-1}$. 
Using the metric ansatz  (\ref{3metric}) and the stress-energy tensor
(\ref{3source}), the Einstein equations become 
\begin{align}
\label{3solnset1}
    &\frac{3}{2} \frac{\sigma^{\prime\prime}} {\sigma} 
   +\frac{3}{4}(n-1)\frac{\sigma^\prime}{\sigma}\frac{\gamma^\prime}{\gamma}
   +\frac{1}{8}(n-1)(n-4) \frac{\gamma^{\prime 2}}{\gamma^2}
   +\frac{1}{2}(n-1)\frac{\gamma^{\prime\prime}}{\gamma}
   -\frac{1}{2\gamma}(n-1)(n-2) \nonumber \\
    &\qquad = -\frac{1}{M_D^{n+2}}\left[\Lambda_D - f_0(\rho)\right] +
    \frac{\Lambda_{phys}}{M_P^2}\frac{1}{\sigma}~,
\end{align}
\begin{align}
\label{3solnset2}
    &\frac{3}{2}\frac{\sigma^{\prime 2}}{\sigma^2}
   +(n-1)\frac{\sigma^\prime}{\sigma}\frac{\gamma^\prime}{\gamma}
   +\frac{1}{8}(n-1)(n-2) \frac{\gamma^{\prime 2}}{\gamma^2}
   -\frac{1}{2\gamma}(n-1)(n-2) \nonumber\\
    &\qquad = -\frac{1}{M_D^{n+2}}\left[\Lambda_D - f_\rho(\rho)\right] +
    \frac{2\Lambda_{phys}}{M_P^2}\frac{1}{\sigma}~,\\
 \label{3solnset3}
    &2\frac{\sigma^{\prime\prime}} {\sigma} 
    +\frac{1}{2}\frac{\sigma^{\prime 2}}{\sigma^2}
   +(n-2)\frac{\sigma^\prime}{\sigma}\frac{\gamma^\prime}{\gamma}
   +\frac{1}{8}(n-2)(n-5) \frac{\gamma^{\prime 2}}{\gamma^2}
   +\frac{1}{2}(n-2)\frac{\gamma^{\prime\prime}}{\gamma}
    -\frac{1}{2\gamma}(n-2)(n-3)\nonumber\\
   &\qquad  = -\frac{1}{M_D^{n+2}}\left[\Lambda_D - f_\theta(\rho)\right] 
    +\frac{2\Lambda_{phys}}{M_P^2}\frac{1}{\sigma}~,
\end{align}
where $^\prime$ denotes differentiation with respect to $\rho$ and the Einstein
equations arising from all the angular components simply reduce to 
one angular equation (\ref{3solnset3}). The constant
$\Lambda_{phys}$ represents the physical 4-dimensional
cosmological constant, where
\begin{equation}
     R_{\mu\nu}^{(4)} - \frac{1}{2} g_{\mu\nu} R^{(4)} + 
     \frac{\Lambda_{phys}}{M_P^2} g_{\mu\nu}=0~.
\end{equation}
The system of equations (\ref{3solnset1})--(\ref{3solnset3}) for $f_i=0$
was first derived in \cite{Randjbar-Daemi:1985wg} and describes
the  generalization of the setup considered in 
\cite{Rubakov:1983bz,Randall:1999vf,Gherghetta:2000qi,Olasagasti:2000gx}, to
the case where there are $n$ transverse dimensions, together with a
nonzero cosmological constant in 4-dimensions and stress-energy tensor
in the bulk. If we eliminate two of
the equations in (\ref{3solnset1})--(\ref{3solnset3}) then the source
functions satisfy
\begin{equation}
     f_\rho^\prime = 2 \frac{\sigma'}{\sigma} (f_0-f_\rho) 
      + \frac{n-1}{2} \frac{\gamma'}{\gamma} (f_\theta-f_\rho)~,
\end{equation}
which is simply a consequence of the conservation of the stress-energy
tensor $\nabla_M T^M_{\;\; N} = 0$. In general, the Ricci scalar corresponding to 
the metric ansatz (\ref{3metric}) is
\begin{align}
\label{3ricci}
      R&=-4\frac{\sigma''}{\sigma}- \frac{\sigma^{\prime2}}{\sigma^2}
         -2(n-1) \frac{\sigma'}{\sigma}\frac{\gamma'}{\gamma}  
         -(n-1)\frac{\gamma''}{\gamma}-
       \frac{1}{4}(n-1)(n-4)\frac{\gamma^{\prime2}}{\gamma^2} \nonumber\\
       &+ (n-1)(n-2)\frac{1}{\gamma} +
       \frac{4}{\sigma}\frac{\Lambda_{\rm phys}}{M_P^2}~.
\end{align}
The boundary conditions at the origin of the transverse space
are assumed to be
\begin{equation} \label{3boundsigmagamma}
     \sigma^\prime\big|_{\rho=0} = 0~, \quad (\sqrt{\gamma})^\prime
     \big|_{\rho=0} = 1  \quad {\rm and} 
     \quad \gamma\big|_{\rho=0} = 0~,
\end{equation}
which is consistent with the usual regular solution in flat space.
We also set $\sigma(0) =A $, where $A$ is constant.
We can now define various components of 
the brane tension per unit length as
\begin{equation}
     \mu_i = -\int\limits_0^\infty d\rho\, \sigma^2 \gamma^{(n-1)/2}\, 
     f_i(\rho)~.
\end{equation}
where $i = 0,\rho,\theta$. Using the system of equations
(\ref{3solnset1})--(\ref{3solnset3}) we obtain the following relations 
\begin{align} 
  \label{3junc1}
  (n-2) \mu_0-\mu_\rho-(n-1) \mu_\theta &=
  \left[2 \Lambda_D- \frac{(n+2)\Lambda_{phys} M_D^{n+2}}{M_P^2} \right] \, 
   \int\limits_0^\infty \sigma^2 \sqrt{\gamma^{n-1}} d\rho 
\end{align}
and
\begin{align}
  \label{3junc2}
  4 \mu_0 +\mu_\rho-3 \mu_\theta =&
  (n+2) M_D^{n+2} A^2 \delta_{n 2} + (n+2)(n-2) M_D^{n+2} \int\limits_0^\infty \sigma^2 \sqrt{\gamma^{n-3}} d\rho 
  \nonumber \\
  &-2 \Lambda_D \int\limits_0^\infty \sigma^2 \sqrt{\gamma^{n-1}} d\rho \, .
\end{align}
In the derivation of relations (\ref{3junc1}) and (\ref{3junc2}) we used the boundary conditions 
(\ref{3boundsigmagamma}) together with the assumptions that all integrals converge and that furthermore 
\begin{equation}
  \lim_{\rho \to \infty} \sigma \sigma' \sqrt{\gamma^{n-1}}=
  \lim_{\rho \to \infty} \sigma^2 \sqrt{\gamma^{n-2}} \left(\sqrt\gamma\right)'=0 \, .
\end{equation} 
Since we are mainly interested in solutions that localize gravity, these assumptions are 
physically reasonable. As we will see below, in most of the cases where the condition of localization 
of gravity is satisfied also the above assumptions are valid.
The equations (\ref{3junc1}) and (\ref{3junc2}) are the general
conditions relating the brane tension components to the metric
solution of the Einstein equations
(\ref{3solnset1})--(\ref{3solnset3}) and lead to nontrivial
relationships between the components of the brane tension per unit
length. 
Furthermore, by analogy with the solution for local strings we can
identify (\ref{3junc1}) as the gravitational mass  per unit length and
(\ref{3junc2}) as the angular deficit per unit length. Thus the source
for the 3-brane, in general curves the transverse space.

From the Einstein term in the D-dimensional Lagrangian we can obtain
the effective 4-dimensional reduced Planck mass. Using the spherically
symmetric  metric ansatz (\ref{3metric}), the 4-dimensional reduced
Planck mass is given by
\begin{equation}
        M_P^2 = {\cal A}_{n-1} M_D^{n+2} \int\limits_0^\infty d\rho\, \sigma\, 
        \gamma^{(n-1)/2}~.
\end{equation}
where ${\cal A}_{n-1}=2\pi^{\frac{n}{2}}/\Gamma[\frac{n}{2}]$ is the surface area of 
an $n-1$ dimensional unit sphere. We are interested in obtaining solutions to the Einstein
equations  (\ref{3solnset1})--(\ref{3solnset3}) such that a finite
four-dimensional Planck mass is obtained. This leads to various
possible asymptotic behavior for the metric factors $\sigma$
and $\gamma$ in the  limit $\rho\rightarrow\infty$. Below, we will
concentrate only on the case of vanishing  $4$-dimensional cosmological 
constant, $\Lambda_{phys}=0$.

\section{Strictly local defect solutions}

First, let us assume that the functions $f_i(\rho)$ vanish sufficiently rapid outside
the core of the topological defect. In order to obtain a finite
4-dimensional Planck scale, we require a solution of the system of
equations (\ref{3solnset1})--(\ref{3solnset3}) for which the function
$\sigma \gamma^{(n-1)/2}$ goes to zero when  $\rho\rightarrow\infty$.
For $n=1$ and $n=2$ the solutions are known to exist, see \cite{Randall:1999vf}
and \cite{Gherghetta:2000qi} correspondingly. However, when $n \geq 3$ the structure
of the equations is qualitatively different because of the presence of the 
$1/\gamma$ terms in (\ref{3solnset1})--(\ref{3solnset3}). Thus, there is no simple generalization of the
solutions found for $n=1$ and $2$.

To neutralize the effect of the $1/\gamma$ term, one can look for
asymptotic solutions for which $\gamma$ is a positive constant.  However, one
can easily check that the system of equations 
(\ref{3solnset1})--(\ref{3solnset3}) does not allow a solution for
which $\gamma$ tends to a constant when $\rho \rightarrow \infty$,
and $\sigma$ is a negative exponential.  

Alternatively, we can assume that there is an asymptotic solution for
which $\gamma \rightarrow \infty$ but $\sigma$ tends to zero faster
than $\gamma^{(n-1)/2}$ and omit the troublesome $1/\gamma$ term from
the equations of motion. In this case  the set of
equations~(\ref{3solnset1})--(\ref{3solnset3}) can  be simply reduced
to a single equation, as in $6$-dimensional case \cite{Rubakov:1983bz,Gherghetta:2000qi}:
\begin{equation}
\label{3zeqn}
     z'' = -\frac{d U(z)}{dz}~,
\end{equation}
where the potential $U(z)$ is given by
\begin{equation}
\label{3upot}
     U(z) = \frac{(n+3)}{4(n+2)}\,\frac{\Lambda_D}{M_D^{n+2}}\, z^2~.
\end{equation}
With this parametrization the metric functions $\sigma(\rho)$ and
$\gamma(\rho)$  can be written in terms of $z(\rho)$ as
\begin{eqnarray}
     \sigma &=& |z'|^{(2-\sqrt{(n+2)(n-1)})/(n+3)}\,
        |z|^{(2+\sqrt{(n+2)(n-1)})/(n+3)} \\
     \gamma &=& |z'|^{6/(1-n+2\sqrt{(n+2)(n-1)})}\,
        |z|^{6/(1-n-2\sqrt{(n+2)(n-1)})}~.
\end{eqnarray}
Solving equation (\ref{3zeqn}) with the potential (\ref{3upot}) gives 
the general solution
\begin{equation}
\label{3zsoln}
           z(\rho) = d_1\, e^{-\frac{1}{4}(n+3) c \rho} + d_2\, 
           e^{\frac{1}{4}(n+3) c \rho}~,
\end{equation}
where $c^2=-8\Lambda_D/((n+2)(n+3) M_D^{n+2})$, 
$d_1, d_2$ are constants and we take $\Lambda_D<0$. This
solution is a generalization of the pure exponential solution
considered earlier and in ref.~\cite{Olasagasti:2000gx}, see below.  In this picture
we can think of particle motion under the influence of the potential
(\ref{3upot}) with position $z(\rho)$ and ``time'' $\rho$. 

Since $1-n+2\sqrt{(n+2)(n-1)}>0$ and $1-n-2\sqrt{(n+2)(n-1)} < 0$,
the metric factor $\gamma$ can be large in two cases. In the  first
case $z(\rho)$ is zero for some $\rho_0$. However, this point is only
a coordinate singularity (the Ricci scalar is regular  at this point)
and the metric can be extended beyond $\rho_0$, leading  then to an
exponentially rising solution for both $\sigma$ and $\gamma$. This,
unfortunately, is not interesting for compactification.  In the
second case both $z'$ and $z$ are non-zero and increase exponentially
for large $\rho$. Thus, there is no possibility of a finite Planck
scale in this case either.

Similarly, we were unable to find solutions in the  reverse case when
$\gamma$ vanishes at infinity. Moreover, even if such solutions were
to exist, they would likely lead to  a singular geometry (naked
singularity), because the Ricci scalar contains a $1/\gamma$ term,
see eq. (\ref{3ricci}). This was indeed shown to be the case for solutions
with regular geometries at $\rho=0$ in \cite{Randjbar-Daemi:1985wg}.

\section{Bulk scalar field}
\subsection{Global topological defects}

The other possibility is to consider defects with different types of
``hair'' i.e. with non-zero stress-energy tensor outside the core of
the defect.  We start with global topological defects. This case 
has been extensively studied in \cite{Gregory:1999gv,Olasagasti:2000gx} and we
list here a number of explicit
solutions (see also~\cite{Oda:2000zc,Dvali:2000ty,Chen:2000at}).  
For simplicity we will restrict ourselves to the case of vanishing 
$4$-dimensional cosmological constant, $\Lambda_{phys} =0$, and
assume that outside the core of the defect we have the following 
asymptotic behavior for $\sigma$:
\begin{equation}
\label{3solnform}
     \sigma(\rho) = e^{-c \rho}~.
\end{equation}

Consider $n$ scalar fields $\phi^a$ with a potential
\begin{equation}
        V(\phi) = \lambda (\phi^a\phi^a - v^2)^2~,
\end{equation}
where $v$ has mass dimension $(n+2)/2$.
Then the potential minimum is at $\phi^a\phi^a = v^2$. 
The defect solution has a 
``hedgehog'' configuration outside the core
\begin{equation}
        \phi^a(\rho) = v \, d^a~,
\end{equation}
$d^a$ being a unit vector in the extra dimensions, 
$d^n=\cos\theta_{n-1},~d^{n-1} = \sin\theta_{n-1} \cos\theta_{n-2},\dots$.

The scalar field gives an additional contribution to the stress-energy 
tensor in the bulk with components
\begin{eqnarray}
         T_\mu^\nu &=& -(n-1) \frac{v^2}{2\gamma} \delta_\mu^\nu~, \\
         T_\rho^\rho &=& -(n-1) \frac{v^2}{2\gamma}~, \\
         T_\theta^\theta &=& -(n-3) \frac{v^2}{2\gamma}~.
\end{eqnarray}
Now, for $\Lambda_D <0$ and $v^2 > (n-2) M_D^{n+2}$ the following
solution leads to the localization of gravity and a regular geometry in
the bulk~\cite{Gregory:1999gv,Olasagasti:2000gx}
\begin{eqnarray}
  \label{3s1} c&=&\sqrt{\frac{2(-\Lambda_D)}{(n+2) M_D^{n+2}}}~,\\
\label{3s2}\gamma &=&\frac{1}{c^2}\left(\frac{v^2}{M_D^{n+2}}-n+2 \right)~.
\end{eqnarray}
The transverse geometry of this solution is that of a cylinder, 
$R_+ \times S_{n-1}$ with $R_+$ being the half-line and $S_{n-1}$ 
being an $n-1$ sphere. 

When $v^2 = (n-2)M_D^{n+2}$ the $1/\gamma$ terms are eliminated
from the system of equations (\ref{3solnset1})--(\ref{3solnset3}) and
the exponential solution to the coupled set of equations 
(\ref{3solnset1})--(\ref{3solnset3}) can 
then be found with 
\begin{equation}
  \gamma(\rho)=R_0^2\sigma(\rho),~~
   c=\sqrt{\frac{8(-\Lambda_D)}{(n+2)(n+3) M_D^{n+2}}}~,
     \label{3sgsol}
\end{equation}
where $R_0$ is an arbitrary length scale. As expected, the negative
exponential solution (\ref{3solnform}) requires that $\Lambda_D < 0$.
Notice that the exponential solution (\ref{3sgsol}) only requires the
``hedgehog'' scalar field configuration in the bulk for transverse
spaces with dimension $n\geq 3$.  No such configuration is needed for
the $5$-dimensional \cite{Randall:1999vf} and $6$-dimensional 
cases~\cite{Gherghetta:2000qi}, which only require
gravity in the bulk.

The Ricci scalar corresponding to the negative exponential solution
with vanishing four-dimensional cosmological constant is 
\begin{equation}
  R=-(n+3)(n+4)\frac{c^2}{4} +(n-1)(n-2)\frac{e^{c \rho}}{R_0^2}~,
\end{equation}
which diverges when $\rho=\infty$.  Thus we see that for $n\geq 3$
the space is no longer  a constant curvature space and in fact has a
singularity at $\rho=\infty$.  This is also confirmed by looking at
the other curvature invariants, $R_{AB}R^{AB}$ and
$R_{ABCD}R^{ABCD}$. Only for the 5d and 6d cases do we obtain a
constant  curvature anti-de-Sitter space. The appearance of a
singularity is similar to the case of the global-string
defect~\cite{Cohen:1999ia}. 



\subsection{Corrections to Newton's Law}

For the solution (\ref{3s1}),(\ref{3s2}) the corrections to Newton's law
are parametrically the same as for 5d case, since $\gamma$ is a constant.
On the other hand, the singular solution (\ref{3sgsol})
will ultimately require that the singularity is smoothed by string 
theory corrections (perhaps similar to the nonsingular deformations 
considered in~\cite{Chodos:2000tf}). Assuming that this is the case, then
the corrections to Newton's law on the 3-brane can be calculated by
generalizing the calculation presented in~\cite{Randall:1999vf,Gherghetta:2000qi} 
(see also~\cite{Csaki:2000fc}).

In order to see that gravity is localized on the 3-brane, let us
now consider the equations of motion for the linearized metric
fluctuations. We will only concentrate on the spin-2 modes and
neglect the scalar modes, which need to be taken into account
together with the bending of the brane~\cite{Garriga:1999yh}. The vector modes
are massive as follows from a simple modification of the results in 
ref.~\cite{Lavrelashvili:1985aa}. For a fluctuation of the form $h_{\mu\nu}(x,z) =
\Phi(z) h_{\mu\nu}(x)$ where $z=(\rho,\theta)$ and $\partial^2
h_{\mu\nu}(x)  = m_0^2 h_{\mu\nu}(x)$, we can separate the variables
by defining $\Phi(z)= \sum_{l_i m} \phi_m(\rho)
Y_{l_i}(\phi,\theta_i)$.  The radial modes satisfy the
equation~\cite{Lavrelashvili:1985aa}
\begin{equation}
\label{3diffop}
     -\frac{1}{\sigma \gamma^{(n-1)/2}}\,\partial_\rho\left[
      \sigma^2\gamma^{(n-1)/2} \,\partial_\rho \phi_m  \right] = 
      m^2 \phi_m~,
\end{equation}
where $m^2=m_0^2 + \tilde\Delta/R_0^2$ contains the contributions from the 
angular momentum modes $l_i$ and $\tilde\Delta$ denotes the Laplacian on $S^{n-1}$. The differential operator (\ref{3diffop}) 
is self-adjoint provided that we impose the boundary conditions
\begin{equation}
\label{3bc}
     \phi_m^\prime(0) = \phi_m^\prime(\infty) = 0~,
\end{equation}
where the modes $\phi_m$ satisfy the orthonormality condition
\begin{equation}
       {\cal A}_{n-1} \int\limits_0^\infty d\rho \, \sigma \,\gamma^{(n-1)/2}\,
       \phi_m^\ast \phi_n = \delta_{mn}~.
\end{equation}
Using the specific solution
(\ref{3solnform}),(\ref{3sgsol}), the differential operator
({\ref{3diffop}) becomes
\begin{equation}
\label{3diffeqn}
      \phi_m^{\prime\prime} -\frac{(n+3)}{2}\, c\, \phi_m^\prime +
     m^2 e^{c\rho}  \phi_m = 0~.
\end{equation}
This equation is the same as that obtained for the 5d domain wall
solution~\cite{Randall:1999vf}, when $n=1$ and the local string-like
solution~\cite{Gherghetta:2000qi} when $n=2$. We see that each extra transverse
coordinate augments this coefficient by $1/2$. When $m=0$ we clearly
see that $\phi_0(\rho)$ = constant is a solution. Thus we have a
zero-mode tensor fluctuation which is localized near  the origin
$\rho =0$ and is normalizable.

The contribution from the nonzero modes will modify Newton's law on the
3-brane. In order to calculate this contribution we need
to obtain the wave-function for the nonzero modes at the origin. The 
nonzero mass eigenvalues can be obtained by imposing the boundary conditions
(\ref{3bc}) on the solutions of the differential equation (\ref{3diffeqn}). 
The  solutions of (\ref{3diffeqn}) are
\begin{equation}
    \phi_m(\rho) = e^{\frac{c}{4} (n+3)\rho}
    \left[ C_1 J_{\frac{1}{2}(n+3)}(\frac{2m}{c} e^{\frac{c}{2}\rho}) 
        + C_2 Y_{\frac{1}{2}(n+3)}(\frac{2m}{c} e^{\frac{c}{2}\rho}) \right]~,
\end{equation}
where $C_1,C_2$ are constants and $J_{\frac{1}{2}(n+3)},
Y_{\frac{1}{2}(n+3)}$ are  the usual Bessel functions. Imposing the
boundary conditions (\ref{3bc}) at a finite radial distance cutoff
$\rho=\rho_{\rm max}$ (instead of $\rho=\infty$) will lead to a
discrete mass spectrum, where for $k=1,2,3,\dots$ we obtain 
\begin{equation}
      m_k \simeq c(k+\frac{n}{4}) \frac{\pi}{2} e^{-\frac{c}{2} \rho_{\rm max}}~.
\end{equation}
With this discrete mass spectrum we find that
in the limit of vanishing mass $m_k$,
\begin{equation}
      \phi_{m_k}^2(0) = \frac{1}{{\cal A}_{n-1} R_0^{n-1}} \frac{\pi c}{8}
     \frac{(n+1)^2}{\Gamma^2[(n+3)/2]} \left(\frac{m_k}{c}\right)^n 
     e^{-\frac{c}{2} \rho_{\rm max}}~,
\end{equation}
where $\Gamma[x]$ is the gamma-function.
On the 3-brane the gravitational potential between two point masses
$m_1$ and $m_2$, will receive a contribution from the discrete, nonzero modes 
given by
\begin{equation}
      \Delta V(r) \simeq G_N  \frac{m_1 m_2}{r}  \frac{\pi
      (n+1)}{4\Gamma^2[(n+3)/2]} \sum_k e^{-m_k r}
      \left(\frac{m_k}{c}\right)^n  e^{-\frac{c}{2} \rho_{\rm max}}~,
\end{equation}
where $G_N$ is Newton's constant.
In the limit that $\rho_{\rm max}\rightarrow \infty$, the spectrum
becomes continuous and the discrete sum is converted into an
integral. The contribution to the gravitational potential then
becomes
\begin{eqnarray}
       V(r)
       &\simeq& G_N \frac{m_1 m_2}{r} \left[1+\frac{1}{2 c^{n+1}}
       \frac{n+1}{\Gamma^2[(n+3)/2]} \int\limits_0^\infty dm\,m^n e^{-m
       r}\right] \\
       &=& G_N \frac{m_1 m_2}{r} \left[1+ \frac{\Gamma[n+2]}{2
       \Gamma^2[(n+3)/2]} \frac{1}{(c r)^{n+1}}\right]
\end{eqnarray}
Thus we see that for $n$ transverse dimensions
the correction to Newton's law from the bulk continuum states grows 
like $1/(c r)^{n+1}$. This correction becomes more suppressed as the number
of transverse dimensions grows, because now the  gravitational field of
the bulk continuum modes spreads out in more dimensions and so
their effect on the 3-brane is weaker.

\section{Bulk $p$-form field} \label{SectionBulkpform}

The global topological defects considered in the previous section
inevitably contain massless scalar fields -- Nambu-Goldstone bosons 
associated with the spontaneous breakdown of the global symmetry. 
Thus, the stability of these configurations is far from being obvious.

We will now consider the possibility of introducing other types of
bulk fields ($p$-form fields),
which directly
lead to a regular geometry. The stability of the corresponding
configurations  may be insured simply by the magnetic flux
conservation. The D-dimensional action is
\begin{equation}
     S= \int d^Dx \sqrt{-g} \left( \frac{1}{2} M_D^{n+2} R - 
     \Lambda_D -\frac{1}{4} 
      F_{A_1\dots A_{p+1}}F^{A_1\dots A_{p+1}} \right) \, .
\end{equation}
The energy-momentum tensor associated with the $p$-form field
configuration is given by
\begin{equation}
     T^A_{\;\;B} = -\left( \frac{1}{4}\delta^A_{\;\;B}
      F_{A_1\dots A_{p+1}}F^{ A_1\dots A_{p+1}} -
       \frac{p+1}{2} F^A_{\quad A_1\dots A_p} F_B^{\;\; A_1\dots A_p}\right) \, .
\end{equation}
A solution to the equation of motion for the $p$-form field when
$p=n-2$ is
\begin{equation} \label{3pFormCharge}
      F_{\theta_1\dots\theta_{n-1}} = Q
(\sin\theta_{n-1})^{(n-2)}\dots \sin\theta_2~,
\end{equation}
where $Q$ is the charge of the field configuration and all other
components of $F$ are equal to zero. In fact, this ``hedgehog''
field configuration is the generalization of the magnetic field of a monopole.
The stress-energy tensor associated with this $p$-form field in the bulk is
\begin{eqnarray}
      T^\mu_\nu &=& -\frac{(n-1)!}{4}\frac{Q^2}{\gamma^{n-1}}
      \delta^\mu_\nu~, \\
      T^\rho_\rho &=& -T^\theta_\theta =
        -\frac{(n-1)!}{4}\frac{Q^2}{\gamma^{n-1}}~.
\end{eqnarray}
Let us again assume a solution with the asymptotic form 
\begin{equation}
\label{3pfansatz}
    \sigma(\rho)= e^{-c\rho} \quad {\rm and} \quad \gamma= {\rm const} \, .
\end{equation}
With this ansatz we see from the Ricci scalar that the transverse
space  will have a constant curvature and the effective
four-dimensional Planck constant will be finite.  If we substitute
this ansatz and also include the contribution of the $p$-form bulk
field to the stress-energy tensor, the Einstein equations 
(\ref{3solnset1})--(\ref{3solnset3}), with $\Lambda_{phys}=0$ are
reduced to the following two equations for the metric factors outside
the 3-brane source
\begin{eqnarray}
\label{3pfsoln1}
    (n-1)!\frac{Q^2}{\gamma^{n-1} M_D^{n+2}}  &-& \frac{1}{2\gamma}(n-2)(n+2)
    +\frac{\Lambda_D}{M_D^{n+2}} =0 \, ,  \\
\label{3pfsoln2}
   c^2 &=& 
   -\frac{1}{2} \frac{\Lambda_D}{M_D^{n+2}}+ \frac{1}{4\gamma}(n-2)^2~.
\end{eqnarray}
We are interested in the solutions of these two equations which 
are exponential, $c^2>0$ and do not change the metric signature, 
$\gamma>0$. Remarkably, solutions to these equations 
exist for which these conditions can be simultaneously satisfied. 
Let us first consider the case $n=2$, where the solutions simply reduce to
\begin{eqnarray}
    \frac{Q^2}{\gamma}  &=& -\Lambda_6 \, , \\
     c^2 &=& -\frac{1}{2} \frac{\Lambda_6}{M_6^4}~.
\end{eqnarray}
Thus, for $\Lambda_6 < 0$ we see that there is a solution that satisfies 
(\ref{3pfansatz}). 

Next we consider the case $n=3$. Only the ``+'' solution to the quadratic 
equation (\ref{3pfsoln1}) gives rise to 
a solution with both $c^2>0$ and $\gamma > 0$. This solution can be written 
in the form
\begin{eqnarray} \label{3pFormAsymptotic1}
    \frac{Q^2}{\gamma M_7^5}  &=& \frac{1}{4}\left[ \frac{5}{2} + 
     \sqrt{\left(\frac{5}{2}\right)^2 - 8 Q^2\frac{\Lambda_7}{M_7^{10}}}\right]\, , \\
   \frac{Q^2 c^2}{M_7^5} &=& \frac{1}{16} \left[ \frac{5}{2} - 8 Q^2
      \frac{\Lambda_7}{M_7^{10}}
     +\sqrt{\left(\frac{5}{2}\right)^2 - 8 Q^2\frac{\Lambda_7}{M_7^{10}}}\right] \, . 
    \label{3pFormAsymptotic2}
\end{eqnarray}
When $Q^2\Lambda_7/M_7^{10} < 25/32 $ we obtain real solutions
which are plotted in Figure~\ref{fig:soln3}.
\begin{figure}[tbp]
\centerline{ \epsfxsize 4.0 truein \epsfbox {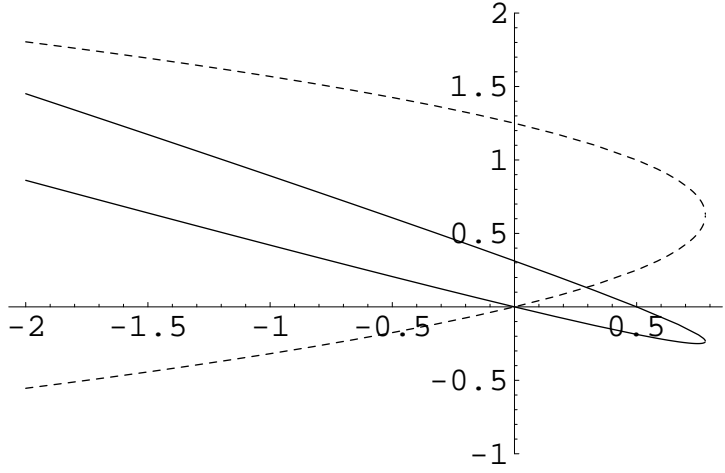}}
\caption{\it The $n=3$ solution for $Q^2/(\gamma M_7^5)$ (dashed line)
and $(Q^2 c^2)/M_7^5$ (solid line), as a function of $Q^2 \Lambda_7/M_7^{10}$.
Only the branches with $c^2 >0$ and $\gamma>0$ lead to solutions
that localize gravity.}
\label{fig:soln3}
\end{figure}
In fact requiring $c^2>0$ we
see that there are solutions not only for $\Lambda_7< 0$, but also 
for $\Lambda_7 \geq0$, provided that $Q^2\Lambda_7/M_7^{10} < 1/2$. 
Thus, the bulk cosmological constant does not need to
be negative in order to localize gravity.

Similarly, solutions for which $c^2>0$ and $\gamma>0$ exist for values
of $n>3$. Again, we find that solutions with
both positive, zero and negative bulk cosmological constants exist.

The nice property of these solutions (\ref{3pfansatz})
is that since $\gamma$ is a constant
the Ricci scalar does not diverge at any point in the transverse space. 
In particular for the $n=2$ solution the Ricci scalar is
$R=5 \Lambda_6/(2 M_6^4)$, while for $n=3$ it is
\begin{equation}
     R= +\frac{M_7^5}{2 Q^2}\left[ \frac{15}{16} + 5 Q^2\frac{\Lambda_7}{M_7^{10}}
     - \frac{3}{8} \sqrt{\left(\frac{5}{2}\right)^2 - 8 Q^2\frac{\Lambda_7}{M_7^{10}}}\right] \, .
\end{equation}
The space is again a constant curvature space, but it is not
necessarily anti-de-Sitter. This is also confirmed by checking
the higher-order curvature invariants $R_{AB}R^{AB}$ and $R_{ABCD}R^{ABCD}$.

Finally, we see from (\ref{3diffop}) that
the equation of motion for the spin-2 radial modes using the solution
(\ref{3pfansatz}) is qualitatively similar to the 5d case. The constant
$\gamma$ factor effectively plays no role in the localization of gravity.
Thus, the corrections to Newton's law will be suppressed by $1/r^2$
for all solutions $n\geq 3$. This is easy to understand since the
geometry of the extra dimensions is simply $R_+\times S_{n-1}$, as in the
global defect case, and there is just one non-compact
dimension for all $n\geq 3$. In the special case when 
$\Lambda_D=0$, the metric solution we have found corresponds to the 
near-horizon metric of a class of extremal non-dilatonic black 
branes~\cite{Gibbons:1994vm}.

Another possible solution for the $p$-form in the bulk includes
\begin{equation}
\label{3osoln}
      F_{\mu_1\dots\mu_n} = \epsilon_{\mu_1\dots\mu_n} \kappa(\rho)~,
\end{equation}
where 
\begin{equation}
        \kappa(\rho)= Q \frac{\gamma^{(n-1)/2}}{\sigma^2}
(\sin\theta_{n-1})^{(n-2)}\dots \sin\theta_2~.
\end{equation}
In this case the contribution to the stress-energy tensor is
\begin{eqnarray}
      T^\mu_\nu &=& -\frac{n!}{4}\frac{Q^2}{\sigma^4} \delta^\mu_\nu~, \\
      T^\rho_\rho &=& T^\theta_\theta = +\frac{n!}{4}\frac{Q^2}{\sigma^4}~.
\end{eqnarray}
This contribution does not appear to make the solution of the 
Einstein equations any easier.

The above two $p$-form solutions have only included components 
in the transverse space. If we also require 
the $p$-form field to transform non-trivially under the 3-brane 
coordinates, then we can have 
\begin{equation}
     F_{0123 \theta_1\dots \theta_{n-1}} = 
     Q(\sin\theta_{n-1})^{(n-2)}\dots \sin\theta_2~,
\end{equation}
where all other components of $F$ are zero. The components
of the stress-energy tensor for this field configuration are
\begin{eqnarray}
      T^\mu_\nu &=&-
      \frac{(n+3)!}{4}\frac{Q^2}{\sigma^4\gamma^{n-1}}
      \delta^\mu_\nu \\ T^\rho_\rho &=& -T^\theta_\theta =
      +\frac{(n+3)!}{4}\frac{Q^2} {\sigma^4\gamma^{n-1}} ~.
\end{eqnarray}
Again, there is no simple solution of the Einstein equations
with the inclusion of this contribution.

Finally, one can also generalize the solution (\ref{3osoln})
\begin{equation}
      F_{\mu\nu\alpha\beta a_1\dots a_n} =
        \epsilon_{\mu\nu\alpha\beta a_1\dots a_n}  \kappa(\rho)~,
\end{equation}
where 
\begin{equation}
     \kappa(\rho)= Q \sigma^2 \gamma^{(n-1)/2}
(\sin\theta_{n-1})^{(n-2)}\dots \sin\theta_2~,
\end{equation}
and the stress-energy tensor
\begin{equation}
      T^M_N = -\frac{(n+4)!}{4} Q^2 \delta^M_N~,
\end{equation}
is a constant. Thus, we see that the addition of a
$n+4$ form field is equivalent to adding a bulk cosmological
constant.

\section{Conclusions}

In summary we can say that the generalization of the
exponential solutions found in \cite{Randall:1999vf} and 
\cite{Gherghetta:2000qi} in the case of one or two extra dimensions which lead 
to the localization of gravity on a 3-brane is nontrivial. In the case of a strictly 
local defect, where the stress-energy tensor vanishes exponentially outside the 
core of the defect, we were unable to find any solution to the Einstein equations.
However, when a scalar field with ``hedgehog'' configuration is added to the bulk, there exist 
solutions which localize gravity. In the case (\ref{3sgsol}) of exponentially vanishing $\gamma$, 
the transverse space no longer has constant curvature, and
furthermore develops a singularity at $\rho=\infty$. The
corrections to Newton's law on the 3-brane are suppressed by 
$1/(c r)^{n+1}$, where we see that each additional extra dimension adds one more power of $1/r$ to 
the correction factor. If however $\gamma$ is a constant (\ref{3s1})--(\ref{3s2}) then regular 
gravity localizing solutions exist.

By adding a $n-1$ form field in the bulk, which can be considered
as the generalization of the magnetic field of a monopole,
gravity localizing solutions can be found for positive, zero
and negative bulk cosmological constant. In this case, the
transverse space has constant curvature but is not an anti-de-Sitter
space. Since $\gamma$ approaches a constant at infinity in transverse space,
the corrections to Newton's law have the same form as in the
original Randall-Sundrum II model~\cite{Randall:1999vf}. Eventually we saw that 
the addition of an $n+4$-form field in the bulk is equivalent to adding a bulk cosmological constant.


In the next chapter we will consider a particular
field theoretical model in $7$ dimensions in which the 
asymptotic solution (\ref{3pfansatz}) found in this section is realized.
More concretely, we will discuss numerical solutions to the coupled 
Einstein and matter-field equations for a 't~Hooft-Polyakov monopole residing in
a $3$-dimensional spherically symmetric transverse space.   
\chapter{A numerical example - the 't~Hooft-Polyakov monopole} \label{cha4}

\section{Introduction}
This chapter is dedicated to a detailed numerical analysis of the coupled Einstein and 
matter field equations for a particular topological defect, a 't~Hooft-Polyakov monopole, residing in 
a $3$-dimensional transverse space of a $7$-dimensional space-time. The motivations for this 
are twofold: first, as we outlined briefly in the last 
chapter, a more involved topological background in the extra dimensional transverse 
space provides richer structure to generate 
chiral fermions on the brane as zero modes of higher-dimensional non-chiral spinor-fields. From this 
point of view, such considerations may help to envisage the construction of realistic models. Second,
a common feature of the solutions to Einstein equations presented in the last chapter was their 
asymptotic nature, valid only outside of the core of the defect. It is not a priori 
clear, whether the fully coupled system of Einstein plus matter field equations allows for 
singularity free solutions that interpolate smoothly between the topological defect at the 
origin of transverse space 
and the asymptotic at infinity in transverse space.
It was shown recently \cite{Giovannini:2001hh} (see also 
\cite{Giovannini:2001uf,Giovannini:2002sb,Giovannini:2002mk})
that gravity can be localized on an abelian Higgs string in 
6 dimensions in the absence 
of any singularity under the assumption of a certain amount of fine tuning between brane tension components. 
The aim of the numerical work presented here is to apply the same idea to the case of a 
monopole in 7~dimensional space-time.
This chapter is organized as follows: in the following section \ref{4TPmonopole4D} we review 
the physics of 't~Hooft-Polyakov monopoles in flat $4$-dimensional Minkowski space-time.
In section \ref{4EQFQ} we present the $SU(2)$ invariant Georgi-Glashow toy model 
(having 't~Hooft-Polyakov monopoles as flat space-time solutions 
\cite{'tHooft:1974qc},\cite{Polyakov:1974ek}) coupled to gravity
in 7 dimensions. The Einstein 
equations and the field equations are obtained in the case of a 
generalized 't~Hooft-Polyakov ansatz for the fields.
While boundary conditions and the desired asymptotic of the solutions are discussed in sections \ref{4BC} 
and \ref{4AS},
we give the relations between the brane tension components necessary for gravity localization in section 
\ref{4FTR}. A 
proof of these relations is given in appendix \ref{appFTR}.
Section \ref{4N} is dedicated to numerics: after presenting the numerical methods used for solving the 
differential equations in section \ref{4NGD}, we discuss the fine-tuning surface (the 
relation between the independent parameters of the model) in section \ref{4NFTS}. 
We give explicit sample solutions in 
section \ref{4samplesol} and discuss their general dependence on the parameters of the model.
Sections \ref{4PSL} and \ref{4FML} finally treat the Prasad-Sommerfield limit and the fine-monopole limit, 
respectively. While in the former case no gravity localizing solutions exist, in the latter case we 
demonstrate the possibility of physical choices of the model parameters which are in agreement 
 with observations. 

\section{The 't~Hooft-Polyakov monopole in flat 4d space-time} \label{4TPmonopole4D}
As we pointed out in the last chapter, from a field theoretical point of view 
monopoles can be characterized as 
non-singular, point-like ($0$-dimensional), finite-energy solutions to classical field 
equations. Their interest for particle physics stems mainly from their 
stability which again can be understood on mere
topological grounds. Excellent reviews and textbooks can be found on this subject, see for 
example \cite{Raj,Vilenkin,Vachaspati:2000cq,Goddard:1977da,Preskill:1986kp,Coleman:1982cx}. 
In the following we summarize the basic properties of 't~Hooft-Polyakov monopoles in 
$4$-dimensional Minkowski space-time.
Monopoles are produced during the process of spontaneous symmetry breaking whenever the vacuum manifold 
$\mathcal{M}=G/H$ of the Higgs field after the symmetry breaking contains non-contractible two surfaces
$\pi_2(G/H) \neq I $. Making use of the second fundamental theorem (\ref{2ndHomTheorem}) of section  
\ref{Homotopy} we see that this can be rephrased as $\pi_1\left(H_0\right) \neq I$ with 
$H_0$ being the component of $H$ connected to the identity. This is the reason why  
monopoles are a general prediction of grand unified theories whenever a single group is to 
encompass all interactions. At some stage of the symmetry breaking $U(1)_{em}$, the gauge group
of electromagnetism has to appear and with it the associated monopoles 
since $\pi_1\left(U(1)\right) \cong Z$.
The existence of monopole solutions was established by the pioneering works of 't~Hooft \cite{'tHooft:1974qc}
 and Polyakov \cite{Polyakov:1974ek} taking the simple example of a model proposed by Georgi and 
Glashow \cite{Georgi:1972cj} in the 
context of electroweak unification. The underlying field theory is a gauge theory based on the 
group $SU(2)$. It contains a scalar triplet $ \Phi^{\tilde{a}}$ transforming in the 
adjoint representation of $SU(2)$ and the corresponding gauge field $W_{\mu}^{\tilde{a}}$. 
Group indices are denoted by 
a tilde : $\tilde{a}, \tilde{b}, \tilde{c} =1 \ldots 3$. The action for this theory is given by
\begin{align} \label{4TPaction}
  S=\int d^4x \left[ -\frac{1}{4} G_{\mu \nu}^{\tilde{a}} G^{\tilde{a} \mu \nu}-\frac{1}{2} \mathcal{D}_\mu 
\Phi^{\tilde{a}} \mathcal{D}^\mu \Phi^{\tilde{a}}-\frac{\lambda}{4} \left( \Phi^{\tilde{a}} 
\Phi^{\tilde{a}} -\eta^2 \right)^2 \right] \, ,
\end{align} 
where
\begin{equation} \label{4cov_der4D}
  \mathcal{D}_\mu \Phi^{\tilde{a}}=\partial_\mu \Phi^{\tilde{a}} + e \, 
  \epsilon^{\tilde{a}\tilde{b}\tilde{c}}\, W_\mu^{\tilde{b}} \Phi^{\tilde{c}} \, 
\end{equation}
and
\begin{equation} \label{4GaugeField4D} 
  G_{\mu \nu }^{\tilde{a}}=\partial_\mu W_\nu^{\tilde{a}}-\partial_\nu W_\mu^{\tilde{a}}+e \, 
  \epsilon^{\tilde{a}\tilde{b}\tilde{c}} 
  \, W_\mu^{\tilde{b}} W_\nu^{\tilde{c}} \, .
\end{equation}
The variation of the action (\ref{4TPaction}) gives the following equations of motion 
for the gauge and scalar fields:
\begin{eqnarray}  
 \mathcal{D}_\mu \mathcal{D}^\mu \Phi^{\tilde{a}}  &=& \lambda \, \Phi^{\tilde{a}} 
	\left( \Phi^{\tilde{b}} \Phi^{\tilde{b}}-\eta^2\right) \, , \label{4EoMscalar4D} \\
  \mathcal{D}_\mu G^{\tilde{a} \mu \nu} &=& -e \, \epsilon^{\tilde{a}\tilde{b}\tilde{c}} 
  \left( \mathcal{D}^\nu \Phi^{\tilde{b}} \right) \Phi^{\tilde{c}} \, . \label{4EoMgauge4D} 
\end{eqnarray}

In this model the $SU(2)$ symmetry is spontaneously broken down to $U(1)$ which can be identified with the 
gauge group of electromagnetism \cite{'tHooft:1974qc}. The monopole corresponds to the simplest, 
topologically non-trivial field configuration (with topological charge 1). 
In terms of the parameters $\eta$ and $\lambda$, the Higgs mass is given by 
$m_H=\eta \, \sqrt{2 \lambda}$. Two of the three vector fields acquire a mass given by $m_W=e \, \eta$.

We are interested in static, spherically symmetric solutions to the set of coupled equations 
(\ref{4EoMscalar4D})--(\ref{4EoMgauge4D}) with finite energy. The energy functional for static configurations
can be written as:
\begin{equation} \label{4EnFunctional}
  E=\int d^3x \left[ 
    \frac{1}{4} G^{\tilde{a}}_{i j} G^{\tilde{a} i j}+
    \frac{1}{2} \mathcal{D}_i \Phi^{\tilde{a}} \mathcal{D}_i \Phi^{\tilde{a}} + 
    \frac{\lambda}{4} \left( \Phi^{\tilde{a}} \Phi^{\tilde{a}} -\eta^2 \right)^2 \right] \, .
\end{equation}
From (\ref{4EnFunctional}) we can see that the condition of finite energy 
implies that the fields fall off at infinity sufficiently fast:
\begin{align} 
  \mathcal{D}_i \Phi^{\tilde{a}} \to 0, \quad \mbox{and} \quad \Phi^{\tilde{a}} \Phi^{\tilde{a}} \to 
  \eta^2  \quad \mbox{for} \; r=\vert x \vert \to \infty \, .
\end{align}
The second condition for the Higgs triplet $\Phi^{\tilde{a}}$ above 
shows that in the limit $r \to \infty$ the field $\Phi$ can be thought of as a function from physical 
infinity of topology $S^2$ to the vacuum manifold $G/H$ also of topology $S^2$. This argument is 
at the base of the homotopy classification of monopoles solutions in this theory since the 
above mappings fall into homotopy classes labeled by an integer $n$ according to $\pi_2(S^2)=Z$.
The so-called 't~Hooft-Polyakov ansatz respects the above boundary conditions for $n=1$:
\begin{align} \label{4TPansatz}
  \Phi^{\tilde{a}}(x^i)&=\delta_{i \tilde{a}} \frac{x^i}{r} \frac{H(r)}{e r} \, , \\
  W_i^{\tilde{a}}(x^i)&=\epsilon_{a i j} \frac{x^i}{r} \frac{1-K(r)}{e r} \, , \qquad W_0^{\tilde{a}}=0 \, ,
\end{align}
provided we have
\begin{equation}
  \lim_{r \to \infty} K(r)=0\, , \quad \lim_{r \to \infty} \frac{H(r)}{e r}=\eta.
\end{equation}
Due to the simplicity of the above ansatz, the field equations (\ref{4EoMscalar4D})-(\ref{4EoMgauge4D}) 
reduce to the following system of ordinary differential equations:
\begin{align}
  r^2 \frac{d^2 K(r)}{dr^2}&=K(r) \left[K(r)^2-1\right] + H(r)^2 K(r) \, ,\label{4TPeq1} \\
  r^2 \frac{d^2 H(r)}{dr^2}&=2 H(r) K(r)^2 + \lambda H(r) \left[\frac{H(r)^2}{e^2}-r^2 \eta^2 \right] \, .
  \label{4TPeq2}
\end{align}
No analytic solutions to the system (\ref{4TPeq1}) and (\ref{4TPeq2}) are known at present 
except for the case $\lambda\to 0$. This limit is known as the Prasad-Sommerfield limit 
\cite{Prasad:1975kr,Bogomolny:1975de} and the solutions to (\ref{4TPeq1}) and (\ref{4TPeq2}) take 
the particular simple form:
\begin{align} \label{4BPSsolution}
  K(r)=\frac{r e \eta}{\sinh(r e \eta)}, \qquad H(r)=\frac{r e \eta}{\tanh(r e \eta)}-1 \, .
\end{align}
Even though 't~Hooft-Polyakov monopoles give rise to plenty of interesting physics, we will stop 
our discussion at this point and come back to our main subject, the gravitating monopole 
in a $7$-dimensional space-time.    

\section{The warped 't~Hooft-Polyakov monopole in 7d curved space-time}
\subsection{Einstein equations - field equations} \label{4EQFQ}

The action for the setup considered in this section is a straightforward generalization of the action 
of a gravitating 't~Hooft-Polyakov monopole in 4 dimensions, objects that have been extensively studied 
in the past \cite{VanNieuwenhuizen:1975tc,Bais:1975gu,Lee:1991vy,Ortiz:1991eu,Breitenlohner:1991aa}, 
to the case of a $7$ dimensional space-time:
\begin{equation} \label{4action}
  S=S_{\mbox{\tiny\it gravity}}+S_{\mbox{\tiny\it brane}}\, .
\end{equation}
$S_{\mbox{\tiny\it gravity}}$ is the 7-dimensional generalization of the Einstein-Hilbert action:
\begin{equation} \label{4act_grav}
  S_{\mbox{\tiny\it gravity}}=\frac{M_7^5}{2} \int d^7 x \sqrt{-g} \left( R - \frac{2 \Lambda_7}{M_7^5} \right).
\end{equation}
Here, $g$ is the determinant of the metric $g_{M N}$ with signature  mostly 
plus $-+\ldots+$. Upper case latin indices $M,N$ will run over $0 \ldots 6$, lower case latin indices $m,n$ 
over $4 \ldots 6$ and greek indices $\mu, \nu$ over $0 \ldots 3$. $M_7$ denotes the fundamental gravity 
scale and $\Lambda_7$ the bulk cosmological constant. $S_{\mbox{\tiny\it brane}}$ is formally 
identical to the Georgi-Glashow action (\ref{4TPaction}) in Minkowski space-time 
apart from the dimensionality:
\begin{equation} \label{4act_brane}
 S_{\mbox{\tiny\it brane}}=\int d^7 x \sqrt{-g} \, \mathcal{L}_m \, ,
\end{equation}
with
\begin{equation}
 \mathcal{L}_m= -\frac{1}{4} G_{M N}^{\tilde{a}} G^{\tilde{a} M N}-\frac{1}{2} 
  \mathcal{D}_M \Phi^{\tilde{a}} \mathcal{D}^M \Phi^{\tilde{a}}-\frac{\lambda}{4} 
  \left( \Phi^{\tilde{a}} \Phi^{\tilde{a}} -\eta^2 \right)^2 \, .
\end{equation}
Here, $\eta$ is the vacuum expectation value of the scalar field.
Following the same conventions as in the last section, we denote group indices by 
a tilde: $\tilde{a}, \tilde{b}, \tilde{c} =1 \ldots 3$. $\mathcal{D}_M \Phi^{\tilde{a}}$ denotes the gauge 
and generally covariant derivative of the scalar field
\begin{equation} \label{4cov_der}
  \mathcal{D}_M \Phi^{\tilde{a}}=\partial_M \Phi^{\tilde{a}} + e \, \epsilon^{\tilde{a}\tilde{b}\tilde{c}}\, 
  W_M^{\tilde{b}} \Phi^{\tilde{c}}.
\end{equation}
Again we have
\begin{equation} \label{4GaugeField}
  G_{M N}^{\tilde{a}}=\partial_M W_N^{\tilde{a}}-\partial_N W_M^{\tilde{a}}+e \, 
  \epsilon^{\tilde{a}\tilde{b}\tilde{c}} \, W_M^{\tilde{b}} W_N^{\tilde{c}}.
\end{equation}
The general coupled system of Einsteins equations and the equations of motion for the scalar field and the gauge field following from the above action are
\begin{eqnarray}  
  R_{M N} -\frac{1}{2} g_{M N} R + \frac{\Lambda_7}{M_7^5} g_{M N} &=& \frac{1}{M_7^5} T_{M N}\, ,\label{4Einstein} \\
  \frac{1}{\sqrt{-g}} \mathcal{D}_M \left( \sqrt{-g} \, \mathcal{D}^M \Phi^{\tilde{a}} \right) &=& \lambda \, \Phi^{\tilde{a}} 
	\left( \Phi^{\tilde{b}} \Phi^{\tilde{b}}-\eta^2\right) \, , \label{4EoMscalar} \\
  \frac{1}{\sqrt{-g}} \mathcal{D}_M \left( \sqrt{-g} \, G^{\tilde{a} M N} \right) &=& -e \, \epsilon^{\tilde{a}\tilde{b}\tilde{c}} \left( \mathcal{D}^N \Phi^{\tilde{b}} \right) \Phi^{\tilde{c}} \, , \label{4EoMgauge} 
\end{eqnarray}
where the stress-energy tensor $T_{M N}$ is given by
\begin{equation} \label{4EnMom}
  T_{MN}\equiv-\frac{2}{\sqrt{-g}} \frac{\delta S_{\mbox{\tiny\it brane}}}{\delta g^{M N}}= G_{M L}^{\tilde{a}} G_{N}^{\tilde{a}\,L}+ \mathcal{D}_M \Phi^{\tilde{a}} \mathcal{D}_N \Phi^{\tilde{a}} + g_{M N} \mathcal{L}_m \, .
\end{equation}
We are interested in static monopole-like solutions to the set of equations (\ref{4Einstein})-(\ref{4EoMgauge}) respecting both $4D$-Poincar\'e invariance on the brane and rotational invariance in the transverse space. The fields $\Phi^{\tilde{a}}$ and $W_M^{\tilde{a}}$ (and as a result $G_{M N}^{\tilde{a}}$) should not depend on the coordinates on the brane $x^{\mu}$. The brane is supposed to be located at the center of the magnetic monopole. A general non-factorizable ansatz for the metric satisfying the above conditions is 
\begin{equation} \label{4metric}
  ds^2=M^2(\rho) g^{(4)}_{\mu\nu} dx^{\mu} dx^{\nu} + d\rho^2+ L^2(\rho) \left( d\theta^2+\sin^2 \theta \, d\varphi^2 \right) \, ,
\end{equation}
where $g^{(4)}_{\mu\nu}$ is interpreted as the metric of $4D$ general relativity which satisfies the $4D$-Einstein equations with an arbitrary cosmological constant $\Lambda_{\it phys}$, also appearing in the $7D$-Einstein equations 
\cite{Rubakov:1983bz}, \cite{Randjbar-Daemi:1985wg}. In the following we will only consider the case $\Lambda_{\it phys}=0$ and we take $g^{(4)}_{\mu\nu}$ 
to be the Minkowski metric $\eta_{\mu \nu}$ with signature $(-+++)$. Since the choice $\Lambda_{\it phys}=0$ is not
motivated by any physical argument, a possible solution to the cosmological constant problem depends on a mechanism explaining why the value $0$ is preferred. 
The 't~Hooft-Polyakov ansatz (\ref{4TPansatz}) (see \cite{'tHooft:1974qc}) 
for the fields is readily generalized to our setup.
After transforming to spherical coordinates this ansatz becomes
\begin{align} 
  &\vec{W}_{\rho}=0, \quad  \vec{W}_{\theta}=-\frac{1-K(\rho)}{e}\,\vec{e}^{\,\,\varphi}, \quad  
  \vec{W}_{\varphi}=\frac{1-K(\rho)}{e}\,\sin\theta \, \vec{e}^{\, \, \theta}, \quad \vec{W}_{\mu}=0 \, , 
  \nonumber \\ \label{4tHooftPolysphere}
  &\qquad \vec{\Phi}=\frac{H(\rho)}{e \rho} \vec{e}^{\, \, \rho} \, ,
\end{align}
with flashes indicating vectors in internal $SU(2)$ space and with
\begin{eqnarray} \label{4intvec}
  \vec{e}^{\,\,\rho}&=&\left( \sin\theta \cos\varphi,\sin\theta \sin\varphi,\cos\theta \right) \, , \\
  \vec{e}^{\,\,\theta}&=&\left( \cos\theta \cos\varphi,\cos\theta \sin\varphi,-\sin\theta \right)\, , \\
  \vec{e}^{\,\,\varphi}&=&\left( -\sin\varphi,\cos\varphi,0 \right) \, .
\end{eqnarray}
Using the ansatz (\ref{4tHooftPolysphere}) for the fields together with the metric (\ref{4metric}) in the coupled system of differential equations (\ref{4Einstein})--(\ref{4EoMgauge}) gives
\begin{eqnarray}
  3\, \frac{M''}{M}+ 2\, \frac{\mathcal{L}''}{\mathcal{L}}+ 6\, \frac{M'\,\mathcal{L}'}{M\,\mathcal{L}} + 
  3\, \frac{{M'}^2}{M^2} + \frac{{\mathcal{L}'}^2}{{\mathcal{L}}^2} -\frac{1}{\mathcal{L}^2} &=& 
   \beta \left(\epsilon_0-\Lambda_7 \right) \, , \label{4EinsteinP0}\\
  8\, \frac{M'\,\mathcal{L}'}{M\,\mathcal{L}} + 6\, \frac{{M'}^2}{M^2} + 
    \frac{{\mathcal{L}'}^2}{{\mathcal{L}}^2} - \frac{1}{\mathcal{L}^2} &=& 
  \beta \left(\epsilon_\rho-\Lambda_7 \right) \, , \label{4EinsteinPrho}\\ 
  4\, \frac{M''}{M}+ \frac{\mathcal{L}''}{\mathcal{L}}+ 4\, \frac{M'\,\mathcal{L}'}{M\,\mathcal{L}} + 
  6\, \frac{{M'}^2}{M^2} &=& 
   \beta \left(\epsilon_\theta-\Lambda_7 \right) \, ,\label{4EinsteinPtheta}
\end{eqnarray}
\begin{eqnarray}
  J''+2\left(2\, \frac{M'}{M}+\frac{\mathcal{L}'}{\mathcal{L}} \right) J'-2\,\frac{K^2 J}{\mathcal{L}^2}&=&
  \alpha J \left(J^2-1\right) \, ,\label{4EqMovphi}\\
  K''+\frac{K \left( 1- K^2 \right)}{\mathcal{L}^2}+4\, \frac{M'}{M} K' &=& J^2 K \, .\label{4EqMovW}
\end{eqnarray}
where primes denote derivatives with respect to the transverse radial coordinate $r$ 
rescaled by the mass of the gauge boson $m_W$:
\begin{equation}
  r=m_W \rho = \eta \, e \, \rho.
\end{equation}
All quantities appearing in the above equations are dimensionless, including $\alpha, \beta$ and $\gamma$. 
We have
\begin{equation}
  \alpha = \frac{\lambda}{e^2}=\frac{1}{2} \left(\frac{m_H}{m_W}\right)^2 \, , \quad
  \beta = \frac{\eta^2}{M_7^5} \, , \quad
  \gamma = \frac{\Lambda_7}{e^2 \eta^4} \, , \quad
  \epsilon_i = \frac{f_i}{e^2 \eta^4} \, ,
\end{equation}
with 
\begin{equation} \label{4f_def}
  T^{\mu}_{\nu}= \delta^{\mu}_{\nu} f_0 \,, \;\;
  T^{\rho}_{\rho}= f_\rho \, , \;\;
  T^{\theta}_{\theta}= f_\theta \,, \;\;
  T^{\varphi}_{\varphi}= f_\varphi \,, \;\; 
  \mathcal{L}=L m_W \, ,\;\;
  J=\frac{H}{\eta \, e \, \rho}.
\end{equation}
The dimensionless diagonal elements of the stress-energy tensor are given by
\begin{eqnarray}\label{4enmomelements1}
  \epsilon_0&=&-\left[ \frac{K'^2}{\mathcal{L}^2}+\frac{\left(1-K^2\right)^2}
  {2 \mathcal{L}^4}+\frac{1}{2} J'^2 + 
  \frac{J^2 K^2}{\mathcal{L}^2}+\frac{\alpha}{4} \left( J^2-1\right)^2  \right]\, , \\ \label{4enmomelements2}
  \epsilon_\rho&=& \frac{K'^2}{\mathcal{L}^2}-\frac{\left(1-K^2\right)^2}{2 \mathcal{L}^4}+\frac{1}{2} J'^2 - 
  \frac{J^2 K^2}{\mathcal{L}^2}-\frac{\alpha}{4} \left( J^2-1\right)^2 \, , \\\label{4enmomelements3}
  \epsilon_\theta&=& \frac{\left(1-K^2\right)^2}{2 \mathcal{L}^4}-\frac{1}{2} J'^2 - 
	 	     \frac{\alpha}{4} \left( J^2-1 \right)^2 \, .
\end{eqnarray}
The rotational symmetry in transverse space implies that the $(\theta\,\theta)$ and the $(\varphi\varphi)$ components 
of the Einstein equations are identical (and that $\epsilon_\varphi=\epsilon_\theta$). Equations 
(\ref{4EinsteinP0})-(\ref{4EinsteinPtheta}) are not functionally independent \cite{Gherghetta:2000jf}. 
They are related by the 
Bianchi identities (or equivalently by conservation of stress-energy $\nabla_M T^{M}_{N}=0$). When solving the system
(\ref{4EinsteinP0})--(\ref{4EqMovW}) numerically, eq. (\ref{4EinsteinPrho}) can be considered as a constraint 
which is automatically satisfied.

\noindent The Ricci scalar and the curvature invariants $R$, $R_{A B}\,R^{A B}$, $R_{A B C D}\,R^{A B C D}$, 
 $C_{A B C D}\,C^{A B C D}$ with  $C_{A B C D}$ the Weyl tensor are given by
\begin{eqnarray}
    \frac{R}{m_W^2}&=&
    \frac{2}{{\mathcal{L}}^2} - 
    \frac{2\,{\mathcal{L}'}^2}{{\mathcal{L}}^2} - 
    \frac{16\,\mathcal{L}'\,M'}{\mathcal{L}\,M} - 
    \frac{12\,{M'}^2}{{M}^2} - 
    \frac{4\,\mathcal{L}''}{\mathcal{L}} - 
    \frac{8\,M''}{M} \, ,\label{4Ricci} \\
   \frac{R_{A B}\,R^{A B}}{m_W^4}&=&
    \frac{2}{{\mathcal{L}}^4} - 
    \frac{4\,{\mathcal{L}'}^2}{{\mathcal{L}}^4} + 
    \frac{2\,{\mathcal{L}'}^4}{{\mathcal{L}}^4} - 
    \frac{16\,\mathcal{L}'\,M'}{{\mathcal{L}}^3\,M} + 
    \frac{16\,{\mathcal{L}'}^3\,M'}{{\mathcal{L}}^3\,M}  \nonumber \\
    &\phantom{=}&+\frac{48\,{\mathcal{L}'}^2\,{M'}^2}{{\mathcal{L}}^2\,{M}^2} +
    \frac{48\,\mathcal{L}'\,{M'}^3}{\mathcal{L}\,{M}^3} + 
    \frac{36\,{M'}^4}{{M}^4} -
    \frac{4\,\mathcal{L}''}{{\mathcal{L}}^3} + 
    \frac{4\,{\mathcal{L}'}^2\,\mathcal{L}''}{{\mathcal{L}}^3} \nonumber \\
    &\phantom{=}&+\frac{16\,\mathcal{L}'\,M'\,\mathcal{L}''}{{\mathcal{L}}^2\,M} + 
    \frac{6\,{\mathcal{L}''}^2}{{\mathcal{L}}^2} +
    \frac{16\,\mathcal{L}'\,M'\,M''}{\mathcal{L}\,{M}^2} + 
    \frac{24\,{M'}^2\,M''}{{M}^3}\nonumber \\
    &\phantom{=}&+\frac{16\,\mathcal{L}''\,M''}{\mathcal{L}\,M} + 
    \frac{20\,{M''}^2}{{M}^2} \, ,\label{4RabRab} \\
   \frac{R_{A B C D}\,R^{A B C D}}{m_W^4}&=&
    \frac{4}{{\mathcal{L}}^4} - 
    \frac{8\,{\mathcal{L}'}^2}{{\mathcal{L}}^4} + 
    \frac{4\,{\mathcal{L}'}^4}{{\mathcal{L}}^4} + 
    \frac{32\,{\mathcal{L}'}^2\,{M'}^2}{{\mathcal{L}}^2\,{M}^2} + 
    \frac{24\,{M'}^4}{{M}^4}  \nonumber \\
    &\phantom{=}&+\frac{8\,{\mathcal{L}''}^2}{{\mathcal{L}}^2} + 
    \frac{16\,{M''}^2}{{M}^2} \, ,\label{4RabcdRabcd} \\
   \frac{C_{A B C D}\,C^{A B C D}}{m_W^4}&=&\frac{8}{3\,{\mathcal{L}}^4} - 
    \frac{16\,{\mathcal{L}'}^2}{3\,{\mathcal{L}}^4} + 
    \frac{8\,{\mathcal{L}'}^4}{3\,{\mathcal{L}}^4} + 
    \frac{128\,\mathcal{L}'\,M'}{15\,{\mathcal{L}}^3\,M} - 
    \frac{128\,{\mathcal{L}'}^3\,M'}{15\,{\mathcal{L}}^3\,M}  \nonumber \\
    &\phantom{=}&-\frac{16\,{M'}^2}{5\,{\mathcal{L}}^2\,{M}^2} + 
    \frac{208\,{\mathcal{L}'}^2\,{M'}^2}{15\,{\mathcal{L}}^2\,{M}^2} - 
    \frac{64\,\mathcal{L}'\,{M'}^3}{5\,\mathcal{L}\,{M}^3} + 
    \frac{24\,{M'}^4}{5\,{M}^4} \nonumber \\
    &\phantom{=}&+\frac{32\,\mathcal{L}''}{15\,{\mathcal{L}}^3} -
    \frac{32\,{\mathcal{L}'}^2\,\mathcal{L}''}{15\,{\mathcal{L}}^3} - 
    \frac{64\,\mathcal{L}'\,M'\,\mathcal{L}''}{15\,{\mathcal{L}}^2\,M} +  
    \frac{32\,{M'}^2\,\mathcal{L}''}{5\,\mathcal{L}\,{M}^2} \nonumber \\
    &\phantom{=}&+\frac{64\,{\mathcal{L}''}^2}{15\,{\mathcal{L}}^2} -  
    \frac{32\,M''}{15\,{\mathcal{L}}^2\,M} + 
    \frac{32\,{\mathcal{L}'}^2\,M''}{15\,{\mathcal{L}}^2\,M} + 
    \frac{64\,\mathcal{L}'\,M'\,M''}{15\,\mathcal{L}\,{M}^2}  \nonumber \\
    &\phantom{=}&-\frac{32\,{M'}^2\,M''}{5\,{M}^3} - 
    \frac{128\,\mathcal{L}''\,M''}{15\,\mathcal{L}\,M} + 
    \frac{64\,{M''}^2}{15\,{M}^2} \, .\label{4CabcdCabcd}
\end{eqnarray}
Following \cite{Gherghetta:2000jf} we can define various brane tension components by 
\begin{equation} \label{4branetensions}
  \mu_i = -\int\limits_0^{\infty} dr M(r)^4 \mathcal{L}(r)^2 \epsilon_i(r) \, .
\end{equation}
\subsection{Boundary conditions} \label{4BC}
The boundary conditions should lead to a regular solution at the origin. Thus we have to impose
\begin{equation} \label{4BCMorigin}
  M\vert_{r=0}=1 \, , \quad   M'\vert_{r=0}=0 \, , \quad  
  \mathcal{L}\vert_{r=0}=0 \, , \quad \mathcal{L}'\vert_{r=0}=1 \, ,
\end{equation}
where the value $+1$ for $M\vert_{r=0}$ is a convenient choice that can be obtained by
rescaling brane coordinates.

Both the vacuum manifold of the scalar triplet $\Phi^{\tilde a}$ and infinity in transverse space 
($r=\infty$) are homeomorphic to $S^2$, so that all possible field configurations with specified boundary 
conditions at infinity can be classified by a topological (winding) number, see 
e.g.~\cite{Raj} or \cite{Vilenkin}.
For the monopole solutions we are interested in, this winding number is $1$. The scalar field given by the 
ansatz (\ref{4tHooftPolysphere}) is radial at infinity in the bulk. Unwinding such a field configuration 
is impossible, as is continuously contracting a smooth ``identity'' mapping $S^2 \to S^2$ to a mapping where
the whole $S^2$ is mapped into a single point. The radial component 
of the scalar field $J(r)$ must have at least one zero \cite{'tHooft:1974qc} which is chosen to be at the
 origin ($r=0$) of transverse space. Thus 
\begin{equation} 
  J(0)=0 \, , \quad  \lim_{r \to \infty} J(r) = 1 \, .
\end{equation} 
For the gauge field the condition of vanishing covariant derivative of $\Phi^{\tilde a}$ at infinity 
and the regularity of the ansatz (\ref{4tHooftPolysphere}) gives
\begin{equation} \label{4Kboundary}
 K(0)=1 \, , \quad  \lim_{r \to \infty} K(r) = 0.
\end{equation}
Up to this point our considerations were similar to 
those of \cite{Giovannini:2001hh}. However, the question of localizing 
gravity differs from the string case in several ways. Most importantly, in order to obtain solutions that
 localize gravity, i.e. solutions for which
\begin{equation} \label{4PlanckMass}
  M_P^2=\frac{4 \pi M_7^5}{m_W^3} \int\limits_{0}^{\infty} M(r)^2 \mathcal{L}(r)^2 dr
\end{equation}
is finite, it is impossible that $M(r)$ and $\mathcal{L}(r)$ should both be exponentially decreasing in the bulk.
No solutions corresponding to a strictly local defect have been found for $n=3$ extra dimensions, see 
\cite{Gherghetta:2000jf} and the discussion in the last chapter. 
This follows from the presence of the $1/\mathcal{L}(r)^2$-term in the Einstein equations.
The solutions we are looking for are concrete (numerical) realizations of the metric asymptotic
(\ref{3pfansatz}), (\ref{3pFormAsymptotic1}), (\ref{3pFormAsymptotic2})
found in 
the last section for the more general case of a p-Form field: 
$M(r)=M_0 e^{-\frac{c}{2} r}\, , \, \mathcal{L}(r)=\mbox{const.}$
Another important difference from the $5D$ case \cite{Randall:1999vf} and the $6D$ case 
\cite{Giovannini:2001hh} is that even though the space-time has constant curvature (for $r \to \infty$), 
it is not necessarily anti-de-Sitter. Finally, both signs of the bulk cosmological constant 
are possible.
\subsection{Asymptotic of the solutions - behavior at the origin} \label{4AS}
Once boundary conditions at the center of the defect are imposed for the fields and 
the metric, the system of equations (\ref{4EinsteinP0})--(\ref{4EqMovW}) can be solved 
in the vicinity of the origin by developing the fields and the 
metric into a power series in the (reduced) transverse radial variable $r$. 
For the given system this can be done up to any desired order. We give the power series 
up to third order in $r$:
\begin{align}
  M(r) &= 1 - \frac{1}{60} r^2\,\beta \,\left( \alpha  + 4\,\gamma  - 6\,{K''(0)}^2 \right) + \mathcal{O}(r^4) \, ,  \label{4asy_origin_M}\\ 
  \mathcal{L}(r)&=  r + \frac{1}{360} r^3\,\beta \,\left( \alpha  + 4\,\gamma - 30\,{J'(0)}^2 - 66\,{K''(0)}^2 \right) + \mathcal{O}(r^5) \, ,  \label{4asy_origin_L}\\
  J(r)&=  r J'(0) + r^3\,J'(0) \frac{-9\alpha  + \alpha \beta  + 4\beta \gamma + 
  6\beta {J'(0)}^2}{90}\nonumber \\
   &\phantom{=}+ \frac{3 K''(0) + \beta {K''(0)}^2}{15} + \mathcal{O}(r^5) \, ,  \label{4asy_origin_J}\\
 K(r) &= 1 + \frac{1}{2} r^2\,K''(0) + \mathcal{O}(r^4) \,  \label{4asy_origin_K}.
\end{align}
It can easily be shown that the power series of  $M(r)$ and $K(r)$ only involve even powers of $r$ 
whereas those of 
$\mathcal{L}(r)$ and $J(r)$ involve only odd ones. The expressions for $\mathcal{L}(r) $ 
and $J(r)$ are therefore valid up 
to $5^{\mathrm th}$ order. One observes that the solutions satisfying the boundary 
conditions at the origin can be parametrized by five parameters 
$\left[\alpha,\beta,\gamma, J'(0), K''(0) \right]$.
For arbitrary combinations of these parameters the corresponding metric solution will not 
satisfy the boundary 
conditions at infinity. Therefore, the numerical task is to find those parameter combinations for which 
(\ref{4PlanckMass}) is finite. For completeness, we give the power series expansion at the origin 
to zero-th order for the stress-energy tensor components 
\begin{eqnarray} \label{4epsorigin}
  \epsilon_0 \vert_{r=0}&=&-\frac{1}{4} \left[ \alpha + 6 \, J'(0)^2 + 6 \, K''(0)^2 \right] \, , \\ 
  \epsilon_{\rho}\vert_{r=0}=\epsilon_{\theta}\vert_{r=0}&=&-\frac{1}{4} \left[ \alpha + 
  2 \, J'(0)^2 - 2 \, K''(0)^2 \right] \, ,
\end{eqnarray}
and the curvature invariants:
\begin{align} \label{4curvatorigin1}
  \frac{R}{m_W^2}&= \frac{\beta}{10} \, \left[ 7\,\alpha  + 28\,\gamma  + 30\,{J'(0)}^2 + 18\,{K''(0)}^2 
  \right]\, ,\\ \label{4curvatorigin2}
  \frac{R_{A B}\,R^{A B}}{m_W^4}&=\frac{{\beta}^2}{100} \,\left[ 7\,{\alpha }^2 \hspace{-.5mm}+\hspace{-.5mm}
    56\,\alpha \,\gamma \hspace{-.5mm} 
   +\hspace{-.5mm} 112\,{\gamma }^2 \hspace{-.5mm}+\hspace{-.5mm} 60\,\alpha \,{J'(0)}^2 
    \hspace{-.5mm}+\hspace{-.5mm} 240\,\gamma \,{J'(0)}^2 +300\,{J'(0)}^4 \right.  \\
  & \left. \qquad \qquad   \hspace{-.5mm}+\hspace{-.5mm} 12\,\left( 3\,\alpha  + 12\,\gamma  + 
    70\,{J'(0)}^2 \right) \,{K''(0)}^2 \hspace{-.5mm}+\hspace{-.5mm} 732\,{K''(0)}^4 \right] \, ,\nonumber 
\end{align}
\begin{align} \label{4curvatorigin3}
  \frac{R_{A B C D}\,R^{A B C D}}{m_W^4}&=\frac{{\beta }^2}{300} \, \left[ 17\,{\alpha }^2 
  \hspace{-.5mm}+\hspace{-.5mm} 136\, 
  \alpha \, \gamma \hspace{-.5mm}+\hspace{-.5mm} 
  272 \, {\gamma}^2 \hspace{-.5mm}-\hspace{-.5mm} 60\, \alpha \, {J'(0)}^2 
   \hspace{-.5mm}-\hspace{-.5mm} 240\,\gamma \, {J'(0)}^2 + 900 \, {J'(0)}^4 \right. \nonumber \\
    & \quad \qquad \left.   - 36\,\left( 9\,\alpha  + 36\,\gamma  - 
            110\,{J'(0)}^2 \right) \,{K''(0)}^2 + 
4932\,{K''(0)}^4 \right]\, ,\\
  \frac{C_{A B C D}\,C^{A B C D}}{m_W^4}&=\label{4curvatorigin4}
  \frac{{\beta }^2}{30}\,{\left[ \alpha  + 4\,\gamma  - 6\,\left( {J'(0)}^2 + 3\,{K''(0)}^2 \right)  
  \right] }^2 \, .
\end{align}

\subsection{Asymptotic of the solutions - behavior at infinity}
The asymptotic of the metric functions $M(r)$ and $\mathcal{L}(r)$ far from the monopole is:
\begin{equation} \label{4asympto}
  M(r)=M_0 \, e^{-\frac{c}{2}r} \quad \mbox{and} \quad \mathcal{L}(r)=\mathcal{L}_0=\mbox{const}\, ,
\end{equation}
where only positive values of $c$ lead to gravity localization. This induces the following asymptotic for 
the stress-energy components and the various curvature invariants:
\begin{eqnarray} \label{4epsinfinity}
  \lim_{r \to \infty} \epsilon_0(r)=\lim_{r \to \infty} \epsilon_{\rho}(r)&=&-\frac{1}{2 {\mathcal{L}_0}^4} \, , \\
  \lim_{r \to \infty} \epsilon_{\theta}(r)&=&\frac{1}{2 {\mathcal{L}_0}^4} \, .
\end{eqnarray}
\begin{eqnarray}
  \lim_{r \to \infty} \frac{R(r)}{m_W^2} &=& -5 c^2 + \frac{2}{\mathcal{L}_0^2} \, , \label{4curvinvinfinity1}\\
  \lim_{r \to \infty} \frac{R_{A B}(r) R^{A B}(r)}{m_W^4} &=& 5 c^4 + \frac{2}{\mathcal{L}_0^4} \, , \label{4curvinvinfinity2}\\
  \lim_{r \to \infty} \frac{R_{A B C D}(r) R^{A B C D}(r)}{m_W^4} &=& \frac{5}{2} c^4 + \frac{4}{\mathcal{L}_0^4} \, , \label{4curvinvinfinity3}\\
  \lim_{r \to \infty} \frac{C_{A B C D}(r) C^{A B C D}(r)}{m_W^4} &=& \frac{c^4}{6} +\frac{8}{3 \, {\mathcal{L}_0^4}}
           -\frac{4 c^2}{\mathcal{L}_0^2} \, .\label{4curvinvinfinity4}
\end{eqnarray}
$c$ and $\mathcal{L}_0$ are determined 
by Einsteins equations for large $r$ and the corresponding asymptotic was derived 
in the last chapter in (\ref{3pFormAsymptotic1})-(\ref{3pFormAsymptotic2}). In the present case
these relations become: 
\begin{eqnarray} 
  c^2&=&\frac{5}{32 \beta} \left( 1-\frac{16}{5} \gamma \beta^2 + \sqrt{1-\frac{32}{25} \gamma \beta^2} \right)\, , 
    \label{4Einsteinasymptocsqr}\\
  \frac{1}{\mathcal{L}_0^2}&=&\frac{5}{8\beta}\left( 1 + \sqrt{1-\frac{32}{25} \gamma \beta^2} \right) 
   \label{4Einsteinasympto1overLsqr} \, .
\end{eqnarray}
The charge of the field configuration can either be determined by comparing the energy-momentum tensor with 
the general expression given in (\ref{3pFormCharge}) of section \ref{SectionBulkpform} or by a 
direct calculation of the magnetic field strength tensor following 't~Hooft \cite{'tHooft:1974qc}:
\begin{equation} \label{4FST}
  \mathcal{G}_{M N}=\frac{\vec{\Phi} \cdot \vec{G}_{M N}}{\vert \vec{\Phi}\vert}-
  \frac{1}{e \vert \vec{\Phi}\vert^3} \, \vec{\Phi} \cdot \left( \mathcal{D}_M \vec{\Phi} \times \mathcal{D}_N 
\vec{\Phi} \right).
\end{equation}
The only nonzero component of $ \mathcal{G}_{M N}$ is $ \mathcal{G}_{\theta \varphi}=-\frac{\sin\theta}{\vert e \vert}$.
Either way gives $Q=\frac{1}{e}$.
In order to obtain some information about the asymptotic behavior of the fields $J(r)$ and $K(r)$ at infinity we insert
relations (\ref{4asympto}) into (\ref{4EqMovphi}) and (\ref{4EqMovW}). Furthermore we use $K(r)=\delta K(r)$ and 
$J(r)=1-\delta J(r)$ with $\delta K \ll 1$ and $\delta J \ll 1$ for $1\ll r$ to linearize these equations:
\begin{eqnarray}
  \delta J'' - 2 \, c \, \delta J' -2 \alpha \delta J &=& 0 \, , \label{4lineqJ}\\
  \delta K'' - 2 \, c \, \delta K' +\left(\frac{1}{\mathcal{L}_0^2}-1\right) \delta K &=& 0\, \label{4lineqK} .
\end{eqnarray}
By using the ansatz $\delta K = A e^{-k r}$ and $\delta J = B e^{-j r}$ we find 
\begin{eqnarray}
  k_{1,2}=-c\pm\sqrt{c^2-\left( \frac{1}{\mathcal{L}_0^2}-1\right)} \label{4kasympt} \, ,\qquad 
  j_{1,2}=-c\pm\sqrt{c^2+2\alpha}\, .
\end{eqnarray} 
To satisfy the boundary conditions we obviously have to impose $k>0$ and $j>0$. We distinguish two cases:
\begin{enumerate}
\item \fbox{$c > 0$} Gravity localizing solutions. In this case there is a unique 
$k$ for $\frac{1}{\mathcal{L}_0^2}>1$ with  the positive sign in (\ref{4kasympt}).
\item \fbox{$c < 0$} Solutions that do not localize gravity.
\begin{enumerate}
  \item \fbox{$\frac{1}{\mathcal{L}_0^2} > 1 $}  
 \raisebox{-5mm}{
  \parbox{12cm}{
  \begin{enumerate}
    \item \fbox{$c^2 \geq \frac{1}{\mathcal{L}_0^2}-1$} \, Both solutions of (\ref{4kasympt}) are positive.
    \item \fbox{$c^2 < \frac{1}{\mathcal{L}_0^2}-1$} \, In this case there are no real solutions.
  \end{enumerate}
  }
  }
  \item \fbox{$\frac{1}{\mathcal{L}_0^2} < 1 $} \, There is a unique solution with the positive sign in (\ref{4kasympt}) .
  \item \fbox{$\frac{1}{\mathcal{L}_0^2} = 1 $} \, The important $k$-value here is $k=-2 c$.
\end{enumerate}
\end{enumerate}
It is easily shown  that for large enough $r$ the linear approximation (\ref{4lineqK}) 
to the equation of motion of the gauge field is always valid. Linearizing the equation for
the scalar field however breaks down for $2k<j$ due to the presence of the term $\propto K^2 J$ in
(\ref{4EqMovphi}). In that case $J(r)$ approaches $1$ as $e^{-2 k r}$. For solutions with $c>0$ (the case of 
predominant interest) a detailed discussion of the validity of $2k>j$ gives:
\begin{enumerate}
\item \fbox{$\alpha=0$} \, Prasad-Sommerfield limit \cite{Prasad:1975kr}. One has $2k > j$. The asymptotic of 
the scalar field is governed by $e^{- j r}$.
\item \fbox{$0<\alpha<2$} \, The validity of $2k > j$ depends on different inequalities between $\alpha$,
 $\frac{1}{\mathcal{L}_0^2}$ and \nolinebreak $c^2$.
  \begin{enumerate}
    \item \fbox{$\alpha+\frac{1}{\mathcal{L}_0^2} \leq 1$}  $ \longrightarrow 2k > j $
    \item  \fbox{$\alpha+\frac{1}{\mathcal{L}_0^2} > 1$} and 
               \fbox{$\frac{\alpha}{2}+\frac{1}{\mathcal{L}_0^2} < 1$}\, 

In this case $ 2k \geq j$ is equivalent to 
     $c^2 \leq \frac{-\left(1-\frac{\alpha}{2}-\frac{1}{\mathcal{L}_0^2}\right)^2}{1-\alpha-\frac{1}{\mathcal{L}_0^2}}$,
     where equality in one of the equations implies equality in the other.  
    \item \fbox {$\frac{\alpha}{2}+\frac{1}{\mathcal{L}_0^2} \geq 1$}  $ \longrightarrow 2k < j $ and the scalar field 
    asymptotic is governed by $e^{- 2 k r}$.
  \end{enumerate}
\item \fbox{$\alpha >2$} $ \longrightarrow 2k < j $. Also in this case the gauge field $K(r)$ determines 
  the asymptotic of the scalar field $J(r)$.
\end{enumerate}
This information is essential for obtaining numerical solutions to the equations (\ref{4EinsteinP0})-(\ref{4EqMovW})
since the integration has to be stopped at some finite distance from the origin where boundary conditions have to be 
imposed.
\subsection{Fine-tuning relations} \label{4FTR}
It is possible to derive analytic relations between the different components of the brane tension valid for 
gravity localizing solutions. Integrating linear combinations of Einsteins equations 
(\ref{4EinsteinP0})--(\ref{4EinsteinPtheta}) between $0$ and $\infty$ gives after multiplication with 
$M^4(r) \mathcal{L}^2(r)$:
\begin{eqnarray}
  \mu_0-\mu_{\theta}&=&\frac{1}{\beta} \int\limits_{0}^{\infty} M^4 d r = 
       \int\limits_{0}^{\infty} \left(1 -K^2 \right)\frac{M^4}{\mathcal{L}^2} d r \, , \label{4FineTunRel1}\\
  \mu_0-\mu_{\rho}- 2 \mu_{\theta}&=&2 \gamma \int\limits_{0}^{\infty} M^4 \mathcal{L}^2 d r \, , \label{4FineTunRel2}\\
  \mu_0+\mu_{\rho}+ 2 \mu_{\theta}&=&  \alpha \int\limits_{0}^{\infty} \left( 1 - J^2 \right) 
        M^4 \mathcal{L}^2 d r \, \label{4FineTunRel3}.
\end{eqnarray} 
To obtain the above relations integration by parts was used where the boundary terms dropped due
to the boundary conditions given above. A detailed derivation of these relations is given in 
appendix \ref{appFTR}.

\section{Numerics} \label{4N}
\subsection{General discussion} \label{4NGD}
As already pointed out, the numerical problem is to find those solutions to the system of differential 
equations (\ref{4EinsteinP0})--(\ref{4EqMovW}) and boundary conditions for which the integral 
defining the $4$-dimensional Planck-scale (\ref{4PlanckMass}) is finite. This is a two point 
boundary value problem on the interval $r=[0,\infty)$ depending on three independent parameters
$(\alpha,\beta,\gamma)$. Independently of the numerical method employed, the system of equations 
(\ref{4EinsteinP0})--(\ref{4EqMovW}) was rewritten in a different way in order for the integration
to be as stable as possible. By introducing the derivatives of the unknown functions $J(r)$, $K(r)$,
$M(r)$ and $\mathcal{L}(r)$ as new dependent variables one obtains a system of ordinary first order equations.
In the case of $M(r)$ it turned out to be convenient to define $y_7=M'(r)/M(r)$ as a new unknown function
(rather than $M'(r)$) since the boundary condition for $y_7$ at infinity then simply reads
\begin{equation}
  \lim_{r\to \infty} y_7=-\frac{c}{2} \, .
\end{equation}
With the following definitions, we give the form of the 
equations which is at the base of several numerical methods employed:
\begin{equation} \label{4numsys}
  \begin{array}{lclrcl}
    y_1(r) &=& J(r)\, ,& \quad y'_1 &=& y_5 \, , \\  
    y_2(r) &=& K(r)\, ,&  y'_2 &=& y_6 \, , \\  
    y_3(r) &=& M(r)\, ,&  y'_3 &=& y_3 \, y_7 \, , \\  
    y_4(r) &=& \mathcal{L}(r)\, ,& y'_4 &=& y_8 \, , \\  
    y_5(r) &=& J'(r)\, ,& y'_5 &=& -2 \left( 2 \, y_7+\frac{y_8}{y_4}\right) y_5+2\, \frac{y_1 y_2^2}{y_4^2}+
               \alpha \, y_1 \left( y_1^2-1 \right) \, , \\  
    y_6(r) &=& K'(r)\, ,& y'_6 &=& -\frac{y_2 \left( 1- y_2^2 \right)}{y_4^2} - 4 y_6 y_7 + y_1^2 y_2 \, , \\  
    y_7(r) &=& \frac{M'(r)}{M(r)}\, ,& y'_7 &=& -4 y_7^2 - 2 y_7 \frac{y_8}{y_4} -
               \frac{\beta}{5} \left( 2 \gamma + \epsilon_0 - \epsilon_{\rho}-2 \epsilon_{\theta} \right) \, , \\  
    y_8(r) &=& \mathcal{L}'(r)\, ,& y'_8 &=& -\frac{y_8^2}{y_4}-4 y_7 y_8 + \frac{1}{y_4}-
 		\frac{\beta y_4}{5} \left( 2 \gamma -4 \epsilon_0 - \epsilon_{\rho}+3 \epsilon_{\theta} \right) \, ,
  \end{array}
\end{equation}
with 
\begin{eqnarray} \label{4epscomb}
  \epsilon_0 - \epsilon_{\rho}-2 \epsilon_{\theta} &=& 
       \frac{\alpha}{2} \left( y_1^2 - 1 \right)^2 
      -\frac{\left(1-y_2^2\right)^2}{y_4^4}-2 \, \frac{y_6^2}{y_4^2} \, , \\
  -4 \epsilon_0 - \epsilon_{\rho}+3 \epsilon_{\theta} &=&
       \frac{\alpha}{2} \left( y_1^2 - 1 \right)^2 
   + 4 \, \frac{\left(1-y_2^2\right)^2}{y_4^4}
   + 5 \, \frac{y_1^2 \, y_2^2}{y_4^2} + 3 \, \frac{y_6^2}{y_4^2} \, . 
\end{eqnarray}
This is an autonomous, ordinary system of coupled differential equations depending on the parameters
$(\alpha,\,\beta,\,\gamma)$. The boundary conditions are 
\begin{equation} 
  \begin{array}{rcl} y_1(0)&=&0 \, , \\ \lim\limits_{r\to\infty} y_1(r) &=& 1 \, , \end{array} \quad
  \begin{array}{rcl} y_2(0)&=&1 \, , \\ \lim\limits_{r\to\infty} y_2(r) &=& 0 \, , \end{array} \quad
  \begin{array}{rcl} y_3(0)&=&1 \, , \\ y_7(0) &=& 0 \, , \end{array}\quad
  \begin{array}{rcl} y_4(0)&=&0 \, , \\ y_8(0) &=& 1 \, .  \end{array}\quad
\end{equation}
In order to find solutions with the desired metric asymptotic at infinity, it is useful to define either 
one (or more)
of the parameters $(\alpha,\,\beta,\,\gamma)$ or the constants $J'(0)$ and $K''(0)$ as additional 
dependent variables, 
e. g. $y_9(r)=\alpha$ with $y'_9(r)=0$, see \cite{numrec}.
Before discussing the different methods that were used we give some common numerical problems encountered.
\begin{itemize}
\item Technically it is impossible to integrate to infinity. The possibility of compactifying the independent
variable $r$ was not believed to simplify the numerics. Therefore, the integration has to be 
stopped at some upper 
value of $r=r_{\mbox{\tiny max}}$. For most of the solutions this was about $20 \sim 30$.
\item Forward integration with arbitrary values of $(\alpha,\,\beta,\,\gamma,\, J'(0)\, ,K''(0))$ turned
out to be very unstable. This means that even before some integration routine (e.g. the Runge-Kutta method \cite{numrec})
reached $r_{\mbox{\tiny max}}$, the values of some $y_i$ went out of range which was due to the presence of terms 
$\frac{1}{\left(\mathcal{L}\right)^n}$ or quadratic and cubic (positive coefficient) terms in $y_i$.
\item Some right hand sides of (\ref{4numsys}) contain terms singular at the origin, such that their sums remain
regular. Starting the integration at $r=0$, is therefore impossible. To overcome this problem the solution in
terms of the power series (\ref{4asy_origin_M})--(\ref{4asy_origin_K}) was used within the interval $\left[ 0, \,  \epsilon \right]$ (for $\epsilon=0.01$).
\item The solutions are extremely sensitive to initial conditions which made it unavoidable to pass from single precision to double precision (from about $7$ to about $15$ significant digits). However this didn't completely solve the 
problem. Even giving initial conditions (corresponding to a gravity localizing solution) at the origin to machine 
precision is in general not sufficient to obtain satisfactory precision at $r_{\mbox{\tiny max}}$ in a single 
Runge-Kutta forward integration step from $\epsilon$ to $r_{\mbox{\tiny max}}$.
\end{itemize}

\subsection{Discussion of solutions - fine-tuning surface} \label{4NFTS}
The method used for exploring the parameter space $(\alpha,\,\beta,\,\gamma)$ for gravity localizing solutions 
was a generalized version of the shooting method called the multiple shooting method 
\cite{numrec,acton,keller,BulirschStoer}. 
In the shooting method, a boundary value problem is solved by combining a
 root-finding method (e.g. Newton's
method \cite{numrec}) with forward integration. In order to start the integration at one boundary, the root finding 
routine specifies particular values for the so-called shooting parameters and compares the results of the integration
with the boundary conditions at the other boundary. This method obviously fails whenever the ``initial guess'' for the 
shooting parameters is too far from a solution such that the forward integration does not reach the second boundary.
For this reason, the multiple shooting method was used in which the interval 
$\left[ \epsilon \, , r_{\mbox{\tiny max}} \right]$ was divided into a variable number of subintervals in 
each of which the shooting method was applied. This of course drastically increased the number of shooting 
parameters and as a result the amount of computing time. However, it resolved two problems:
\begin{itemize}
\item Since the shooting parameters were specified in all subintervals the precision of the parameter values 
at the origin was no longer crucial for obtaining high precision solutions.
\item Out of range errors can be avoided by increasing the number of shooting intervals (again at the cost of 
increasing computing time).
\end{itemize}
\begin{figure}[htb]
\centerline{ \epsfxsize = 90mm \epsfbox {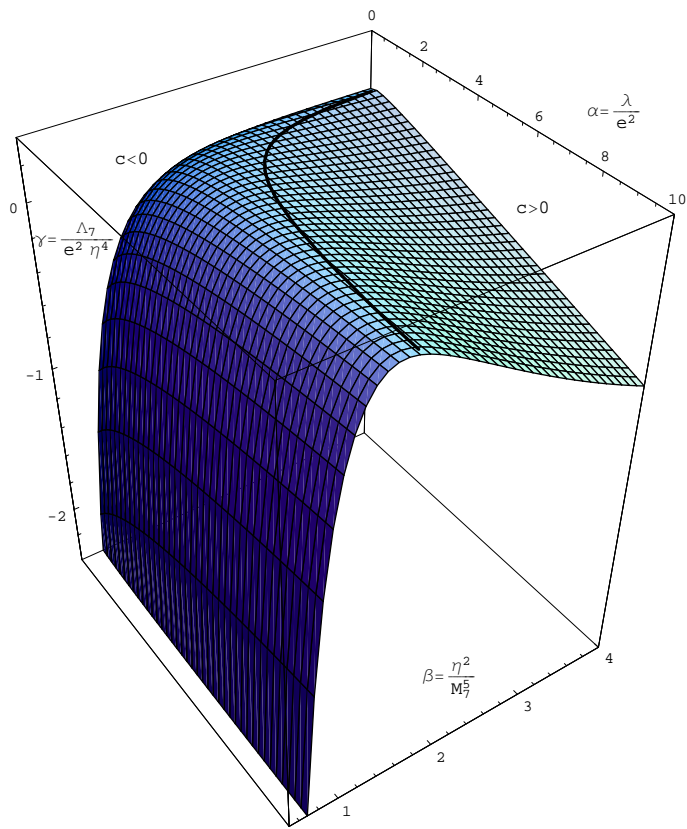}}
\caption[a]{\it Fine-tuning surface for solutions with the metric asymptotic (\ref{4asympto}). The bold 
line separates solutions
that localize gravity $(c>0)$ from those that do not $(c<0)$. Numerically obtained values of 
$\gamma=\frac{\Lambda_7}{e^2 \eta^4}$ are plotted as a function of $\alpha = \frac{\lambda}{e^2}$ and 
$\beta = \frac{\eta^2}{M_7^5}$.}
\label{4FTS}
\end{figure}
Despite all these advantages of multiple shooting, a first combination of parameter values 
$(\alpha,\,\beta,\,\gamma,\, J'(0)\, ,K''(0))$ leading to gravity localization could not be found by this
 method, since convergence depends 
strongly on how close initial shooting parameters are to a real solution. This first solution was found
by backward integration combined with the simplex method for finding zeros of one real function 
of several real variables. For a discussion of the simplex method see e.g. \cite{numrec}. This solution 
corresponds to the parameter values $(\alpha=1.429965428 , \, \beta=3.276535576 , \, \gamma=0.025269415) $. 
Once this solution was known, it was straightforward to investigate with the multiple shooting method, 
which subset of the $(\alpha,\,\beta,\,\gamma)$-space leads to the desired metric asymptotic. We used a 
known solution $(\alpha_1,\,\beta_1,\,\gamma_1)$ to obtain starting values for the shooting parameters of a  
closeby other solution $(\alpha_1+\delta\alpha,\,\beta_1+\delta\beta,\,\gamma_1+\delta\gamma)$. This leads 
in general
to rapid convergence of the Newton-method. Nevertheless, $\delta\alpha,\,\delta\beta,\,\delta\gamma$
still had to be small. By this simple but time-consuming operation the so-called fine-tuning surface in 
parameter space was found, see Fig.~\ref{4FTS}. 
A point on this surface $(\alpha_0,\,\beta_0,\,\gamma_0)$ corresponds to a particular
solution with the metric asymptotic (\ref{4asympto}). The bold line separates gravity localizing 
solutions $c>0$ from
solutions that do not localize gravity $c<0$. The parameter space has been thoroughly explored within 
the rectangles 
$\Delta\alpha \times \Delta\beta = [0,\,10]\,  \times \, [0.5,\, 4.0]$ and 
$\Delta\alpha \times \Delta\beta = [0,\,1]\,  \times \, [3.5,\, 10]$ 
in the $(\alpha,\beta)$-plane.
To illustrate the dependence on the parameter $\beta$ (strength of gravity), we present in the 
following subsection 
a series of solutions corresponding to decreasing values of $\beta$ for a fixed value of $\alpha=1$. 
 
\subsection{Examples of numerical solutions} \label{4samplesol}

Fig.~\ref{4sol155} and Fig.~\ref{4sol135} show two different numerical solutions corresponding to positive 
and negative bulk cosmological constant, respectively. Both solutions localize gravity $c>0$.
The corresponding parameter values are 
$(\alpha=1.0000000 , \, \beta=5.50000000 , \, \gamma=-0.05434431) $ and 
$(\alpha=1.0000000 , \, \beta=3.50000000 , \, \gamma=0.02678351) $.
Figs.~\ref{4sol155_e} and~\ref{4sol135_e} and Figs.~\ref{4sol155_c} and~\ref{4sol135_c} show the corresponding
components of the stress-energy tensor and the curvature invariants.
For high values of $\beta$ the metric function $\mathcal{L}(r)$ develops a maximum before attaining
its boundary value. Gravity dominates and the volume of the transverse space stays finite. 
Furthermore, it can be seen from Fig.~\ref{4FTS} that for every fixed $\alpha$ there is a
particular value of $\beta$ such that $c$ equals zero, 
which is the case for all points on the solid line shown in Fig.~\ref{4FTS}.
By looking at eq.~(\ref{4Einsteinasymptocsqr}), we immediately see that $c=0$ is equivalent to
 $\beta^2 \gamma=\frac{\Lambda_7}{e^2 \, M_7^{10}}=\frac{1}{2}$. 
We will discuss the limit $c \ll 1$ in more detail in section \ref{4FML}.
For lower values of $\beta$ the transverse space has infinite volume and gravity can not be localized. 
See Figs.~\ref{4sol118} and ~\ref{4sol118_e} for a solution that does not localize gravity $(c<0)$ and
corresponds to $(\alpha=1.0000000 , \, \beta=1.80000000 , \, \gamma=0.04053600) $.
\begin{figure}
\centerline{ \epsfxsize = 13cm \epsfbox {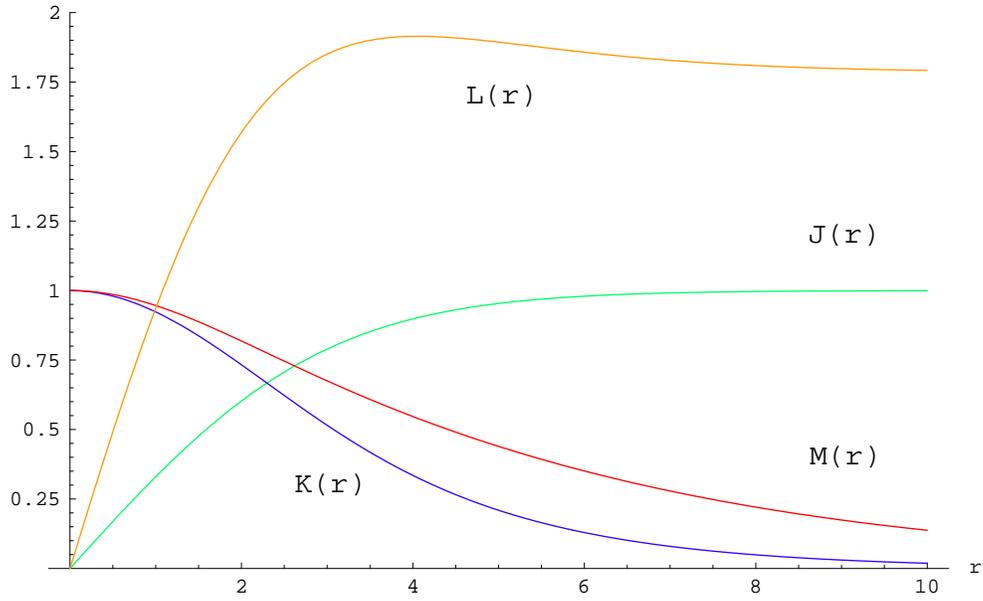}}
\caption[a]{\it Gravity-localizing solution with negative bulk cosmological constant corresponding to the parameter values  
$(\alpha=1.00000000, \, \beta=5.50000000, \, \gamma=-0.05434431) $}\label{4sol155}
\end{figure}
\begin{figure}
\centerline{ \epsfxsize = 13cm \epsfbox {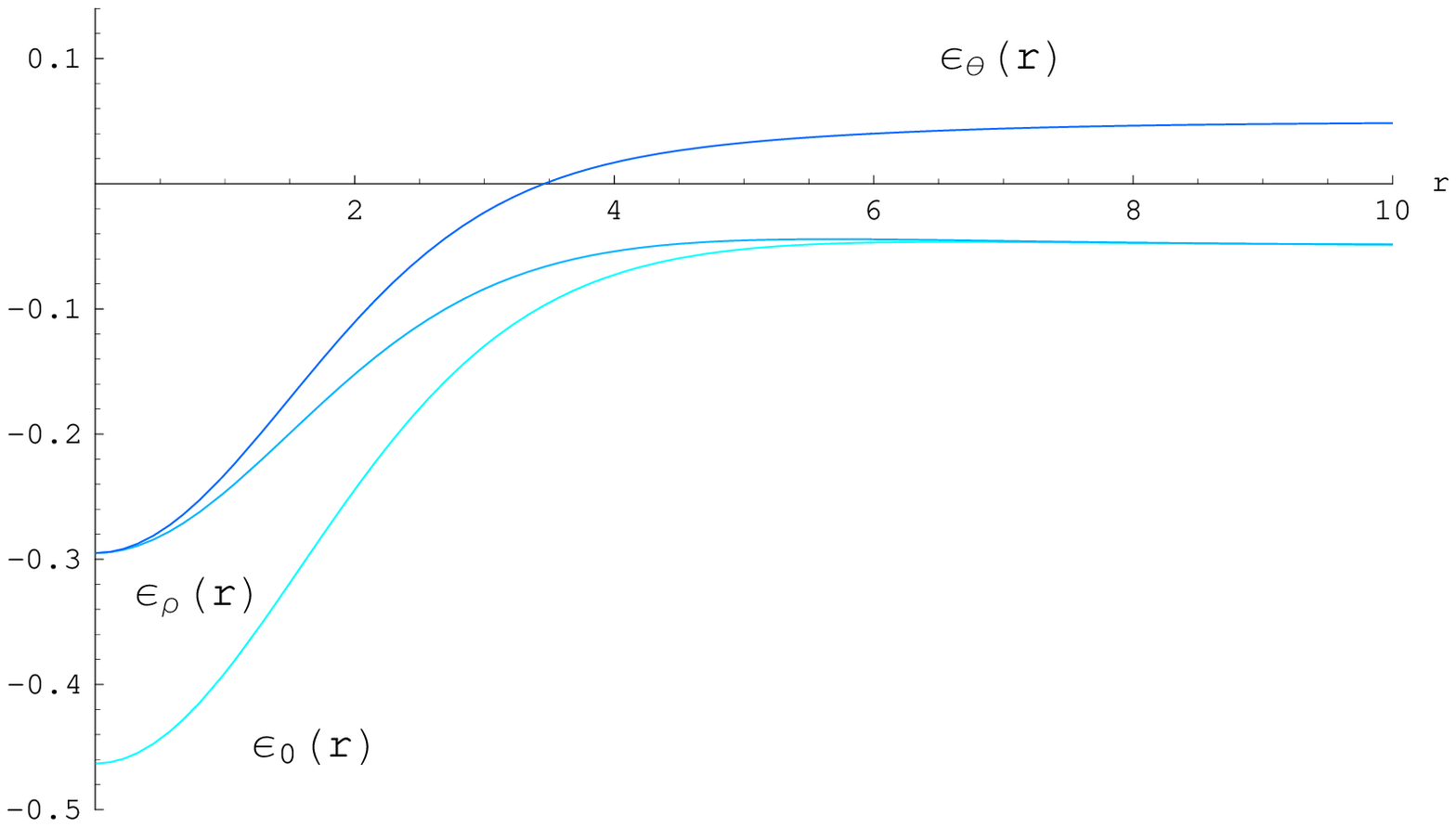}}
\caption[a]{\it Stress-energy components for the solution given in Fig.~\ref{4sol155}. \\
$(\alpha=1.00000000, \, \beta=5.50000000, \, \gamma=-0.05434431)$} \label{4sol155_e}
\end{figure}
\clearpage
\begin{figure}
\centerline{ \epsfxsize = 13cm \epsfbox {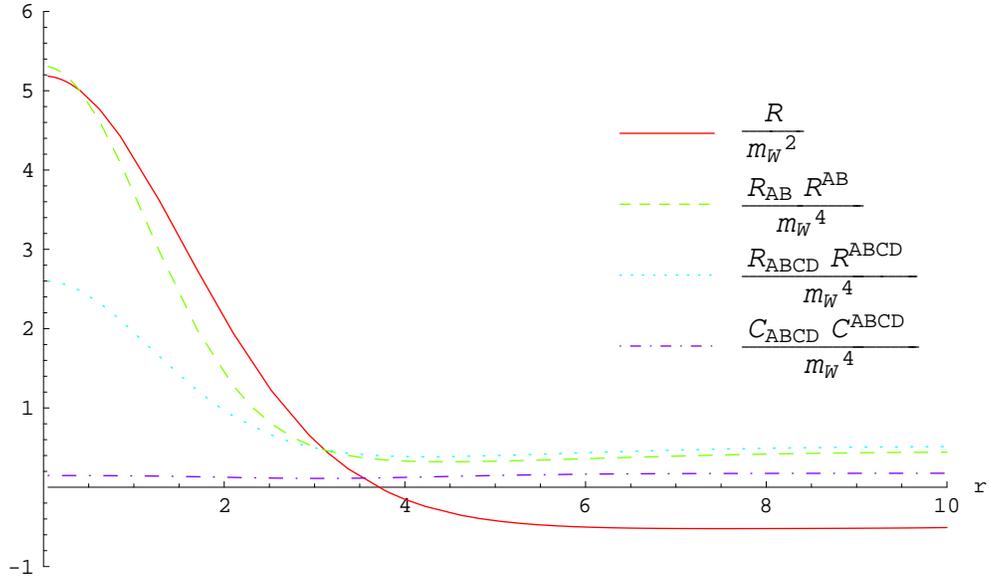}}
\caption[a]{\it Curvature invariants for the solution given in Fig.~\ref{4sol155}. \\
$(\alpha=1.00000000, \, \beta=5.50000000, \, \gamma=-0.05434431) $} \label{4sol155_c}
\end{figure}
\begin{figure}
\centerline{ \epsfxsize = 13cm \epsfbox {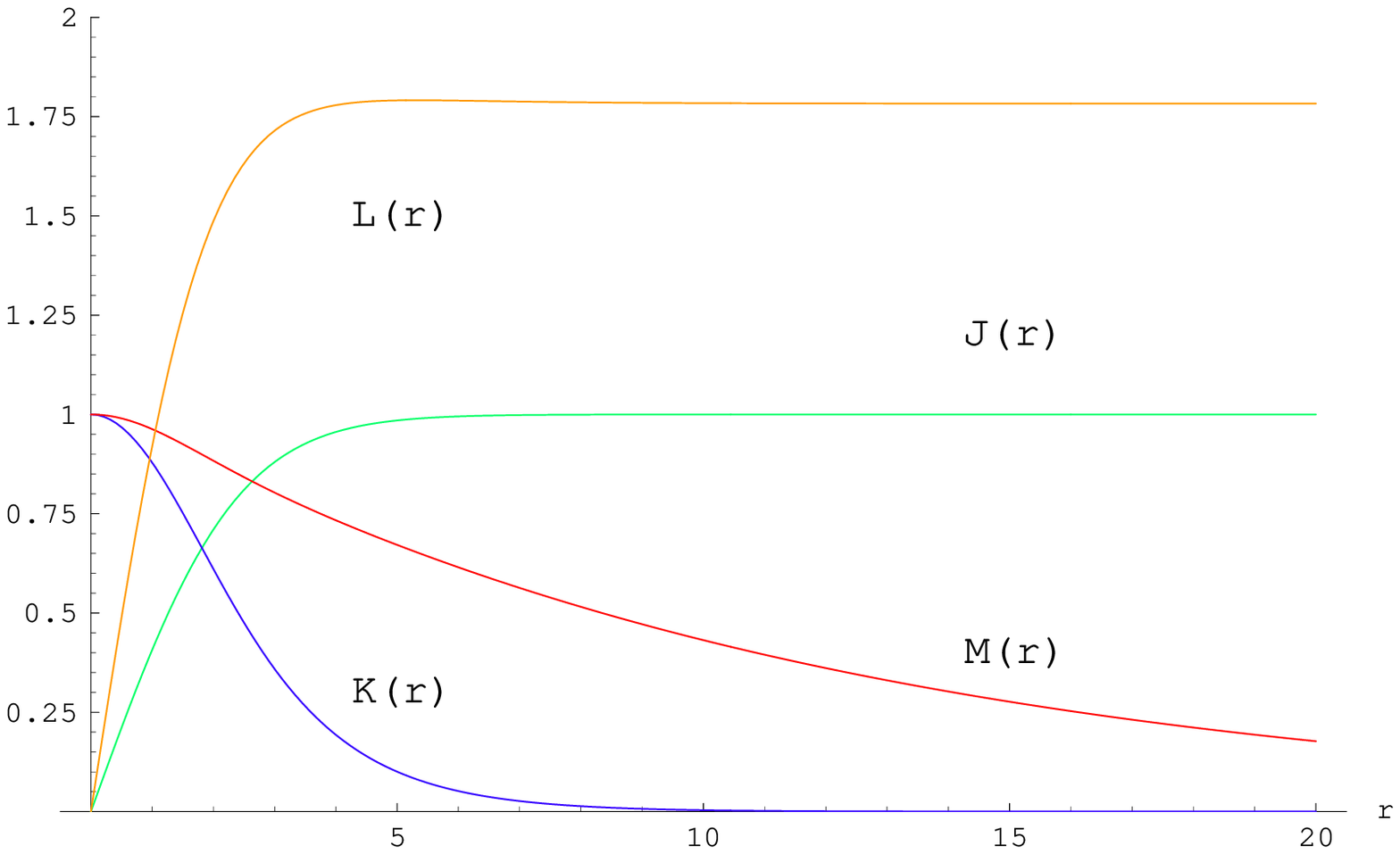}}
\caption[a]{\it Gravity-localizing solution with positive bulk cosmological constant corresponding to the parameter values
$(\alpha=1.00000000, \, \beta=3.50000000, \, \gamma=0.02678351) $} \label{4sol135}
\end{figure}
\clearpage
\begin{figure}
\centerline{ \epsfxsize = 13cm \epsfbox {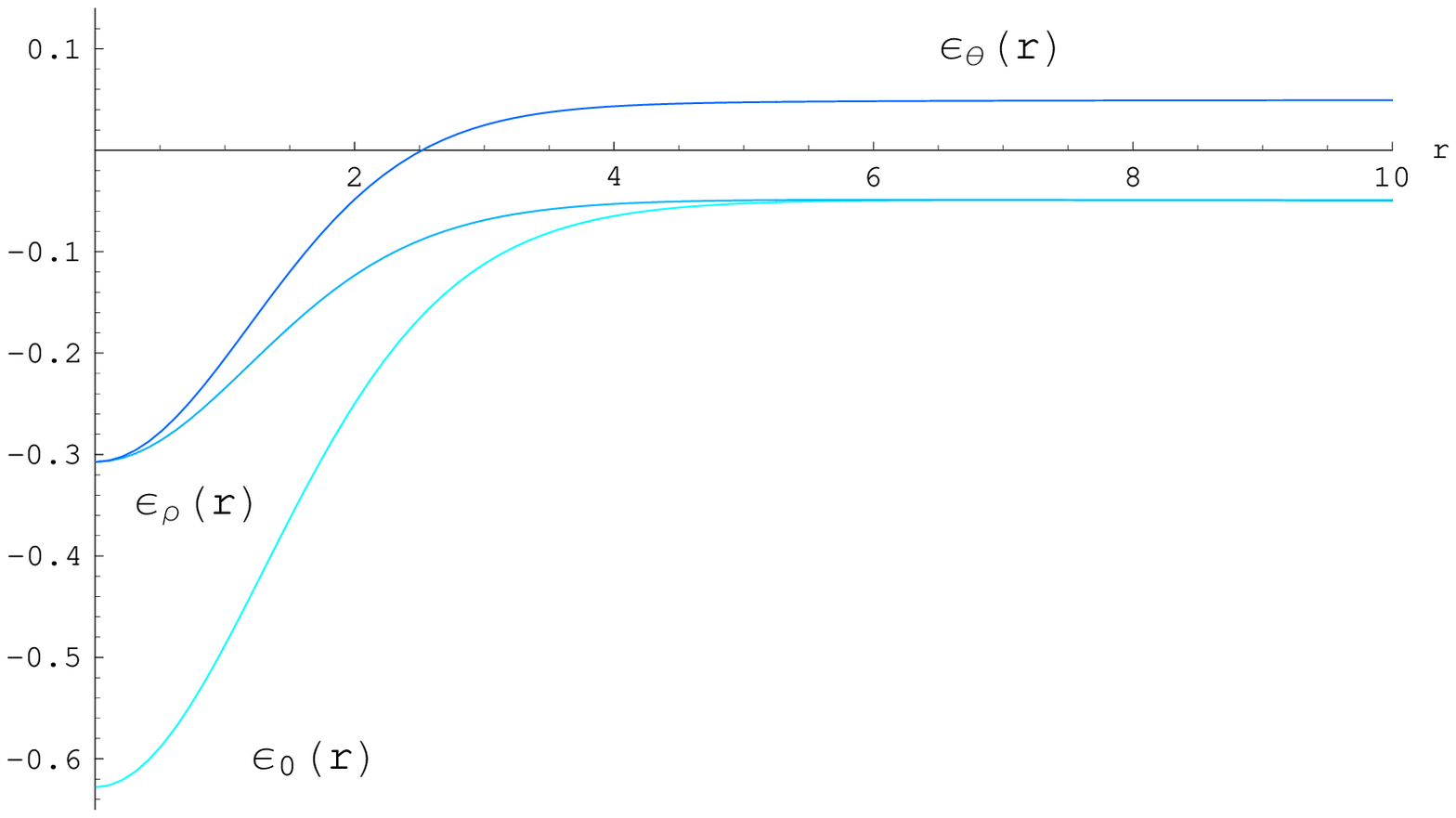}}
\caption[a]{\it Stress-energy components for the solution given in Fig.~\ref{4sol135}. \\
$(\alpha=1.00000000, \, \beta=3.50000000, \, \gamma=0.02678351)$} \label{4sol135_e}
\end{figure}
\begin{figure}
\centerline{ \epsfxsize = 13cm \epsfbox {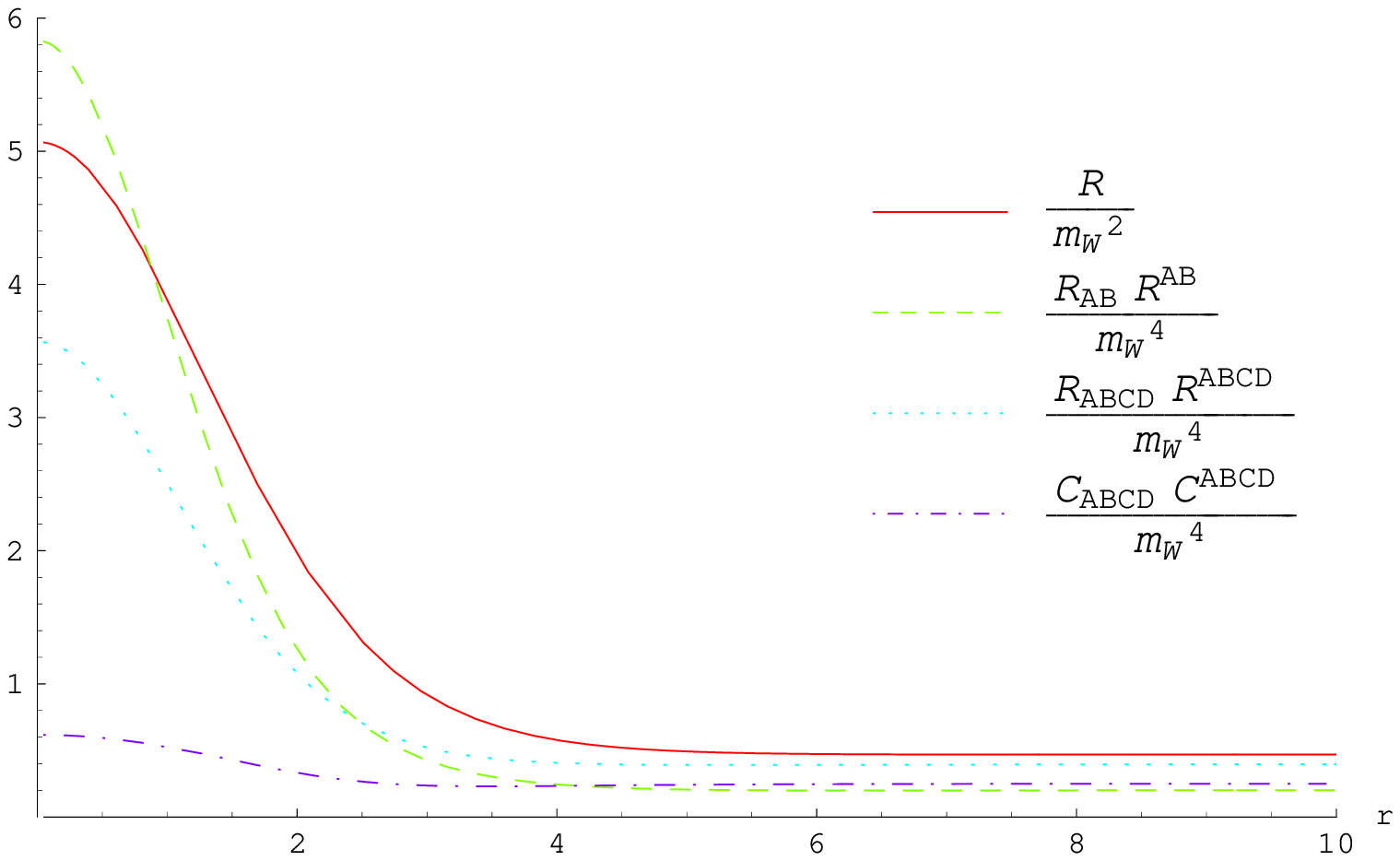}}
\caption[a]{\it Curvature invariants for the solution given in Fig.~\ref{4sol135}. \\
$(\alpha=1.00000000, \, \beta=3.50000000, \, \gamma=0.02678351) $}\label{4sol135_c}
\end{figure}
\clearpage
\begin{figure}
\centerline{ \epsfxsize = 13cm \epsfbox {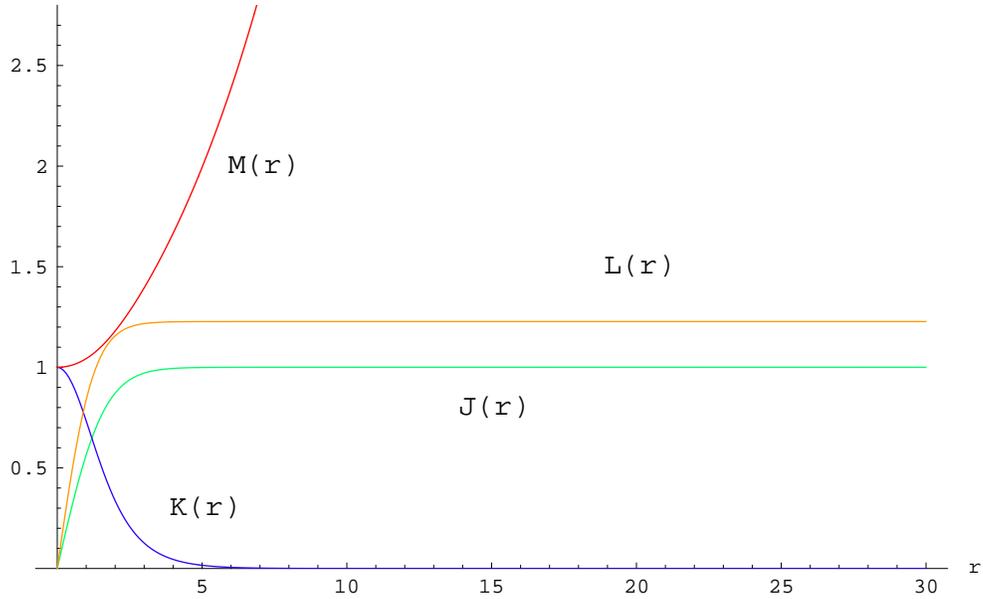}}
\caption[a]{\it Example of a solution that does not localize gravity $(c<0)$ corresponding to the parameter values
$(\alpha=1.00000000, \, \beta=1.80000000, \, \gamma=0.04053600) $} \label{4sol118}
\end{figure}
\begin{figure}
\centerline{ \epsfxsize = 13cm \epsfbox {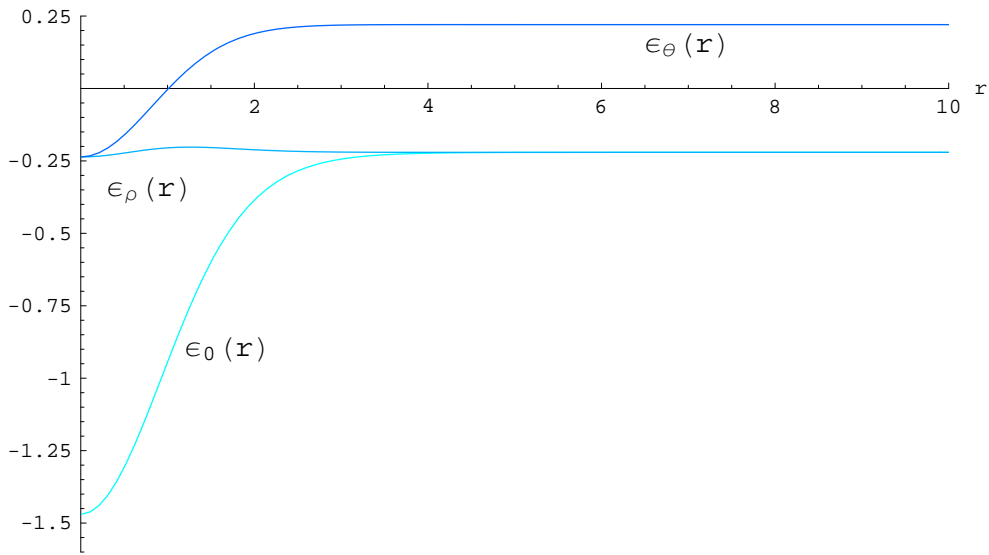}}
\caption[a]{\it Stress-energy components for the solution given in Fig.~\ref{4sol118}. \\
$(\alpha=1.00000000, \, \beta=1.80000000, \, \gamma=0.04053600)$} \label{4sol118_e}
\end{figure}
\clearpage
\newpage

\section{Prasad-Sommerfield limit $(\alpha=0)$} \label{4PSL}
The Prasad-Sommerfield limit $(\alpha=0)$ was explored numerically for $\beta$-values ranging from $0.4$ to 
about $70$. The corresponding intersection of the fine-tuning surface Fig.~\ref{4FTS} and the plane $\alpha=0$ 
is given in Fig.~\ref{4PSlimit}. There exist no gravity localizing solutions as the separating line in 
Fig.~\ref{4FTS} indicates. The point in Fig.~\ref{4PSlimit} corresponds to the solution shown in Figs.~\ref{4sol036} and
\ref{4sol036_e}. One sees that $\gamma$
tends to zero for $\beta$ going to infinity and that $\gamma$ tends to $-\infty$ for $\beta$ going to zero. 
\begin{figure}[h]
\centerline{ \epsfxsize = 13cm \epsfbox {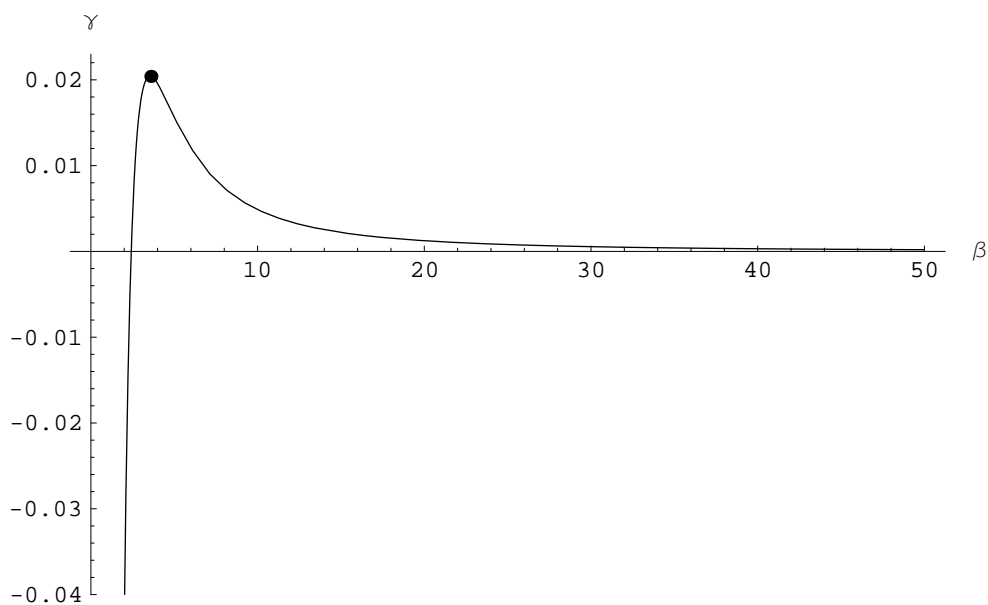}}
\caption[a]{\it Section of the fine-tuning surface Fig.~\ref{4FTS} corresponding to the Prasad-Sommerfield limit $\alpha=0$. 
None of the shown combinations of $\beta$ and $\gamma$-values correspond to solutions that lead to a finite
transverse volume. The point indicates the sample solution given below.}
\label{4PSlimit}
\end{figure}
\begin{figure}[htbp]
\centerline{ \epsfxsize = 13cm \epsfbox {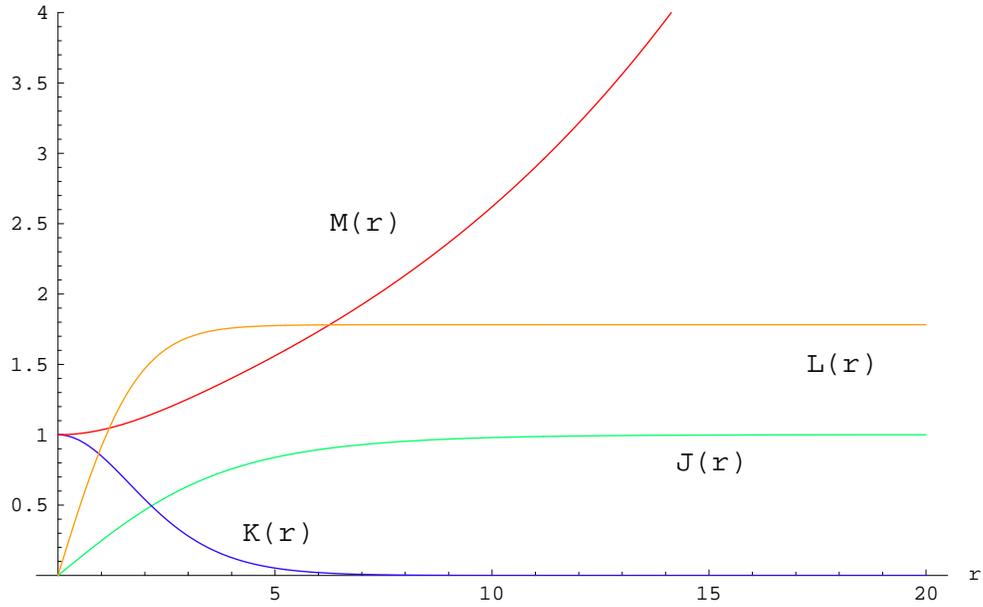}}
\caption[a]{\it Sample solution for the Prasad-Sommerfield limit corresponding to the parameter values  
$(\alpha=0.00000000, \, \beta=3.60000000, \, \gamma=0.02040333) $}
\label{4sol036}
\end{figure}

\begin{figure}[htbp]
\centerline{ \epsfxsize = 13cm \epsfbox {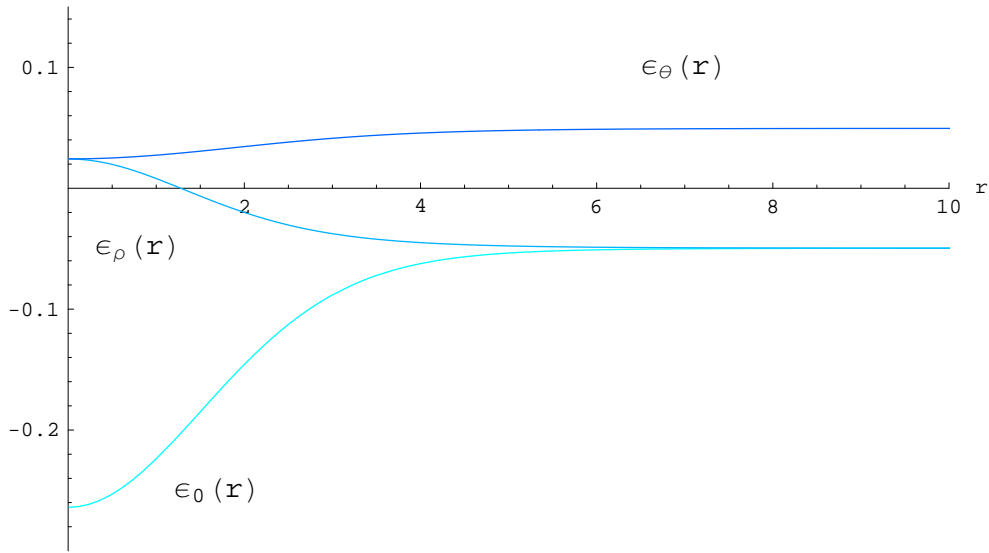}}
\caption[a]{\it Stress-energy components for the solution given in Fig.~\ref{4sol036}.\\
$(\alpha=0.00000000, \, \beta=3.60000000, \, \gamma=0.02040333)$}
\label{4sol036_e}
\end{figure}
\clearpage

\section{Fine-monopole limit - physical requirements on solutions}\label{4FML}
The fine monopole can be characterized by $c \ll 1$. As anticipated in section \ref{4samplesol}, $c=0$ 
in eq.~(\ref{4Einsteinasymptocsqr}) immediately leads to  
$\beta^2 \gamma=\frac{\Lambda_7}{e^2 \, M_7^{10}}=\frac{1}{2}$. Hence, we deduce that the 
fine-monopole limit can
not be realized for a negative bulk cosmological constant. In addition, from
(\ref{4Einsteinasympto1overLsqr}) it follows that $\mathcal{L}_0 = \sqrt{\beta}$. 
The $c\ll 1$ limit is qualitatively different from its analogue in the $6D$-string case 
\cite{Giovannini:2001hh}. The 
solutions do not correspond to strictly local defects. Einstein equations never 
decouple from the field equations.
Due to the particular metric asymptotic (\ref{4asympto}), the stress-energy tensor components tend to constants at infinity in transverse space. In the $6$-dimensional string case the fine-string limit was realized as a 
``strictly local'' defect (in the language of  chapter \ref{cha3}) having 
stress-energy vanishing exponentially outside the string core.
Despite this difference, the discussion of the physical requirements which we have to impose 
on our solutions is very similar. In the following we show that
in the fine-monopole limit the dimensionful parameters of the system 
$\Lambda_7, \,  M_7, \, m_W, \, \lambda, \, e $ can be chosen in such a way that
all of the following physical requirements are simultaneously satisfied:
\begin{enumerate}
  \item $M_P^2$ equals $ (1.22 \cdot 10^{19} \, \mbox{GeV})^2$. \label{4R1}
  \item The corrections to Newtons law do not contradict latest measurements.\label{4R2}
  \item Classical gravity is applicable in the bulk.\label{4R3}
  \item Classical gravity is applicable in the monopole core $(r=0)$.\label{4R4}
\end{enumerate}
To find solutions with the above mentioned properties, it is possible to restrict oneself to a particular
value of $\alpha$, e.g. $\alpha=\frac{1}{2}$. This choice corresponds to equal vector and Higgs masses $m_W=m_H$.
Even though extra dimensions are infinite, the fact that $M(r)$ decreases exponentially permits the definition of 
an effective ``size'' $r_0$ of the extra dimensions:
\begin{equation} \label{4sizeofexdim}
  M=M_0 \, e^{-\frac{c}{2}r}=M_0 \, e^{-\frac{c \, m_W}{2}\frac{r}{m_W}} \quad \Rightarrow \quad r_0 \equiv \frac{2}{c \, m_W} \, .
\end{equation}
In order to solve the hierarchy problem in similar lines to \cite{Arkani-Hamed:1998rs,Antoniadis:1998ig} 
we parametrize the fundamental gravity 
scale as follows:
\begin{equation} \label{4fundgravscale}
  M_7= \kappa \, 10^3 \, \mbox{GeV} \, .
\end{equation}
$\kappa=1$ then sets the fundamental scale equal to the electroweak scale.
\begin{enumerate}
  \item The expression (\ref{4PlanckMass}) for the square of the Planck mass $M_P^2$ can be approximated in the fine-monopole 
  limit by using the asymptotic (\ref{4asympto}) for the metric in the integral (\ref{4PlanckMass}) rather than
 the exact (numerical) solutions. This gives
  \begin{equation}
     M_P^2 \approx \frac{4 \pi M_7^5}{m_W^3}\, M_0^2 \, \mathcal{L}_0^2 \, \frac{1}{c} \, . 
  \end{equation}
  By using one of Einstein's equations at infinity 
\begin{equation} \label{4EinsteininfI}
  \mathcal{L}_0^2=\frac{1}{4 c^2+2 \beta \gamma} \, ,
\end{equation}
  and by developing to lowest order in $c$ one finds
  \begin{equation}
   M_P^2 \approx \frac{4 \pi M_7^5 M_0^2}{m_W^3} \frac{1}{2 \beta \gamma} \left(\frac{1}{c}+ \mathcal{O}(c) \right) \, .  
  \end{equation}
Numerical solutions for $c \to 0$ and $\alpha=\frac{1}{2}$ converge to the following approximate
parameter values 
\begin{eqnarray} \label{4finemononum}
  \beta &=& 3.266281 \, ,\nonumber \\
  \gamma &=& 0.0468665 \, ,\nonumber \\
  \beta^2 \gamma &=& 0.4999995 \, ,\nonumber\\
  M_0 &=& 0.959721 \, .
\end{eqnarray}
The above values were obtained by extrapolation of solutions to $c=0$, see Figs.~\ref{4fmlbeta} to \ref{4fmlM0}. 
Therefore, the errors are of order $10^{-6}$ which is considerably higher than average errors from the integration which were smaller than $10^{-8}$.  

\begin{figure}[htbp]
\centerline{\epsfxsize = 12.5cm \epsfbox {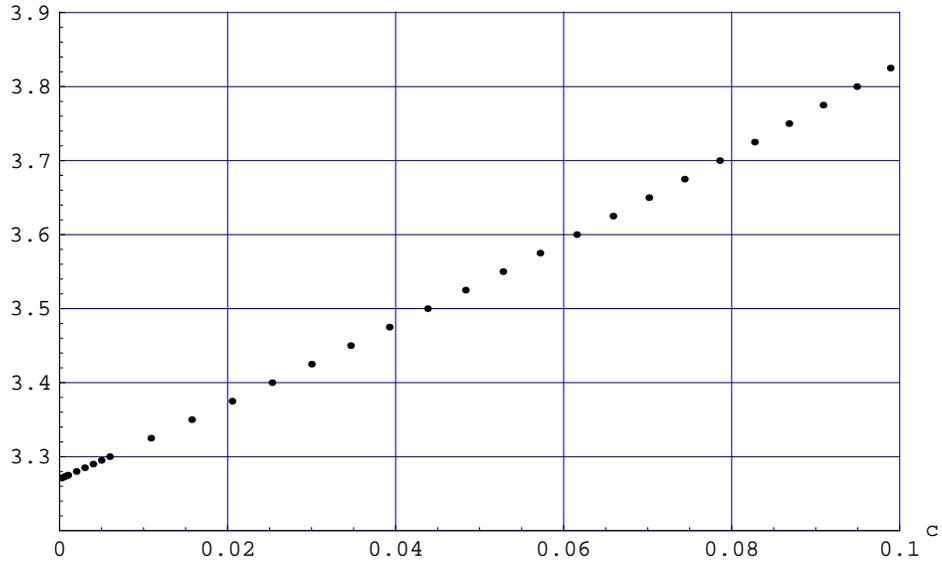}}
\caption[a]{\it Behavior of $\beta$ in the fine-monopole limit $c \to 0$ for $\alpha=\frac{1}{2}$.}
\label{4fmlbeta}
\end{figure}
\begin{figure}[htbp]
\centerline{\epsfxsize = 12.5cm \epsfbox {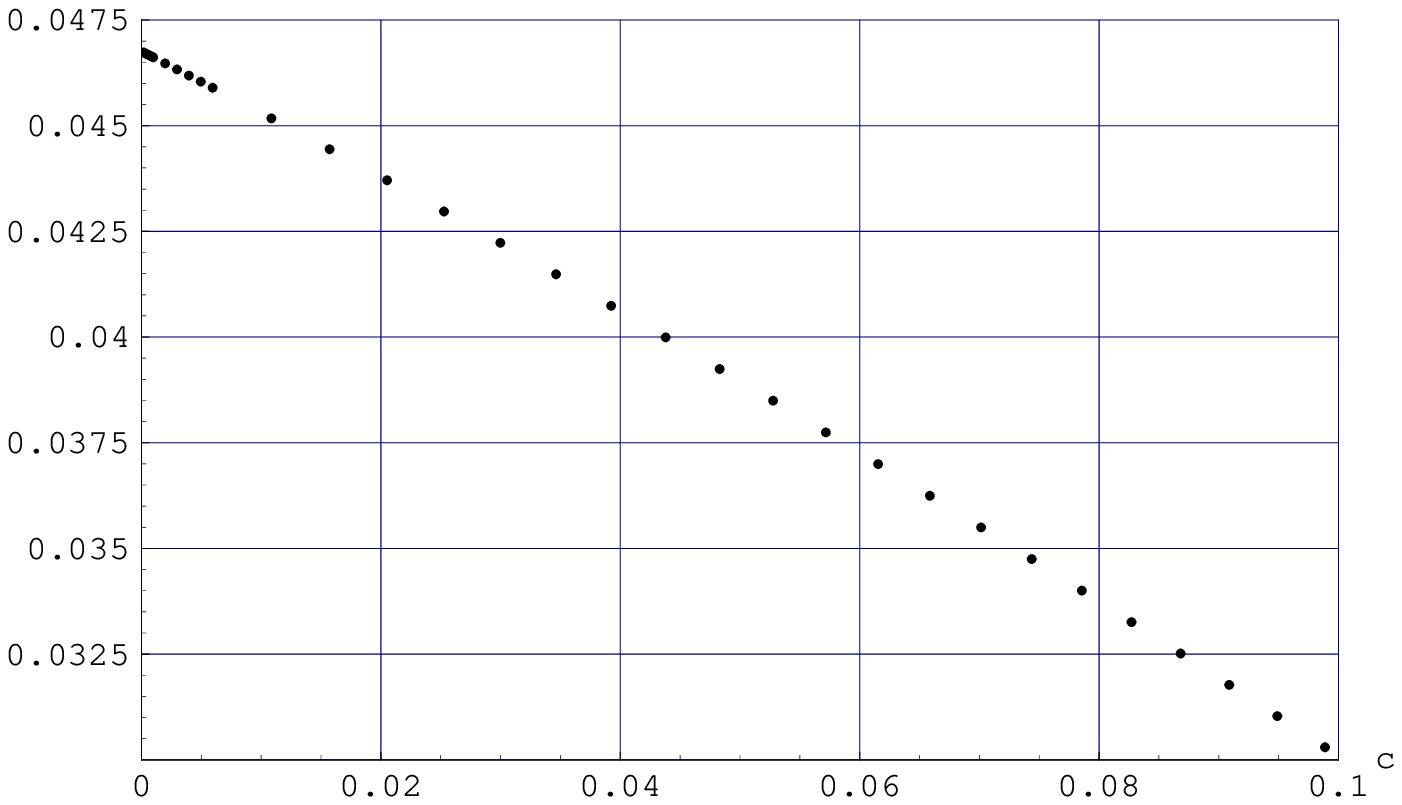}}
\caption[a]{\it Behavior of $\gamma$ in the fine-monopole limit $c \to 0$ for $\alpha=\frac{1}{2}$.}
\label{4fmlgamma}
\end{figure}
\begin{figure}[htbp]
\centerline{\epsfxsize = 12.5cm \epsfbox {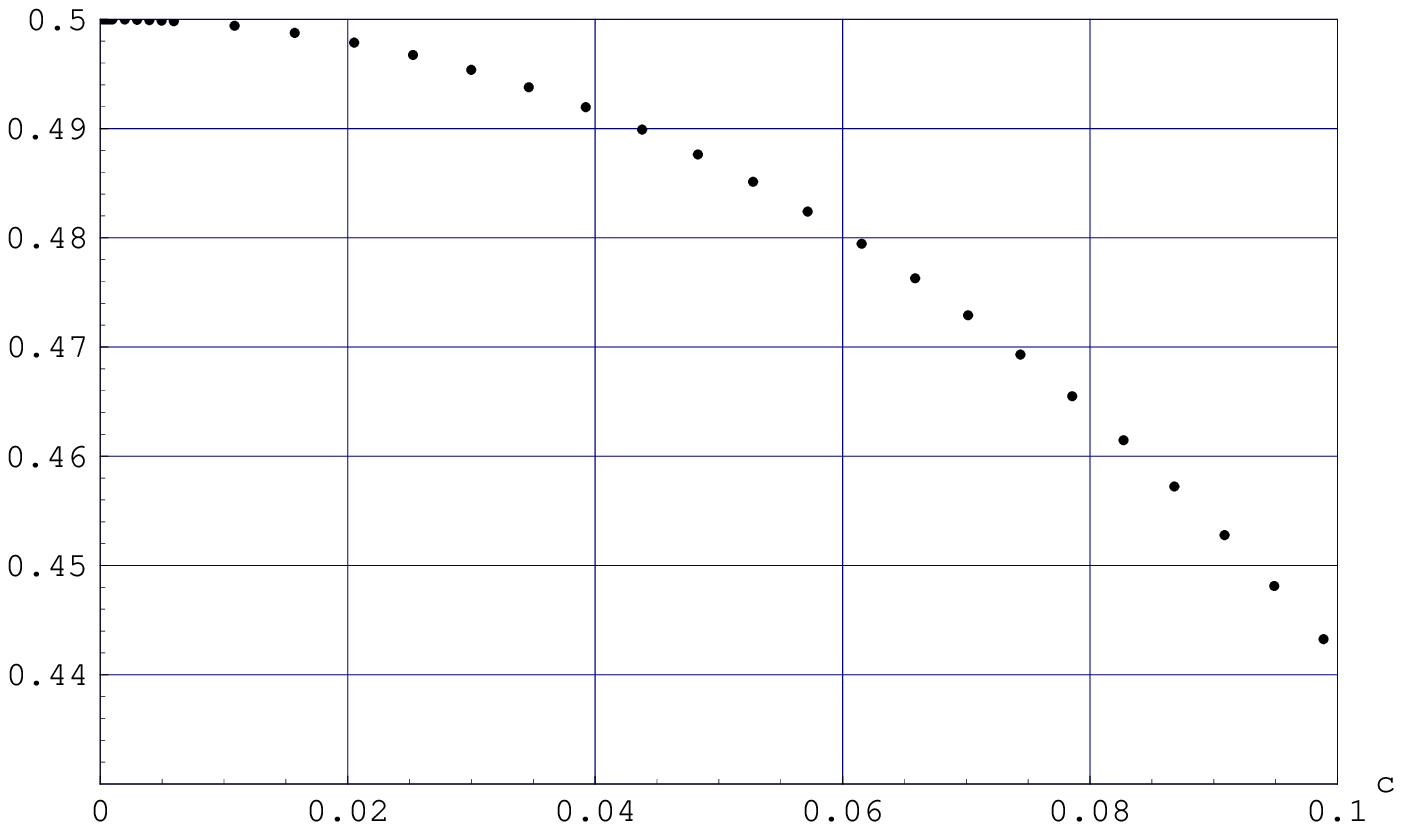}}
\caption[a]{\it Behavior of $\beta^2 \gamma=\frac{\Lambda_7}{e^2 \, \eta^4}$ in the fine-monopole limit $c \to 0$ for
 $\alpha=\frac{1}{2}$.}
\label{4fmlbgsqr}
\end{figure}
\begin{figure}[htbp]
\centerline{\epsfxsize = 12.5cm \epsfbox {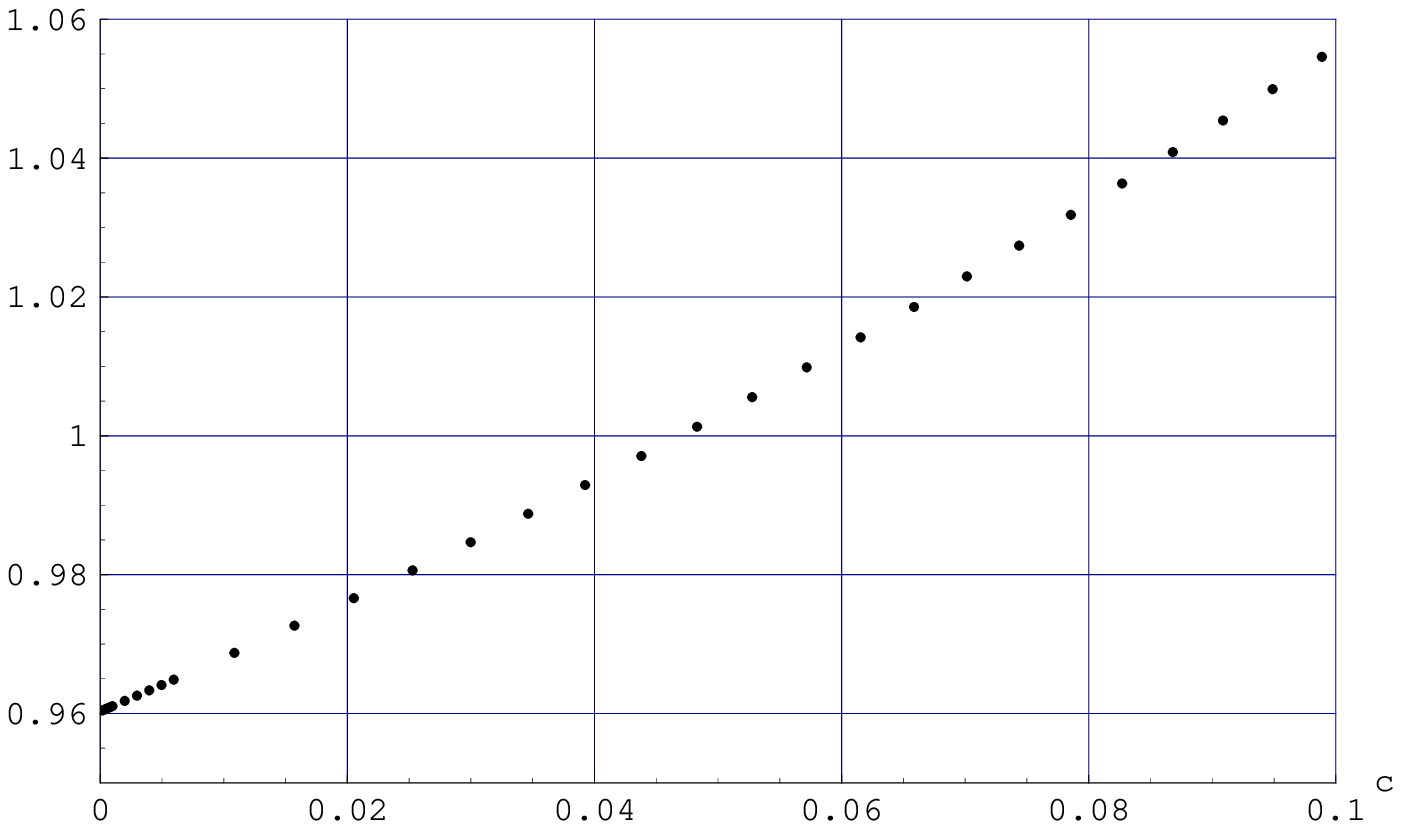}}
\caption[a]{\it Behavior of $M_0$ in the fine-monopole limit $c \to 0$ for $\alpha=\frac{1}{2}$.}
\label{4fmlM0}
\end{figure}

Note that $\beta^2 \gamma$ tends to $\frac{1}{2}$ in the fine-monopole limit (see Fig.~\ref{4fmlbgsqr}). Neglecting all orders different from $\frac{1}{c}$ 
in the above expression for $M_P^2$ and using (\ref{4sizeofexdim}) one has
  \begin{equation} \label{4Wmass}
    m_W=1.13 \cdot \, 10^{-5} \, \frac{\kappa^{5/2}}{\xi^{1/2}} \mbox{GeV} \, ,
  \end{equation}
  where $r_0$ has been parametrized by $r_0=\frac{0.2 \, \mbox{\tiny mm}}{\xi}\approx 10^{12}\, \mbox{GeV}^{-1} 
   \cdot \frac{1}{\xi}$.
  \item Since Newton's law is established down to $0.2 \, \mbox{mm}$ \cite{Hoyle:2000cv}, 
we simply need to have $\xi>1$.
  \item In order to have classical gravity applicable in the bulk, we require that curvature at infinity
is negligible with respect to the corresponding power of the fundamental scale. By looking at 
eqs.~(\ref{4curvinvinfinity1})--(\ref{4curvinvinfinity4}), we see that we have to impose 
  \begin{equation} 
    c^2 \, m_W^2 \ll M_7^2 \quad \mbox{and} \quad \frac{m_W^2}{\mathcal{L}_0^2} \ll M_7^2 \, .
  \end{equation}
Using (\ref{4EinsteininfI}) and the first of the above relations, the second one can immediately be transformed into 
\begin{equation}
  \beta \, \gamma \, m_W^2 \ll M_7^2 \, .
\end{equation}
One then finds 
  \begin{equation} \label{4reqiii}
    2 \cdot 10^{-15} \frac{\xi}{\kappa} \ll 1 \quad \mbox{and} \quad
    4 \cdot 10^{-3} \left( \frac{\kappa^3}{\xi} \right)^{1/2} \ll 1 \, ,
  \end{equation}
  where $\beta$ and $\gamma$ have again been replaced by their limiting values (\ref{4finemononum}) 
  for $c \to 0$. 
  \item \label{4} Classical gravity is applicable in the monopole core whenever the curvature invariants
  $R^2$, $R_{A B} R^{A B}$, $R_{A B C D} R^{A B C D}$ and $C_{A B C D} C_{A B C D}$ are small compared to the forth 
power of the fundamental gravity scale. Since these quantities are of the order of the mass $m_W^4$ 
(see (\ref{4curvatorigin1})--(\ref{4curvatorigin4})) we have
\begin{equation} \label{4reqiv}
  m_W^4 \ll M_7^4 \, .
\end{equation} 
Using (\ref{4fundgravscale}) and (\ref{4Wmass}) one finds
\begin{equation}
  1.1 \cdot 10^{-8} \left( \frac{\kappa^3}{\xi} \right)^{1/2} \ll 1 \, .
\end{equation}
\end{enumerate}
Finally, the fine-monopole limit requires 
\begin{equation} \label{4thincondition}
  c=\frac{2}{r_0 \, m_W} \ll 1 \, .
\end{equation}
Again (\ref{4Wmass}) implies
\begin{equation} \label{4thinmonolimit}
  1.8 \cdot 10^{-7} \left( \frac{\xi^3}{\kappa^5} \right)^{1/2} \ll 1 \, .
\end{equation}
It is now easy to see that for a wide range of parameter combinations
all these requirements on $\xi$ and $\kappa$ can simultaneously be satisfied.  
One possible choice is $\xi=100$ and $\kappa=1$. This shows that already in the
case $\alpha=\frac{1}{2}$ there are physical solutions corresponding to a fine monopole 
in the sense of eq.~(\ref{4thincondition}), respecting all of the above requirements (\ref{4R1})-(\ref{4R4}).

\section{Conclusions}
As we have seen in this chapter, it is possible to realize the idea of warped compactification on 
a topological defect in a higher-dimensional space-time with $n=3$ transverse dimensions by 
considering a specific field theoretical model. Numerical solutions were found in the case of a 
't~Hooft-Polyakov monopole residing in the $3$-dimensional transverse space of a $7$-dimensional 
bulk space-time. As discussed in the general case in section \ref{SectionBulkpform}
of chapter \ref{cha3}, the 
transverse space approaches a constant curvature space at spatial infinity, however, not necessarily 
an anti-de-Sitter space. Both signs of the bulk cosmological constant are possible in order to 
localize gravity. 
We also considered a fine-monopole limit in the case $\alpha=1/2 \;\; (m_W=m_H)$ and verified 
explicitly that the model proposed is not in conflict with Newton's law, that it leads to a possible 
solution of the hierarchy problem and that classical gravity is applicable in the bulk and in the core 
of the defect.

In the next and last chapter of this work, we are going to study a brane-world scenario which in many 
respects resembles the Randall-Sundrum II model. We are going to treat the case of a singular 
$3$-brane embedded in a $5$-dimensional anti-de-Sitter bulk space-time. Contrary to the 
Randall-Sundrum II model, however, we consider a brane with induced metric given by 
Einstein's static universe rather than by $4$-dimensional Minkowski space-time. One of the advantages 
of such a setup, compared to the Randall-Sundrum II case, is that the corresponding space-time manifold 
turns out to be geodesically complete.  
\chapter{Einstein static universe as a brane in extra dimensions} \label{cha5}

\section{Introduction}
\label{5ESUintro}

Recent proposals of large \cite{Antoniadis:1990ew}--\cite{Antoniadis:1998ig} and infinite 
\cite{Rubakov:bb,Randall:1999ee,Randall:1999vf} extra dimensions provide new opportunities for 
addressing several outstanding problems of modern theoretical 
physics like for example the hierarchy problem 
\cite{Antoniadis:1990ew}--\cite{Cohen:1999ia} 
or the cosmological constant problem \cite{Rubakov:1983bz}--\cite{Randjbar-Daemi:1985wg}.
These proposals triggered an immense research activity in theories involving $3$-branes with
interests ranging from elucidating the global space-time structure of brane world scenarios, properties 
of gravity, cosmology and brane cosmological perturbations, generalizations to higher dimensions etc.
While the possibilities are rich, realistic scenarios remain rare. For example, the simple Randall-Sundrum II
model \cite{Randall:1999vf} faces the problem of being geodesically incomplete 
\cite{Rubakov:2001kp,Muck:2000bb,Gregory:2000rh}. 
It is therefore reasonable to look for adequate alternatives or generalizations 
to the Randall-Sundrum II model which avoid amongst others the 
above mentioned problem while sharing its pleasant feature of an effective $4$-dimensional low
energy gravity on the brane.

In this final chapter, we present a $5$-dimensional brane-world model which solves the geodesic
incompleteness of the Randall-Sundrum II model while preserving its phenomenological 
properties concerning the localization of gravity. 

We try to illustrate our motivation for considering a particular geometry by using the simple 
picture of a domain structure in extra dimensions, resulting from a spontaneously broken 
discrete symmetry. The associated Higgs-field takes different values in regions separated by domain walls, 
which restricts the possibilities of combining domain walls depending on the global topology of the 
space-time under consideration. For simplicity we restrict ourselves to five dimensions.
Let us analyze a few simple cases:
\begin{itemize}
{\item Non-compact extra dimension}: $y \in (-\infty, + \infty)$.
In this case, by choosing the origin $y=0$ to coincide with the position of the brane, we 
obtain two non-overlapping regions
$(-\infty, 0) $ and $(0, + \infty)$ and we can thus consistently have one brane in such a theory. 
An example for this configuration is provided by the Randall-Sundrum II model \cite{Randall:1999vf}.
{\item Compact extra dimension}: $y \in [0 , 2 \pi]$. We now consider two possibilities, depending on
the spatial topology of our manifold:
\begin{itemize}
{\item{\bf (a)}} $\mathbb{R}^3 \times S^1$: If the ordinary dimensions are supposed to be 
non-compact, it is not possible
to consistently put only one brane in the extra dimensions. At least two branes are needed.
{\item{\bf (b)}} $S^4$: If {\it all} spatial coordinates are part of a compact
manifold, the simple picture shown in Fig.~\ref{5ct} seems to suggest that it is possible to consistently put 
a single brane in the bulk space-time.
\end{itemize}
\end{itemize}
To the best of our knowledge, case {\bf (b)} has not yet been considered in the literature and 
our aim is to present such a construction.


This chapter is organized as follows: in Section \ref{5ESUone} we discuss the basic geometric and 
topological properties of our brane-world scenario like  Einstein equations, junction
conditions and the distance hierarchy between the extra dimension $R$ and the observable universe 
$R_U$. Section \ref{5ESUtwo} is dedicated to the study of geodesics and the demonstration, that our model 
does not suffer from being geodesically incomplete. In section \ref{5ESUthree} we present a detailed 
computation 
of the propagator of a massless scalar field in the given background, serving as a easy, phenomenological 
approach to the study of gravity. We discuss the behavior of the two-point function in three different 
distance regimes. It turns out that the computations are rather technical and we therefore collect large 
parts of it in four appendices. We eventually draw conclusions in section \ref{5ESUfour}.


\begin{figure}[htpb]
\centerline{\epsfxsize=3.0in\epsfbox{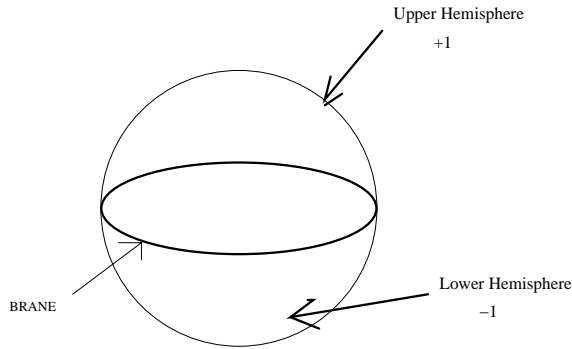}}
\caption{$S^3$-brane embedded in $S^4$. $+1$ and $-1$ symbolically 
represent the different vacuum expectation values taken by the Higgs field in different regions of space.}
\label{5ct}
\end{figure}

\section{The background equations}
\label{5ESUone}
The aim of this section is to present the topology and geometry of the brane-world model 
which we set out to study but also to motivate how this particular model emerged through 
imposing several physical conditions on the more general set of solutions.

\subsection{Einstein equations}

We would like to construct a space-time with all spatial dimensions being part of the 
same compact manifold. One more motivation for this lies in a possible solution of the strong 
CP-problem within theories with extra dimensions \cite{Khlebnikov:1987zg}.  
The overall topology would then be given by $\mathbb{R}\times \mathbb{K}$,
where $\mathbb{R}$ represents the time-coordinate and $\mathbb{K}$ any compact manifold. 
For simplicity we will restrict ourselves to the case of one extra dimension and an induced metric 
on the brane characterized by a spatial component of geometry $S^3$. The idea is to combine two 
$5$-dimensional regions of space-time dominated by a cosmological constant $\Lambda$ in such a 
way, that the border of the two regions can be identified with a $3$-brane, constituting 
our observable universe. As pointed out in \cite{Khlebnikov:1987zg} the manifold $\mathbb{K}$ has to be highly 
anisotropic in order to single out the small extra dimension from the three usual ones. It
is not a priory clear whether the Einstein equations allow for such solutions at all and if so whether 
gravity can be localized on the brane in such a setup.

We choose the following ansatz for the $5$-dimensional metric consistent with the above 
requirements:
\begin{eqnarray} \label{5metric}
  ds^2 = g_{M N} dx^M dx^N=-\sigma^2(\theta)\, dt^2 + R_U^2 \, \gamma ^2(\theta) 
         \, d\Omega^2_3 + R^2 \, d\theta^2 \, ,
\end{eqnarray}
where $\theta \in [-\frac{\pi}{2},\frac{\pi}{2}]$ denotes the extra coordinate and 
$R$ and $R_U$ are constants representing the size 
of the extra dimension and the size of the observable universe at present, respectively.
$d\Omega^2_3$ denotes the line element of a $3$-sphere:
\begin{equation} \label{5S3metric}
  d\Omega^2_3 =d\varphi^2_1 + \sin ^2 \varphi_1 \, d\varphi^2_2 + \sin ^2 \varphi_1\, \sin ^2 \varphi_2\,  
  d\varphi^2_3 \, ,
\end{equation}
where $\varphi_1, \varphi_2$ belong to the interval $\left[ 0, \pi \right]$, $ \varphi_3 $
to $\left[ 0, 2 \pi \right]$.
Capital latin letters $M,N,..$ will range from $0$ to $4$.
In order to obtain a compact space, we require $\gamma(\pm\frac{\pi}{2})=0$.

Metrics which can locally be put into the form (\ref{5metric}) are sometimes referred to as 
{\em asymmetrically warped} metrics due to two different functions ($\sigma$ and $\gamma$ in our 
notations) multiplying the temporal and spatial part of the $4$-dimensional coordinate 
differentials. Their relevance in connection with $4$-dimensional Lorentz-violation at 
high energies was first pointed out in \cite{Visser:1985qm}. More recent discussions of this subject 
can be found in \cite{Csaki:2000dm},\cite{Dubovsky:2001fj} or \cite{Rubakov:2001kp} and 
references therein. A common prediction of theories of this kind is that dispersion 
relations get modified at high energies. 
For a field-theoretical discussion of Lorentz violating effects 
in the context of the Standard Model of particle physics see \cite{Coleman:1998ti}.

The Einstein equations in 5 dimensions with a bulk cosmological constant 
$\Lambda$ and a stress-energy tensor $T_{M N}$ take the following form:
\begin{eqnarray} \label{5einsteineq}
  R_{M N} -\frac{1}{2} \, R \, g_{M N} + \Lambda \, g_{M N}
  = \frac{8 \pi}{M^3} \, T_{M N} \, ,
\end{eqnarray}
where $M$ is the fundamental scale of gravity.
We choose to parametrize the stress-energy tensor in the following way, consistent with the symmetries
of the metric:
\begin{eqnarray} \label{5stress}
  T^0_{\;\;0}=\epsilon_0(\theta) < 0\, , \quad
  T^i_{\;\;j}= \delta^i_j\,\epsilon(\theta)\, , \quad
  T^\theta_{\;\;\theta}=\epsilon_\theta(\theta) \, ,
\end{eqnarray}
where, as indicated, the above diagonal components depend only on the extra coordinate~$\theta$.
Lower case latin indices $i,j$ label the coordinates on $S^3$. 

Using the metric ansatz (\ref{5metric}) together with the stress-energy tensor (\ref{5stress}) 
the Einstein equations (\ref{5einsteineq}) become
\begin{eqnarray}
\frac{3}{R^2} \left[ \left( \frac{\gamma^{\prime}}{\gamma}\right)^2 + \frac{\gamma^{\prime \prime}}{\gamma} -
 \left( \frac{R}{R_U} \right)^2 \frac{1}{\gamma ^2} \right] + \Lambda &=& \frac{8 \pi}{M^3} 
  \epsilon_0 \, , \label{500} \\
\frac{1}{R^2} \left[ \left( \frac{\gamma ^{\prime}}{\gamma}\right)^2 +
2 \frac{\gamma^{\prime}}{\gamma} \frac{\sigma^{\prime}}{\sigma}
+2 \frac{\gamma ^{\prime \prime}}{\gamma} + \frac{\sigma^{\prime \prime}}{\sigma} -
 \left( \frac{R}{R_U} \right)^2 \frac{1}{\gamma^2} \right] + \Lambda &=& \frac{8 \pi}{M^3} 
  \epsilon \, , \label{5ii} \\
\frac{3}{R^2} \left[ \left( \frac{\gamma^{\prime}}{\gamma}\right)^2 +
\frac{\gamma^{\prime}}{\gamma} \frac{\sigma^{\prime}}{\sigma} -
 \left( \frac{R}{R_U} \right)^2 \frac{1}{\gamma ^2} \right] + \Lambda &=& \frac{8 \pi}{ M^3} 
  \epsilon_{\theta} \, , \label{5thetatheta}
\end{eqnarray}
where $^{\prime}$ denotes differentiation with respect to $\theta$.
The conservation of stress-energy, or equivalently, the Bianchi-identities lead to the following constraint 
relating the three independent components $\epsilon_0$, $\epsilon$ and $\epsilon_\theta$:
\begin{eqnarray}\label{5constraint}
  \epsilon^\prime_{\theta} + \left( \frac{\sigma^{\prime}}{\sigma} +
  3 \frac{\gamma^{\prime}}{\gamma}\right) \, \epsilon_{\theta} -
  \frac{\sigma^{\prime}}{\sigma}\, \epsilon_{0} - 3 \, \frac{\gamma^{\prime}}{\gamma} \epsilon =0 \, .
\end{eqnarray}

\subsection{Vacuum solution}
We do not intend to provide a field theoretical model which could 
generate the geometry we are about the describe. Our aim is to study a singular brane, located at
$\theta_b=0$ separating two vacuum regions in the bulk. Of course the solutions to the Einstein 
equations in vacuum with a cosmological constant term are nothing but the familiar 
de-Sitter ($\Lambda>0$), Minkowski ($\Lambda=0$) and anti-de-Sitter ($\Lambda<0$) space-times.
To see how these geometries can be recovered using the line element (\ref{5metric}) we set
the right hand sides of (\ref{500})--(\ref{5thetatheta}) equal to zero and subtract eq.~(\ref{5thetatheta}) 
from eq.~(\ref{500}) to obtain $\sigma = c \, \gamma^{\prime}$, where $c$ is a constant.
Putting this back in eqs.~(\ref{500})--(\ref{5thetatheta}), we are left with only one independent differential
equation  
\begin{align}\label{500n}
  \frac{3}{R^2} \left[ \left( \frac{\gamma^{\prime}}{\gamma}\right)^2 + 
  \frac{\gamma^{\prime \prime}}{\gamma} -
   \left( \frac{R}{R_U} \right)^2 \frac{1}{\gamma^2} \right] + \Lambda = 0 \, .
\end{align}
The solution of (\ref{500n}) is trivial and by means of $\sigma = c \, \gamma^{\prime}$ we find 
\begin{equation} \label{5sigmaandgamma}
  \gamma(\theta) =
  \frac{\sinh \left[\omega\left(\frac{\pi}{2}-|\theta| \right)\right]}
     {\sinh (\omega \frac{\pi}{2})}\, ,\qquad
  \sigma (\theta) =
  \frac{\cosh \left[\omega\left(\frac{\pi}{2}-|\theta| \right)\right]}
    {\cosh (\omega \frac{\pi}{2})}\, ,
\end{equation}
provided that $R_U = R \sinh (\omega \frac{\pi}{2})/\omega$ and $\omega^2 =  -\Lambda  R^2 /6 $.
Notice that the solution (\ref{5sigmaandgamma}) does not contain any integration constant because we 
already imposed the boundary conditions $\gamma(\pm\frac{\pi}{2})=0$  
and $\gamma(0)=\sigma(0)=1$. Moreover, we chose the whole setup to be symmetric under the transformation
$\theta \to -\theta$.
From (\ref{5sigmaandgamma}) it is now obvious that locally (for $\theta>0$ and $\theta<0$) 
the line element (\ref{5metric}) correctly describes de-Sitter, Minkowski and anti-de-Sitter space-times for 
imaginary, zero and real values of $\omega$, respectively. For our purposes only the $AdS$ solution will be
of any interest as we will see in the following section. A simple change of coordinates starting from
(\ref{5metric}) and (\ref{5sigmaandgamma}) shows that the $AdS$-radius in our notations is given by 
$R_{AdS}=R/\omega=\sqrt{-6/\Lambda}$.\footnote{See also footnote \ref{5changecoords}.} 
At this stage, the validity of the vacuum solution 
(\ref{5sigmaandgamma}) is restricted to the bulk, since it is not even differentiable in the classical 
sense at $\theta=0$. In order to give sense to (\ref{5sigmaandgamma}) for all values of $\theta$ we will 
have to allow for some singular distribution of stress-energy at $\theta=0$ and solve 
(\ref{500})--(\ref{5thetatheta}) in the sense of distributions.

\subsection{The complete solution for a singular brane}
\label{5SingularBrane}
As announced, we now refine our ansatz for the energy-momentum tensor (\ref{5stress}) 
to allow for a solution of eqs.~(\ref{500})--(\ref{5thetatheta}) in the whole interval 
$-\pi/2\leq\theta\leq\pi/2$:
\begin{align} \label{5stressrefined}
  \epsilon_0 (\theta ) = c_0 \, \frac{\delta(\theta)}{R} \, ,\quad  
  \epsilon (\theta ) = c  \, \frac{\delta(\theta)}{R} \, ,\quad 
  \epsilon_{\theta} (\theta ) = c_\theta \, \frac{\delta(\theta)}{R} \, .
\end{align}
After replacing the above components (\ref{5stressrefined}) in eqs.~(\ref{500})--(\ref{5thetatheta}) and 
integrating over $\theta$ from $-\eta$ to $\eta$, followed by the limit $\eta \to 0$ we find:
\begin{align}  
  :\frac{\gamma^\prime}{\gamma}: \;\; &= \frac{8 \pi R}{3 M^3} c_0 \label{5FineTuneI} \, ,\\
  :\frac{\sigma^\prime}{\sigma}: \;+ \;2:\frac{\gamma^\prime}{\gamma}: \;\; &=\frac{8\pi R}{M^3}c 
  \label{5FineTuneII} \, ,\\
  0&=c_\theta \, ,
\end{align}
where the symbol $: \ldots :$ is used to denote the jump of a quantity across the brane defined by:
\begin{equation} \label{5::def}
  : f : \;\; \equiv \lim_{\eta\to0} \left[f(\eta)-f(-\eta)\right].
\end{equation}
Note that in the above step we made use of the identity
\begin{equation}
  \frac{\gamma^{\prime \prime}}{\gamma} = \left(\frac{\gamma^\prime}{\gamma} \right)^\prime + 
   \left( \frac{\gamma^\prime}{\gamma} \right)^2 
\end{equation}
together with the continuity of $\gamma$ on the brane. Specifying (\ref{5FineTuneI}) and (\ref{5FineTuneII}) 
to (\ref{5sigmaandgamma}) we have:
\begin{align}
  c_0 &= - \frac{3}{4\pi} M^3 \frac{\omega}{R} \, \coth \left(\omega \frac{\pi}{2}\right) \, ,\label{5solc0}\\
  c_{\phantom{0}} &= - \frac{1}{4\pi} M^3 \frac{\omega}{ R} \, 
   \left[ \tanh \left(\omega \frac{\pi}{2}\right)+2 \coth \left(\omega \frac{\pi}{2}\right)\right] 
   \, , \label{5solc} \\
  c_\theta &= 0 \, . \label{5solctheta}
\end{align}
Eqs.~(\ref{5solc0}) and (\ref{5solc}) relate the energy-density and the pressure of the singular brane to the bulk cosmological constant $\Lambda$ (via $\omega$), the size of the extra dimension $R$ 
and the fundamental scale of gravity $M$. With the above relations (\ref{5solc0})--(\ref{5solctheta}) we 
can now interpret (\ref{5sigmaandgamma}) as a solution to the Einstein equations 
(\ref{500})--(\ref{5thetatheta}) in the sense 
of distributions. Also the stress-energy conservation constraint 
(\ref{5constraint}) is satisfied based on the identity $\delta(x) \, \mbox{sign}(x)=0$ again in the 
distributional sense. 

Note that $c_0/c \to 1$ in the limit $\omega \to \infty$. 
Moreover, for larger and larger values of $\omega$, $c_0$ and $c$ 
approach the brane tension of the Randall-Sundrum II model and the above eqs.~ (\ref{5solc0}) and 
(\ref{5solc}) merge to the equivalent relation in the Randall-Sundrum II case. This is no surprise since
taking the limit $\omega \to \infty$ corresponds to inflating and flattening the 3-brane 
so that we expect to recover the case of the flat Randall-Sundrum II brane.

We finish this section by the discussion of some physical properties of our manifold. We first  
observe that its spatial part is homeomorphic to a $4$-sphere $S^4$. 
This is obvious from the metric (\ref{5metric}) and the explicit expression for $\gamma$ given in 
(\ref{5sigmaandgamma}). Geometrically, however, our manifold differs from $S^4$ due to the high
anisotropy related to the smallness of the extra dimension.
The ratio of typical distance scales in the bulk and on the brane is given by
\begin{eqnarray} \label{5hierarchy}
  \frac{R}{R_{U}}= \frac{\omega}{\sinh \left( \frac{\omega\pi}{2}\right)} \, .
\end{eqnarray}
It is now clear that the above ratio (\ref{5hierarchy}) can only be made very small in the case of 
real $\omega$ ($AdS$-space-time). For the size of the observable universe we take the lower bound 
$R_U > 4 \mbox{Gpc}\sim 10^{28}\,\mbox{cm}$ while the size of the extra dimension is 
limited from above 
\cite{Hoyle:2000cv}: $R<10^{-2}\,\mbox{cm}$, leaving us with $\omega > 50$.

Finally we would like to point out a very peculiar property of the manifold under consideration: 
as it can immediately be deduced from the line element (\ref{5metric}) and (\ref{5sigmaandgamma}),
any two points on the brane are separated by not more than a distance of the order of $R$, regardless of their 
distance as measured by an observer on the brane using the induced metric.

\section{Geodesics}
\label{5ESUtwo}

It is well known, that the Randall-Sundrum II model is time-like and light-like geodesically incomplete 
\cite{Rubakov:2001kp,Muck:2000bb,Gregory:2000rh} which means that there exist inextendible time-like and light-like geodesics.\footnote{For a precise
definition of a geodesically incomplete space see e.g. \cite{Wald}.} An inextendible geodesic is a geodesic
parametrized by an affine parameter $\tau$ such that by using up only a finite amount of affine
parameter the geodesic extends over infinite coordinate distances. In a more physical language one could 
reformulate the above statement by saying that it takes only a finite amount of affine parameter $\tau$ in 
order to reach the infinities of the incomplete space-time. As we will illustrate later in this 
chapter, the reason why the Randall-Sundrum II setup ceases to be geodesically complete is simply due to
a specific way of gluing two patches of $AdS_5$. 
One of the main motivations for considering this setup was to provide an alternative to the Randall-Sundrum II model that 
has the advantage of being geodesically complete while conserving the pleasant phenomenological features of 
the latter.
We divide the discussion of geodesics in two parts: in subsection \ref{5RSgeod} we illustrate the effects of 
incomplete geodesics in the Randall-Sundrum II setup for time-like and light-like geodesics. In the following
subsection \ref{5ourgeod} we demonstrate why our setup is geodesically complete by looking at corresponding 
geodesics. 
Finally, we complement the discussions by illustrating the physics 
with the use of the Penrose-diagram of (the universal covering space-time of) $AdS_5$.
\subsection{Geodesics in the Randall-Sundrum II setup}
\label{5RSgeod}

Our discussion of geodesics in this chapter is in no sense meant to be complete.
Without going into the details of the computations we merely intend to present the solutions of 
the geodesic equations in certain cases. For more general and more complete discussions of this issue we
refer to the literature, see e.g.~\cite{Muck:2000bb,Youm:2001qc} and references therein.
We first consider light-like geodesics in the Randall-Sundrum II background metric given in appendix 
\ref{appParallel}, eq.~(\ref{5RSmetric}). 
Let us suppose that a photon is emitted at the brane at $y=0$ in the positive $y$-direction at coordinate 
time $t=0$ then reflected at $y=y_1$ at the time $t=t_1$ to be observed by an 
observer on the brane at time $t=t_2$. In this situation $t$ corresponds to the proper time 
of an observer on the brane at rest. A simple calculation reveals
\begin{equation} \label{5RSgeodint}
   t_2=2 t_1=\frac{2}{k} \left( e^{k y_1}-1\right).
\end{equation}
A brane bound observer will therefore note that it takes an infinite time for a photon to escape 
to $y=\infty$. However, parameterizing the same geodesic by an affine parameter $\tau$ reveals the
light-like incompleteness of the Randall-Sundrum II space-time. Let the events of emission, reflection (at
$y=y_1$) and arrival on the brane again be labeled by $\tau=0$, $\tau=\tau_1$ and $\tau=\tau_2$, 
respectively. Using the geodesic equation 
\begin{equation} \label{5Geodesicequation}
  \frac{d^2 x^\mu}{d \tau^2}+\Gamma^\mu_{\nu \rho} \frac{d x^\nu}{d \tau}
  \frac{d x^\rho}{d\tau}=0 \, ,
\end{equation}
an easy computation shows:
\begin{equation} \label{5RSgeodintau}
  \tau_2=2 \tau_1=\frac{2}{c k} \left(1- e^{-k y_1}\right).
\end{equation}
The constant $c$ is a remnant of the freedom in the choice of an affine parameter.\footnote{In general, two affine parameters $\lambda$ and $\tau$ are related by $\lambda=c \tau + d$, since 
this is the most general transformation leaving the geodesic equation (\ref{5Geodesicequation}) invariant.
The choice of the origin of time to coincide with the emission of the photon only fixes $d$ but does not 
restrict $c$.}
From the last equation we see that now
\begin{equation} 
  \lim_{y_1 \to \infty} \tau_1 = \frac{1}{c k},
\end{equation}
meaning that in order to reach infinity ($y=\infty)$ in the extra dimension it takes only 
a finite amount $1/(c k)$ of affine parameter $\tau$, the expression of incompleteness 
of the Randall-Sundrum II space-time with respect to affinely parametrized light-like geodesics.

In the case of time-like geodesics, the inconsistency is even more striking. As shown in e.g. 
\cite{Gregory:2000rh}, a massive particle starting at the brane with vanishing initial velocity 
travels to $y=\infty$ in finite proper time given by $\tau_p=\pi/(2 k)$ while
for a brane-bound observer this happens in an infinite coordinate time. 
To summarize, the Randall-Sundrum II brane-world model is geodesically incomplete both for null
and for time-like geodesics. As we will see in the next section, the geodesic incompleteness
is a direct consequence of the use of a particular coordinate system in $AdS_5$, the so-called 
Poincar\'e coordinate system, which covers only a part of the full $AdS_5$ space-time.
We will also see that the problem of incomplete geodesics is absent in the setup we propose here. 
  
\subsection{Geodesics in the background (\ref{5metric}) and Penrose-diagram}
\label{5ourgeod}

We would now like to answer similar questions to the ones considered in the previous section for the
background (\ref{5metric}).
For example, we would like to know what time it takes for light to travel from the brane (at 
$\theta$=0) in the $\theta$-direction to a given point in the ``upper hemisphere'' with $\theta$ 
coordinate $\theta_1$, to be reflected and to return to the brane. As in the last section, 
$t$ and $\tau$ will denote the coordinate time (proper time of a stationary observer on the brane)
and the affine parameter used for parameterizing the geodesics, respectively. Again we choose 
$t=0$ ($\tau=0$) for the 
moment of emission, $t_1$ ($\tau_1$) for the reflection and $t_2$ ($\tau_2$) for the time where the photon 
returns to the brane. Due to the enormous hierarchy of distance scales in our model, one might wonder 
whether photons (or gravitons) are able to carry information from an arbitrary point on the brane to 
any other point on the brane connected to the first one by a null geodesic in the extra dimension. 
For an observer on the brane such a possibility would be interpreted as 4-dimensional 
causality violation. However, as we will see in the following, none of these possibilities exist in our model.
Omitting all details we find
\begin{equation} \label{5Ourgeodint}
   t_2=2 t_1=4 R_U \coth\left(\frac{\omega\pi}{2}\right) 
   \arctan \left[\frac{\sinh\left( 
  \frac{\omega\theta_1}{2}\right)}{\cosh\left[\frac{\omega}{2}\left(\pi-\theta_1\right)\right]} \right],
\end{equation}
so that an observer on the brane will see that the photon reaches the ``north pole'' $\theta_1=\pi/2$ at 
finite time
\begin{equation} 
  t_1=R_U \coth\left(\frac{\omega\pi}{2}\right) \arctan \left[ \sinh\left( \frac{\omega \pi}{2}\right)\right] 
  \approx R_U \frac{\pi}{2}\, .
\end{equation} 
Note that due to the warped geometry, it is $R_U$ entering the last relation and not $R$, so that 
even though the physical distance to the ``north pole'' is of the order of $R$ it takes a time of the 
order of $R_U$ for photons to reach it, excluding causality violation on the brane as 
discussed above. If the same geodesic is parametrized using an affine parameter we obtain 
\begin{equation} 
  \tau_2=2 \tau_1=\frac{2}{c}\frac{R}{\omega}\left[\tanh \left(\frac{\omega\pi}{2}\right)-
   \frac{\sinh\left[ \omega \left(\frac{\pi}{2}-\theta_1\right)\right]}{\cosh \frac{\omega\pi}{2}}\right]\, ,
\end{equation}
such that the amount of affine parameter needed to reach $\theta_1=\pi/2$ starting from the 
brane is:
\begin{equation} 
  \Delta \tau = \frac{1}{c}\frac{R}{\omega} \tanh \left(\frac{\omega\pi}{2}\right) \approx
  \frac{1}{c} \frac{R}{\omega} \, .
\end{equation}
Here again $c$ reflects the freedom in the choice of the affine parameter.

The results formally resemble those of the Randall-Sundrum II case. However, the important difference is that
in our case each geodesic can trivially be extended to arbitrary values of the affine parameter,
a simple consequence of the compactness of our space. Once the photon reaches the point $\theta=0$ it 
continues on its geodesic, approaching the brane, entering the ``southern hemisphere'', etc. 
It is clear that it needs an infinite amount of affine parameter in order to travel infinite coordinate 
distances. Therefore, the null geodesics in our setup which are the analogues of the incomplete 
null geodesics in the Randall-Sundrum II setup turn out to be perfectly complete due to the 
compactness of our space. The situation for time-like geodesics is fully analog to the case of the
null geodesics.

To end this section about the geometric properties of our model, we would like to discuss the conformal 
structure of our space-time and point out differences to the Randall-Sundrum II setup. Let us review briefly 
the basic properties of $AdS_5$ space-time to the extend that we will need it in the following 
discussion.\footnote{We mainly follow \cite{Aharony:1999ti}.}
$AdS_5$ space-time can be thought of as the hyperboloid defined by 
\begin{equation} \label{5hyperboloid}
  X_0^2+X_5^2-X_1^2-X_2^2-X_3^2-X_4^2=a^2
\end{equation}
embedded in a flat space with metric
\begin{equation} \label{5AdSDefineMetric}
  ds^2=-dX_0^2-dX_5^2+dX_1^2+dX_2^2+dX_3^2+dX_4^2,
\end{equation}
$a$ being the so-called $AdS$-radius. The {\em global coordinates} of $AdS_5$ are defined by 
\begin{align}\label{5GlobalCoordsDef}
  X_0&=a \cosh \chi \cos \tau,          &X_5=&a \cosh \chi \sin \tau, \nonumber \\
  X_1&=a \sinh \chi \cos \varphi_1,     &X_2=&a \sinh \chi \sin \varphi_1 \cos \varphi_2, \nonumber  \\
  X_3&=a \sinh \chi \sin \varphi_1 \sin \varphi_2 \cos \varphi_3,  
  &X_4=&a \sinh \chi \sin \varphi_1 \sin \varphi_2 \sin \varphi_3,
\end{align}
where the coordinates are confined by $0\leq \chi$, $-\pi \leq \tau \leq \pi$, $0\leq\varphi_1\leq \pi$,
$0\leq\varphi_2\leq \pi$, $0\leq\varphi_3\leq 2\pi$ and $\tau=-\pi$ is identified with $\tau=\pi$. 
These coordinates cover the full hyperboloid exactly once. Allowing $\tau$ to take values on the real line
without the above identification of points gives the universal covering space $CAdS_5$ of $AdS_5$.\footnote{Whenever we used the word $AdS_5$ so far in this chapter we actually meant $CAdS_5$. 
For reasons of clarity, however, we will brake with this common practice 
in the rest of this section.}
In these coordinates, the line element (\ref{5AdSDefineMetric}) can be written:\footnote{\label{5changecoords}It is obvious that the metric (\ref{5metric}) for $0\leq \theta \leq \pi/2$ 
($-\pi/2 \leq \theta \leq 0$) reduces to the above line-element (\ref{5AdSGlobalCoords}) under the 
following coordinate transformation: $\chi=\omega (\frac{\pi}{2}\mp \theta)$, 
$\tau= t \omega/\left[R \cosh \left(\frac{\omega\pi}{2}\right)\right]$, together with $a=R/\omega$.}
\begin{equation} \label{5AdSGlobalCoords}
  ds^2=a^2\left(-\cosh^2\chi \; d\tau^2+d\chi^2+\sinh^2\chi \; d\Omega_3^2\right) \, .
\end{equation}

Another coordinate system can be defined by 
\begin{align}\label{5PoincareCoordsDef}
  X_0&= \frac{1}{2 u} \left[1+u^2\left(a^2+\vec{x}^2-\bar{t}^{\,2}\right)\right] ,
 &X_5=&a \, u \, \bar{t} , \nonumber \\
  X^i&=a \, u \, x^i, \; i=1,2,3\; ,  
 &X^4=& \frac{1}{2 u} \left[1-u^2\left(a^2-\vec{x}^2+\bar{t}^{\, 2}\right)\right],
\end{align}
with $u>0$, $\bar{t} \in (-\infty, \infty)$ and $x^i \in (-\infty, \infty)$.
In these {\em Poincar\'e coordinates} the line element (\ref{5AdSDefineMetric}) takes the form
\begin{equation} \label{5AdsPoincareCoords}
  ds^2= a^2 \left[ \frac{du^2}{u^2} + u^2 \left(-d\bar{t}^{\, 2}+d\vec{x}^2\right)\right].
\end{equation}
In contrast to the global coordinates, the Poincar\'e coordinates do not cover the whole of 
the $AdS_5$ and $CAdS_5$ space-times \cite{Aharony:1999ti}. From (\ref{5AdsPoincareCoords}), after changing coordinates 
according to $dy=-a \, du/u$ and rescaling $t$ and $x^i$ by the $AdS$-radius $a$, we recover the original 
Randall-Sundrum II coordinate system given in (\ref{5RSmetric}). 
The restrictions on $y$ in the Randall-Sundrum II setup further 
limit the range covered by their coordinate system to the $0<u \leq 1$ domain of the Poincar\'e coordinates.

Coming back to the global coordinates, we introduce $\rho$ by
\begin{equation} 
  \tan \rho = \sinh \chi \;\; \mbox{with} \;\; 0\leq \rho < \frac{\pi}{2},
\end{equation}
so that (\ref{5AdSGlobalCoords}) becomes 
\begin{equation} \label{5PenroseCoords}
  ds^2=\frac{a^2}{\cos^2 \rho} \left(-d\tau^2+d\rho^2+\sin^2\rho \; d\Omega_3^2 \right).
\end{equation}
\begin{figure}[htb]
\begin{center}
\input{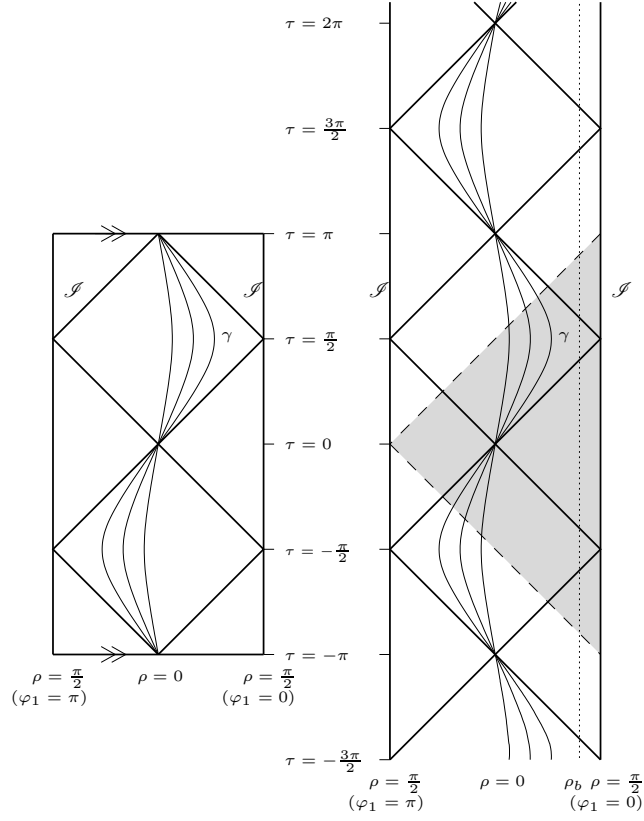}
\end{center}
\caption{Penrose diagram of $AdS_5$ space and its universal covering space $CAdS_5$}
\label{5penrose}
\end{figure}
The Penrose-diagrams of $AdS_5$ space-time and its universal covering space-time $CAdS_5$ are shown in 
Fig.~\ref{5penrose}, see \cite{HawkingEllis,Avis:1977yn}.
While the $AdS_5$ space-time contains closed time-like curves (denoted by $\gamma$ in the figure), 
its universal covering space-time $CAdS_5$ does not.
The arrows in the left diagram indicate that the lines $\tau=-\pi$ and $\tau=\pi$ should be identified.
The symbol $\mathscr{I}$ stands for the time-like surface $\rho=\pi/2$ (spatial infinity). It is this 
surface which is responsible for the absence of a Cauchy surface in $AdS$-space. 
We will concentrate in the following 
on the $CAdS_5$ diagram. First we note that each point in the diagram corresponds to a $3$-sphere.
The shaded region indicates the patch covered by the Poincar\'e (and Randall-Sundrum) coordinates.
Note that the position of the Randall-Sundrum brane cannot be represented in a simple way in the
Penrose diagram of $CAdS_5$. The reason is that the $u=\mbox{const.}$ hypersurfaces of 
the Poincar\'e coordinates 
generate a slicing of flat $4$-dimensional Minkowski space-times while the points in the diagram represent
(curved) $3$-spheres.
From the Penrose diagram it is immediately clear that the Randall-Sundrum II space-time is geodesically 
incomplete. The time-like curves denoted by $\gamma$ emanating from the origin ($\rho=0,\tau=0$) 
will all eventually exit the shaded region after a finite coordinate time $\tau$. These geodesics appear
inextendible from the point of view of the Randall-Sundrum space-time. The problems arise due to the 
arbitrary cutting of a space-time along the borders of a given coordinate patch which covers only a part of 
the initial space-time.
Similar conclusions can be drawn for null geodesics, represented by lines making angles of $45$ degrees 
with the vertical lines in Fig.~\ref{5penrose}.
The vertical dotted line with topology $\mathbb{R} \times S^3$ at coordinate 
$\rho_b=\arctan \left[ \sinh\left( \frac{\omega\pi}{2}\right) \right]$ close to $\rho=\pi/2$ in the 
right diagram 
corresponds to the location of our curved $3$-brane, $\tau$ being 
proportional to our time coordinate $t$ (see footnote \ref{5changecoords}). We note that since 
$\omega > 50$, this line is drawn by far
too distant from $\rho=\pi/2$ as a simple expansion shows:
\begin{equation} 
  \rho_b=\arctan \left[ \sinh\left( \frac{\omega \pi}{2}\right)\right] \sim \frac{\pi}{2} -2 e^{-\frac{\omega\pi}{2}}+ 
  \mathcal{O}(e^{-\frac{3 \omega\pi}{2}}) \, .
\end{equation}
The slice of $CAdS_5$ to the right of $\rho_b$ 
($\rho_b \leq \rho \leq \pi/2$) is discarded and replaced by another copy of the slice to the left 
of $\rho_b$ ($0\leq\rho\leq \rho_b$).

From the Penrose diagram we can also deduce that our space-time is geodesically complete. Time-like 
geodesics (emanating from ($\rho=0,\tau=0$)) cross the brane $\rho_b$ at some point or return to the origin
depending on the initial velocities of the particles that define them. In both cases (as in the case of 
null geodesics) there is no obstacle to extend the affine parameter to larger and larger values. 

As we already mentioned, the absence of a Cauchy surface in $CAdS$ space-time is due to the existence 
of the time-like surface $\mathscr{I}$ at spatial infinity. Even though the family of space-like 
hypersurfaces $\tau=\mbox{const.}$ cover the whole $CAdS$ space-time, none of these hypersurfaces 
is a Cauchy surface.
By considering the two null-geodesics which cross the origin ($\tau=\rho=0$) in the right diagram of 
Fig.~\ref{5penrose} we note that they will never reach for example the hypersurface $\tau=\pi$. Since 
the affine parameter $\lambda$ corresponding to these geodesics lies within the full interval $(-\infty,\infty)$, 
the hypersurface $\tau=\pi$ can not be a 
Cauchy surface. Put in another way, there are points on $\mathscr{I}$ which lie neither in the future 
nor in the past Cauchy development of the hypersurface $\tau=\pi$. Similar arguments hold for any other 
$\tau=\mbox{const.}$ hypersurface. 

By construction, our space-time excludes $\mathscr{I}$. A corresponding null-geodesics in our 
setup which starts for example at the ``north pole'' $\theta=\pi/2$ will reach the brane after using up 
only a finite amount of affine parameter and can be extended to the ``southern hemisphere'' where it will 
eventually reach the ``south pole'' $\theta=-\pi/2$ (again by using up only a finite amount of affine 
parameter), etc. It is therefore clear that all inextendible, non space-like lines in our manifold 
eventually have to cross any given $\tau=\mbox{const.}$ hypersurface. We conclude that the family 
of $\tau=\mbox{const.}$ hypersurfaces indeed constitutes a family of Cauchy surfaces for the manifold
considered in this chapter. 

\section{Gravity localization on the brane}
\label{5ESUthree}

There are at least two equivalent ways of addressing the problem of the localization of gravity
in brane-world scenarios. Both ways have their advantages and disadvantages. The first way is
based on a detailed study of the Kaluza-Klein excitations of the graviton. After integrating out the 
extra dimension(s) one obtains an effective 4 dimensional Lagrangian involving the full tower of 
Kaluza-Klein gravitons. Considering the low energy scattering process of two test particles on the brane via 
exchange of Kaluza-Klein excitations allows to relate the non-relativistic scattering amplitude to the
static potential felt by the two test particles. Since the coupling of each individual Kaluza-Klein 
particle to matter on the brane is proportional to the value of its transverse wave-function at the position
of the brane, this approach necessitates a proper normalization of all Kaluza-Klein modes. An advantage of
this approach is the possibility of distinguishing between the contributions to the potential coming from 
the zero mode (Newton's law) and from the higher modes (corrections thereof).

The second approach is based on a direct calculation of the graviton two-point function in the space-time
under consideration, having the obvious advantage of bypassing all technicalities related to the Kaluza-Klein
spectrum and the wave-function normalization. However, a physical interpretation of the effects of individual 
Kaluza-Klein modes from the point of view of a 4-dimensional observer is, to say the least, not 
straightforward. Finally, due to the equivalence of the two approaches, it is clearly possible to fill this 
gap by reading of the Kaluza-Klein spectrum and the (normalized) wave-functions from the two-point function 
by locating its poles and determining corresponding residues.   

In the setup considered in this chapter, we choose to work in the second approach, the direct evaluation of 
the Green's functions, due to additional difficulties arising from the non-Minkowskian nature of the 
induced metric on the brane. The lack of Poincar\'e invariance on the brane clearly implies the absence of 
this symmetry in the effective 4-dimensional Lagrangian as well as non-Minkowskian dispersion relations for 
the Kaluza-Klein modes. 

\subsection{Perturbation equations and junction conditions}

We would like to study the fluctuations $H_{ij}$ of the metric in the background (\ref{5metric}) 
defined by:
\begin{equation} 
  ds^2=-\sigma^2 dt^2+R_U^2 \gamma^2 \left[\eta_{ij}+2 H_{ij}\right] d\varphi^i d\varphi^j +R^2 d\theta^2\, .
\end{equation}
$H_{ij}$ transform as a transverse, traceless second-rank tensor with respect to coordinate transformations
on the maximally symmetric space with metric $\eta_{ij}$, a $3$-sphere in our case:
\begin{equation} 
  \eta^{ij} H_{ij}=0\, , \quad \eta^{i j} \tilde{\nabla}_i H_{j k}=0\, ,
\end{equation}
where $\tilde{\nabla}$ denotes the covariant derivative associated with the metric $\eta_{ij}$ on $S^3$.
The gauge invariant symmetric tensor $H_{ij}$ does not couple to the vector and scalar perturbations. 
Its perturbation
equation in the bulk is obtained from the transverse, traceless 
component of the perturbed Einstein equations in $5$ dimensions:
\begin{equation} 
  \delta R_{M N}-\frac{2}{3} \Lambda \delta g_{M N}=0 \, .
\end{equation}
Specifying this equation to the case of interest of a static background we  find:
\begin{equation} \label{5Hfluct}
  \frac{1}{\sigma^2} \ddot{H}-\frac{1}{R_U^2\gamma^2}\tilde{\Delta} H_{ij} -
  \frac{1}{R^2}H_{ij}^{\prime \prime}-
  \frac{1}{R^2} \left( \frac{\sigma^\prime}{\sigma}  
  +3\frac{\gamma^\prime}{\gamma}\right)H_{ij}^{\prime}
  +\frac{2}{R_U^2\gamma^2}H_{ij} = 0 \, ,
\end{equation}
where $\tilde{\Delta}$ denotes the Laplacian on $S^3$. The last equation is valid in the bulk in the 
absence of sources. We can now conveniently expand $H_{ij}$ in the basis of the symmetric, transverse, 
traceless tensor harmonics $\hat{T}^{(l \lambda)}_{ij}$ on $S^3$ 
\cite{Mukohyama:2000ui,Rubin:tc,Allen:1986tt}:
\begin{equation} \label{5Hexpand}
  H_{ij}=\sum_{l=2}^\infty \sum_\lambda \Phi^{(l \lambda)}(t,\theta) \hat{T}^{(l \lambda)}_{ij}\, ,
\end{equation}
where the $\hat{T}^{(l \lambda)}_{ij}$ satisfy
\begin{align} \label{5TTHarmonics}
  \tilde{\Delta} \hat{T}^{(l \lambda)}_{ij} + k_l^2 \hat{T}^{(l \lambda)}_{ij} =0\, , 
  \qquad k_l^2= l(l+2)-2\, , \qquad l=2,3,... \nonumber  \\
  \eta^{ij}\tilde{\nabla}_i \hat{T}^{(l \lambda)}_{jk}=0\, , \qquad 
  \eta^{ij} \hat{T}^{(l \lambda)}_{ij} =0\, , \qquad 
  \hat{T}^{(l \lambda)}_{[ij]}=0\, .
\end{align}
The sum over $\lambda$ is symbolic and replaces all eigenvalues needed to describe the 
full degeneracy of the subspace of solutions for a given value of $l$.
Introducing the expansion (\ref{5Hexpand}) into (\ref{5Hfluct}) and using the orthogonality 
relation of the tensor harmonics \cite{Allen:1986tt} 
\begin{equation} \label{5TTTOrthogonality}
  \int \sqrt{\eta } \, \eta^{ik} \, \eta^{jl} \, \hat{T}^{(l \lambda)}_{ij} 
  \, \hat{T}^{(l' \lambda')}_{kl} d^3\varphi = 
   \delta^{l l'} \, \delta^{\lambda \lambda'} \, ,
\end{equation}
we obtain:
\begin{equation} \label{5Hfluctexpansion}
  \frac{1}{\sigma^2} \ddot{\Phi}^{(l \lambda)}- \frac{1}{R^2}\Phi^{(l \lambda) \prime \prime}-
  \frac{1}{R^2} \left( \frac{\sigma^\prime}{\sigma} + 3 \frac{\gamma^\prime}{\gamma}\right)
   \Phi^{(l \lambda) \prime}+\frac{l(l+2)}{R_U^2\gamma^2} \Phi^{(l \lambda)}=0 \, .
\end{equation}
This equation has to be compared to the equation of a massless scalar field in the background 
(\ref{5metric}). After expanding the massless scalar in the corresponding scalar harmonics on
$S^3$, we recover (\ref{5Hfluctexpansion}) with the only difference in the eigenvalue parameters
due to different spectra of the Laplacian $\tilde{\Delta}$ for scalars and for tensors.
Motivated by this last observation, and in order to avoid the technicalities related to the tensorial nature
of the graviton $H_{ij}$, we confine ourselves to the study of the Green's functions 
of a massless scalar field in the background (\ref{5metric}).

The differential equation (\ref{5Hfluctexpansion}) alone does not determine $H_{ij}$
uniquely. We have to impose proper boundary conditions for $H_{ij}$. While imposing square 
integrability will constitute one boundary condition at $\theta=\pm \pi/2$, 
the behavior of $H_{ij}$ on the brane will be dictated by the Israel junction condition \cite{Israel:rt}:
\begin{equation} \label{5israelcondition}
  : K_{\mu \nu} : \;\;\; =-\frac{8 \pi}{M^3}\left(T_{\mu\nu}-\frac{1}{3} T_\kappa^{\;\kappa} \bar{g}_{\mu \nu}
  \right).
\end{equation}
Here $K_{\mu\nu}$ and $\bar{g}_{\mu\nu}$ denote the extrinsic curvature and the induced metric on the brane, 
respectively. $T_{\mu\nu}$ is the $4$-dimensional stress-energy tensor on the 
brane\footnote{\; $T^0_{\;\;0}=c_0$, and $T^i_{\;\;j}=\delta^i_{\;\;j} c$ in the notation of section 
\ref{5SingularBrane}.} and the symbol $: \ldots :$ is defined in (\ref{5::def}).  
In our coordinate system the extrinsic curvature is given by:
\begin{equation} \label{5extrCurvature}
  K_{\mu\nu}=\frac{1}{2 R} \frac{\partial{\bar{g}_{\mu \nu}}}{\partial \theta} \, .
\end{equation}
Hence, the non-trivial components of the Israel condition become:
\begin{align} 
  : \frac{\sigma^\prime}{\sigma} : &= -\frac{8 \pi R}{M^3} \left(\frac{2}{3} c_0-c 
     \right) \, , \label{5Israel00BG} \\
  : \frac{\gamma^\prime}{\gamma} \left( \eta_{ij}+2 H_{ij} \right) + H^\prime_{ij} : 
   &= \frac{8 \pi R}{3 M^3} c_0 \left(\eta_{ij}+2 H_{ij} \right). \label{5Israelij}
\end{align}
Separating the background from the fluctuation in (\ref{5Israelij}) we find:
\begin{align} 
  : \frac{\gamma^\prime}{\gamma} : &= \frac{8 \pi R}{3 M^3} c_0 \, ,\label{5IsraelijBG}\\
  : H^\prime_{ij} : &= 0 \label{5IsraelijFluct}\, ,
\end{align}
where we only used the continuity of $H_{ij}$ on the brane. While eq.~(\ref{5IsraelijBG}) directly coincides 
with eq.~(\ref{5FineTuneI}) of section \ref{5SingularBrane}, eq.~(\ref{5Israel00BG}) turns out 
to be a linear combination of (\ref{5FineTuneI}) and (\ref{5FineTuneII}).

Eq.~(\ref{5IsraelijFluct}) taken alone implies the continuity of $H^\prime_{ij}$ on the brane. If 
in addition the fluctuations are supposed to satisfy the $\theta \to-\theta$ symmetry, this condition
reduces to a Neumann condition, as for example in our treatment of the scalar two-point function 
in the Randall-Sundrum II case (see appendix \ref{appParallel}).

In the next section we are going to find the static Green's functions of a massless scalar field in the 
Einstein static universe background. This will serve as a preparatory step for how to handle the more 
complicated case of a non-invertible Laplacian. More importantly, it will provide the necessary reference
needed for the interpretation of the effect of the extra dimension on the potential between two test masses 
on the brane.

\subsection{Gravity in the Einstein static universe}
As announced in the previous section, we will concentrate on the massless scalar field. 
Our aim is to solve the analog of Poisson equation in the Einstein static universe background.
Due to its topology $\mathbb{R} \times S^3$ we will encounter a difficulty related to the existence 
of a zero eigenvalue of the scalar Laplacian on $S^3$ necessitating the introduction of a modified 
Green's function.\footnote{For an elementary introduction to the concept of modified Green's functions see e.g
\cite{Stakgold}.} Note that in the more physical case where the full tensor structure of the 
graviton is maintained no such step is necessary since all eigenvalues of the tensorial Laplacian
are strictly negative on $S^3$, see (\ref{5TTHarmonics}).

\subsubsection{Definition of a modified Green's function}
We take the metric of Einstein's static universe in the form:
\begin{equation} \label{5esu_metric}
  ds^2=g_{\mu\nu}dx^\mu dx^\nu = - dt^2+A^2 d \Omega_3^2
\end{equation}
with $d\Omega_3^2$ being the line element of a $3$-sphere given in (\ref{5S3metric}) and $A$ being its 
constant radius. 
The equation of a massless scalar field then becomes:
\begin{equation} \label{5ScaFieldEqESU}
  \frac{1}{\sqrt{-g}}\partial_\mu \left[\sqrt{-g} \, g^{\mu\nu}\partial_\nu u(t,\vec{x}) \right]= j(t,\vec x)
\end{equation}
which in the static case reduces to
\begin{equation} \label{5StaticEqESU}
  \mathcal{D} u(\vec x) \equiv \frac{1}{A^2} \tilde{\Delta} u(\vec x) = j(\vec x) \, ,
\end{equation}
where $\tilde{\Delta}$ is the scalar Laplacian on $S^3$ and we choose to write $\vec x$ for the collection
of the three angles on $S^3$.
All functions involved are supposed to obey periodic boundary condition on $S^3$ so that in
Green's identity all boundary terms vanish:
\begin{equation} \label{5GreensIdentityESU}
  \int \sqrt{-g} \, \left(\mathcal{D} \, v(\vec x)\right) \, \overline{u}(\vec x) \, d^3 \vec{x} = 
  \int \sqrt{-g} \, v(\vec x) \, \overline{\mathcal{D} \, u(\vec x)} \, d^3 \vec{x}=
  \int \sqrt{-g} \, v(\vec x) \, \overline{j}(\vec x) \, d^3\vec{x} \, .
\end{equation}
Here and in the following, bars denote complex conjugate quantities.
The operator $\mathcal{D}$ trivially allows for an eigenfunction with zero eigenvalue (the constant
function on $S^3$):
\begin{equation}
  \mathcal{D} u_0(\vec x) = 0 \, ,\quad \int \sqrt{-g} \, u_0(\vec x) \, \overline{u}_0(\vec x) \, 
  d^3 \vec x=1 \, .
\end{equation}  
Then, since the homogeneous equation has not only the trivial solution ($u=0$) which satisfies 
the periodic boundary conditions in the angular variables, the operator $\mathcal{D}$ cannot be 
invertible. Therefore, there 
is no solution to the equation (\ref{5StaticEqESU}) for an arbitrary source.  
In order to still define a ``modified'' Green's function we have to restrict the possible 
sources to sources that satisfy the following solvability condition:
\begin{equation} \label{5solvabilityESU}
  \int \sqrt{-g} \, j(\vec x) \, \bar{u}_0(\vec x) \, d^3\vec x=0 \, .
\end{equation}
Formally this condition can be obtained by replacing $v(\vec x)$ by the non-trivial solution 
of the homogeneous equation $u_0(\vec x)$  in (\ref{5GreensIdentityESU}).
We now define the modified Green's function as a solution of
\begin{equation} \label{5ModGreenFunctionESU}
  \mathcal{D}_x \mathcal{G}(\vec x,\vec x')=\frac{\delta^3(\vec x-\vec x')}{\sqrt{-g}}-u_0(\vec x)
  \, \bar{u}_0(\vec x') \, .
\end{equation}
Replacing $v(\vec x)$ by $\mathcal{G}(\vec x,\vec x')$ in Green's identity (\ref{5GreensIdentityESU}) we obtain (after complex conjugation) the desired integral representation for the solution of (\ref{5StaticEqESU}):
\begin{eqnarray} \label{5esuIntReprESU}
  u(\vec x) = C u_0(\vec x)+\int \sqrt{-g} \, j(\vec x') \, \overline{\mathcal{G}}(\vec x',\vec x) \, 
  d^3 \vec x' \, ,
\end{eqnarray}
where $C$ is given by
\begin{equation} 
  C=\int \sqrt{-g} \, u(\vec x') \, \overline{u}_0(\vec x') \, d^3\vec x' \, .
\end{equation}
Note that the source 
\begin{equation}
  j_0(\vec x)=\frac{\delta^3(\vec x-\vec x')}{\sqrt{-g}}-u_0(\vec x)\bar{u}_0(\vec x')
\end{equation}
trivially satisfies the solvability condition (\ref{5solvabilityESU}) and represents a point source at 
the location $\vec x=\vec x'$ compensated by a uniform, negative mass density. To fix the 
normalization  of $u_0(\vec x)$ we write:
\begin{equation}
  u_0(\vec x)=N_0 \Phi_{100}(\vec x) \,\,\, \mbox{with} \,\,\, \int \sqrt{\eta} \, \Phi_{100}(\vec x) 
  \overline{\Phi}_{100}(\vec x) d^3 \vec x=1,
\end{equation}
where $\Phi_{100}=\frac{1}{\pi \sqrt{2}}$ and so
\begin{align}
  \int \sqrt{-g} \, u_0(\vec x) \, \overline{u}_0(\vec x) \, d^3\vec x &= \vert N_0 \vert^2 A^3 
  \underbrace{\int \sqrt{\eta} \, \Phi_{100}(\vec x) \, \overline{\Phi}_{100}(\vec x) \, d^3 \vec x}_{=1} 
   = \vert N_0 \vert^2 A^3=1 \, ,
\end{align}
so that
\begin{equation}
  u_0(\vec x)=\frac{1}{A^{3/2}}\Phi_{100}(\vec x).
\end{equation}
In order to solve the differential equation (\ref{5ModGreenFunctionESU}) defining the modified Green's 
function $\mathcal{G}(\vec x, \vec x')$, we expand in eigenfunctions of the Laplace operator on $S^3$, 
the so-called scalar harmonics $\Phi_{\lambda l m}$ with properties 
(see e.g.~\cite{Kodama:2000fa,GribMamaMost}):
\begin{equation} \label{5ScalarHarmonicsESU}
  \tilde{\Delta} \Phi_{\lambda l m} =(1-\lambda^2 )\Phi_{\lambda l m} \, , 
  \qquad \lambda=1,2,\ldots; l=0,\ldots,\lambda-1;m=-l,\ldots,l \, .
\end{equation} 
Our ansatz therefore reads:
\begin{equation}
  \mathcal{G}(\vec x, \vec x')=\sum_{\lambda=1}^\infty \sum_{l=0}^{\lambda-1}\sum_{m=-l}^l 
  \Phi_{\lambda l m}(\vec x) \, c_{\lambda l m}(\vec x') \, .
\end{equation}
Introducing this in (\ref{5ModGreenFunctionESU}) we obtain 
\begin{equation} 
  \sum_{\lambda=1}^{\infty}\sum_{l=0}^{\lambda-1}\sum_{m=-l}^l \left[-\frac{\lambda^2-1}{A^2}
  c_{\lambda l m}(\vec x')\right] \Phi_{\lambda l m}(\vec x)= 
   \frac{\delta^3(\vec x-\vec x')}{\sqrt{-g}}-u_0(\vec x)\, \overline{u}_0(\vec x')\, .
\end{equation} 
If we now multiply by $\Phi_{\lambda' l' m'}(\vec x) \, \sqrt{\eta}$ and integrate over $S^3$ we find
\begin{equation}
 -\frac{\lambda'^2-1}{A^2}c_{\lambda' l' m'}(\vec x') =\frac{1}{A^3} 
  \overline{\Phi}_{\lambda' l' m'}(\vec x')-\bar{u}_0(\vec x') \int\limits_{S^3} \sqrt{\eta} \,  
  u_0(\vec x) \, \overline{\Phi}_{\lambda' l' m'}(\vec x) \, d^3 \vec x,
\end{equation}
where we made use of the orthogonality relation of the scalar harmonics
\begin{equation} \label{5Harmonicsorthogonality}
 \int \sqrt{\eta} \, \overline{\Phi}_{\lambda l m}(\vec x) \, \Phi_{\lambda' l' m'}(\vec x) \, d^3 \vec x=
 \delta_{\lambda \lambda'}   \delta_{l l'} \delta_{m m'}.
\end{equation}
In the case $\lambda'=1$ (and vanishing $l'$ and $m'$) the above equation is identically
satisfied for all values of $c_{1 0 0}(\vec x')$.\footnote{This means that we have the freedom to choose 
$c_{1 0 0}(\vec x')$ freely. Our choice is $c_{1 0 0}(\vec x')=0$ without restricting generality 
since from eq. (\ref{5esuIntReprESU}) we immediately conclude that for sources satisfying 
(\ref{5solvabilityESU}) there will never be any contribution to $u(\vec x)$ coming from $c_{1 0 0}(\vec x')$.}

In the case $\lambda'\neq 1$ the coefficient $c_{\lambda' l' m'}(\vec x')$ follows to be
\begin{equation}
 c_{\lambda' l' m'}(\vec x')=\frac{\overline{\Phi}_{\lambda' l' m'}(\vec x')}{A(1-\lambda'^2)}
\end{equation}
so that the formal solution for the modified Green's function can be written as
\begin{equation} 
  \mathcal{G}(\vec x,\vec x') =
  \sum_{\lambda=2}^{\infty} \sum_{l=0}^{\lambda-1} \sum_{m=-l}^{l} 
   \frac{\Phi_{\lambda l m}(\vec x) \bar{\Phi}_{\lambda l m}(\vec x')}{A(1-\lambda^2)}\, .
\end{equation}
Due to the maximal symmetry of the $3$-sphere, the Green's function 
$\mathcal{G}(\vec x,\vec x')$ can only depend on 
the geodesic distance $s(\vec x, \vec x') \in [0,\pi]$ between the two points 
$\vec x$ and $\vec x'$:
\begin{eqnarray} \label{5DistanceOnS3}
  \cos s &=& \cos \varphi_1 \cos \varphi'_1+\sin\varphi_1\sin\varphi'_1 \cos\beta \, ,\nonumber\\
  \cos\beta&=& \cos\varphi_2 \cos \varphi'_2+ \sin\varphi_2\sin\varphi'_2 \cos(\varphi_3-\varphi'_3) \, ,
\end{eqnarray}
which in the case $\varphi_2=\varphi'_2$ and $\varphi_3=\varphi'_3$ clearly reduces to 
$s=\varphi_1-\varphi'_1$. Indeed, the sum over $l$ and $m$ can be performed 
using\footnote{See e.g.~\cite{GribMamaMost}.}
\begin{equation} \label{5SumOverS3}
  \sum_{l=0}^{\lambda-1} \sum_{m=-l}^{l} \bar{\Phi}_{\lambda l m}(\vec x) 
   \Phi_{\lambda l m}(\vec x')=\frac{\lambda}{2 \pi^2} \frac{\sin \left(\lambda s\right)}{\sin s}
\end{equation}
such that even the remaining sum over $\lambda$ can be done analytically:
\begin{equation} \label{5ESUresult}
  \tilde{\mathcal{G}}(s) \equiv \mathcal{G}(\vec x,\vec x')=
   \frac{1}{8 \pi^2 A}-\frac{1}{4\pi A}\left[(1-\frac{s}{\pi})\cot s \right]\, , \qquad s \in [0,\pi]\, .
\end{equation}
To interpret this result, we develop $\tilde{\mathcal{G}}(s)$ around $s=0$ 
obtaining\footnote{Note that the constant term in the expansion (\ref{5expansioninsESU}) is specific 
to our choice of $c_{1 0 0}(\vec x')$.}
\begin{equation} \label{5expansioninsESU}
  \tilde{\mathcal{G}}(s)=\frac{1}{A}\left[-\frac{1}{4\pi s} +\frac{3}{8\pi^2}+ \frac{s}{12 \pi}
   +\mathcal{O}(s^2) \right] \, .
\end{equation}
By introducing the variable $r=s A$ and by treating $\tilde{\mathcal{G}}(s)$ as a
gravitational potential we find 
\begin{equation}
  \frac{1}{A} \frac{d\tilde{\mathcal{G}}(s)}{ds}=\frac{1}{4\pi r^2}+\frac{1}{12\pi A^2}+\mathcal{O}(r/A^3).
\end{equation}
We notice that for short distances, $r \ll A$, we find the expected flat result whereas the corrections to 
Newton's law become important at distances $r$ of the order of $A$ in the form of a constant attracting
force.\footnote{There exists numerous articles treating the gravitational potential of a point source 
in Einstein's static universe (see \cite{Astefanesei:2001cx,Nolan:1999wf} and references therein). 
We only would like to point out here similarities between our  
result (\ref{5ESUresult}) and the line element of a Schwarzschild metric in an Einstein static universe
background given in \cite{Astefanesei:2001cx}.}

\subsection{Green's function of a massless scalar in the background (\ref{5metric})}
We are now prepared to address the main problem of this chapter namely the computation
of the modified Green's function of a massless scalar field in the background space-time
(\ref{5metric}). Since our main interest focuses again on the low energy properties of the two-point
function, we will limit ourselves to the static case. The main 
line of reasoning is the same as in the previous section. Due to the fact that our space has the
global topology of a 4-sphere $S^4$, we again are confronted with a non-invertible differential operator.
We would like to solve 
\begin{equation} \label{5ScalarEq}
  \mathcal{D} u(\varphi_i,\theta) 
  \equiv \frac{\tilde{\Delta} u(\varphi_i,\theta)}{R_U^2 \gamma(\theta)^2}+
  \frac{1}{R^2} \frac{1}{\sigma(\theta)\gamma(\theta)^3} \frac{\partial}{\partial\theta}
  \left[ \sigma(\theta)\gamma(\theta)^3 \frac{\partial u(\varphi_i,\theta)}{\partial \theta}\right]=
  j(\varphi_i,\theta)  \, ,
\end{equation}
where the independent angular variable ranges are  
$0\leq\varphi_1\leq\pi;\,0\leq\varphi_2\leq\pi;\, 0\leq\varphi_3\leq 2 \pi;\,-\pi/2\leq\theta\leq\pi/2$. 
Every discussion of Green's functions is based on Green's identity relating the differential 
operator under consideration to its adjoint operator. Since $\mathcal{D}$ is formally self-adjoint we have
\begin{align} \label{5GreensIdentity}
  &\int\limits_0^\pi d\varphi_1 \int\limits_0^\pi d\varphi_2 \int\limits_0^{2\pi} d\varphi_3 
  \int\limits_{-\frac{\pi}{2}}^{\frac{\pi}{2}} d\theta \sqrt{-g}
  \left[(\mathcal{D} v) \overline{u} -v \overline{(\mathcal{D} u)}\right]= \\ \nonumber
  &\;\;\;
  \int\limits_0^\pi d\varphi_1 \int\limits_0^\pi d\varphi_2 \int\limits_0^{2\pi} d\varphi_3 \frac{R_U^3}{R} 
  \sqrt{\eta}\left[ \sigma(\theta)\gamma(\theta)^3 
  \left( \bar{u} \frac{\partial v}{\partial\theta}-
     \frac{\partial \bar{u}}{\partial\theta} v \right) 
   \right]_{-\frac{\pi}{2}}^{\phantom{-}\frac{\pi}{2}}+\ldots \, ,
\end{align}
where we dropped the arguments of $u$ and $v$ for simplicity.
The dots in (\ref{5GreensIdentity}) refer to boundary terms in the variables $\varphi_i$ and since 
we again employ an eigenfunction expansion in scalar harmonics on $S^3$, these 
boundary terms will
vanish. In order to find an integral representation of the solution $u(\varphi_i,\theta)$ of 
(\ref{5ScalarEq}) we have to impose appropriate boundary conditions on $u$ and $v$ at 
$\theta=\pm \pi/2$. For the time being we assume this to be the case such that all boundary
terms in (\ref{5GreensIdentity}) vanish and proceed with the formal solution of (\ref{5ScalarEq}).
We will address the issue of boundary conditions in $\theta$ in detail in appendices \ref{appParallel} and
\ref{appDEQ}.

In the following we will collectively use $x$ instead of $(\varphi_i, \theta)$.
The homogeneous equation $\mathcal{D} \, u(x)=0$ does not have a unique solution under the 
assumption of periodic boundary conditions. In addition to the trivial solution $(u=0)$ we also find 
\begin{equation} \label{5ZeroMode}
  \mathcal{D} \, u_0(x) = 0\, , \qquad \int \sqrt{-g} \, u_0(x) \, 
   \overline{u}_0(x) \, d^4x=1 \, .
\end{equation} 
Therefore, the corresponding inhomogeneous equation (\ref{5ScalarEq}) does not have a solution unless
we again restrict the space of allowed sources: 
\begin{equation} \label{5solvability}
  \int \sqrt{-g} \, j(x) \, \overline{u}_0(x) \, d^4x=0 \, .
\end{equation}
As in the last section, this condition can be obtained by replacing $v$ by the non-trivial solution 
of the homogeneous equation $u_0$  in (\ref{5GreensIdentity}).
We now define the modified Green's function by
\begin{equation} \label{5ModGreenFunction}
  \mathcal{D}_x \mathcal{G}(x,x')=\frac{\delta^4(x-x')}{\sqrt{-g}}-u_0(x)\overline{u}_0(x') \, .
\end{equation}
From Green's identity (\ref{5GreensIdentity}), with $v(x)$ given by $\mathcal{G}(x,x')$, 
we again obtain after complex conjugation the desired integral representation:
\begin{eqnarray} \label{5esuIntRepr}
  u(x) = C u_0(x)+\int \sqrt{-g} \, j(x') \overline{\mathcal{G}(x',x)} \, d^4x' \, ,
\end{eqnarray}
with 
\begin{equation} 
  C=\int \sqrt{-g} \, u(x') \, \overline{u}_0(x') \, d^4x' \, .
\end{equation}
As before, the source
\begin{equation}
  j_0(x)=\frac{\delta^4(x-x')}{\sqrt{-g}}-u_0(x)\, \overline{u}_0(x')
\end{equation}
satisfies the solvability condition (\ref{5solvability}) by construction.
The normalization of the constant mode $u_0(x)$ is slightly more involved than 
before due to the nontrivial measure $\sigma(\theta)\gamma(\theta)^3$ in the $\theta$ 
integration. By inserting  
\begin{equation} 
  u_0(x)=N_0 \Phi_{1 0 0}(\varphi_i) \chi_1(\theta) \, \, \mbox{with}\,\, \chi_1(\theta)=1
\end{equation}
in the integral in (\ref{5ZeroMode}) we obtain 
\begin{equation}
  N_0 =\left[\frac{2\omega}{R R_U^3 \tanh \left(\frac{\omega \pi}{2}\right)}\right]^{\frac{1}{2}} \,.
\end{equation}
For the solution of eq. (\ref{5ModGreenFunction}) we use the ansatz
\begin{equation}
  \mathcal{G}(\varphi_i,\varphi'_i,\theta,\theta')=\sum_{\lambda=1}^{\infty}\sum_{l=0}^{\lambda-1}
  \sum_{m=-l}^l \Phi_{\lambda l m}(\varphi_i) \, c_{\lambda l m}(\varphi'_i,\theta,\theta')  \, ,
\end{equation}
where from now on we decide to write all arguments explicitly.
After inserting this in (\ref{5ModGreenFunction}) we find
\begin{align}
  &\sum_{\lambda=1}^{\infty}\sum_{l=0}^{\lambda-1} \sum_{m=-l}^l \left[
    -\frac{\lambda^2-1} {R_U^2 \gamma(\theta)^2}c_{\lambda l m}(\varphi'_i,\theta,\theta') 
  \right.\nonumber\\  & \qquad\qquad \qquad \left. 
    + \frac{1}{R^2} \frac{1}{\sigma(\theta)\gamma(\theta)^3}\frac{\partial}{\partial\theta}\left(
   \sigma(\theta)\gamma(\theta)^3 
   \frac{\partial c_{\lambda l m}(\varphi'_i,\theta,\theta')}{\partial \theta}\right)\right] 
   \Phi_{\lambda l m}(\varphi_i)\nonumber \\
  &\qquad\qquad\qquad 
   =\frac{\delta^3(\varphi_i-\varphi'_i) \, \delta(\theta-\theta')}
     {\sqrt{-g}}-u_0(\varphi_i,\theta)\, \overline{u}_0(\varphi'_i,\theta') \, .
\end{align}
After multiplication by $\sqrt{\eta} \, \overline{\Phi}_{\lambda' l' m'}(\varphi_i)$ and integration over 
$S^3$ we obtain
\begin{align}\label{5TransverseEquation}
  &-\frac{\lambda'^2-1} {R_U^2 \gamma(\theta)^2}c_{\lambda' l' m'}(\varphi'_i,\theta,\theta')+
   \frac{1}{R^2}\frac{1}{\sigma(\theta)\gamma(\theta)^3} \frac{\partial}{\partial\theta}\left[
   \sigma(\theta)\gamma(\theta)^3 \frac{\partial c_{\lambda' l' m'}(\varphi'_i,\theta,\theta')}
    {\partial \theta}\right]\\ \nonumber 
  & \qquad =\frac{1}{R_U^3 R}\frac{1}{\sigma(\theta)\gamma(\theta)^3} \delta(\theta-\theta')
  \overline{\Phi}_{\lambda' l' m'}(\varphi'_i) 
  -\overline{u}_0(\varphi'_i,\theta') \int\limits_{S^3}\sqrt{\eta} 
  \, u_0(\varphi_i,\theta) \, 
  \overline{\Phi}_{\lambda' l' m'}(\varphi_i) \, d^3\varphi_i \, .
\end{align}
We now have to distinguish the cases $\lambda'=1$ and $\lambda'\neq 1$.
\clearpage
\begin{enumerate}
  \item $( \lambda' l' m')=(1 0 0)$. 

In this case, the last term on the right hand side of 
(\ref{5TransverseEquation}) will give a non-vanishing contribution:
\begin{align}
     &\frac{1}{R^2}\frac{1}{\sigma(\theta)\gamma(\theta)^3} \frac{\partial}{\partial\theta}\left[
     \sigma(\theta)\gamma(\theta)^3 \frac{\partial c_{1 0 0}(\varphi'_i,\theta,\theta')}
      {\partial \theta}\right] \\
     &\quad =\frac{1}{R_U^3 R} \frac{1}{\sigma(\theta)\gamma(\theta)^3} \delta(\theta-\theta')
    \overline{\Phi}_{1 0 0}(\varphi'_i)\nonumber 
 \\     &\quad \quad 
   -\vert N_0 \vert^2 \bar{\Phi}_{1 0 0}(\varphi'_i) 
     \chi_1(\theta) \overline{\chi}_1(\theta') \underbrace{\int\limits_{S^3} \sqrt{\eta} \, \Phi_{1 0 0}(\varphi_i) 
     \overline{\Phi}_{1 0 0}(\varphi_i) \, d^3 \varphi_i}_{=1} \nonumber \, .
\end{align}
By defining $g^{(1)}(\theta,\theta')$ by the relation
\begin{equation}
  c_{1 0 0}(\varphi'_i,\theta,\theta')=\frac{1}{R_U^3 R} \bar{\Phi}_{1 0 0}(\varphi'_i) 
   g^{(1)}(\theta,\theta')
\end{equation}
we obtain the following differential equation for $g^{(1)}(\theta,\theta')$:
\begin{equation} \label{5TransverseEq1}
  \frac{1}{R^2} \frac{1}{\sigma(\theta)\gamma(\theta)^3} \frac{\partial}{\partial\theta}
    \left[ \sigma(\theta)\gamma(\theta)^3 \frac{\partial g^{(1)}(\theta,\theta')}{\partial \theta}
    \right]=\frac{\delta(\theta-\theta')}{\sigma(\theta)\gamma(\theta)^3}-
    \tilde{\chi}_1(\theta) \bar{\tilde{\chi}}_1(\theta') \, ,
\end{equation}
where we used 
\begin{equation} 
  \tilde{\chi}_1(\theta)\equiv \left[ \frac{2\omega}{\tanh \left(\frac{\omega\pi}{2}\right)} 
  \right]^{\frac{1}{2}} \chi_1(\theta)\, , \qquad (\chi_1(\theta)\equiv1) \,.
\end{equation}
Note that we defined $\tilde{\chi}_1(\theta)$ in such a way that 
\begin{equation} 
  \int\limits_{-\frac{\pi}{2}}^{\frac{\pi}{2}} \sigma(\theta)\gamma(\theta)^3 \tilde{\chi}_1(\theta)
   \bar{\tilde{\chi}}_1(\theta) d\theta=1.
\end{equation}

\item $(\lambda' l' m')\neq(1 0 0)$. 

Due to the orthogonality of $\Phi_{1 0 0}$ and 
$\Phi_{\lambda' l' m'}$ on $S^3$, the last term on the right hand side of 
(\ref{5TransverseEquation}) vanishes. We therefore have
  \begin{align}
     &\frac{1-\lambda'^2}{R_U^2 \gamma(\theta)^2} c_{\lambda' l' m'}(\varphi'_i,\theta,\theta')+
     \frac{1}{R^2}\frac{1}{\sigma(\theta)\gamma(\theta)^3} \frac{\partial}{\partial\theta}\left[
     \sigma(\theta)\gamma(\theta)^3 \frac{\partial c_{\lambda' l' m'}(\varphi'_i,\theta,\theta')}
      {\partial \theta}\right]\qquad \qquad \nonumber \\
     &\qquad =\frac{1}{R_U^3 R} \frac{1}{\sigma(\theta)\gamma(\theta)^3} \delta(\theta-\theta')
    \bar{\Phi}_{\lambda' l' m'}(\varphi'_i)\, .
  \end{align}
Introducing $g^{(\lambda')}(\theta,\theta')$ again via
\begin{equation}
  c_{\lambda' l' m'}(\varphi'_i,\theta,\theta')=\frac{1}{R_U^3 R} \bar{\Phi}_{\lambda' l' m'}(\varphi'_i)
  g^{(\lambda')}(\theta,\theta')\, ,
\end{equation}
we see that $g^{(\lambda')}(\theta,\theta')$ has to satisfy
\begin{align} \label{5TransverseEq2}
    &\frac{1-\lambda'^2}{R_U^2 \gamma(\theta)^2} g^{(\lambda')}(\theta,\theta')+
    \frac{1}{R^2} \frac{1}{\sigma(\theta)\gamma(\theta)^3} \frac{\partial}{\partial\theta}
    \left[ \sigma(\theta)\gamma(\theta)^3 \frac{\partial g^{(\lambda')}(\theta,\theta')}{\partial \theta}
    \right]
=\frac{\delta(\theta-\theta')}{\sigma(\theta)\gamma(\theta)^3} \, .
\end{align}
\end{enumerate}

Combining the above results for $\lambda=1$ and $\lambda\neq 1$ we are able to write the formal solution
 of (\ref{5ModGreenFunction}) as
\begin{equation} 
  \mathcal{G}(\varphi_i,\varphi'_i,\theta,\theta')=\frac{1}{R_U^3 R}\sum_{\lambda=1}^{\infty}
  \sum_{l=0}^{\lambda-1}\sum_{m=-l}^l \Phi_{\lambda l m}(\varphi_i)\bar{\Phi}_{\lambda l m}(\varphi'_i)
  g^{(\lambda)}(\theta,\theta') \, .
\end{equation}

We obtain a further simplification of this formal solution by 
employing the spherical symmetry on $S^3$, see eq.~(\ref{5SumOverS3}), leaving us with a 
representation of the 
two-point function by a Fourier sum, which is natural for a compact space without boundaries:
\begin{equation} \label{5FormalSolution}
  \tilde{\mathcal{G}}(s,\theta,\theta')=\mathcal{G}(\varphi_i,\varphi'_i,\theta,\theta')
  =\frac{1}{R R_U^3} \sum_{\lambda=1}^{\infty} \frac{\lambda}{2 \pi^2} \frac{\sin(\lambda s)}{\sin s}
  g^{(\lambda)}(\theta,\theta') \, .
\end{equation}

Finding the formal solution (\ref{5FormalSolution}) was straightforward apart from minor complications 
inherent to the use of a
modified Green's function. To solve the differential equations (\ref{5TransverseEq1}) and 
(\ref{5TransverseEq2}) by imposing appropriate boundary conditions again is a routine task without 
any conceptual difficulties, though slightly technical in nature. Replacing back this solution in 
(\ref{5FormalSolution}) we are left with a Fourier sum which at first sight looks intractable, 
anticipating the fact that the  solutions $g^{(\lambda)}(\theta,\theta')$ (for $\lambda\neq1$) are
given by hypergeometric functions. Nevertheless it is possible to 
extract the desired asymptotic information from the sum (\ref{5FormalSolution}). 

In order not to disturb the transparency and fluidity of the main discussion, we provide large parts 
of the technical calculations in four appendices. 
In appendix \ref{appParallel} we report similarities and 
differences in the evaluation of the two-point functions between our case and the case of Randall-Sundrum, 
since the latter served as a guideline for handling the more difficult case under consideration.
The solutions of eqs. (\ref{5TransverseEq1}) and (\ref{5TransverseEq2}) are presented in 
appendix \ref{appDEQ} 
and the evaluation of the Fourier sum (\ref{5FormalSolution}) at distances exceeding the size of the extra dimension in 
appendix \ref{appSum}. 
Eventually, appendix \ref{appGreenUltraShort} contains the evaluation of the Fourier sum 
(\ref{5FormalSolution}) for distances smaller than the extra dimension.
In this way, we can offer the reader less interested in the details of the computations to have
the main results at hand.

From its definition (\ref{5ModGreenFunction}) we understand that the Green's function (\ref{5FormalSolution}) 
can be considered as the response of the scalar field to the combination of a point-like source located 
at coordinates $(\varphi'_i, \theta')$ and a delocalized, compensating negative contribution.   
Since we would like to see the response to a point-like particle on the brane, we put $\theta'=0$ in eq.
(\ref{5FormalSolution}) and explicitly write the $\lambda=1$ term:
\begin{equation} \label{5GreenSolBrane}
  \tilde{\mathcal{G}}(s,\theta,0)=\frac{g^{(1)}(\theta,0)}{2 \pi^2 R R_U^3}+
  \frac{R}{4 \pi^2 R_U^3}\frac{1}{\sin s} S[s,\theta,\omega] \, ,
\end{equation}
with $S[s,\theta,\omega]$ given by (\ref{5TheSumB}) of appendix \ref{appSum} 
(see also (\ref{5glambdaratio}) of appendix \ref{appDEQ}).
The general result for the sum $S[s,0,\omega]$ obtained in appendix \ref{appSum} is 
\begin{align} \label{5SresApp}
  S[s,0,\omega] &\equiv  \lim_{\theta \to 0} S[s,\theta,\omega] 
  = -\frac{2}{\omega} \frac{z(0)^\frac{1}{2}}{1-z(0)}
  \left(\frac{\pi-s}{2}\cos s -\frac{1}{4}\sin s \right)\\ \nonumber
  &\;\;-\frac{1}{2 \omega} z(0)^{-\frac{1}{2}} \ln\left[1-z(0)\right] \sin s
  -\frac{1}{\omega} z(0)^\frac{1}{2} \lim_{\theta \to 0} R[s,\theta,\omega]\;,
\end{align} 
where we have $z(0) = \tanh^2 \left(\frac{\omega\pi}{2} \right)$ and 
where we refer to (\ref{5defR}) for the definition of $R[s,\theta,\omega]$. The first term in 
(\ref{5SresApp}) is the zero mode contribution\footnote{At this point we have to explain what we mean by ``zero mode'' in this context, since 
due to the asymmetrically warped geometry, the spectrum of Kaluza-Klein excitations will not 
be Lorentz-invariant and strictly speaking, different excitations cannot be characterized by different 
$4$-dimensional masses. It would be more accurate to say that for a given value of the 
momentum eigenvalue $\lambda$, there exists a tower of corresponding Kaluza-Klein excitations 
with energies given by $E_n^\lambda$ with $n=0,1,\ldots$. In our use of language the ``zero mode branch'' 
of the spectrum or simply the ``zero mode'' is defined to be the collection of the lowest energy excitations corresponding to 
all possible values of $\lambda$, 
that is by the set $\left\{\left(E_0^\lambda,\lambda\right),\, \lambda=1,2,\ldots\right\}$. The definition 
is readily generalized to higher branches of the spectrum.} 
and we see that it reproduces exactly the $4$-dimensional 
static Green's function of Einstein's static universe given in (\ref{5ESUresult}). 
The other two terms are the contributions from the higher Kaluza-Klein modes.

We first concentrate on the case where $s \sim 1$ or what is equivalent $r\sim R_U$. Since one 
can easily convince oneself that $R[s,0,\omega]\sim \left[1-z(0)\right]^0$ in this regime, the 
contributions of the higher Kaluza-Klein modes are strongly suppressed with respect to the 
zero mode contribution. This means that as in the case of the Randall-Sundrum II model it 
is the zero mode which dominates the behavior of gravity at distances much larger than the extra
dimensions $R$. The main difference to the Randall-Sundrum II case is that the zero mode of our model 
not only gives rise to the 
typical $4$-dimensional $1/r$ singularity but also accounts for the compactness of space by
reproducing the Einstein static universe behavior (\ref{5ESUresult}). This result is somewhat 
surprising given the extreme anisotropy of our manifold. Due to the fact that the distances 
between two arbitrary points on the brane are of the order of $R$, one might intuitively 
expect that the extra dimension can be effective in determining also the large distance 
behavior of gravity (on the brane). As we could show by direct calculation the above expectation 
turns out to be incorrect. 
 
Next, we consider physical distances $r$ much larger than the extra dimension $r\gg R$ and much smaller 
than the observable universe $r \ll R_U$
in which case the results for $R[s,0,\omega]$ can be seen to be:
\begin{align} 
  R[s,0,\omega]&\sim\frac{\pi}{2 s^2}+
  \frac{\pi}{2}\frac{1-z(0)}{s^4} \left\{ 8-6\ln 2 -6 \ln\left[s \left[1-z(0)\right]^{-1/2}\right]\right\} 
  \\ \nonumber 
  &+\mathcal{O}\left[\left[1-z(0)\right]^2 \frac{\ln\left[s \left[1-z(0)\right]^{-1/2}\right]}{s^6}\right] 
  \, ,
\end{align}
valid for $\left[1-z(0)\right]^{1/2} \ll s \ll 1$.
The zero mode contribution (to $S[s,0,\omega]$) in this regime is simply the constant obtained by 
setting $s=0$ in the first term of (\ref{5SresApp}) so that we obtain:
\begin{equation} 
  S[s,0,\omega] \sim -\frac{\pi}{\omega} \frac{z(0)^{\frac{1}{2}}}{1-z(0)} 
  \left\{1+\frac{1}{2 \bar{s}^2}+\frac{1}{\bar{s}^4}\left[4-3 \ln 2 - 3 \ln \bar{s} \right] +
  \mathcal{O} \left( \frac{\ln \bar{s}}{\bar{s}^6}\right)\right\} \, ,
\end{equation}
where we introduced $\bar{s}=s \left[1-z(0)\right]^{-1/2}$. Inserting this result in (\ref{5GreenSolBrane})
and using physical distance $r=R_U s$ instead of $s$ we obtain
\begin{align} \label{5FullCorrections}
  &\tilde{\mathcal{G}}(s,0,0)=\\
  &=\frac{g^{(1)}(0,0)}{2 \pi^2 R R_U^3}\hspace{-0.5mm}-\hspace{-0.5mm}
  \frac{1}{4 \pi r} \frac{\omega \coth\left( \frac{\omega \pi}{2}\right)}{R} 
  \left\{1\hspace{-0.5mm}+\hspace{-0.5mm}\frac{1}{2 \bar{s}^2}\hspace{-0.5mm}+\hspace{-0.5mm}
  \frac{1}{\bar{s}^4}\left[4\hspace{-0.5mm}-\hspace{-0.5mm}3 \ln 2 \hspace{-0.5mm}-\hspace{-0.5mm}
    3 \ln \bar{s} \right] \hspace{-0.5mm}+\hspace{-0.5mm}
  \mathcal{O} \left( \frac{\ln \bar{s}}{\bar{s}^6}\right)\right\}\nonumber \\
  &=\frac{g^{(1)}(0,0)}{2 \pi^2 R R_U^3}-
  \frac{1}{4 \pi r} \frac{\omega \coth\left( \frac{\omega \pi}{2}\right)}{R} 
  \left\{1+\frac{\tanh^2\left(\frac{\omega\pi}{2}\right)}{2 \bar{r}^2}+
   \frac{\tanh^4\left(\frac{\omega\pi}{2}\right)}{\bar{r}^4} \, \times \right. \nonumber \\
  & \;\; \left. \times \, \left[4-3 \ln 2 - 
  3 \ln \left(\frac{\bar{r}}{\tanh \left(\frac{\omega \pi}{2}\right)}\right) \right]
  +  \mathcal{O} \left[ \frac{\tanh^6\left(\frac{\omega\pi}{2}\right)}{\bar{r}^6} \nonumber 
    \ln \left(\frac{\bar{r}}{\tanh \left(\frac{\omega \pi}{2}\right)}\right) \right]\right\} \, .
\end{align}
Since we would like to compare our result with the corresponding correction in the Randall-Sundrum II
case, we introduced the dimensionless distance variable $\bar{r}=r \omega/R$, 
the physical distance measured in units of the AdS-radius, in the last line of the above result. 
We see that 
in complete agreement with the Randall-Sundrum II scenario, our setup reproduces 
$4$-dimensional gravity at large distances with extremely suppressed corrections. The only remnant 
effect from the different global topology manifest itself through the factors of 
$\tanh \left(\frac{\omega \pi}{2}\right)$ and $\coth \left(\frac{\omega \pi}{2}\right)$ which
are very close to $1$. 
We furthermore emphasize that
apart from these  deviations, the asymptotic we obtained coincides exactly with the 
asymptotic for the case of a massless scalar field 
in the Randall-Sundrum II background, see (\ref{5I1Asymptotic})
and e.g.~\cite{Giddings:2000mu,Kiritsis:2002ca,Callin:2004py,Ghoroku:2003bs}. 
From the factor $\omega \coth \left(\frac{\omega \pi}{2}\right)/R$ 
in eq. (\ref{5FullCorrections}) we see that also the relation between the fundamental scale $M$ and 
the Planck-scale $M_{Pl}$ gets modified only by the same factor of 
$\tanh\left(\frac{\omega \pi}{2}\right)$:
\begin{equation} 
  M_{Pl}^2 = M^3 \frac{R}{\omega} \tanh \left(\frac{\omega \pi}{2} \right) \, ,
\end{equation}
where we remind that $R/\omega$ is nothing but the AdS-Radius.

Eventually, we treat the case of distances inferior to the extra dimension $r\ll R$. 
After using the result (\ref{55DNewton}) of appendix \ref{appGreenUltraShort} in (\ref{5GreenSolBrane}), a 
short calculation reveals 
\begin{align} \label{5Short5DCorrections}
  \tilde{\mathcal{G}}(s,0,0)&=\frac{g^{(1)}(0,0)}{2 \pi^2 R R_U^3}-\frac{1}{4\pi^2}
  \frac{1}{R_U^2 s^2+R^2 \theta^2}, 
\end{align}
a result that has to be compared to the characteristic solution of the Poisson equation in 4-dimensional
flat space. In $n$-dimensional flat space one has:
\begin{equation} \label{5CharLaplSolnDspace}
  \Delta \left[ -\frac{1}{(n-2) V_{S^{n-1}} r^{n-2}}\right]=\delta(r) \, , \qquad 
  r=\left(\sum_{i=1}^n x_i^2\right)^{\frac{1}{2}} \, ,
\end{equation}
with $V_{S^{n-1}}=2\pi^{\frac{n}{2}}/\Gamma[\frac{n}{2}]$ denoting the volume of the $n-1$ sphere. Specifying
to $n=4$, we recover the correct prefactor of $-1/4\pi^2$ in (\ref{5Short5DCorrections}) multiplying the 
$1/r^2$ singularity.

Finally, we mention that in none of the considered cases we payed attention to the additive constant 
in the two-point function on the brane. The arbitrary constant entering the solution 
$g^{(1)}(\theta,\theta')$ can always be chosen in such a way that $g^{(1)}(0,0)$ vanishes on the brane (see
appendix~\ref{appDEQ}).

\section{Conclusions}
\label{5ESUfour}

In this chapter we considered a particular brane world model in 5 dimensions with the 
characteristic property that the spatial part of the space-time manifold (including the 
extra dimension) is compact and has the topology of a $4$-sphere $S^4$. Similar to the 
original Randall-Sundrum~II model, the $3$-brane is located at the boundary between two
regions of $AdS_5$ space-time. The coordinates of $AdS_5$ used by Randall and Sundrum 
are closely related to the so-called Poincar\'e coordinates of $AdS_5$. While the extra 
dimension in this set of coordinates provides a slicing of $AdS_5$ along flat $4$-dimensional
Minkowski sections (resulting in a flat Minkowskian induced metric on the brane), 
their disadvantage is that they do not cover the whole of $AdS_5$ space-time. 

The coordinates we used are the global coordinates of $AdS_5$ known to provide 
a global cover of the $AdS_5$ space-time. In this case the ``extra'' dimension labels different
sections with intrinsic geometry $\mathbb{R} \times S^3$, the geometry of 
Einsteins static universe. The induced metric on the $3$-brane in our setup is therefore also 
given by $\mathbb{R} \times S^3$.

As we illustrated with the use of the Penrose-diagram of $AdS_5$, the incompleteness of the 
Poincar\'e patch is at the origin of the incompleteness of the Randall-Sundrum II 
space-time with respect to time-like and light-like geodesics. Moreover, we were able to demonstrate 
that the setup considered in this chapter provides an alternative to the Randall-Sundrum II model 
which does not suffer from the drawback of being geodesically incomplete. The latter point was 
part of the main motivations for the study of this setup.

The spatial 
part of our manifold is characterized by an extreme anisotropy with respect to one of the 
coordinates (the extra coordinate) accounting for thirty orders of magnitude between the
size of the observable universe and present upper bounds for the size of extra dimensions. 

Another interesting property of our manifold, related to the anisotropy of its spatial part, is the
fact that {\em any two points} on the brane are separated by a distance of the order of the 
size of the extra dimension $R$ regardless of their distance measured by means of the 
induced metric on the brane. Despite the difference in the global topology, the properties of 
gravity localization turned out to be very similar to the Randall-Sundrum~II model, though much more 
difficult to work out technically. We computed the static (modified) Green's function of a massless 
scalar field in our background and could show that in the intermediate distance regime 
$R \ll r \ll R_U$ the $4$-dimensional Newton's law is valid for two test particles on the brane, 
with asymptotic corrections terms identical to the Randall-Sundrum~II case up to tiny 
factors of $\tanh\left(\frac{\omega\pi}{2}\right)$. We could also 
recover the characteristic $5$-dimensional behavior of the Green's function for distances smaller than
the extra dimension $r \ll R$. Eventually we saw that in the regime of cosmic distances 
$r \sim R_U$, somewhat 
counterintuitive given our highly anisotropic manifold, the Green's function is dominated 
by the behavior of the corresponding static (modified)
Green's function in Einstein's static universe.

In the simple setup considered in this chapter, the $3$-brane is supposed to be motionless.
In the light of recent progress in the study of $4$-dimensional cosmic evolution
induced by the motion of the brane in the bulk, it would be interesting to explore this possibility and 
see what kind of modifications of our results we would have to envisage. Finally, a related, 
important question which would be interesting to address would be the question of stability of our setup.
\chapter*{Concluding remarks}  \label{cha6}

Topological defects or topological solitons have important applications not only in $4$-dimensional
condensed matter physics, particle physics and cosmology, but also more recently in theories with 
extra dimensions, as sketched in this work. There are various reasons to study topological defects in 
the latter context: to start with, we mention their classical stability. Strictly speaking, 
the topological arguments in 
favor of classical stability are valid only in the case of a flat space-time. 
However, in certain cases of topological defects weakly coupled to gravity, we can expect to obtain 
good candidates for stable configurations. Another reason why topological defects are interesting for
model building in theories involving higher dimensions is that they explain the chiral nature of 
the observed $4$-dimensional fermions. The higher dimensional 
Dirac operator in a topologically non-trivial background contains a certain number of zero modes 
which form a $4$-dimensional point of view 
can be identified as Standard Model fermions. It would be desirable to reproduce in this way the field 
content of the Standard Model from a higher dimensional theory.

The introduction of extra dimensions in any theory always has to be accompanied by 
an explanation for why our world appears to be $4$-dimensional at low energies. One crucial 
test of such theories when gravity is concerned is whether they are able to predict the 
correct $4$-dimensional Newton's law 
in the distance regime explored by latest observations. At this point, non-factorizable metrics 
play an important role since it has been show recently that backgrounds involving such metrics 
do pass the test of reproducing the correct $4$-dimensional behavior of the gravitational potential. 
It is the interplay between these two ingredients, topological defects and non-factorizable geometries in
higher dimensions, which lies at the center of interest of this work. After an 
introductory chapter on non-factorizable metrics, we presented some possibilities of topological defect 
solutions to the higher dimensional Einstein equations. The results presented in chapter \ref{cha3} were
mainly of an asymptotic nature in the region of transverse space far from the core of the defects. 
In chapter \ref{cha4} we gave a numerical realization of one of the metric asymptotics found in chapter 
\ref{cha3} in the case of a magnetic monopole residing in transverse space, as a particular example of a 
bulk $p$-form field. The main motivation for the study presented in chapter \ref{cha5} was 
the incompleteness of the Randall-Sundrum II space-time with respect to time-like and light-like 
geodesics. As we showed, the brane-world model we discussed is geodesically complete and still 
reproduces the correct $4$-dimensional behavior of gravity on the brane. 

Needless to say that numerous open questions remain. As mentioned above, the stability of all solutions
of the Einstein equations given in this work has to be analyzed carefully. A related problem would be
the study of the dynamics, in a cosmological context for example, since all cases considered correspond 
to static solutions.
The verification of Newton's law on the brane is only one necessary consistency check.
More generally it should be verified that $4$-dimensional general relativity can be recovered on the brane.

To summarize, higher dimensional brane-world scenarios provide numerous opportunities for addressing 
many unresolved problems of theoretical particle physics and cosmology. However, the introduction of extra 
dimensions always has to be accompanied by an explanation for why they are not observed at 
accessible energies. Despite numerous attempts in this direction, no theoretically and phenomenologically
satisfying model exists at present. It is interesting to see in which way new collider 
experiments will change our present view of extra dimensions.

\chapter*{Acknowledgments}
{\em 
I am very grateful to many people for their continuous help and support during this work.
First, I would like to thank my supervisor Prof. Mikhail Shaposhnikov for countless discussions 
and for helpful advice throughout this work. Second, I am also grateful to Prof. Jean-Jacques Loeffel 
for competent help in mathematical questions. Then, I have to thank Prof. Peter Tinyakov 
for the time he sacrificed in numerous discussions of physical and mathematical nature. 
Eventually I am deeply indebted to Prof. Bernhard Schnizer for his continuous support
and advice.
Next, I would like to thank my friends: first of all Sylvain Wolf for 
his support and help during the first two years of my stay in Lausanne. Second, Alessandro Gruppuso for 
his kind hospitality during my stay in Italy and for teaching me the proper meaning of the 
word collaboration. Third, Elmar Teufl for his valuable advice in mathematical issues of various kinds.
Finally, my thanks goes to all my colleagues in ITP inside and outside the cosmology group for the 
friendly working environment. I also wish to express my gratitude to our librarian Madame 
Josiane Moll for her kindness and for her help in ordering books from outside our library. 

Last, and probably most importantly, I wish to deeply thank my wife for her everlasting understanding, 
care and support throughout the last two years and in particular during the period of typesetting of the 
manuscript.}
\thispagestyle{empty}
\appendix
\chapter{A generalization of the metric (\ref{WarpedMetric})} \label{appDoubleWarp}
In this appendix we plan to consider a particular metric ansatz which can be considered as
a generalization of the metric (\ref{WarpedMetric}) given in chapter \ref{cha2}. As we will see,
the Einstein equations with a cosmological constant term in otherwise empty space-time are very restrictive
in the case we are going to study. The metric generalization which we will briefly
describe is used in theories of higher dimensions characterized by two scale-factors (in the 
cosmological sense). In this context of {\em dynamical compactification} 
\cite{Chodos:1979vk}-\cite{Darabi:2003st}, the cosmological expansion
of our $4$-dimensional universe is thought to be accompanied by a shrinking of the size of an internal 
manifold.

The form (\ref{WarpedMetric}) of the metric can still be generalized in the following sense
\begin{equation} \label{WarpedMetricGeneralized}
  ds^2=\hat{g}_{M N} \, dx^M \, dx^N \, = \, \sigma\left(x^a \right) \, g_{\mu \nu}\left(x^\rho\right) \, 
   dx^\mu \, dx^\nu +
  \eta\left(x^\mu \right) \,\tilde{g}_{a b}\left(x^c\right) dx^a \, dx^b \, ,
\end{equation}
introducing a second warp factor $\eta\left(x^\mu \right)$ depending only on the coordinates
$x^\mu$. In this case the connections become:
\begin{align} \label{WarpedConnectionsGeneralized}
  \hat{\Gamma}^c_{a b}&=\tilde{\Gamma}^c_{a b} \, , \quad
  \hat{\Gamma}^a_{b \gamma}=\frac{1}{2\, \eta} \nabla_\gamma \, \eta \, \delta^a_{\; b} \, , \quad
  \hat{\Gamma}^\gamma_{a b}=-\frac{1}{2\, \sigma} \nabla^\gamma \, \eta \, \tilde{g}_{a b} \, , \quad \\
  \hat{\Gamma}^c_{\mu \nu}&=-\frac{1}{2\, \eta} g_{\mu \nu} \tilde{\nabla}^c \, \sigma \, , \quad 
  \hat{\Gamma}^\nu_{\mu b}=\frac{1}{2 \, \sigma}\delta^\nu_{\; \mu} \tilde{\nabla}_b \, \sigma \, , \quad
  \hat{\Gamma}^\rho_{\mu \nu}=\Gamma^\rho_{\mu \nu} \, . \nonumber 
\end{align}
Similarly, the Ricci tensors $\hat{R}_{M N}$, $R_{\mu \nu}$, $\tilde{R}_{a b}$ and 
Ricci scalars $\hat{R}$, $R$ and $\tilde{R}$ corresponding to  
$\hat{g}_{M N}$, $g_{\mu \nu}$ and $\tilde{g}_{a b}$ generalize to:
\begin{align} \label{WarpedRicciTensorGeneralized}
  \hat{R}_{\mu \nu} &= R_{\mu \nu} -\frac{1}{2 \, \eta} g_{\mu \nu} 
  \left[ \tilde{\nabla}_a \, \tilde{\nabla}^a \, \sigma + \frac{d_1-2}{2 \, \sigma} 
         \tilde{\nabla}_a \, \sigma \, \tilde{\nabla}^a \, \sigma \right] \\
  & \quad \qquad - \frac{d_2}{2 \, \eta} \nabla_\mu \, \nabla_\nu \, \eta  
  + \frac{d_2}{4 \, \eta^2} \nabla_\mu \, \eta \, \nabla_\nu \, \eta \, ,  \nonumber \\
  \hat{R}_{a b} &= \tilde{R}_{a b} -\frac{1}{2 \, \sigma} \tilde{g}_{a b} 
  \left[ \nabla_\mu \, \nabla^\mu \, \eta + \frac{d_2-2}{2 \, \eta} 
         \nabla_\mu \, \eta \, \nabla^\mu \, \eta \right] \\
  & \quad \qquad - \frac{d_1}{2 \, \sigma} \tilde{\nabla}_a \, \tilde{\nabla}_b \, \sigma  \, 
  + \frac{d_1}{4 \, \sigma^2} \tilde{\nabla}_a \, \sigma \, \tilde{\nabla}_b \, \sigma \, ,  \nonumber
\end{align}
\begin{align}
  \hat{R}_{\mu a} &= \frac{d_1+d_2-2}{4 \, \sigma \, \eta} \, \nabla_\mu \, \eta \, \tilde{\nabla}_a \, 
   \sigma\, , \label{WarpedRicciTensorOffDiagonal} \\
  \hat{R}&=\frac{1}{\eta}\, \tilde{R} -\frac{d_1}{\sigma \, \eta} \label{WarpedRicciScalarGeneralized}
            \tilde{\nabla}_a \, \tilde{\nabla}^a \, \sigma  - 
    \frac{d_1\left(d_1-3\right)}{4\, \sigma^2 \, \eta} 
    \tilde{\nabla}_a \, \sigma \, \tilde{\nabla}^a \, \sigma  \\
   &+ \frac{1}{\sigma}\, R -\frac{d_2}{\eta \, \sigma} 
            \nabla_\mu \, \nabla^\mu \, \eta  - 
    \frac{d_2\left(d_2-3\right)}{4\, \eta^2 \, \sigma} 
    \nabla_\mu \, \eta \, \nabla^\mu \, \eta \nonumber \, .
\end{align}
In the case $\eta=1$ the relations (\ref{WarpedConnectionsGeneralized})-(\ref{WarpedRicciScalarGeneralized}) 
obviously reduce to (\ref{WarpedConnections})-(\ref{WarpedRicciScalar}).

For the generalized metric (\ref{WarpedMetricGeneralized}) the  Einstein equations (\ref{FullEinstein}) 
turn out to be more restrictive, at least in the case of vanishing energy-momentum tensor, 
see \cite{Katanaev:1998ry} for related work in $4$ dimensions. 
Apart from the physically uninteresting case $d_1+d_2=2$, the 
off-diagonal components $_{\mu a}$ of (\ref{FullEinstein}) imply that either $\sigma$ or $\eta$ has to be 
constant, as a direct consequence of (\ref{WarpedRicciTensorOffDiagonal}).
To some extend, this brings us 
back to the original metric (\ref{WarpedMetric}). In particular with the choice 
$\eta=1$, (\ref{WarpedMetricGeneralized}) reduces to (\ref{WarpedMetric}). 
The case $\sigma=1$ has been studied in the context of dynamical 
compactification of extra dimensions \cite{Chodos:1979vk}-\cite{Darabi:2003st}. 
Due to the formal invariance of the metric under the following interchange 
of ``conventional'' and ``internal'' dimensions
\begin{equation} \label{ExternalInternalInvariance}
  x^\mu \leftarrow \hspace{-3mm} \rightarrow x^a \, \quad 
  g_{\mu \nu}\left(x^\rho\right) \leftarrow \hspace{-3mm}\rightarrow \tilde{g}_{a b} \left(x^c\right) \, \quad
  \sigma(x^a) \leftarrow \hspace{-3mm}\rightarrow \eta(x^\mu) \, ,
\end{equation}
the Einstein equations for this case can be obtained from the Einstein equations 
(\ref{FullEinsteinExplicit1})--(\ref{FullEinsteinExplicit3})
by generalized replacements of this kind. However, it is more useful to present the equations 
(\ref{FullEinstein}) in a somewhat more suggestive way:
\begin{align} 
  &R_{\mu \nu}-\frac{1}{2} \, g_{\mu \nu} \, R+\Lambda \, g_{\mu \nu} = 
   \frac{d_2}{2\,\eta} \nabla_\mu \, \nabla_\nu \, \eta 
  -\frac{d_2}{4\, \eta^2} \nabla_\mu \, \eta \, \nabla_\nu \, \eta\, , 
   \label{FullEinsteinExplicitGeneralizedmunu}\\
  &\tilde{R}_{a b}-\frac{1}{2} \, \tilde{g}_{a b} \, \tilde{R}+\tilde{\Lambda} \, \tilde{g}_{a b} =0 \, 
  \label{FullEinsteinExplicitGeneralizedab},
\end{align}
where 
\begin{align}
  \Lambda&=\frac{d_2\left(d_2-3\right)}{8\,\eta^2} \, \nabla^\lambda \, \eta \, \nabla_\lambda \, \eta +
           \frac{d_2}{2 \, \eta} \nabla_\lambda \, \nabla^\lambda \, \eta + \Lambda_B 
           -\frac{1}{2\,\eta}\, \tilde{R} \, , \label{Lambdad1} \\
  \tilde{\Lambda}&=\Lambda_B \, \eta +\frac{\left(d_2-1\right)\left(d_2-4\right)}{8\,\eta}
   \, \nabla^\lambda \, \eta \, \nabla_\lambda \, \eta
   +\frac{d_2-1}{2}\nabla_\lambda \, \nabla^\lambda \, \eta -\frac{\eta}{2}\, R \, . \label{Lambdad2}
\end{align}
Due to the choice $\sigma=1$ the off-diagonal components $_{\mu a}$ are identically satisfied.
We are now interested in solutions to the above equations (\ref{FullEinsteinExplicitGeneralizedmunu}) and 
(\ref{FullEinsteinExplicitGeneralizedab}) in which 
the dependencies on the coordinates are separated. In other words, we would like to eliminate the 
dependence of $\Lambda$ in (\ref{Lambdad1}) on the ``internal'' coordinates $x^a$ via $\tilde{R}$. This is 
equivalent to imposing that the internal space should be a constant curvature space 
$\tilde{R} \neq \tilde{R}\left(x^a\right)$. By taking the trace of 
eq.~(\ref{FullEinsteinExplicitGeneralizedab}), this by itself implies that also $\tilde{\Lambda}$ has to
be a constant. It is now an algebraic exercise to take the trace also of 
eq.~(\ref{FullEinsteinExplicitGeneralizedmunu}) in order to eliminate the dependencies of $\Lambda$ 
and $\tilde{\Lambda}$ on $\tilde{R}$ and $R$, respectively. The result is 
\begin{align} 
  \Lambda&=\frac{d_1-2}{d_1+d_2-2} \, \Lambda_B+\frac{d_2}{4\, \eta}
        \nabla_\lambda \, \nabla^\lambda \, \eta 
   -\frac{d_2}{8 \, \eta^2} \, \nabla_\lambda \, \eta \nabla^\lambda \, \eta
  \, , \label{LambdaResult} \\
  \tilde{\Lambda}&=\frac{d_2-2}{d_1+d_2-2} \Lambda_B \, \eta + 
   \frac{\left(d_2-2\right)^2}{8 \, \eta} \, \nabla_\lambda \, \eta \nabla^\lambda \, \eta +
   \frac{\left(d_2-2\right)}{4} \, \nabla_\lambda \, \nabla^\lambda \, \eta \label{LambdatildeResult} \, .
\end{align}
The inverse powers of $\eta$ in (\ref{LambdaResult}) suggest the following redefinition of the warp-factor:
\begin{equation} \label{warpRedefine}
  \eta\left(x^\mu\right)=e^{2 \, \rho\left(x^\mu\right)} \, ,
\end{equation}
so that the ``effective'' $4$-dimensional Einstein equations (\ref{FullEinsteinExplicitGeneralizedmunu}) 
take the form:
\begin{align} \label{EinsteinEffective}
  R_{\mu \nu} - \frac{1}{2}\, g_{\mu \nu} \, R + \frac{d_1-2}{d_1+d_2-2}\, \Lambda_B g_{\mu \nu} = 
  T_{\mu \nu} \, ,
\end{align}
where we introduced $T_{\mu \nu}$ by 
\begin{equation} \label{IntroTmunu}
  T_{\mu \nu} = d_2 \left\{ \left[ \nabla_\mu \, \rho \nabla_\nu \, \rho + 
     \nabla_\mu \, \nabla_\nu \, \rho \right]- \frac{1}{2} \, g_{\mu \nu} 
   \left[\nabla_\mu \, \rho \, \nabla^\mu \rho + \nabla_\mu \, \nabla^\mu \rho \right]\right\} \, .
\end{equation} 
Eq.~(\ref{EinsteinEffective}) has to be supplemented by the highly non-linear ``field''-equation 
(\ref{LambdatildeResult}) 
for the warp factor $\rho$:
\begin{align} \label{effectiveWarpFactorEquation}
  \nabla_\lambda \, \nabla^\lambda \, \rho + 
  d_2 \, \nabla_\lambda \, \rho \, \nabla^\lambda \, \rho
  -\frac{2}{d_2-2} \, e^{-2 \rho} \, \tilde{\Lambda} + \frac{2}{d_1+d_2-2} \, \Lambda_B =0 \, ,
\end{align}
where $\tilde{\Lambda}$ is related to the Ricci scalar $\tilde{R}$ of the internal constant curvature 
space by
\begin{equation}
  \tilde{\Lambda}=\frac{2-d_2}{2 \, d_2} \tilde{R} \, .
\end{equation}
From eq.~(\ref{EinsteinEffective}) we see that the effective $4$-dimensional cosmological constant is a
fraction of the bulk cosmological constant and that from a $4$-dimensional point of view 
the warp-factor plays the role of a scalar field coupled to Einstein-equations via the effective 
energy-momentum tensor (\ref{IntroTmunu}). Note that in addition to the standard 
energy-momentum tensor of a massless scalar field given by the two first terms in the square brackets
of eq.~(\ref{IntroTmunu}) there are also two non-standard contributions given by the two last terms in 
the square brackets. Since we consider a pure gravity theory it is needless to say that there is no
Planck-scale involved in the coupling of gravity to the scalar field $\rho$.
The constant $\tilde{\Lambda}$ is a free parameter of the theory.

\chapter{Fine-tuning relations for the 't~Hooft-Polyakov monopole} \label{appFTR}
By taking linear combinations of Einstein equations (\ref{4EinsteinP0}) to (\ref{4EinsteinPtheta}) 
one can easily derive the following relations:
\begin{eqnarray}
  \frac{\left[M(r)^3 M'(r) \mathcal{L}(r)^2 \right]'}{M(r)^4 \mathcal{L}(r)^2}&=&
    -\frac{\beta}{5} \left( 2 \gamma + \epsilon_0 - \epsilon_\rho - 2 \epsilon_\theta \right) \, , 
\label{4Einsteincomb1}\\
  \frac{\left[M(r)^4 \mathcal{L}(r) \mathcal{L}'(r)\right]'}{M(r)^4 \mathcal{L}^2(r)}-\frac{1}{\mathcal{L}(r)^2}&=&
    -\frac{\beta}{5} \left( 2 \gamma - 4 \epsilon_0 - \epsilon_\rho + 3 \epsilon_\theta \right) \, . \label{4Einsteincomb2}
\end{eqnarray}
By multiplying with $M(r)^4 \mathcal{L}(r)^2$, integrating from $0$ to $\infty$ and using the definition of 
the brane tension components (\ref{4branetensions}) we obtain
\begin{eqnarray}
  M(r)^3 M'(r) \mathcal{L}(r)^2 \vert_{0}^{\infty}&=& 
    -\frac{2\beta\gamma}{5} \int\limits_{0}^{\infty} M(r)^4 \mathcal{L}(r)^2 dr  \label{4A1} \\ \nonumber
   &&+ \frac{\beta}{5} \left( \mu_{0}-\mu_{\rho}-2 \mu_{\theta}\right) \, , \\
  M(r)^4 \mathcal{L}(r) \mathcal{L}'(r) \vert_{0}^{\infty}&=&\hspace{-2mm}\int\limits_{0}^{\infty} 
    \hspace{-2mm} M(r)^4 dr - \label{4A2}
   \frac{2 \beta \gamma}{5} \int\limits_0^{\infty} \hspace{-2mm} M(r)^4 \mathcal{L}(r)^2 dr \\ \nonumber
    &&- \frac{\beta}{5} 
      \left(4 \mu_0 + \mu_\rho - 3 \mu_\theta \right) \, . 
\end{eqnarray}
Using the boundary conditions for the metric functions (\ref{4BCMorigin}) and (\ref{4asympto}), and taking the difference 
of the eqs.~(\ref{4A1}) and (\ref{4A2}) then establishes 
the first part of eq.~(\ref{4FineTunRel1}). To prove the second part of (\ref{4FineTunRel1}), we start 
 from the general expressions of the stress-energy components $\epsilon_i$, 
relations~(\ref{4enmomelements1})--(\ref{4enmomelements3}):
\begin{align}
  \mu_0-\mu_{\theta}&=\int\limits_0^{\infty} dr M(r)^4 \mathcal{L}(r)^2 
   \left( \epsilon_\theta - \epsilon_0 \right)  \\ \nonumber
   & = \int\limits_0^{\infty} dr M(r)^4 \mathcal{L}(r)^2 \left\{ 
    \frac{K'(r)^2}{\mathcal{L}(r)^2}+\frac{\left[1-K(r)^2\right]^2}{\mathcal{L}(r)^4} + 
    \frac{J(r)^2 K(r)^2}{\mathcal{L}(r)^2} \right\} \, . 
\end{align}
Multiplying the equation of motion for the gauge field (\ref{4EqMovW}) by $K(r)$ and substituting 
the $J(r)^2 K(r)^2$ term gives 
\begin{eqnarray}
  \mu_0-\mu_\theta=\int\limits_0^{\infty} dr M(r)^4 \left\{ K'(r)^2+\frac{1- K(r)^2}{\mathcal{L}(r)^2}
     + \frac{\left[ M(r)^4 K'(r) \right]' K(r)}{M(r)^4} \right\}  \, .
\end{eqnarray}
Integration by parts in the last term of the above equation leads to
\begin{eqnarray}
   \mu_0-\mu_\theta=\int\limits_0^{\infty} dr M(r)^4 \left[\frac{1-K(r)^2}{\mathcal{L}(r)^2}\right]+ 
   \left[ K(r) M(r)^4 K'(r) \right]_0^{\infty} \, ,
\end{eqnarray} 
which together with the behavior of the gauge field at the origin $K'(0) =0$ (see (\ref{4asy_origin_K})) finishes
 the proof of relation (\ref{4FineTunRel1}). 

\noindent
The proof of relation (\ref{4FineTunRel2}) is simply obtained by rewriting (\ref{4A1}).

\noindent
To establish (\ref{4FineTunRel3}) we start directly from the definitions of the stress-energy tensor components,
relations~(\ref{4enmomelements1})-(\ref{4enmomelements3}):
\begin{equation} \label{4FTR3int}
  \mu_0+\mu_{\rho}+2\mu_{\theta}= \int\limits_{0}^{\infty} dr M(r)^4 \mathcal{L}(r)^2 
    \left[ J'(r)^2+\frac{2 \, J(r)^2 K(r)^2}{\mathcal{L}(r)^2}+\alpha \left( J(r)^2-1\right)^2
    \right] \, . 
\end{equation}
Collecting derivatives in the equation of motion for the scalar field (\ref{4EqMovphi}) and multiplying by $J(r)$ gives
\begin{equation}
  \frac{2 \, J(r)^2 K(r)^2}{\mathcal{L}(r)^2} = 
    \frac{\left[ M(r)^4 \mathcal{L}(r)^2 J'(r) \right]'}{M(r)^4 \mathcal{L}(r)^2} J(r) 
    -\alpha J(r)^2 \left( J(r)^2-1\right) \, .
\end{equation}
Eliminating now the second term in the equation (\ref{4FTR3int}) leads to 
\begin{align}
  \mu_0+\mu_{\rho}+2\mu_{\theta}&= \int\limits_{0}^{\infty} dr M(r)^4 \mathcal{L}(r)^2
   \left[ J'(r)^2+\frac{\left[ M(r)^4 \mathcal{L}(r)^2 J'(r) \right]'}{M(r)^4 \mathcal{L}(r)^2} J(r) + 
  \alpha \left( 1- J(r)^2 \right) \right] \, .
\end{align}
If we now expand and integrate the second term by parts we are left with
\begin{align} 
 \mu_0+\mu_{\rho}+2\mu_{\theta}=& \alpha \int\limits_{0}^{\infty} dr \left( 1- J(r)^2 \right) M(r)^4 
  \mathcal{L}(r)^2 + 
  \left[M(r)^4 \mathcal{L}(r)^2 J(r) J'(r) \right]_0^{\infty} \, ,
\end{align}
which reduces to (\ref{4FineTunRel3}) when the boundary conditions for 
the metric  (\ref{4BCMorigin}) and (\ref{4asympto}) are used.
\thispagestyle{empty}
\chapter{Computational parallels in the corrections to Newton's law} \label{appParallel}

The purpose of this appendix is to review briefly the calculations of the static two-point function of 
a scalar field in the Randall-Sundrum II background \cite{Giddings:2000mu,Kiritsis:2002ca} and to 
compare each stage with the corresponding 
stage of calculations in the background considered in chapter \ref{cha5}. This serves mainly for underlining
similarities and differences between the two calculations. Let us begin by writing down the metric of 
the Randall-Sundrum II model
\begin{equation} \label{5RSmetric}
  ds^2=e^{-2 k \vert y\vert} \eta_{\mu \nu} dx^\mu dx^\nu + dy^2\, , \qquad -\pi r_c \leq y \leq \pi r_c \, .
\end{equation}
Here, $y$ stands for the extra dimension while $k$ and $\eta_{\mu \nu}$ denote the inverse radius of $AdS_5$
space and the (4-dimensional) Minkowski metric with signature $-+++$. Due to the orbifold symmetry, the
allowed range of $y$ is $0 \leq y \leq \pi r_c$.\footnote{We allow for a finite $r_c$ only to impose 
boundary conditions in a proper way. Eventually we are interested in the limit $r_c \to \infty$.}
As it is well known, each $4$-dimensional graviton mode in this background satisfies the
 equation of a massless scalar field. 
Therefore, for the study of the potential between two test masses on the brane we confine ourselves 
to solving the equation of a massless scalar field with an arbitrary time-independent source:
\begin{equation} \label{5RSScalarEq}
  \mathcal{D} u(\vec x,y) = j(\vec x,y) \qquad \mbox{with} \; \; \mathcal{D}=e^{2 k y} \Delta_x -
   4 k\frac{\partial}{\partial y}+\frac{\partial^2}{\partial y^2} \, .
\end{equation}
Since the operator $\mathcal{D}$ is formally self-adjoint, the corresponding Green's function
will satisfy:
\begin{equation} \label{5RSGreendef}
  \mathcal{D} \mathcal{G}(\vec x,\vec x';y,y')=\frac{\delta^3(\vec x-\vec x') \, \delta(y-y')}{\sqrt{-g}} \, .
\end{equation}
We are now able to write down the usual integral representation of the solution of (\ref{5RSScalarEq}):
\begin{equation} \label{5RSSolIntRep}
  u(\vec x,y)=\int \sqrt{-g} \, j(\vec x',y')\,  \overline{G(\vec x,\vec x';y,y')} \, d^3 \vec x' dy',
\end{equation}
where the $y'$-integration extends from $0$ to $\pi r_c$ and the $\vec x'$ integrations from
$-\infty$ to $\infty$.
The absence of boundary terms in (\ref{5RSSolIntRep}) is of course the result of an appropriate choice 
of  boundary conditions for $u(\vec x,y)$ and  $G(\vec x,\vec x';y,y')$. In the above coordinates 
of the Randall-Sundrum II case the orbifold boundary conditions together with the Israel condition 
imposed on the fluctuations of the metric give rise to a Neumann boundary condition at $y=0$. One can easily 
convince oneself that in the limit $r_c \to \infty$ the resulting Green's 
function is independent of the choice of the (homogeneous) boundary condition at $y=\pi r_c$. We therefore
follow \cite{Randall:1999vf} and use also a Neumann boundary condition at $y=\pi r_c$. Another important 
point is that the Green's function we are considering is specific to the orbifold boundary condition and
so describes the situation of a semi-infinite extra dimension.
We decided to carry out the calculations in the semi-infinite case as opposed to \cite{Randall:1999vf}, 
where eventually the orbifold boundary conditions are dropped and the case of a fully infinite extra 
dimension is considered.\footnote{However, as long as matter on the 
brane is considered, the Green's function for the two setups differ only by a factor of $2$.}
An immediate consequence of this will be that the constant factors of the 
characteristic short distance singularities of the solutions of Laplace 
equations also will be modified by a factor of $2$. Finally, we note that this factor of $2$ can be 
accounted for by an overall redefinition of the 5-dimensional Newton's constant leaving the two theories
with equivalent predictions. We conclude the discussion of boundary conditions by noting that
at infinity in $\vec x'$ we suppose that $u(\vec x',y')$ and $G(\vec x, \vec x'; y,y')$ vanish 
sufficiently rapidly.

The solution of eq.~(\ref{5RSGreendef}) can be found most conveniently by Fourier expansion 
in the $\vec x$ coordinates and by direct solution of the resulting differential equation for the 
 ``transverse'' Green's function. After performing the integrations over the angular coordinates in
Fourier space we obtain:
\begin{equation} \label{5RSGreenFormalSolution}
  \mathcal{G}(\vec x-\vec x',y,y')=\frac{1}{2 \pi^2}\int\limits_0^\infty 
  \frac{p \sin\left(p \vert \vec x-\vec x' \vert \right)}{\vert \vec x-\vec x' \vert} g^{(p)}(y,y') dp \, ,
\end{equation}
where $g^{(p)}(y,y')$ satisfies
\begin{equation} \label{5RSTransGreen}
  \frac{1}{e^{-4 k y}} \frac{\partial}{\partial y}
   \left[e^{-4 k y} \frac{\partial}{\partial y} g^{(p)}(y,y')\right]-p^2 e^{2 k y} g^{(p)}(y,y')=
  \frac{\delta(y-y')}{e^{-4 k y}}\, , \quad (0 \leq y \leq \pi r_c)
\end{equation}
together with Neumann boundary conditions at $y=0$ and $y=\pi r_c$.
The solution of (\ref{5RSTransGreen}) is straightforward with the general result
\begin{align}
  g^{(p)}(y,y') = & \frac{e^{2 k \left(y_>+y_< \right)}}
     {k \left[ I_1\left( \frac{p}{k}\right) K_1\left( \frac{p}{k} e^{k \pi r_c} \right)-
     I_1\left( \frac{p}{k} e^{k \pi r_c} \right) K_1\left( \frac{p}{k}\right) \right]} \, \times \\
     &\times \left[I_1\left(\frac{p}{k}\right) K_2\left(\frac{p}{k} e^{k y_<}\right) \nonumber 
        + K_1\left(\frac{p}{k}\right) I_2\left(\frac{p}{k} e^{k y_<}\right) \right] \, \times \\
        \nonumber 
        & \times \left[I_1\left(\frac{p}{k} e^{k \pi r_c} \right) K_2\left(\frac{p}{k} e^{k y_>}\right)
        + K_1\left(\frac{p}{k} e^{k \pi r_c} \right) I_2\left(\frac{p}{k} e^{k y_>}\right) \right] \, ,
\end{align}
where $y_>$ ($y_<$) denote the greater (smaller) of the two numbers $y$ and $y'$ and 
$I_1(z), I_2(z)$, $K_1(z), K_2(z)$ are modified Bessel functions.
Since we want to study sources on the brane, we now set $y'=0$ and take the well-defined 
limit $r_c\to \infty$, as can easily be verified from the asymptotic behavior of $I_n$ and $K_n$
for large values of $z$. We are therefore able to write the static scalar two-point function 
as
\begin{eqnarray} \label{5RSGreenSol}
  \mathcal{G}(\vec x-\vec x';y,0)=-\frac{e^{2 k y}}{2 \pi^2} \int\limits_0^\infty
  \frac{\sin\left(p \vert \vec{x}-\vec x' \vert \right)}{\vert \vec x-\vec x' \vert}
  \frac{K_2 \left(\frac{p}{k} e^{k y} \right)}{K_1 \left(\frac{p}{k} \right)} \, dp \, .
\end{eqnarray}
We would like to point out a particularity of the Fourier integral (\ref{5RSGreenSol}). By using the 
asymptotic expansions for the modified Green's functions $K_1(z)$ and $K_2(z)$ we have \cite{AbrStegun}:
\begin{equation} \label{5KoverKAsymptotic}
  \frac{K_2 \left(\frac{p}{k} e^{k y} \right)}{K_1 \left(\frac{p}{k} \right)} \sim 
  e^{-\frac{1}{2} k y} e^{-\frac{p}{k}\left(e^{k y}-1\right)}\left[1+\mathcal{O}\left(\frac{1}{p}
  \right)\right],
\end{equation}
showing that for $y=0$ the integral over $p$ in (\ref{5RSGreenSol}) does not exist. The correct 
value of the Green's function on the brane is therefore obtained by imposing continuity at $y=0$:
\begin{equation} 
  \mathcal{G}(\vec{x}-\vec{x}';0,0)\equiv\lim_{y \to 0} \mathcal{G}(\vec{x}-\vec{x}';y,0).
\end{equation}

Note that this subtlety is still present in our case of the Fourier sum representation of the modified
 Green's function (\ref{5FormalSolution}), as discussed in appendix \ref{appSum}. 

The representation (\ref{5RSGreenSol}) is very suitable for obtaining the short distance behavior of the
Green's function. By inserting the expansion (\ref{5KoverKAsymptotic}) into (\ref{5RSGreenSol}) we can evaluate the 
integral which will give reasonable results for distances much smaller than $1/k$:
\begin{equation} 
  \mathcal{G}(\vec x-\vec x';y,0)\sim-\frac{e^{\frac{3}{2} k y}}{2 \pi^2}
  \frac{1}{\vert \vec x-\vec x' \vert^2+y^2} \, .
\end{equation}
We see that after replacing the exponential factor by $1$ (valid for $y \ll 1/k$) we 
recover (up to a factor of $2$) the correct $5$-dimensional behavior of the Green's function in flat 
4-dimensional space (\ref{5CharLaplSolnDspace}).\footnote{As discussed above, the factor of $2$ is 
a direct consequence of the boundary conditions corresponding to a semi-infinite extra dimension.} 

For large distances the Fourier representation (\ref{5RSGreenSol}) is less suited for obtaining corrections 
to Newton's law. This is due to the fact that all but the first term in the expansion of 
$K_2/K_1$ in powers of $p$ around $p=0$ lead to divergent contributions upon inserting 
in (\ref{5RSGreenSol}). We therefore seek another method which will allow us to obtain the corrections 
to Newton's law by term-wise integration. 

The idea is to promote (\ref{5RSGreenSol}) to a contour integral in the complex $p$-plane and to
shift the contour in such a way that the trigonometric function is transformed into 
an exponential function. First we introduce dimensionless quantities by rescaling with $k$ according to
$X=\vert \vec x - \vec x' \vert k$, $Y = y k$, $z=p \vert \vec x - \vec x' \vert$:
\begin{equation} 
  \mathcal{\bar{G}}(X,Y,0)=-\frac{k^2 e^{2 Y}}{2 \pi^2 X^2}
  \underbrace{\int\limits_0^\infty \sin z \frac{K_2\left(\frac{z}{X} e^Y \right)}{K_1\left(\frac{z}{X} 
   \right)} dz}_{\equiv I\left[X,Y\right]} \, .
\end{equation}
Following \cite{Kiritsis:2002ca} (see also \cite{AbrStegun}) we now use the relation 
\begin{equation} \label{5BesselKRelation}
  K_2[w]=K_0[w]+\frac{2}{w}K_1[w] 
\end{equation}
to separate the zero mode contribution to the Green's function (Newton's law) from the contributions 
coming from the higher Kaluza-Klein particles (corrections to Newton's law).
Using (\ref{5BesselKRelation}) in $I[X,Y]$ we find 
\begin{equation} 
  I[X,Y]=\underbrace{\int\limits_0^\infty \sin z \frac{K_0\left(\frac{z}{X} e^Y \right)}
    {K_1\left(\frac{z}{X}\right)} dz}_{\equiv I_1[X,Y]}+
    \underbrace{\frac{2 X}{e^Y} \int\limits_0^\infty \frac{\sin z}{z} 
   \frac{K_1\left(\frac{z}{X} e^Y \right)}{K_1\left(\frac{z}{X} \right)} dz}_{\equiv I_2[X,Y]} \, .
\end{equation}
The additional factor of $1/z$ in the integrand of $I_2[X,Y]$ allows us to take the limit $Y \to 0$ with the 
result:
\begin{equation} 
  \lim_{Y \to 0} I_2[X,Y]=\pi X \, .
\end{equation} 
We still need $Y>0$ for convergence in $I_1[X,Y]$. The next step is to use 
$\sin z=\Im \left\{e^{\imath z}\right\}$ and to exchange the operation $\Im$ with the integration 
over z:\footnote{This is justified since both the real and the imaginary part of the 
resulting integral converge.}
\begin{equation} \label{5RSI1FourierRep}
  I_1[X,Y]=\Im\left\{\int\limits_0^\infty e^{\imath z} 
  \frac{K_0\left(\frac{z}{X} e^Y \right)}{K_1\left(\frac{z}{X} \right)} dz\right\} \, .
\end{equation}
The integrand in (\ref{5RSI1FourierRep}) is a holomorphic function of $z$ in the first 
quadrant (see \cite{AbrStegun}, p.377 for details). We can therefore apply \textsc{Cauchy}'s 
theorem to the contour depicted in Fig.~\ref{5intcontour}.
\begin{figure}[tbp]
\begin{center}
\input{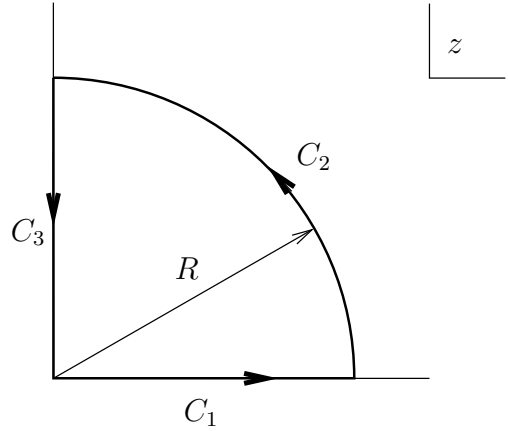}
\end{center}
\caption{Contour used in \textsc{Cauchy}'s theorem to evaluate the integral (\ref{5RSI1FourierRep}).}
\label{5intcontour}
\end{figure}
As we will show shortly, the contribution from the arc 
$C_2=\left\{z \, |\,  z=R e^{\imath \varphi},\;0\leq \varphi \leq \frac{\pi}{2} \right\}$ 
vanishes in the limit $R \to \infty$ (the limit we are interested in).
Symbolically we therefore have
\begin{equation} 
  \lim_{R \to \infty} \left\{\int_{C_1} \ldots + \int_{C_3} \ldots  \right\}=0\, ,
\end{equation}
where $\ldots$ replace the integrand in (\ref{5RSI1FourierRep}) and the sense of integration is as
indicated in the figure. This means that we can 
replace the integration over the positive real axis in (\ref{5RSI1FourierRep}) by an integration 
over the positive imaginary axis without changing the value of $I_1[X,Y]$. Substituting now $z$ in
favor of $n$  according to $n=-\imath z$ we obtain:
\begin{align} \label{5I1closetoresult}
  I_1[X,Y]&=\Im \left\{ \int\limits_0^\infty e^{-n} 
  \frac{K_0\left(\frac{\imath n}{X} e^Y \right)}{K_1\left(\frac{\imath n}{X} \right)} \imath dn \right\}=
  \Im \left\{ \int\limits_0^\infty e^{-n} 
  \frac{-H_0^{(2)} \left(\frac{n}{X} e^Y \right)}{H_1^{(2)}\left(\frac{n}{X} \right)} dn \right\}\nonumber \\
  &=\int\limits_0^\infty e^{-n} \frac{Y_0\left(\frac{n}{X} e^Y\right) J_1\left(\frac{n}{X}\right)-
                               J_0\left(\frac{n}{X} e^Y\right) Y_1\left(\frac{n}{X}\right) }
                               {\left[J_1\left(\frac{n}{X} \right)\right]^2 +
                                \left[Y_1\left(\frac{n}{X} \right)\right]^2} \, dn \, .
\end{align}
In the first line in (\ref{5I1closetoresult}) we replaced modified Bessel functions with imaginary arguments
by Hankel functions of real argument. In the second line we explicitly took the imaginary part after replacing
the identities relating the Hankel functions and the Bessel functions $J$ and $Y$ of the first kind. 
The benefit from the rotation of the contour of integration is obvious at this stage. First, the last 
integral in (\ref{5I1closetoresult}) converges also for $Y=0$ due to the presence of the exponential
function in the integrand. In this case we have
\begin{equation} \label{5RSMainResult}
  I_1[X,0]=\frac{2 X}{\pi} \int\limits_0^\infty \frac{e^{-n}}{n}
  \frac{1}{\left[J_1\left(\frac{n}{X} \right)\right]^2 + \left[Y_1\left(\frac{n}{X} \right)\right]^2} 
  \, dn \, .
\end{equation}
Note that we used the Wronskian relation for Bessel functions to simplify the numerator, the general
 reference being once more \cite{AbrStegun}. 
Second, it is straightforward to obtain the large $X$ asymptotic ($1\ll X $) of the integral 
(\ref{5RSMainResult}) by a simple power series expansion in $n$ around $n=0$ of the integrand
(apart from the exponential factor), followed by term-wise integration. For the results see 
\cite{Callin:2004py}, 
where the above integral (\ref{5RSMainResult}) has been found in the wave-function approach:
\begin{equation} \label{5I1Asymptotic}
  I_1[X,0]\sim\frac{\pi}{2 X}+\frac{\pi}{X^3}\left(4-3 \ln 2 - 3\ln X \right) + 
  \mathcal{O}\left[\frac{\ln X}{X^5}\right].
\end{equation}
The result for $\mathcal{\bar{G}}(X,0,0)$ is therefore:
\begin{equation} 
  \mathcal{\bar{G}}(X,0,0)=-\frac{k^2}{2 \pi X}
  \left[1+\frac{1}{2 X^2}+\frac{1}{X^4} \left(4-3 \ln 2 - 3\ln X \right) + 
  \mathcal{O}\left(\frac{\ln X}{X^6}\right)\right].
\end{equation} 

However, we still have to verify that the contribution from the arc $C_2$ in 
Fig.~\ref{5intcontour} vanishes in the limit $R \to \infty$. Parametrizing $z$ by 
$z=R e^{\imath \varphi}, \; 0 \leq \varphi \leq \pi/2$ we find:
\begin{align} \label{5KoverKestimate}
  \left| \int_{C_2} e^{\imath z} 
  \frac{K_0\left(\frac{z}{X} e^Y \right)}{K_1\left(\frac{z}{X} \right)} dz \right|
  &= \left| \int\limits_0^{\frac{\pi}{2}} e^{\imath R e^{\imath \varphi}} 
     \frac{K_0\left(\frac{R e^{\imath \varphi}}{X} e^Y \right)}
      {K_1\left(\frac{R e^{\imath \varphi}}{X} \right)} \left( \imath R e^{\imath \varphi} \right) 
      d\varphi \right| \nonumber \\
  &\leq R \int\limits_0^\frac{\pi}{2} \left| e^{\imath R \left(\cos \varphi+ \imath \sin \varphi\right)}\right|
      \cdot \left| \frac{K_0\left(\frac{R e^{\imath \varphi}}{X} e^Y \right)}
      {K_1\left(\frac{R e^{\imath \varphi}}{X} \right)}\right|  d\varphi \, .
\end{align}
Making now use of the asymptotic properties of the modified Bessel functions to estimate the second modulus 
in the last integrand in (\ref{5KoverKestimate}) we can write:
\begin{equation} 
  \left| \frac{K_0\left(\frac{R e^{\imath \varphi}}{X} e^Y \right)}
      {K_1\left(\frac{R e^{\imath \varphi}}{X} \right)}\right| 
   \leq C \left| e^{-\frac{Y}{2}} e^{-R e^{\imath \varphi}\frac{e^Y-1}{X}}\right| \, ,
\end{equation}
so that
\begin{align} 
  \left| \int_{C_2} e^{\imath z} 
  \frac{K_0\left(\frac{z}{X} e^Y \right)}{K_1\left(\frac{z}{X} \right)} dz \right| <
  C R e^{-\frac{Y}{2}} \int\limits_0^{\frac{\pi}{2}} e^{-R\left[ \sin\varphi + 
  \left( \frac{e^Y-1}{X} \right) \cos \varphi \right]} d\varphi \, ,
\end{align}
with $C$ being a constant of the order of unity, independent of $R$ and $\varphi$.
To finish the estimate we observe that 
\begin{equation} 
  \sin \varphi + \left( \frac{e^Y-1}{X} \right) \cos \varphi \geq \epsilon
  \equiv \min \left[1, \frac{e^Y-1}{X} \right] > 0 \, , \, \forall\, \varphi \in [0,\frac{\pi}{2}]\, ,
  \;\; (Y>0)
\end{equation}
and obtain:
\begin{equation} \label{5arcresult}
  \left| \int_{C_2} e^{\imath z} 
  \frac{K_0\left(\frac{z}{X} e^Y \right)}{K_1\left(\frac{z}{X} \right)} dz \right| <
  \frac{C R \pi}{2} e^{-\frac{Y}{2}}  e^{-R \epsilon} \; \to 0 \quad \mbox{for} \; R \to \infty.
\end{equation}
This establishes the result that in the limit $R \to \infty$ the arc $C_2$ does not contribute to the 
integral in (\ref{5RSI1FourierRep}) and completes our discussion of the two-point function in the 
Randall-Sundrum II setup. 

We are now going to summarize briefly the complications arising when 
the general scheme of computations outlined above for the Randall-Sundrum II case are applied to the brane
setup considered in chapter \ref{cha5}. First and foremost, the main difference is due to the fact that the 
induced metric on our brane is not the flat Minkowski metric but the metric of the Einstein static universe.
This implies the use of a modified Green's function. In the eigenfunction expansion, the discrete 
scalar-harmonics on $S^3$ will replace the continuous plane-wave eigenfunctions. 
We therefore expect the formal solution for the two-point function (the analog of 
(\ref{5RSGreenFormalSolution})) to take the form of a Fourier sum.

The solution to the \textsl{transverse} Green's function (the analog of eq. (\ref{5RSTransGreen})) as 
presented in appendix \ref{appDEQ} turns out to be again straightforward and completes the formal solution. 

When trying to distinguish between the zero mode contribution and the contribution coming from 
higher modes (in the language of the Kaluza-Klein approach) a relation similar to (\ref{5BesselKRelation}) 
proves useful. While the evaluation of the zero mode contribution poses no problems, the corrections 
coming from higher modes are much more involved this time. We face the following 
major technical difficulties: there is no analog of \textsc{Cauchy}'s 
theorem in the discrete case of the Fourier sum. One solution to this problem is to use a variant of 
\textsc{Euler-Maclaurin}'s sum rule in order to replace the Fourier sum by an analog Fourier integral and  
a remainder term also in the form of an integral. Now the procedure again is similar to the one outlined for 
the Randall-Sundrum case. Another technical problem, however, in connection with large distances 
corrections has its origin in the simple fact that by large distances in our setup 
we mean distances large with respect to the extra dimension $R$ but also small with respect to 
the size of the observable universe $R_U$.
The detailed computations of the corrections in this distances regime can be found 
in appendix \ref{appSum}.

What concerns the computation of the two-point function at distances smaller than the extra dimension, there
is no conceptual difference to the Randall-Sundrum case. The Fourier sum can be calculated 
analytically after inserting the corresponding large momentum asymptotic.  We give the details in appendix 
\ref{appGreenUltraShort}.
\chapter{Solutions to the differential equations (\ref{5TransverseEq1}) and (\ref{5TransverseEq2})} 
\label{appDEQ}
The aim of this appendix is to obtain the solutions to the eqs.~(\ref{5TransverseEq1}) and 
(\ref{5TransverseEq2}) and 
hence to complete the formal computation of the modified Green's function. 
We will concentrate mainly on (\ref{5TransverseEq2}) for two reasons: first and foremost, because
in the formal expansion (\ref{5FormalSolution}) $g^{(1)}(\theta,\theta')$ multiplies a constant function
in $s$ and since we are mainly interested in the behavior of the Green's function on the brane, this constant
is of no relevance for our considerations. Secondly, because the construction of the Green's functions 
$g^{(1)}(\theta,\theta')$ and $g^{(\lambda)}(\theta,\theta')$ follows the standard procedure, so it is not 
necessary to go into details twice. We will merely state the result for $g^{(1)}(\theta,\theta')$. 
The equation we would like to solve is 
\begin{align} \label{5TransvGreen}
    &\frac{1-\lambda^2}{R_U^2 \gamma(\theta)^2} g^{(\lambda)}(\theta,\theta')+
    \frac{1}{R^2} \frac{1}{\sigma(\theta)\gamma(\theta)^3} \frac{\partial}{\partial\theta}
    \left[ \sigma(\theta)\gamma(\theta)^3 \frac{\partial g^{(\lambda)}(\theta,\theta')}{\partial \theta}
    \right] 
=\frac{\delta(\theta-\theta')}{\sigma(\theta)\gamma(\theta)^3} \, ,
\end{align}
where $\lambda=2,3,4\ldots$ and the independent variables $\theta$ and $\theta'$ are restricted to the 
intervals $-\pi/2\leq\theta\leq\pi/2$, 
and $-\pi/2\leq\theta'\leq\pi/2$. 
The homogeneous equation associated with (\ref{5TransvGreen}) can be reduced to the hypergeometric 
equation. Its general formal solution in the interval $0\leq\theta\leq\pi/2$ is given by
\begin{eqnarray} \label{5TransvGreenGenSol}
  \varphi^{(+)}(\lambda,\theta)=a_1^{(+)} \varphi_1(\lambda,\theta)+a_2^{(+)} \varphi_2(\lambda,\theta) 
  \, , \qquad   0\leq\theta\leq\frac{\pi}{2}
\end{eqnarray}
where 
\begin{align} 
  \varphi_1(\lambda,\theta)&\hspace{-.5mm}=\hspace{-.5mm}\frac{\tanh^{\lambda-1}\left[\omega\left(\frac{\pi}{2}-\theta\right)\right]}
  {\cosh^4\left[\omega\left(\frac{\pi}{2}-\theta\right)\right]} \hspace{-2mm}
  \phantom{1}_2 F_1\left[\frac{\lambda+3}{2},\frac{\lambda+3}{2};\lambda+1;
        \tanh^2\left[\omega\left(\frac{\pi}{2}-\theta\right)\right]\right] , \nonumber \\ 
  \varphi_2(\lambda,\theta)&\hspace{-.5mm}=\hspace{-.5mm}\frac{\tanh^{\lambda-1}\left[\omega\left(\frac{\pi}{2}-\theta\right)\right]}
  {\cosh^4\left[\omega\left(\frac{\pi}{2}-\theta\right)\right]}\hspace{-2mm}
  \phantom{1}_2 F_1\left[\frac{\lambda+3}{2},\frac{\lambda+3}{2};3;
        \cosh^{-2}\left[\omega\left(\frac{\pi}{2}-\theta\right)\right]\right] \, .
\end{align}
Due to the symmetry of the background under $\theta\to -\theta$ we can immediately write down the 
general solution of the homogeneous equation obtained from (\ref{5TransvGreen}) in the interval
$-\pi/2\leq\theta\leq 0$:
\begin{eqnarray} \label{5TransvGreenGenSolNeg}
  \varphi^{(-)}(\lambda,\theta)=a_1^{(-)} \varphi_1(\lambda,-\theta)+a_2^{(-)} \varphi_2(\lambda,-\theta) \, ,
  \qquad -\frac{\pi}{2}\leq\theta\leq 0 \, .
\end{eqnarray}
We now have to decide whether we want to restrict our Green's function to describe perturbations 
which also possess the symmetry $\theta\to-\theta$ of the background (as in the Randall-Sundrum case) or not. 
In the first case one can limit the solution of (\ref{5TransvGreen}) to $\theta>0 $ and continue the 
result symmetrically into $\theta<0$. Note that imposing the symmetry $\theta \to -\theta$ on the 
perturbations goes 
hand in hand with imposing $\theta \to -\theta$ for the source which means that one should add a corresponding
delta-function $\delta(\theta+\theta')$ to the right hand side of (\ref{5TransvGreen}). The 
Israel-condition (\ref{5IsraelijFluct}) on the brane with $\theta=0$ then gives 
$\frac{\partial g^{(\lambda)}}{\partial \theta}=0$, as in the Randall-Sundrum case. 

Since in our setup we do not have any convincing argument for imposing the $\theta\to-\theta$ symmetry also on the
metric fluctuations (the scalar field), the second case, which allows a breaking of this symmetry will be the
case of our choice. The Israel-condition (\ref{5IsraelijFluct}) now merely indicates that the perturbation
(the scalar field) should have a continuous first derivative at the brane.

In order to obtain a uniquely defined solution we have to impose two boundary conditions 
on $g^{(\lambda)}(\theta,\theta')$ one at $\theta=\pi/2$ the other at $\theta=-\pi/2$. It turns out that
the requirement of square integrability of our solutions (with the correct weight-function) forces us to 
discard one of the two fundamental solutions at the boundaries $\theta=\pm\pi/2$, namely 
$\varphi_2(\lambda,\theta)$.

Before going into details, we introduce the following abbreviations:
\begin{align} 
  p(\theta)=\frac{\sigma(\theta)\gamma(\theta)^3}{R^2}\, ,  \qquad 
  s(\theta)=\frac{\sigma(\theta)\gamma(\theta)}{R_U^2} \, .
\end{align}
Eq. (\ref{5TransvGreen}) then becomes
\begin{equation} 
  \left[p(\theta) {g^{(\lambda)}}{'}(\theta,\theta')\right]'-(\lambda^2-1) \,
  s(\theta) g^{(\lambda)}(\theta,\theta')=\delta(\theta-\theta') \, ,
\end{equation}
where $'$ denotes derivatives with respect to $\theta$ here and in the following. 
We start by taking $\theta'>0$. The case $\theta'<0$ is fully analogous to the one considered 
up to minus signs.
Our ansatz for $g^{(\lambda)}(\theta,\theta')$ is
\begin{equation} \label{5GreenAnsatz}
  g^{(\lambda)}(\theta,\theta')=\left\{
  \begin{array}{lrrl} A(\theta') \, \varphi_1^{(\lambda)}(-\theta)&-\frac{\pi}{2}&\leq \theta \leq& 0 \, ,\\
                    B(\theta') \, \varphi_1^{(\lambda)}(\theta)+C(\theta') \, \varphi_2^{(\lambda)}(\theta)\quad
                             &0 &\leq \theta \leq& \theta' \, ,\\
                    D(\theta') \, \varphi_1^{(\lambda)}(\theta)&\theta' &\leq \theta \leq& \frac{\pi}{2} \, .
  \end{array} \right.
\end{equation}
We impose the following boundary and matching conditions on $g^{(\lambda)}(\theta,\theta')$ which will uniquely 
determine the coefficients $A, B, C, D$:
\begin{enumerate} 
  \item{continuity of $g^{(\lambda)}(\theta,\theta')$ at $\theta=0$.}
  \item{continuity of ${g^{(\lambda)}}{'}(\theta,\theta')$ at $\theta=0$.}
  \item{continuity of $g^{(\lambda)}(\theta,\theta')$ at $\theta=\theta'$.}
  \item{jump condition of ${g^{(\lambda)}}{'}(\theta,\theta')$ at $\theta=\theta'$.}
\end{enumerate}
Using the ansatz (\ref{5GreenAnsatz}) we obtain from the above conditions:
\begin{equation} 
  \begin{array}{rclcc}
  \left[A(\theta')-B(\theta')\right] \varphi_1^{(\lambda)}(0) &-&C(\theta') \varphi_2^{(\lambda)}(0)&=&0\, ,\\
  \left[A(\theta')+B(\theta')\right] {\varphi_1^{(\lambda)}}{'}(0) &+&C(\theta') 
   {\varphi_2^{(\lambda)}}{'}(0)&=&0\, , \\
  \left[B(\theta')-D(\theta')\right] \varphi_1^{(\lambda)}(\theta') &+&C(\theta') 
   \varphi_2^{(\lambda)}(\theta')&=&0\, ,\\
  \left[B(\theta')-D(\theta')\right] {\varphi_1^{(\lambda)}}{'}(\theta') &+&C(\theta') 
  {\varphi_2^{(\lambda)}}{'}(\theta')&=&-\frac{1}{p(\theta')}   \, ,
  \end{array}
\end{equation}
the solution of which is easily found to be:
\begin{align} \label{5solABCD}
  A(\theta')&=\frac{R^2}{2}\frac{\varphi_1^{(\lambda)}(\theta')}
   {\varphi_1^{(\lambda)}(0) {\varphi_1^{(\lambda)}}{'}(0)}
  \, , \nonumber \\
  B(\theta')&=\frac{R^2}{2}\frac{\varphi_1^{(\lambda)}(\theta')}
         {\mathcal{W}\left[\varphi_1^{(\lambda)},\varphi_2^{(\lambda)},0\right]}
         \frac{\varphi_1^{(\lambda)}(0) {\varphi_2^{(\lambda)}}{'}(0)+
         \varphi_2^{(\lambda)}(0) {\varphi_1^{(\lambda)}}{'}(0)}
         {\varphi_1^{(\lambda)}(0) {\varphi_1^{(\lambda)}}{'}(0)} \, ,\nonumber \\
  C(\theta')&=-R^2 \frac{\varphi_1^{(\lambda)}(\theta')}{\mathcal{W}
          \left[\varphi_1^{(\lambda)},\varphi_2^{(\lambda)},0\right]} \, , \nonumber \\
  D(\theta')&=B(\theta')- R^2 \frac{\varphi_2^{(\lambda)}(\theta')}{\mathcal{W}
          \left[\varphi_1^{(\lambda)},\varphi_2^{(\lambda)},0\right]} \, ,
\end{align}
where $\mathcal{W}\left[\varphi_1^{(\lambda)},\varphi_2^{(\lambda)},0\right]$ denotes the Wronskian 
of $\varphi_1^{(\lambda)}$ and $\varphi_2^{(\lambda)}$ at $\theta=0$. In obtaining (\ref{5solABCD}) we 
used the relation
\begin{equation} 
  \mathcal{W}\left[\varphi_1^{(\lambda)},\varphi_2^{(\lambda)},\theta\right]=
  \frac{\mathcal{W}\left[\varphi_1^{(\lambda)},\varphi_2^{(\lambda)},0\right]}{R^2\, p(\theta)} \, .
\end{equation}
As already mentioned, the case $\theta'<0$ can be treated in complete analogy to the case $\theta'>0$. 
Combining the two results gives the final expression for the solution $g^{(\lambda)}(\theta,\theta')$ of 
(\ref{5TransvGreen}) (for  $\lambda=2,3,\ldots$): 
\begin{equation} \label{5GreenfullResult}
  g^{(\lambda)}(\theta,\theta')=\left\{
  \begin{array}{lrrl} A(\vert \theta'\vert) \, \varphi_1^{(\lambda)}(\vert \theta\vert)&-\frac{\pi}{2}
                         &\leq \theta \, \mbox{sign}(\theta') \leq& 0\, , \\
                 B(\vert \theta'\vert) \, \varphi_1^{(\lambda)}(\vert \theta\vert)+C \, (\vert \theta'\vert) 
                     \varphi_2^{(\lambda)}(\vert \theta\vert)\;
                         &0 &\leq \theta \, \mbox{sign}(\theta') \leq& \vert \theta' \vert \, , \\
                    D(\vert \theta'\vert) \, \varphi_1^{(\lambda)}(\vert \theta \vert )
                      & \vert \theta' \vert & \leq \theta \, \mbox{sign}(\theta') \leq& \frac{\pi}{2} \, .
  \end{array} \right.
\end{equation}
It is easily verified that the Green's function (\ref{5GreenfullResult}) has the property
\begin{equation} 
   g^{(\lambda)}(\theta,\theta') = g^{(\lambda)}(\theta',\theta) \, ,
\end{equation}
as expected for a self-adjoint boundary value problem. We will not need the result 
(\ref{5GreenfullResult}) in its full generality. We focus on the case where sources are located at
the brane ($\theta'=0$) so that (\ref{5GreenfullResult}) reduces to:
\begin{equation} \label{5glambdaratio}
  g^{(\lambda)}(\theta,0)=\frac{R^2}{2}
  \frac{\varphi_1^{(\lambda)}(\vert \theta\vert)}{{\varphi_1^{(\lambda)}}{'}(0)} \, , 
  \qquad (\lambda=2,3,\ldots) \, .
\end{equation}

For the sake of completeness, we also give the solution $g^{(1)}(\theta,\theta')$ to (\ref{5TransverseEq1}) 
\begin{equation} \label{5Transverse1Repeat}
  \frac{1}{R^2} \frac{1}{\sigma(\theta)\gamma(\theta)^3} \frac{\partial}{\partial\theta}
    \left[ \sigma(\theta)\gamma(\theta)^3 \frac{\partial g^{(1)}(\theta,\theta')}{\partial \theta}
    \right]=\frac{\delta(\theta-\theta')}{\sigma(\theta)\gamma(\theta)^3} -
    \tilde{\chi}_1(\theta) \bar{\tilde{\chi}}_1(\theta') \, .
\end{equation}
The general procedure of finding the solution is fully analogous to the case of 
$g^{(\lambda)}(\theta,\theta')$ (for $\lambda=2,3,\ldots$), the only difference being 
that in regions where $\theta \neq \theta'$, (\ref{5Transverse1Repeat}) reduces to an
inhomogeneous differential equation. In the interval $0 \leq \theta \leq \pi/2$ the general solution
is given by
\begin{eqnarray} 
  \psi(\theta)=\psi_0(\theta) + A \psi_1(\theta)+B \psi_2(\theta),
\end{eqnarray}
with
\begin{align} 
  \psi_0(\theta)&=-\frac{R^2}{2 \omega \tanh \left(\frac{\omega\pi}{2}\right)} 
  \ln\left[ \cosh\left[\omega \left(\frac{\pi}{2}-\theta\right)\right]\right]\, , \\ \nonumber
  \psi_1(\theta)&=1\, , \qquad \psi_2(\theta)=2 \ln\left[\tanh\left[\omega \left(\frac{\pi}{2}-\theta\right)\right]\right]+
   \coth^2\left[\omega \left(\frac{\pi}{2}-\theta\right)\right]. 
\end{align}
With these definitions, it is easy to verify that $g^{(1)}(\theta,\theta')$ is given by:
\begin{equation} \label{5resg1}
  g^{(1)}(\theta,\theta')=\mbox{const.}+\psi_0(\vert \theta\vert)+\psi_0(\vert \theta'\vert)+C 
  \left\{\begin{array}{lrcl} \psi_2(0) & -\frac{\pi}{2} \leq \theta \, \mbox{sign}(\theta')\leq & 0 \\
                             \psi_2(\vert \theta\vert) & 0 \leq \theta \, \mbox{sign}(\theta') \leq &\vert \theta' \vert\\
                             \psi_2(\vert \theta' \vert) & \vert \theta' \vert \leq \theta \, \mbox{sign}(\theta') \leq& 
                              \frac{\pi}{2}\end{array} \right. \, ,
\end{equation}
with 
\begin{equation}
   C=-\frac{R^2}{\mathcal{W}\left[\psi_1,\psi_2,0 \right]} \, .
\end{equation}
By $\mathcal{W}\left[\psi_1,\psi_2,0 \right]$ we again mean the Wronskian of $\psi_1$ and $\psi_2$ 
evaluated at $\theta=0$.
\thispagestyle{empty}
\chapter{Detailed computation of the corrections to Newton's law} \label{appSum}
In this appendix we give a detailed computation of the sum
\begin{equation} \label{5TheSumB}
  S[s,\theta,\omega]=\sum_{\lambda=2}^{\infty}\lambda \sin \left(\lambda s \right)
  \frac{\varphi_1^{(\lambda)}(\theta)}{{\varphi'}_1^{(\lambda)}(0)} \, ,\qquad 0<\theta \leq \frac{\pi}{2}
\end{equation}
with 
\begin{align} \label{5RatioOf2Hyps}
  & \frac{\varphi_1^{(\lambda)}(\theta)}{{\varphi'}_1^{(\lambda)}(0)}= 
  -\frac{z(\theta)^\frac{\lambda-1}{2} \left[1-z(\theta)\right]^2 
  \phantom{1}_2 F_1\left[\frac{\lambda+3}{2},\frac{\lambda+3}{2};\lambda+1;z(\theta)\right]}
  {\omega(\lambda-1) z(0)^\frac{\lambda-2}{2} \left[1-z(0)\right]^2 
  \phantom{1}_2 F_1\left[ \frac{\lambda+1}{2},\frac{\lambda+3}{2};\lambda+1;z(0) \right]} \, ,
\end{align}
where we explicitly excluded $\theta=0$ since the Fourier sum (\ref{5TheSumB}) does
not converge for this value of $\theta$.\footnote{Note that this mathematical delicacy about the 
divergence of the Green's function on the brane also applies to the Randall-Sundrum case of the 
Fourier integral representation of the two-point function.
The obvious remedy is to keep $\theta$ strictly greater than
zero during the whole calculation and eventually take the limit $\theta \to 0$.}
We also used the definition $z(\theta)=\tanh^2\left[\omega\left(\frac{\pi}{2}-\theta\right)\right]$.

Note that the sum $S[s,\theta,\omega]$ contains the contribution to the two-point function
coming from zero mode and higher Kaluza-Klein modes. Inspired by the Randall-Sundrum 
case, we try to separate the two contributions by functional relations between contiguous 
Gauss hypergeometric functions. Using (15.2.15) of reference \cite{AbrStegun} for 
$a=b-1=\left(\lambda+1\right)/2,\,c=\lambda+1$ we obtain
\begin{align} 	
  &\left[1-z(\theta)\right] \phantom{1}_2 F_1\left[ \frac{\lambda+3}{2},\frac{\lambda+3}{2};
  \lambda+1;z(\theta) \right]= \\
  &\frac{2}{\lambda+1}
  \hspace{-.5mm}\phantom{1}_2 F_1\left[ \frac{\lambda+1}{2},\frac{\lambda+3}{2};
  \lambda+1;z(\theta) \right] \nonumber
   \hspace{-.5mm}+\hspace{-.5mm}\frac{\lambda-1}{\lambda+1}\hspace{-.5mm}
  \phantom{1}_2 F_1\left[ \frac{\lambda+1}{2},\frac{\lambda+1}{2};
  \lambda+1;z(\theta) \right]  .
\end{align}
In this way we are able to rewrite $S[s,\theta,\omega]$ in the form:
\begin{equation} 
  S[s,\theta,\omega]=  S_1[s,\theta,\omega] + S_2[s,\theta,\omega]\, ,
\end{equation}
where
\begin{align} \label{5defS1}
  S_1[s,\theta,\omega]&=-\frac{2}{\omega}\frac{z(0)}{z(\theta)^\frac{1}{2}} 
  \frac{1-z(\theta)}{\left[1-z(0)\right]^2} 
  \, \times \nonumber \\ & \quad \times \, 
  \sum_{\lambda=2}^{\infty}\frac{\lambda \sin \left(\lambda s\right)}{\lambda^2-1} 
  \left[\frac{z(\theta)}{z(0)}\right]^\frac{\lambda}{2}
  \frac{\phantom{1}_2 F_1\left[ \frac{\lambda+1}{2},\frac{\lambda+3}{2};
  \lambda+1;z(\theta) \right]}{\phantom{1}_2 F_1\left[ \frac{\lambda+1}{2},\frac{\lambda+3}{2};
  \lambda+1;z(0) \right]}\, , \\
  S_2[s,\theta,\omega]&=-\frac{1}{\omega}\frac{z(0)}{z(\theta)^\frac{1}{2}}
  \frac{1-z(\theta)}{\left[1-z(0)\right]^2}  \, \times \nonumber \\
  & \quad 
  \times \, \sum_{\lambda=2}^{\infty}\frac{\lambda \sin\left( \lambda s\right)}{\lambda+1} \label{5defS2}
  \left[\frac{z(\theta)}{z(0)}\right]^\frac{\lambda}{2}
  \frac{\phantom{1}_2 F_1\left[ \frac{\lambda+1}{2},\frac{\lambda+1}{2};
  \lambda+1;z(\theta) \right]}{\phantom{1}_2 F_1\left[ \frac{\lambda+1}{2},\frac{\lambda+3}{2};
  \lambda+1;z(0) \right]} \, .
\end{align}
From the asymptotic formula (\ref{5GeneralAsymptoticOfHyp}) we infer that now the sum $S_1[s,\theta,\omega]$ 
is convergent even for $\theta=0$ due to the additional power of $\lambda$ in the denominator. 
We therefore obtain
\begin{align} 
  S_1[s,0,\omega]&=-\frac{2}{\omega} \frac{z(0)^\frac{1}{2}}{1-z(0)}
  \sum_{\lambda=2}^{\infty}\frac{\lambda \sin \left(\lambda s\right)}{\lambda^2-1} 
  =-\frac{2}{\omega} \frac{z(0)^\frac{1}{2}}{1-z(0)}
  \left(\frac{\pi-s}{2}\cos s -\frac{1}{4}\sin s \right)\, .
\end{align}
Since we recognized in the last sum the two-point function of Einstein's static universe, we attribute
the contribution coming from $S_1[s,\theta,\omega]$ to the zero mode
in the Kaluza-Klein spectrum.
We also should mention that in $S_2[s,\theta,\omega]$ we still need $\theta>0$ for convergence since 
only then the factor $\left[z(\theta)/z(0)\right]^\frac{\lambda-1}{2}$ provides an exponential cutoff 
in $\lambda$ for the sum.

The main challenge in the evaluation of the two-point function is therefore to tame the sum 
$S_2[s,\theta,\omega]$. Our strategy is the following: first we extend it from $\lambda=0$ to
$\infty$ by adding and subtracting the $\lambda=0$ and $\lambda=1$ terms. Next, we replace the sum
over $\lambda$ by the sum of two integrals employing a variant of the \textsc{Euler-Maclaurin} sum 
rule called the
\textsc{Abel-Plana} formula (see. e.g. \cite{Olver}, p. 289-290). Finally, we shall see that the
 resulting (exact) integral representation will allow us to extract the desired asymptotic of the 
two-point function on the brane in the distance regime of interest ($R\ll r \ll R_U$). 

As announced, we start by extending the range of the sum from $0$ to $\infty$. Since the addend with
$\lambda=0$ vanishes, we only need to subtract the $\lambda=1$ term with the result: 
\begin{equation} \label{5S2extended}
  S_2[s,\theta,\omega]=-\frac{1}{2 \omega}\frac{z(0)^\frac{1}{2}}{z(\theta)} \frac{1-z(\theta)}{1-z(0)} 
  \ln\left[1-z(\theta)\right] \sin s 
  -\frac{1}{\omega} \frac{z(0)}{z(\theta)^\frac{1}{2}} \frac{1-z(\theta)}{1-z(0)}
  R[s,\theta,\omega]\; ,
\end{equation}
where $R[s,\theta,\omega]$ is defined by 
\begin{equation} \label{5defR}
  R[s,\theta,\omega]=\sum_{\lambda=0}^{\infty}\frac{\lambda \sin \left(\lambda s\right)}{\lambda+1}
  \left[\frac{z(\theta)}{z(0)}\right]^\frac{\lambda}{2}
  \frac{\phantom{1}_2 F_1\left[ \frac{\lambda+1}{2},\frac{\lambda+1}{2};
  \lambda+1;z(\theta) \right]}{\phantom{1}_2 F_1\left[ \frac{\lambda+1}{2},\frac{\lambda-1}{2};
  \lambda+1;z(0) \right]} \, .
\end{equation}
We used the identity 
\begin{equation} \label{5HypIdentity}
  \phantom{1}_2 F_1\left[ \frac{\lambda+1}{2},\frac{\lambda+3}{2};\lambda+1;z(0) \right]=
  \left[1-z(0)\right]^{-1} 
  \hspace{-2mm}
  \phantom{1}_2 F_1\left[ \frac{\lambda+1}{2},\frac{\lambda-1}{2};\lambda+1;z(0) \right]
\end{equation}
in the last step (see 15.3.3 of \cite{AbrStegun}).
The next step consists of extending the $\sin$ function in
(\ref{5defR}) to an exponential and taking the imaginary part out of the sum (as in the 
Randall-Sundrum case). Now we make use of the \textsc{Abel-Plana} formula, 
first considering the partial sums:
\begin{align} 
  &R^{(n)}[s,\theta,\omega]=\Im \left[ \frac{1}{2} f(0,s,\theta,\omega)+
        \frac{1}{2} f(n,s,\theta,\omega)+\int\limits_0^n f(\lambda,s,\theta,\omega) \, d\lambda \right. 
         \nonumber \\
	&+\imath \left. \int\limits_0^{\infty}\frac{f(\imath y,s,\theta,\omega)\hspace{-1mm}-\hspace{-1mm}
          f(n+\imath y,s,\theta,\omega)\hspace{-1mm}-\hspace{-1mm}
        f(-\imath y,s,\theta,\omega)\hspace{-1mm}+\hspace{-1mm}
        f(n-\imath y,s,\theta,\omega)}{e^{2\pi y}-1} dy  \right] \, ,
\end{align}
where for the sake of clarity we introduced 
\begin{equation} \label{5fdef}
  f(\lambda,s,\theta,\omega)\equiv \frac{\lambda e^{\imath\lambda s}}{\lambda+1}
  \left[\frac{z(\theta)}{z(0)}\right]^\frac{\lambda}{2}
  \frac{\phantom{1}_2 F_1\left[ \frac{\lambda+1}{2},\frac{\lambda+1}{2};
  \lambda+1;z(\theta) \right]}{\phantom{1}_2 F_1\left[ \frac{\lambda+1}{2},\frac{\lambda-1}{2};
  \lambda+1;z(0) \right]} \, .
\end{equation}
In the limit $n\to\infty$ we find:
\begin{equation} \label{5TIntRep}
  R[s,\theta,\omega]=\Im \Big[ \underbrace{\int\limits_0^\infty f(\lambda,s,\theta,\omega) d\lambda}_
	{\equiv R_1[s,\theta,\omega]}
	+\underbrace{\imath \int\limits_0^{\infty}\frac{f(\imath y,s,\theta,\omega)-
	f(-\imath y,s,\theta,\omega)}{e^{2\pi y}-1} dy}_{\equiv R_2[s,\theta,\omega]} \Big] \, .
\end{equation}
\section{Leading order asymptotic of $\Im\left\{R_1[s,\theta,\omega]\right\}$}
We focus first on $R_1[s,\theta,\omega]$ and observe that the function 
$f(\lambda,s,\theta,\omega)$ is holomorphic (in $\lambda$) in the first 
quadrant.\footnote{Everything but the ratio of the hypergeometric functions is clearly holomorphic. 
The dependence
of Gauss's hypergeometric function on the parameters is also holomorphic so the only danger comes from
the zeros of $\phantom{1}_2 F_1\left[ \frac{\lambda+1}{2},\frac{\lambda-1}{2};\lambda+1;z(0) \right]$. These,
however, turn out to be outside of the first quadrant.}
Therefore, in perfect analogy with the Randall-Sundrum case treated in appendix 
\ref{appParallel}, we can use \textsc{Cauchy}'s theorem to replace the integration over the positive real 
axis by an 
integration over the positive imaginary axis.\footnote{To be mathematically correct, one has to apply
\textsc{Cauchy}'s theorem to the quarter of the disk of radius $R$  bounded by the positive real axis, the 
positive imaginary axis and the arc joining the points $R$ and $\imath R$, see Fig.~\ref{5intcontour} in 
appendix \ref{appParallel}. 
The above statement will then be correct if in the limit $R\to \infty$ the contribution from 
the arc tends to zero. We will verify this in section \ref{5arcjust} of this appendix.}
After substituting $\lambda$ with $\imath y$ we obtain :
\begin{equation} \label{5R1shifted}
  R_1[s,\theta,\omega]=-\int\limits_0^\infty \frac{y (1-\imath y)}{1+y^2} e^{-y s}
  \left[\frac{z(\theta)}{z(0)}\right]^\frac{\imath y}{2}
  \frac{\phantom{1}_2 F_1\left[ \frac{1+\imath y}{2},\frac{1+\imath y}{2};
  1+\imath y;z(\theta) \right]}{\phantom{1}_2 F_1\left[ \frac{1+\imath y}{2},\frac{-1+\imath y}{2};
  1+\imath y;z(0) \right]} dy \, .
\end{equation}
The upper deformation of the path of integration is advantageous for at least two reasons. Firstly, 
through the appearance of the exponential in the integrand, convergence on the brane does no longer rely
on $\theta>0$ and we can set $\theta$ equal to zero in (\ref{5R1shifted}) to obtain:
\begin{equation} \label{5R1shiftedBrane}
  R_1[s,0,\omega]=-\int\limits_0^\infty \frac{y (1-\imath y)}{1+y^2} e^{-y s}
  \frac{\phantom{1}_2 F_1\left[ \frac{1+\imath y}{2},\frac{1+\imath y}{2};
  1+\imath y;z(0) \right]}{\phantom{1}_2 F_1\left[ \frac{1+\imath y}{2},\frac{-1+\imath y}{2};
  1+\imath y;z(0) \right]} dy \, .
\end{equation}   
Secondly, the form (\ref{5R1shifted}) is perfectly suited for obtaining asymptotic expansions
for large distances. Note that since $s\in[0,\pi]$, we carefully avoided saying 
``for large s'' since this would not correspond to the case of our interest. We try to compute
the two-point function in a distance regime $R \ll r \ll R_U$, distances clearly far beyond 
the size of the fifth dimension but far below the size of the observable universe. Rewritten in
the geodesic distance coordinate $s$ this becomes 
$\omega/\sinh\left(\frac{\omega\pi}{2}\right)\ll s\ll 1$. It is this last relation which complicates
the computation of the asymptotic considerably. We are forced to look at an asymptotic 
evaluation of (\ref{5R1shiftedBrane}) (or (\ref{5R1shifted})) for intermediate $s$ values such that
 we cannot make direct use of Laplace's method. The correct way of extracting the above described
asymptotic is to expand the ratio of hypergeometric functions in (\ref{5R1shiftedBrane}) in a power 
series of the small parameter $1-z(0)$. Using rel. $(15.3.10)$ and $(15.3.11)$ (with $m=1$) of 
\cite{AbrStegun} we find after some algebra
\begin{align} \label{5HypRatioExpansion}
  &\frac{\phantom{1}_2 F_1\left[ \frac{1+\imath y}{2},\frac{1+\imath y}{2};
  1+\imath y;z(0) \right]}{\phantom{1}_2 F_1\left[ \frac{1+\imath y}{2},\frac{-1+\imath y}{2};
  1+\imath y;z(0) \right]}= 
  -\frac{1}{2}(1+\imath y)\left[2\gamma+2\psi(\frac{1+\imath y}{2})+\ln\left[1-z(0)\right]\right]  
	\nonumber \\
  &+\left\{\left(\frac{1+\imath y}{2}\right)^3\left[2 \psi(2)-2 \psi\left(\frac{3+\imath y}{2}\right)
     -\ln\left[1-z(0)\right]\right] \right.\nonumber \\
  &\qquad-\left(\frac{1+\imath y}{2}\right)^2 \left(\frac{-1+\imath y}{2}\right)
    \left[2 \psi(1)-2 \psi\left(\frac{1+\imath y}{2}\right)-\ln\left[1-z(0)\right]\right]\times\nonumber \\
  &\;\;\;\;\;\;\;\left.\times \left[\ln\left[1-z(0)\right]-\psi(1)-\psi(2)
     +\psi\left(\frac{3+\imath y}{2}\right)+
     \psi\left(\frac{1+\imath y}{2}\right)\right] \right\} \left[1-z(0)\right]\nonumber\\
  &+\mathcal{O}\left\{\ln\left[1-z(0)\right] \left[1-z(0)\right]^2 \right\} \, .
\end{align}
In the above relation $\gamma$ denotes \textsc{Euler-Mascheroni}'s constant and 
$\psi(z)\equiv\Gamma'[z]/\Gamma[z]$ the so-called Digamma-function. Note that we 
developed up to linear order in
$1-z(0)$ since we also want to compute the next to leading order term in the asymptotic later
in this appendix. 

We are only interested in the imaginary part of $R_1[s,0,\omega]$ and it turns out that the first term in
the expansion (\ref{5HypRatioExpansion}) after insertion in (\ref{5R1shiftedBrane}) can be integrated 
analytically:
\begin{align} \label{5R1_0_Integral}
  \Im\left\{R_1^{(0)}[s,0,\omega]\right\}\hspace{-.5mm}=\hspace{-.5mm}
  \Im \left\{\int\limits_0^{\infty} y e^{-y s} 
  \psi\left(\frac{1+\imath y}{2}\right)
  dy \right\}
  \hspace{-.5mm}=\hspace{-.5mm}
  \int\limits_0^{\infty} y e^{-y s} \Im\left[\psi\left(\frac{1+\imath y}{2}\right)\right] dy \, .
\end{align}
The last step is justified since also the real part of the first integral in (\ref{5R1_0_Integral}) 
converges, as follows 
immediately from the asymptotic $\psi(z)\sim\ln z-\frac{1}{2 z}+\mathcal{O}(\frac{1}{z^2})$. 
Using
\begin{eqnarray} \label{5ImofDiGamma}
  \Im\left[\psi(\frac{1+\imath y}{2})\right]=\frac{\pi}{2}\tanh\left( \frac{\pi y}{2}\right)
\end{eqnarray}
(\cite{AbrStegun}, p.259, 6.3.12) and introducing 
the $\beta$-function (see \cite{Gradshteyn}, p.331, 3.311,~2.) 
\begin{equation} \label{5defbeta}
  \beta(x)\equiv\frac{1}{2}\left[\psi\left(\frac{x+1}{2}\right)-\psi\left(\frac{x}{2}\right)\right]
\end{equation}
we find
\begin{align} \label{5R1_0Result}
  &\Im\left\{R_1^{(0)}[s,0,\omega]\right\} = \nonumber \\ \nonumber
  &\quad=\frac{\pi}{2}\int\limits_{0}^\infty y e^{-y s} 
   \tanh \left(\frac{\pi y}{2} \right) dy 
   =-\frac{\pi}{2} \frac{d}{ds}\left[\int\limits_{0}^\infty e^{-y s} 
  \tanh\left(\frac{\pi y}{2}\right) dy\right] \\ \nonumber 
  &\quad=-\frac{\pi}{2} \frac{d}{ds}\left[\int\limits_{0}^\infty e^{-y s} 
    \left(\frac{2}{1+e^{-\pi y}}-1\right) dy\right]
  = -\frac{\pi}{2}\frac{d}{ds}
     \left[-\frac{1}{s}+2\int\limits_0^\infty \frac{e^{-y s}}{1+e^{-\pi y}} dy \right] \\ 
  &\quad=-\frac{\pi}{2 s^2}-\frac{d}{ds}\left[\int\limits_0^\infty
   \frac{e^{-\frac{z s}{\pi}}}{1+e^{-z}} dz\right]=-\frac{\pi}{2 s^2}-\frac{1}{\pi}\beta'(\frac{s}{\pi})\, .
\end{align}
By $\beta'$ in (\ref{5R1_0Result}) we mean the derivative of $\beta$ with respect to its argument.

\section{Leading order asymptotic of $\Im\left\{R_2[s,\theta,\omega]\right\}$}
We can treat $R_2[s,\theta,\omega]$ starting from (\ref{5TIntRep}) in very much the same way as 
$R_1[s,\theta,\omega]$. Noting that we can again put $\theta$ equal to zero since the exponential 
factor $e^{2\pi y}$ guarantees the convergence of the integral, $R_2[s,0,\omega]$ becomes explicitly:
\begin{align} \label{5R2explicit}
  R_2[s,0,\omega]=\int\limits_0^\infty \frac{1}{e^{2\pi y}-1}&\left[ \frac{-y+\imath y^2}{1+y^2}
   e^{-y s} \frac{\phantom{1}_2 F_1\left[ \frac{1+\imath y}{2},\frac{1+\imath y}{2};
  1+\imath y;z(0) \right]}{\phantom{1}_2 F_1\left[ \frac{1+\imath y}{2},\frac{-1+\imath y}{2};
  1+\imath y;z(0) \right]} \right. \nonumber \\
   & \left. +\frac{-y-\imath y^2}{1+y^2}
   e^{y s} \frac{\phantom{1}_2 F_1\left[ \frac{1-\imath y}{2},\frac{1-\imath y}{2};
  1-\imath y;z(0) \right]}{\phantom{1}_2 F_1\left[ \frac{1-\imath y}{2},\frac{-1-\imath y}{2};
  1-\imath y;z(0) \right]} \right] dy \, .
\end{align}
If we now employ the expansion (\ref{5HypRatioExpansion}) two times in (\ref{5R2explicit}) we find 
after some algebra:
\begin{align}
  \Im\left\{R_2^{(0)}[s,0,\omega]\right\}=\Im \left\{ \int\limits_0^\infty \frac{y}{e^{2\pi y}-1} \left[
  e^{-y s} \psi\left(\frac{1+\imath y}{2}\right)+e^{y s} \psi\left(\frac{1-\imath y}{2}\right) \right] 
  dy \right\}
\end{align}
and using (\ref{5ImofDiGamma}) we have
\begin{align} \label{5R2IntRes}
  &\Im\left\{R_2^{(0)}[s,0,\omega]\right\} = \\ \nonumber
  &=\int\limits_0^\infty \frac{\pi}{2} \frac{y}{e^{2\pi y}-1}
  \left[ e^{-y s} \tanh \left(\frac{\pi y}{2}\right)+e^{y s} \tanh \left(-\frac{\pi y}{2}\right) \right] dy\\
  &=-\pi\int\limits_0^\infty \frac{y}{e^{2\pi y}-1}\tanh\left(\frac{\pi y}{2}\right)\sinh(y s) \, dy
  =-\pi\int\limits_0^\infty \frac{y}{\left(1+e^{\pi y}\right)^2}\sinh (y s) \, dy
   \nonumber \\
  &=-\pi\frac{d}{ds}\left[\int\limits_0^\infty \frac{\cosh(y s)}{\left(1+e^{\pi y}\right)^2} dy\right] 
  =-\frac{1}{2}\frac{d}{ds}\left[\int\limits_0^1\frac{u^{1-\frac{s}{\pi}}}{(1+u)^2} \, du +
                                \int\limits_0^1\frac{u^{1+\frac{s}{\pi}}}{(1+u)^2} \, du \right] \nonumber \\
  &=\hspace{-.5mm}-\frac{1}{2}\frac{d}{ds} \left\{\frac{1}{2\hspace{-.5mm}-\hspace{-.5mm}\frac{s}{\pi}} 
    \hspace{-1mm}
     \phantom{1}_2 F_1\left[2,2\hspace{-.5mm}-\hspace{-.5mm}\frac{s}{\pi};
     3\hspace{-.5mm}-\hspace{-.5mm}\frac{s}{\pi};-1 \right]\hspace{-.5mm}+\hspace{-.5mm}
   \frac{1}{2\hspace{-.5mm}+\hspace{-.5mm}\frac{s}{\pi}} 
   \hspace{-1mm}\phantom{1}_2 F_1\left[2,2\hspace{-.5mm}+\hspace{-.5mm}\frac{s}{\pi};
   3\hspace{-.5mm}+\hspace{-.5mm}\frac{s}{\pi};-1 
   \right]\right\} \, . \nonumber 
\end{align}
While we substituted $y$ by $u$ according to $u=e^{-\pi y}$ in the third line, we used (3.194, p.313) of
\cite{Gradshteyn} with $\nu=2$, $u=\beta=1$ and $\mu=2\mp s/\pi$ (valid for $s<2 \pi$)
in the last line.
In order to simplify the hypergeometric functions, we first use another relation between 
contiguous functions, namely eq. (15.2.17) of \cite{AbrStegun} with $a=1$, $b=k$, $c=k+1$ and $z=-1$ and 
then the formulae (15.1.21) and (15.1.23) of \cite{AbrStegun} together with the duplication formula 
for the $\Gamma$-function to obtain:
\begin{align}
  \frac{1}{k}\phantom{1}_2 F_1\left[2,k;k+1;-1 \right]&=\phantom{1}_2 F_1\left[1,k;k;-1 \right]-
  \frac{k-1}{k} \phantom{1}_2 F_1\left[1,k;k+1;-1 \right] \nonumber \\
  &=\;\; 2^{-k} \sqrt{\pi}\frac{\Gamma\left[k\right]}{\Gamma\left[\frac{k}{2}\right]
  \Gamma\left[\frac{k+1}{2}\right]}-\left(k-1\right) \beta(k)\nonumber \\
  &=\;\; \frac{1}{2}+(1-k)\beta(k).
\end{align}
After making use of this in (\ref{5R2IntRes}) we finally obtain for the contribution to lowest 
order in $1-z(0)$:
\begin{eqnarray} 
  \Im\left\{R_2^{(0)}[s,0,\omega]\right\}=\frac{1}{2}\frac{d}{ds}\left[\left(1-\frac{s}{\pi}\right)
  \beta(2-\frac{s}{\pi})+\left(1+\frac{s}{\pi}\right) \beta(2+\frac{s}{\pi}) \right].
\end{eqnarray}
Summarizing the results to lowest order in $1-z(0)$, we therefore have
\begin{align} \label{50orderResult}
  \Im\left\{R_1^{(0)}[s,0,\omega]\right\}&=-\frac{\pi}{2 s^2}-\frac{1}{\pi}\beta'(\frac{s}{\pi}),\\ \nonumber
  \Im\left\{R_2^{(0)}[s,0,\omega]\right\}&=\frac{1}{2}\frac{d}{ds}\left[\left(1-\frac{s}{\pi}\right)
  \beta(2-\frac{s}{\pi})+\left(1+\frac{s}{\pi}\right) \beta(2+\frac{s}{\pi}) \right].
\end{align}
Expanding (\ref{50orderResult}) around $s=0$ we find:
\begin{align}
  \Im\left\{R_1^{(0)}[s,0,\omega]\right\}&=\frac{\pi}{2 s^2}-\frac{\pi}{12}+\mathcal{O}(s) \, ,\\ \nonumber 
  \Im\left\{R_2^{(0)}[s,0,\omega]\right\}&=\left( \frac{1}{6}-\frac{3\zeta(3)}{2\pi^2}\right) s +
  \mathcal{O}(s^3) \, ,
\end{align}
with $\zeta$ denoting \textsc{Riemann}'s $\zeta$-function.
As expected the contribution from $R_2^{(0)}[s,0,\omega]$ is sub-leading with respect to the one 
from $R_1^{(0)}[s,0,\omega]$.

\section{Next to leading order asymptotic of $\Im\left\{R_1[s,\theta,\omega]\right\}$}
We would now like to obtain the next to leading order term in the asymptotic of 
$\Im\left\{R_1[s,\theta,\omega]\right\}$, that is the term obtained from (\ref{5R1shiftedBrane})
by taking into account the linear contribution in $1-z(0)$ of the expansion (\ref{5HypRatioExpansion}).
Since the calculation of the integral obtained in this way is rather cumbersome and anyway 
we do not need the full $s$-dependence, we do not intend to evaluate it fully. Extracting the leading 
$s$-divergence at $s=0$ is sufficient for our purposes here. To start, we insert the linear
 term of 
(\ref{5HypRatioExpansion}) in (\ref{5R1shiftedBrane}):
\begin{align}
  &\Im\left\{R_1^{(1)}[s,0,\omega]\right\}= \\ \nonumber 
  &=-\Im \Bigg\{ \int\limits_0^\infty dy y \frac{1-\imath y}{1+y^2} e^{-y s}
   \left\{ \left(\frac{1\hspace{-.5mm}+\hspace{-.5mm}\imath y}{2}\right)^3 
   \left[2 \psi(2)-2 \psi\left(\frac{3\hspace{-.5mm}+\hspace{-.5mm}\imath y}{2}
   \right) -\ln\left[1\hspace{-.5mm}-\hspace{-.5mm}z(0)\right]\right]\right. \nonumber \\
  &\, \quad-\left(\frac{1+\imath y}{2}\right)^2 \left(\frac{-1+\imath y}{2}\right) \nonumber 
  \left[2 \psi(1)-2 \psi\left(\frac{1+\imath y}{2}\right)-\ln\left[1-z(0)\right]\right]\times \\
  &\times  \left[\ln\left[1-z(0)\right]-\psi(1)-\psi(2)+\psi\left(\frac{3+\imath y}{2}\right)+
     \psi\left(\frac{1+\imath y}{2}\right)\right] \Bigg\} \left[1-z(0)\right] \Bigg\} 
   \, .\nonumber
\end{align}
After some algebra and by using the functional relation of the Digamma-function $\psi(z)$ 
(see \cite{AbrStegun})
\begin{equation} 
  \psi\left(z+1\right)=\psi\left(z\right)+\frac{1}{z}
\end{equation}
we find
\begin{align} \label{5R1_1storder}
  \Im\left\{R_1^{(1)}[s,0,\omega]\right\}=-\frac{1-z(0)}{2} \int\limits_0^{\infty}y e^{-y s} \Im \left\{
  \ldots\right\} dy
\end{align}
with
\begin{align} \label{5R1_1storderIntegrand}
  \Im \left\{ \ldots\right\} &=\Im\left\{\left(\frac{1+\imath y}{2}\right)^2 
  \left[2 \psi(2)-2 \psi\left(\frac{1+\imath y}{2}\right)-\ln\left[1-z(0)\right]\right] -(1+\imath y)\right.
  \nonumber \\
  &\quad\qquad+\frac{1+y^2}{4}\left[2\psi\left(1\right)-2\psi\left(\frac{1+\imath y}{2}\right)-
    \ln\left[1-z(0)\right]\right]\times\nonumber\\
  &\quad\qquad\quad\qquad\times\left[\ln\left[1-z(0)\right]-\psi\left(1\right)-\psi\left(2\right)+
    2\psi\left(\frac{1+\imath y}{2}\right)\right] \nonumber \\
  &\quad\qquad+ \frac{1-\imath y}{2}\left[2\psi\left(1\right)-2\psi\left(\frac{1+\imath y}{2}\right)
    -\ln\left[1-z(0)\right]\right] \Bigg\} \, .
\end{align}
We will now keep only the terms proportional to $y^3$ in (\ref{5R1_1storder}) 
($y^2$ in (\ref{5R1_1storderIntegrand})) since only these will contribute to the leading
$1/s^{4}$ singular behavior. After a few lines of algebra we find
\begin{align} \label{5R1_1HighestContribution}
  &\Im\left\{R_1^{(1)}[s,0,\omega]\right\}= \\ 
  &\quad= -\frac{1-z(0)}{2}\left[1-2\gamma-\ln\left[1-z(0)\right]\right]
    \int\limits_0^{\infty} y^3 e^{-y s} \Im \left[\psi\left(\frac{1+\imath y}{2}\right)\right] dy \nonumber\\ 
  &\quad\quad+\left[1-z(0)\right] \int\limits_0^\infty y^3 e^{-y s} 
    \Im\left[\psi\left(\frac{1+\imath y}{2}\right)\right] 
    \Re\left[\psi\left(\frac{1+\imath y}{2}\right)\right] dy \nonumber \\
  &\quad\quad+\; \mbox{terms involving lower powers of }y. \nonumber 
\end{align}
The first integral can be reduced to the integral (\ref{5R1_0_Integral}) simply by replacing each power 
of $y$ by a derivative with respect to $s$ and by taking the derivatives out of the integral: 
\begin{align} 
  \int\limits_0^{\infty} y^3 e^{-y s} \Im \left[\psi\left(\frac{1+\imath y}{2}\right)\right] dy&=
  -\frac{d^3}{ds^3}\left\{\int\limits_0^\infty e^{-y s} \Im \left[\psi\left(\frac{1+\imath y}{2}\right)\right]
  dy \right\}\nonumber \\
  &=-\frac{\pi}{2}\frac{d^3}{ds^3}\left[-\frac{1}{s}+\frac{2}{\pi}\beta\left(\frac{s}{\pi}\right)\right] .
\end{align}
We could not find an analytic expression for the second integral in (\ref{5R1_1HighestContribution}). 
However, all we need is the 
first term of its small $s$  asymptotic\footnote {In the sense $s\ll 1$.} and this can be easily obtained
 from the large $y$ asymptotic of $\Re\left[\psi\left(\left(1+\imath y\right)/2\right)\right]$. Since
asymptotically 
\begin{equation} 
  \psi\left(z\right)\sim\ln z -\frac{1}{2 z}-\frac{1}{12 z^2}+\mathcal{O}(\frac{1}{z^4})\, , \quad 
  (z \to \infty \; \mbox{in} \; \vert \arg z\vert<\pi)
\end{equation}
we expect that this term will contribute logarithmic terms in $s$ and we find after straightforward
expansions
\begin{eqnarray} \label{5RePartAssympt}
  \Re\left[\psi\left(\frac{1+\imath y}{2}\right)\right] \sim \ln\frac{y}{2}-\frac{1}{6 y^2}+
  \mathcal{O}\left(\frac{1}{y^4}\right) \, .
\end{eqnarray}
From the last formula (\ref{5RePartAssympt}) we learn that within our current approximation of keeping 
only highest (cubic) power terms in $y$, it is sufficient to take into account only the 
contribution coming from $\ln\frac{y}{2}$. Therefore, we need to calculate the following integral
\begin{equation}
  \frac{\pi}{2} \int\limits_0^\infty y^3 e^{-y s} \tanh \left(\frac{\pi y}{2} \right)\ln y \, dy=
  -\frac{\pi}{2}\frac{d^3}{ds^3}\Big[\underbrace{\int\limits_0^\infty e^{-y s} 
  \tanh\left( \frac{\pi y}{2} \right)
  \ln y \, dy}_{\equiv \mathcal{J}(s)} \Big] \, ,
\end{equation}
where we used (\ref{5ImofDiGamma}) once more. 
Note that the term proportional to $\ln 2$ can be accounted for by adding a contribution
of the type of the first integral in (\ref{5R1_1HighestContribution}).
One way to find the asymptotic of $\mathcal{J}(s)$ for small $s$ is to integrate the following
asymptotic expansion of $\tanh x$ 
\begin{eqnarray} \label{5tanhexp}
  \tanh x = 1-2 e^{-2 x}+2 e^{-4 x}-2 e^{-6 x}+\ldots=1+2 \, \sum_{\nu=1}^\infty (-1)^\nu e^{-2 \nu x}  
\end{eqnarray}
which converges for all $x>0$. Inserting this in the definition of $\mathcal{J}(s)$ we find:
\begin{align} \label{5calIIntRes}
  \mathcal{J}(s)&=\int\limits_0^\infty e^{-y s} \ln y \, 
  \left[1+2\sum_{\nu=1}^\infty (-1)^\nu e^{-\pi y \nu} \right] dy 
  \\
  &=-\frac{\gamma + \ln s}{s} +2 \sum_{\nu=1}^\infty (-1)^{\nu+1} 
      \frac{\gamma+\ln\left(s+\nu \pi\right)}{s+\nu\pi}\nonumber\\
  &= -\frac{\gamma + \ln s}{s} +\frac{2}{\pi}\left(\gamma+\ln\pi\right) \beta\left(\frac{s}{\pi}+1\right)
      +\frac{2}{\pi} \sum_{\nu=1}^\infty (-1)^{\nu+1}
    \frac{\ln\left(\nu+\frac{s}{\pi}\right)}{\nu+\frac{s}{\pi}} \nonumber \, ,
\end{align}
where we used the series expansion of the $\beta$-function given e.g. in \cite{Gradshteyn}.
The last two terms in the above formula are finite in the limit $s\to0$ 
and therefore will play no role in the asymptotic (since they are multiplied by a factor of $1-z(0)$).

We are now in a position to assemble all contributions to the small $s$ asymptotic of 
$\Im\left\{R_1^{(1)}[s,0,\omega]\right\}$:
\begin{align} \label{51OrderResult}
  &\Im\left\{R_1^{(1)}[s,0,\omega]\right\}\sim \\
  &\quad \sim \frac{1-z(0)}{2}
   \left[1-2\gamma-\ln\left[1-z(0)\right]+2\ln 2\right]
   \frac{\pi}{2}\frac{d^3}{ds^3}\left[-\frac{1}{s}+\frac{2}{\pi}\beta\left(\frac{s}{\pi}\right)\right]
   \nonumber \\
  &\qquad -\left[ 1-z(0) \right] \frac{\pi}{2}\frac{d^3}{ds^3}\Bigg[-\frac{\gamma+\ln s}{s}
    + \frac{2}{\pi}\left(\gamma+\ln\pi\right) \beta\left(\frac{s}{\pi}+1\right) \nonumber \\
  &\qquad\qquad\qquad\;\;\qquad\qquad +\frac{2}{\pi} \sum_{\nu=1}^\infty
    \frac{ (-1)^{\nu+1}\ln\left(\nu+\frac{s}{\pi}\right)}{\nu+\frac{s}{\pi}} \Bigg] 
    +\mathcal{O}(\frac{\ln s}{s^3}) \nonumber \\
  & \quad= \frac{\pi}{2}\frac{1-z(0)}{s^4} \left\{ 8-6\ln 2 -6 
  \ln\left[\frac{s}{\sqrt{1-z(0)}}\right]\right\} +\mathcal{O}(\frac{\ln s}{s^3})\, , \nonumber 
\end{align}
where we used the following series expansions for the $\beta$-function and for 
the last term in (\ref{5calIIntRes}):\footnote{At this point we would like to thank Elmar Teufl for the 
evaluation of the leading order term of the sum in (\ref{5ElmarSum}).}
\begin{align} 
  \beta\left(\frac{s}{\pi}\right)&=\frac{\pi}{s}-\ln 2 +\mathcal{O}(s)\, , \\
  \beta\left(\frac{s}{\pi}+1\right)&=\ln 2+\mathcal{O}(s)\, , \\
  \sum_{\nu=1}^\infty
    \frac{(-1)^{\nu+1}\ln\left(\nu+\frac{s}{\pi}\right)}{\nu+\frac{s}{\pi}} 
    &=\frac{1}{2} \left[\ln(2)\right]^2 - \gamma \ln(2) + \mathcal{O}(s) \label{5ElmarSum} \, .
\end{align}
Note that only the term $\beta\left(s/\pi\right)$ gave a contribution to the leading 
$1/s^4$ divergence.
Relations (\ref{50orderResult}) and (\ref{51OrderResult}) constitute the main result of this 
appendix. They provide the desired asymptotic behavior of the two-point function in the regime of 
intermediate distances $\omega/\sinh\left(\frac{\omega\pi}{2}\right) \ll s\ll 1$.
A couple of remarks are in place. Clearly, (\ref{50orderResult}) is not only an asymptotic result, but 
is valid for all $s \in [0,\pi]$. Secondly, we were writing $\mathcal{O}(\ln s/s^3)$ in 
(\ref{51OrderResult}) to indicate that we dropped all contributions coming from terms in the integrand 
(\ref{5R1_1storderIntegrand}) lower than cubic order. Finally we point out that the development of 
the ratio (\ref{5HypRatioExpansion}) in power of $1-z(0)$ indeed generates an asymptotic in the correct 
distance regime. This is clear from the fact that each new power in $1-z(0)$ is paired with an 
additional power of $y^2$ in the integrand of $\Im\left\{R_1[s,\theta,\omega]\right\}$. This by itself
implies upon multiplication by $e^{-y s}$ and integration over $y$ an additional power of 
$1/s^2$. We therefore effectively generate an expansion in powers of 
$\sqrt{1-z(0)}/s=1/\left[s \cosh \left(\frac{\omega\pi}{2}\right)\right]$. It
is also easy to understand why this rough way of counting powers works. The reason is that the 
asymptotic properties of our integrals $\Im\left\{R_1^{(n)}[s,0,\omega]\right\}$ are determined 
only by the behavior of the corresponding integrands for large values of the integration variable $y$.
The terms $\Re\left[\psi\left(\left(1+\imath y\right)/2\right)\right]$ and 
$\Im\left[\psi\left(\left(1+\imath y\right)/2\right)\right]$ do not interfere with the above power counting
since their behavior for large $y$ is either logarithmic in $y$ (for $\Re$) or constant in $y$ (for $\Im$). 

Finally we want to point out that the expansion in powers of
$1/\left[s \cosh \left(\frac{\omega\pi}{2}\right)\right]$ is 
not useful for obtaining information about the Green's function for distances 
$s \ll \omega/\cosh\left(\frac{\omega\pi}{2}\right)$. In this limit the asymptotic expansion of the 
hypergeometric functions in (\ref{5RatioOf2Hyps}) for large values of $\lambda$ proves the 
most efficient way to recover the properties of a Green's function in a $4$-dimensional space. 
See appendix \ref{appGreenUltraShort} for the detailed form of the Green's function at distances 
smaller than the extra dimension.

\section{Estimate of the arc contribution to $R_1[s,\theta,\omega]$} \label{5arcjust}
This final subsection is dedicated to the verification that the arc denoted $C_2$ in 
Fig.~\ref{5intcontour} of appendix \ref{appParallel} gives a vanishing contribution to the integral
$R_1[s,\theta,\omega]$ and thus justifying the representation (\ref{5R1shiftedBrane}).
We need to know the large $\lambda$ asymptotic behavior of the hypergeometric functions entering 
the definition of $R_1[s,\theta,\omega]$ in (\ref{5fdef}). The relevant formula 
(\ref{5GeneralAsymptoticOfHyp}) can be found in appendix \ref{appGreenUltraShort} where we 
discuss the short distance behavior of the Green's function. After canceling all common factors from the
ratio of the two hypergeometric functions we find:
\begin{align} \label{5FFratioInR1asymptotic}
  &\frac{\phantom{1}_2 F_1\left[ \frac{\lambda+1}{2},\frac{\lambda+1}{2};
  \lambda+1;z(\theta) \right]}{\phantom{1}_2 F_1\left[ \frac{\lambda+1}{2},\frac{\lambda-1}{2};
  \lambda+1;z(0) \right]}= \\ \nonumber 
  & \qquad =\frac{\lambda+1}{\lambda-1} \left[ \frac{z(\theta)}{z(0)}\right]^{-\frac{\lambda+1}{2}} 
  \left[\frac{e^{-\nu(\theta)}}{e^{-\nu(0)}}\right]^\frac{\lambda+1}{2}
  \frac{\left[1+e^{-\nu(\theta)}\right]^{-\frac{1}{2}} \left[1-e^{-\nu(\theta)}\right]^{-\frac{1}{2}}}
  {\left[1+e^{-\nu(0)}\right]^{\frac{3}{2}} \left[1-e^{-\nu(0)}\right]^{\frac{1}{2}}}
  \left[1+\mathcal{O}\left(\frac{1}{\lambda}\right)\right] \, ,
\end{align} 
where $e^{-\nu(\theta)}$ is defined in (\ref{5nudefinition}) of appendix \ref{appGreenUltraShort} 
(see also (\ref{5zdef}) and (\ref{5nuresult})).

Substituting $\lambda=R e^{\imath \varphi}$, we now estimate $R_1[s,\theta,\omega]$:
\begin{align} 
  &\left| \int_{C_2} \frac{\lambda e^{\imath\lambda s}}{\lambda+1}
  \left[\frac{z(\theta)}{z(0)}\right]^\frac{\lambda}{2}
  \frac{\phantom{1}_2 F_1\left[ \frac{\lambda+1}{2},\frac{\lambda+1}{2};
  \lambda+1;z(\theta) \right]}{\phantom{1}_2 F_1\left[ \frac{\lambda+1}{2},\frac{\lambda-1}{2};
  \lambda+1;z(0) \right]} d \lambda\right| \\
  &\qquad\leq
  \int_{C_2} \left| \frac{\lambda e^{\imath\lambda s}}{\lambda+1}
  \left[\frac{z(\theta)}{z(0)}\right]^\frac{\lambda}{2}
  \frac{\phantom{1}_2 F_1\left[ \frac{\lambda+1}{2},\frac{\lambda+1}{2};
  \lambda+1;z(\theta) \right]}{\phantom{1}_2 F_1\left[ \frac{\lambda+1}{2},\frac{\lambda-1}{2};
  \lambda+1;z(0) \right]} \right| d \lambda \nonumber \\
  &\qquad\leq \underbrace{C \left[ \frac{z(0)}{z(\theta)}\right]^{\frac{1}{2}}
  \left| \frac{e^{-\nu(\theta)}}{e^{-\nu(0)}} \right|^\frac{1}{2}
  \frac{\left[1+e^{-\nu(\theta)}\right]^{-\frac{1}{2}} 
  \left[1-e^{-\nu(\theta)}\right]^{-\frac{1}{2}}}{\left[1+e^{-\nu(0)}\right]^{\frac{3}{2}} 
  \left[1-e^{-\nu(0)}\right]^{\frac{1}{2}}}}_{K} R \, \times \nonumber \\
  &\qquad\quad  \times \int\limits_0^{\frac{\pi}{2}} 
  \left| e^{\imath R s e^{\imath \varphi}} \right|
  \left| \frac{R e^{\imath \varphi}}{R e^{\imath \varphi }-1} \right| 
  \left| \left[\frac{e^{-\nu(\theta)}}{e^{-\nu(0)}}\right]^{\frac{R e^{\imath \varphi}}{2}}\right| 
   d\varphi \nonumber \\
  &\qquad=K R \int\limits_0^{\frac{\pi}{2}} 
  \frac{e^{-R\left(s\sin\varphi-\ln a \cos \varphi\right)}}{\sqrt{1-\frac{2 \cos \varphi}{R}+\frac{1}{R^2}}} 
   d\varphi
  \leq K' R \int\limits_0^{\frac{\pi}{2}} e^{-R s\left(\sin\varphi+\frac{\ln a^{-1}}{s}
   \cos \varphi\right)} d\varphi \nonumber \, ,
\end{align}
where $C$ is a constant independent of $R$ and the estimate of the ratio of the two hypergeometric 
functions is valid for large enough $R$. 
We also introduced $a$ as in appendix \ref{appGreenUltraShort} by (\ref{5adef}). Note that $a<1$ for 
$0<\theta$. In the last line we estimated the inverse of the square root by an arbitrary constant (e.g. 2),
certainly good for large enough $R$ and for all $\varphi$.
All what is left is to observe that
\begin{equation} 
  0 < \epsilon \equiv \min[1,\frac{\ln a^{-1}}{s}] \leq \sin\varphi+\frac{\ln a^{-1}}{s} \cos\varphi \, ,
  \quad s>0, \;\; a^{-1} >1 
\end{equation}
to obtain the desired result:
\begin{equation} 
  \left| \int_{C_2} \ldots d\lambda \right| \leq K' R \int\limits_0^{\frac{\pi}{2}}e^{-R s \epsilon}d\varphi=
  \frac{\pi}{2} K' R e^{-R s \epsilon} \to 0 \quad \mbox{for} \quad R \to \infty.
\end{equation}
\thispagestyle{empty}
\chapter{The two-point function at very short distances ($r \ll R$)} \label{appGreenUltraShort}
The aim of this appendix is to calculate the behavior of the two-point function at distances
smaller than the extra dimension $r\ll R$. We will do so by using an appropriate asymptotic expansion 
for large parameter values of the hypergeometric functions in (\ref{5RatioOf2Hyps}). Such an asymptotic can be 
found for example in \cite{Erdelyi} or \cite{Luke}. 
The general formula is\footnote{Since we noticed that the formulae given by Luke, p. 452 eq.~(20)--(23) 
\cite{Luke} and the formula given by Erd\'elyi \cite{Erdelyi} differ by a factor of $2^{a+b}$ 
we performed a numerical check. Its outcome clearly favored Erd\'elyi's formula which we reproduced
in \ref{5GeneralAsymptoticOfHyp}.}
\begin{align} \label{5GeneralAsymptoticOfHyp}
  &\phantom{1}_2 F_1\left[a+n,a-c+1+n;a-b+1+2 n;z \right] = \frac{2^{a+b} \Gamma[a-b+1+2 n] \nonumber
   \left(\frac{\pi}{n}\right)^{\frac{1}{2}}}{\Gamma[a-c+1+n]\Gamma[c-b+n]} \times \\
  &\qquad\qquad\times\frac{e^{-\nu(n+a)}}{z^{(n+a)}}
  \left(1+e^{-\nu}\right)^{\frac{1}{2}-c} \left(1-e^{-\nu}\right)^{c-a-b-\frac{1}{2}}
  \left[1+\mathcal{O}\left(\frac{1}{n}\right)\right] 
\end{align}
with $\nu$ defined via the relation
\begin{equation} \label{5nudefinition}
  e^{-\nu}=\frac{2-z-2(1-z)^{\frac{1}{2}}}{z} \, .
\end{equation}
We need to calculate
\begin{equation} 
  S[s,\theta,\omega]=\sum_{\lambda=2}^{\infty} \lambda \sin \left(\lambda s \right)
  \frac{\varphi_1^{(\lambda)}(\theta)}{{\varphi'}_1^{(\lambda)}(0)}
\end{equation}
with the ratio $\varphi_1^{(\lambda)}(\theta)/{\varphi'}_1^{(\lambda)}(0)$ given
in (\ref{5RatioOf2Hyps}). Using the above expansion (\ref{5GeneralAsymptoticOfHyp}) two times 
we find after numerous cancellations
\begin{align} \label{5Sasymptotic}
  S[s,\theta,\omega]&\sim-\frac{4}{\omega}\left(\frac{1-z(\theta)}{1-z(0)}\right)^2
  \frac{\left[1+e^{-\nu(\theta)}\right]^{-\frac{1}{2}} \left[1-e^{-\nu(\theta)}\right]^{-\frac{5}{2}}}
  {\left[1+e^{-\nu(0)}\right]^{\frac{1}{2}} \left[1-e^{-\nu(0)}\right]^{-\frac{3}{2}}}
  \frac{z(0)^\frac{3}{2}}{z(\theta)^2}\frac{e^{-\frac{3\nu(\theta)}{2}}}{e^{-\frac{\nu(0)}{2}}} 
  \times \nonumber\\
  &\quad \times \sum_{\lambda=2}^{\infty}\frac{\lambda \sin \left(\lambda s\right)}{\lambda-1}
  \left[\frac{e^{-\nu(\theta)}}{e^{-\nu(0)}}\right]^{\frac{\lambda}{2}} \, ,
\end{align}
where we used the following abbreviations
\begin{align}
  z(\theta)&=\tanh^2\left[\omega\left(\frac{\pi}{2}-\theta\right)\right] \, ,  \label{5zdef}\\
  e^{-\nu(\theta)}&=\tanh^2\left[\frac{\omega}{2}\left(\frac{\pi}{2}-\theta\right)\right]
   \, .\label{5nuresult}
\end{align}
Note that (\ref{5nuresult}) is obtained after some lines of algebra by inserting (\ref{5zdef}) in 
(\ref{5nudefinition}). Still the sum in (\ref{5Sasymptotic}) diverges for $\theta=0$ but it turns out that 
it can be performed analytically for $\theta>0$ so that after the summation the limit
$\theta \to 0$ exists. For the sake of a lighter notation we introduce the symbol $a$ by
\begin{equation} \label{5adef}
  a\equiv\left[\frac{e^{-\nu(\theta)}}{e^{-\nu(0)}}\right]^{\frac{1}{2}}=
  \frac{\tanh\left[\frac{\omega}{2}\left(\frac{\pi}{2}-\theta\right)\right]}
       {\tanh\left(\frac{\omega\pi}{4}\right)} \, .
\end{equation}
It is now straightforward to evaluate the sum over $\lambda$:
\begin{align} 
  \sum_{\lambda=2}^{\infty}\frac{\lambda \sin\left(\lambda s\right)}{\lambda-1}
  a^\lambda=&\sum_{\lambda=2}^{\infty} \sin(\lambda s)\, a^\lambda+
  \sum_{\lambda=2}^{\infty}\frac{\sin \left(\lambda s\right)}{\lambda-1} a^\lambda \nonumber \\
  =&-a \sin s + \sum_{\lambda=1}^{\infty} \sin(\lambda s)\, a^\lambda+
  \sum_{\lambda=1}^{\infty}\frac{\sin \left[\left(\lambda+1\right) s\right]}{\lambda} a^{\lambda+1} 
  \nonumber \\
  =&\hspace{-.5mm}-\hspace{-.5mm}a \sin s \hspace{-.5mm}+\hspace{-.5mm} 
   \frac{1}{2}\frac{\sin s}{\frac{1}{2}\left(a+\frac{1}{a}\right)\hspace{-.5mm}-\hspace{-.5mm}\cos s}
  \hspace{-.5mm}-\hspace{-.5mm}\frac{a}{2} \sin s \, 
  \ln \left( 1\hspace{-.5mm}-\hspace{-.5mm}2a\cos s\hspace{-.5mm}+\hspace{-.5mm}a^2\right) \nonumber \\
  &+ a\cos s \arctan\left[\frac{a \sin s}{1-a \cos s}\right] \, .
\end{align}
The full result can therefore be written as
\begin{align} \label{5SasymptoticResult}
  S[s,\theta,\omega]&\sim-\frac{4}{\omega}\left(\frac{1-z(\theta)}{1-z(0)}\right)^2
  \frac{\left[1+e^{-\nu(\theta)}\right]^{-\frac{1}{2}} \left[1-e^{-\nu(\theta)}\right]^{-\frac{5}{2}}}
  {\left[1+e^{-\nu(0}\right]^{\frac{1}{2}} \left[1-e^{-\nu(0)}\right]^{-\frac{3}{2}}}
  \frac{z(0)^\frac{3}{2}}{z(\theta)^2}\frac{e^{-\frac{3\nu(\theta)}{2}}}{e^{-\frac{\nu(0)}{2}}} 
  \times \nonumber \\
  &\qquad\times a \sin s \Bigg\{ \frac{1}{2 a} \frac{1}{\frac{1}{2}\left(a+\frac{1}{a}\right)-\cos s}
   -\frac{1}{2} \ln \left( 1-2 a \cos s+a^2\right) \nonumber \\
  &\;\;\qquad\qquad\qquad + \cot s \, \arctan \left[\frac{a \sin s}{1-a \cos s}\right]
   -1 \Bigg\} \, .
\end{align}
Although  the last result is given in closed form, we have to emphasize that its validity is restricted
to distances smaller than the size of the extra dimension~$R$. 

It is now save to take the limit $\theta\to 0$ in (\ref{5SasymptoticResult}). The result is 
\begin{align} \label{5SasymptoticResultBrane}
  \lim_{\theta\to 0} S[s,\theta,\omega]&=\hspace{-.5mm}-\hspace{-.5mm}\frac{\sinh\left( \frac{\omega\pi}{2}\right)}{\omega}
   \sin s \left\{\frac{1}{4\sin^2(\frac{s}{2})}\hspace{-.5mm}- \hspace{-.5mm}
    \ln \left[2 \sin \left(\frac{s}{2}\right)\right]\hspace{-.5mm}+\hspace{-.5mm}
    \frac{\pi\hspace{-.5mm}-\hspace{-.5mm}s}{2} \cot s \; 
    \hspace{-.5mm}-\hspace{-.5mm}1\right\} \, .
\end{align}
We note that it is the first term in the curly brackets on the right hand side of 
(\ref{5SasymptoticResult}) which is responsible for reproducing the 
characteristic short distance singularity of the two-point function. The main result of 
this appendix is therefore:
\begin{align} 
  \lim_{\theta\to 0} S[s,\theta,\omega]&\sim-\frac{\sinh \left(\frac{\omega\pi}{2}\right)}{\omega}
  \left\{\frac{1}{s}+\frac{\pi}{2}+\mathcal{O}\left(s \ln s\right)\right\} \, .
\end{align}
The $5$-dimensional short distance singularity can even be obtained including the
extra dimension. Apart from the logarithmic term in the curly brackets of (\ref{5SasymptoticResult}), 
all but the first one are finite at short distances (in $\theta$ and $s$). Developing
the denominator of this term, we recover the usual euclidean metric in $\mathbb{R}^4$:
\begin{equation} \label{55DNewton}
  \frac{1}{2}\left(a+\frac{1}{a}\right)-\cos s \sim 
  \frac{1}{2} \left[ \frac{\omega^2}{\sinh^2 \left(\frac{\omega\pi}{2}\right)}\theta^2 +s^2 \right]+ \ldots .
\end{equation}
where the dots denote higher terms in $\theta$ and $s$.
\thispagestyle{empty}

\end{document}